\documentclass[aps,prb,twocolumn,superscriptaddress,floatfix,letterpaper]{revtex4-1}

\pdfoutput=1

\newcommand{\rw}{{\rm w}}
\newcommand{\Tor}{\text{Tor}}
\newcommand{\Ext}{\text{Ext}}
\newcommand{\Hom}{\text{Hom}}
\renewcommand{\mod}{\text{mod}}
\newcommand{\Sq}{\text{Sq}}
\newcommand{\Bs}{\cB}
\newcommand{\RZ}{{\mathbb{R}/\mathbb{Z}}}
\newcommand\se[1]{\stackrel{#1}{=}}

\begin{document}

\begin{titlepage}

\title{Exactly soluble local bosonic cocycle models, statistical transmutation,\\
and simplest time-reversal symmetric topological orders in 3+1D 
%anomalous (fractionalized) symmetry on string-like excitations
}

\author{Xiao-Gang Wen}
\affiliation{Department of Physics, Massachusetts Institute of
Technology, Cambridge, Massachusetts 02139, USA}

\begin{abstract} 
We propose a generic construction of exactly soluble \emph{local bosonic
models} that realize various topological orders with gappable boundaries.  In
particular, we construct an exactly soluble bosonic model that realizes a 3+1D
$Z_2$ gauge theory with emergent fermionic Kramer doublet. We show that the
emergence of such a fermion will cause the nucleation of certain topological
excitations in space-time without pin$^+$ structure.  The exactly soluble model
also leads to a statistical transmutation in 3+1D. In addition, we construct
exactly soluble bosonic models that realize 2 types of time-reversal symmetry
enriched $Z_2$-topological orders in 2+1D, and 20 types of simplest
time-reversal symmetry enriched topological (SET) orders which have only one
non-trivial point-like and string-like topological excitations. 
%Those 20 SETs come from 9 types of time-reversal symmetric $Z_2$-gauge theory
%stacked with 4 types of time-reversal symmetry protected topological states.
%We have also constructed $3$ types of 2+1D double-semion theories that have
%time-reversal symmetry.  
Many physical properties of those topological states are calculated using the
exactly soluble models.  We find that some time-reversal SET orders have
point-like excitations that carry Kramer doublet -- a fractionalized
time-reversal symmetry.  We also find that some $Z_2$ SET orders have
string-like excitations that carry anomalous (non-on-site) $Z_2$ symmetry,
which can be viewed as  a fractionalization of $Z_2$ symmetry on strings.  Our
construction is based on cochains and cocycles in algebraic topology, which is
very versatile.  In principle, it can also realize emergent topological field
theory beyond the twisted gauge theory.  
%It may become an universal construction which can realize all topological
%orders with gappable boundary in 2+1D and 3+1D.

\end{abstract}


\maketitle

\end{titlepage}

{\small \setcounter{tocdepth}{1} \tableofcontents }

\section{Introduction} 

A sign of a comprehensive understanding of a type of phases of matter is being
able to classify all of them.  We understand that the crystal orders are due to
spontaneous symmetry breaking \cite{L3726} of the translation and the rotation
symmetry.  This leads to the classification of all 230 crystal orders in
3-dimensions using group theory. Now we realized that the phases of matter
beyond symmetry breaking theory are due to long range entanglement
\cite{KP0604,LW0605,CGW1038} for topologically ordered phases
\cite{Wtop,Wrig,KW9327}, and due to symmetry-protected short-range entanglement
\cite{GW0931,PBT1039} for symmetry-protected  trivial (SPT) phases
\cite{GW0931,CLW1141,CGL1314}. This leads to complete classification of many
topological phases.  Using projective representations \cite{PBT1039}, we can
classify all 1+1D gapped phases for bosonic and fermionic systems with any
symmetry \cite{CGW1107,FK1103,SPC1139,CGW1128}. We can also classify all 2+1D
gapped liquid\cite{ZW1490,SM1403} phases for bosonic and fermionic systems with
any finite unitary symmetry using unitary modular tensor categories
\cite{RSW0777,W150605768}, $G$-crossed unitary modular tensor categories
\cite{BBC1440}, and/or unitary braided fusion categories over Rep$(G)$ or
sRep$(G^f)$ \cite{LW150704673,LW160205946}.  Those phases are symmetry breaking
phases, topologically ordered phases, SPT phases (such as odd-integer-spin
Haldane phase \cite{H8364,AKL8799} and topological insulators
\cite{KM0502,MB0706,R0922,FKM0703,QHZ0824,WS13063238}),  symmetry enriched
topological (SET) orders, \etc.  So far, we still do not have a classification
of 3+1D  gapped liquid phases, although we know that it is closely related to
unitary 4-category theory with one object.\cite{KW1458,C161007628}

With those powerful classification results, we would like to have a systematic
construction of those topological phases.  Ideally, we would like to have a
universal construction that can realize any given topological phases.  There
are very systematic ways to construct exactly soluble models
\cite{DW9093,TV9265,FNS0428,LW0510,GWW1017,WW1132,WW160607144,TF160402145,BK160501640,C161007628}
based on tensor network \cite{KW1458}.  Using unitary fusion categories as
input, Turaev-Viro state-sum \cite{TV9265} and Levin-Wen string-net models
allow us to realize all 2+1D bosonic topological orders with gappable boundary.
Using finite group $G$ and group 4-cohomoly classes $\om_4\in \cH^4(G;/\R/\Z)$
as input, Dijkgraaf-Witten models allow us to realize all 3+1D bosonic
topological orders whose point-like excitations are all bosons
\cite{LW170404221}.  Using premodular categories as input, Walker-Wang models
can also realize a large class of 3+1D bosonic topological orders.  But
Walker-Wang models cannot realize all Dijkgraaf-Witten models.  A further
generalization of Walker-Wang models in \Ref{WW160607144,C161007628} allow us
to include all Dijkgraaf-Witten models as well.  Such systematic construction
were also generalized to fermion systems
\cite{GWW1017,TF160402145,BK160501640,KT170108264,WG170310937}.

The above constructions are very systematic, but also very complicated and hard
to use.  Despite their complexity, it is still not clear if they can realize
all 3+1D topological orders or not. (We already know that they cannot realize
all 2+1D topological orders.) In this paper, we are going to develop a simpler
systematic construction.  Our constructed models are not a subset of any one of
the above mentioned tensor network constructions. But our construction also
does include  any one of the above mentioned tensor network constructions, as a
subset.

We will start with topological invariants for topological orders. Then, we will
use cochain theory and cohomology theory \cite{DW9093,KT1321,GW14125148} to
construct exactly soluble local bosonic models whose ground states have
topological orders described by the corresponding topological invariants.  In
other words, the low energy effective field theory of those local bosonic
models are the desired topological field theory.  (Here a \emph{local bosonic
model} is defined as a quantum model whose total Hilbert space has a tensor
product decomposition $\cH_\text{tot}=\bigotimes_i \cH_i$ where $\cH_i$ is a
finite dimensional local Hilbert space for site-$i$, and the Hamiltonian is
local respect to such a tensor product decomposition.) Many mathematical
techniques developed for  cohomology theory and algebraic topology will help us
to do concrete calculations with our models.

One class of topological invariants are given by volume-independent partition
function $Z^\text{top}(M^d)$ on manifolds with vanishing Euler number and
Pontryagin number \cite{KW1458} $\chi(M^d)=P(M^d)=0$.  For invertible
topological orders \cite{KW1458,F1478} and for SPT orders
\cite{CGL1314,CGL1204} (which have no non-trivial bulk topological
excitations), such topological invariants are pure phases
\cite{KW1458,F1478,K1459,K1467,W1477,SR160905970}
\begin{align}
 Z^\text{top}(M^d,a^\text{sym}) = \ee^{\ii 2\pi \int_{M^d} W(\rw_i, a^G)+ k\om_d}
\end{align}
where $\rw_i$ is the $i^\text{th}$ Stiefel-Whitney class, $a^G$ the flat
connection that describes symmetry $G$ twist
\cite{LG1209,HW1339,W1447,SCR1325}, and $\om_d$ the gravitational Chern-Simons
term.  For example, a 2+1D $Z_n$-SPT state labeled by $k\in\cH^3(Z_n,\RZ)=\Z_n$
is characterized by its SPT invariant\cite{HW1339,W1447,K1459,K1467,W1477}
(see Section \ref{Gver})
\begin{align}
 Z^\text{top}(M^{2+1},a^{Z_n})=\ee^{\ii k \frac{2\pi}{n} \int_{M^{2+1}}
a^{Z_n} \cup \Bs_n a^{Z_n}}
\end{align}
where $a^{Z_2}$ becomes an 1-cochain and $\Bs_n$  is Bockstein homomorphism
\eqn{Bsn}.

For other non-invertible topological orders (which have non-trivial bulk
topological excitations), their topological invariants can be sums of phases
\begin{align}
 Z^\text{top}(M^d,a^\text{sym}) = \sum_{c\in H^*(M^d;\M)}\ee^{\ii 2\pi \int_{M^d} W(c,\rw_i, a^\text{sym})+ k\om_d},
\nonumber
\end{align}
where $c$ are cohomology classes.  Our constructed local bosonic model is
designed to produce such form of topological invariants. The construction is
very versatile and many exactly soluble local bosonic models can be constructed
systematically to produce all the topological invariants of the above form
(with $k=0$).  Some of those models have emergent gauge theories or emergent
Dijkgraaf-Witten theories \cite{DW9093}. Other models have emergent ``twisted''
gauge theories beyond Dijkgraaf-Witten type.  

In this paper, we will discuss many different types of gauge theories. To avoid
confusion, here we will explain the terminology that will be used in this
paper.  We will use untwisted (UT) gauge theory to refer to the usual lattice
gauge theories (without any twist) \cite{K7959}.  We will use all-boson (AB)
gauge theory to refer to the lattice gauge theories (may be twisted) where all
the pure gauge charges are bosons.  We will use emergent-fermion (EF) gauge
theory to refer to the lattice gauge theory (may be twisted) where some  pure
gauge charges are fermions.  We will use the term $G$-gauge theory to refer
gauge theory with $G$ gauge group.  The Dijkgraaf-Witten theories \cite{DW9093}
are AB gauge theories.  This is because the Dijkgraaf-Witten theories can be
viewed as the $G$-SPT states with the gauged symmetry $G$ \cite{LG1209}, all
the gauge charge are bosonic in Dijkgraaf-Witten theories.  

We will also discuss 3+1D topological theories beyond Dijkgraaf-Witten
theories.  Many of those theories do not contain gauge fields, and it is hard
to call them gauge theories. However, the point-like topological excitations in
those theories have the same fusion rule as 3+1D gauge theories, \ie fuse like
the irreducible representations of a group $G$.  So we will still call those
3+1D topological theories as gauge theory, which include EF gauge theories.
Certainly, the EF gauge theories are not  Dijkgraaf-Witten theories in 3+1D.

We would like to mention that there are many related constructions of
topological field theories using 1-form, 2-form gauge fields \etc
\cite{KS14010740,WGW1489,YG14102594,KT13094721,GW14125148,GWW1568,YG150805689,PY161209298}.
In contrast to those work, the cocycle models constructed in this paper are
defined on lattice instead of continuous manifold.  Also cocycle models are not
gauge theories. They are local bosonic models without any gauge redundancy.  In
other words, the emergent topological field theories studied in this paper are
free of all anomalies. In comparison, some 1-form, 2-form gauge field theories
defined on continuous manifold can be anomalous since they may not be emergable
from local lattice theories \cite{W1313,KW1458,K1467}.

In this paper, we will use $\se{n}$ to mean equal up to a multiple of $n$, and
use $\se{\dd }$ to mean equal up to $\dd f$ (\ie up to a coboundary).  We will
use $[f]_n$ to mean $\mod(f,n)$ and $\<l,m\>$ to mean the greatest common
divisor of $l$ and $m$ ($\<0,m\>\equiv m$).  We also introduce some modified
$\del$-functions
\begin{align}
 \del_n(x)&=
\begin{cases}
 1 & \text{ if } x\se{n}0,\\
 0 & \text{ otherwise }.\\
\end{cases}
&
 \bar\del(x)&=
\begin{cases}
 1 & \text{ if } x\se{\dd}0,\\
 0 & \text{ otherwise }.\\
\nonumber 
\end{cases}
\end{align}
\begin{align}
 \bar\del_n(x)&=
\begin{cases}
 1 & \text{ if } x\se{n,\dd}0,\\
 0 & \text{ otherwise }.\\
\end{cases}
\nonumber 
\end{align}

\section{A summary of results}

\def\arraystretch{1.4} \setlength\tabcolsep{3pt}
\begin{table*}[t]
\caption{
The 2+1D time-reversal ($T$) symmetric topological orders from four 1-cocycle
models in \eqn{Z2Z4T1}.  They have 3 or 4 types of point-like topological
excitations.  $d_i$'s and $s_i$'s are the quantum dimensions and spins of those
excitations.  A quantum dimension $d=2$ means that the excitation has 2
internal degrees of freedom. $2_\pm$ means that the 2 internal degrees of
freedom form a  $T^2=1$ time-reversal doublet or a  $T^2=-1$ Kramer doublet.
Spin $s=\frac 12$ corresponds to a fermion, 
and $s=\frac 14$ a semion.
Spin $s=\frac 34$ is the time-reversal conjugate of a semion.
The fourth and fifth columns are volume-independent partition functions
$Z^\text{top}_{M^3}$ with $M^3=S^1\times \Si_g, S^1\times \Si^\text{non}_g$,
where $\Si_g$ is the genus $g$ Riemannian surface and $\Si^\text{non}_g$ is the
genus $g$ non-orientable surface.
} \label{Tsymm3}
 \centering
 \begin{tabular}{ |c|c|c|c|c|c| }
 \hline
$k_0k_1k_2$  & $(d_1,d_e,\cdots)$ &  $(s_1,s_e,\cdots)$ & $Z^\text{top}_{S^1\times \Si_g}$ &  $Z^\text{top}_{S^1\times \Si^\text{non}_g}$ &Comments\\
 \hline
$000$  & $(1,1,1,1)$ & $(0,0,0,\frac12)$ & $4^g$ & $2^g$ & $Z_2$-gauge theory (three bosons and one fermion)\\
$001$  & $(1,1,2_-,2_-)$ & $(0,0,0,\frac12)$ & $4^g$ & $2^{g-1}[1+(-)^g]$ & A boson and a fermion are Kramer doublets \\
$10*$  & $(1,2_-,1,2_-)$ & $(0,0,0,\frac12)$ & $4^g$ & $2^{g-1}[1+(-)^g]$ & The same SET order as above \\
 \hline
$010$  & $(1,1,2_+)$ & $(0,0,[\frac14,\frac34])$ & $4^g$ &$2^{g-1}$  & Two semions form a $T^2=1$ time-reversal doublet \\
$011$  & $(1,1,2_-)$ & $(0,0,[\frac14,\frac34])$ & $4^g$ &$2^{g-1}$  & Two semions form a $T^2=-1$ Kramer doublet\\
$11*$  & $(1,2_-,2_-,2_+)$ & $(0,0,[\frac14,\frac34],[\frac14,\frac34])$ & $4^g$ &$2^{g-2}[1+(-)^g]$  & A boson is Kramer doublet\\
 \hline
 \end{tabular}
\end{table*}

The cocycle models introduced in this paper not only can realize many types of
topological orders, SPT orders, and SET orders, they are exactly soluble in the
sense that that their partition function can be calculated exactly on any
space-time manifold \cite{TV9265}.  Those models are realizable by commuting
projectors.  
%They should have symmetric gapped boundaries.  
Because the models are exactly soluble, we can use them to compute many
physical properties of those topological phases, such as ground state
degeneracies, fractional quantum numbers on point-like and string-like
topological excitations, braiding statistics, topological partition functions,
dimension reduction, \etc.

\subsection{Symmetry fractionalization on string-like defects in SPT
states}

\begin{figure}[tb]
\begin{center}
\includegraphics[scale=0.25]{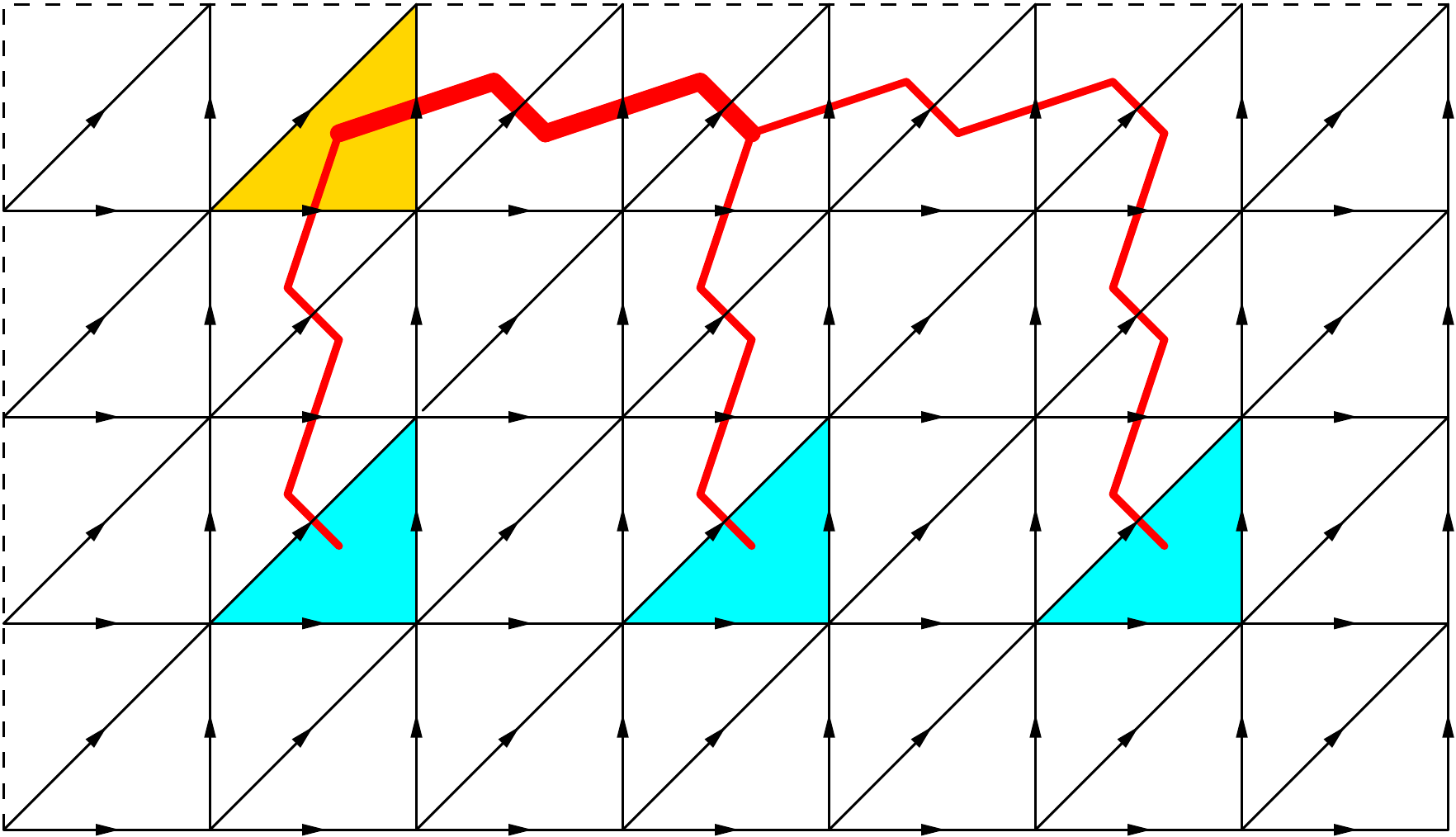} \end{center}
%1
\caption{ (Color online)
Three \emph{identical} $Z_3$-symmetry twist defects
(blue triangles) on a torus.  
The red line are the symmetry twist line.
A symmetry twist defect is an end of symmetry twist line.
}
\label{ZnDefect}
\end{figure}

One way to probe SPT order is to measure fractional quantum number carried by
symmetry twist defect (see Fig. \ref{ZnDefect}).  For example, consider a 2+1D
$Z_n$-SPT state which is labeled by $k$.  In \Ref{W1447} it was shown that a
symmetry-twist defect
% in (see Fig. \ref{ZnDefect}) will 
can carry a  $Z_n$ quantum number $2k$ (\ie each defect will carry a fractional
$Z_n$ quantum number $\frac{2k}{n}$.  We can use this property to measure the
2+1D $Z_n$-SPT order.

%labeled by
%$k\in\cH^3(Z_n,\RZ)=\Z_n$ (or more precisely, characterized by the SPT
%invariant $ Z^\text{top}=\ee^{\ii k \frac{2\pi}{n} \int_{M^{2+1}} a^{Z_n} \cup
%\Bs_n a^{Z_n}} $).  

Similar results also appear in higher dimensions.  Consider a  3+1D $Z_n\times
\t Z_n$-SPT state which is labeled by $k_1=0,\cdots,n-1$ and
$k_2=0,\cdots,n-1$.  A $Z_n$-symmetry-twist defect will be a line defect in
3+1D. We show that such a line defect must be gapless or symmetry breaking,
which behave just like the edge state of some 2+1D SPT state.  This phenomena
can be viewed as symmetry fractionalization on defect lines.

To see the edge state of which 2+1D SPT state that the defect line carries , we
need to specify that the 3+1D $Z_n\times \t Z_n$-SPT state is described by the
following SPT invariant $ Z^\text{top}=\ee^{\ii \frac{2\pi}{n} \int_{M^{3+1}}
k_1 a^{Z_n} \cup a^{\t Z_n} \cup \Bs_n a^{\t Z_n} + k_2 a^{\t Z_n} \cup a^{Z_n}
\cup \Bs_n a^{Z_n} } $.  Then a $Z_n$-symmetry-twist defect line will carry the
edge state of a 2+1D $Z_n\times \t Z_n$-SPT state characterized by the SPT
invariant $ Z^\text{top}=\ee^{\ii \frac{2\pi}{n} \int_{M^{2+1}} k_1 a^{\t Z_n}
\cup \Bs_n a^{\t Z_n} -k_2 a^{Z_n} \cup \Bs_n a^{\t Z_n} } $ (see Section
\ref{DW4Dred}).

To be more precise, the $Z_n$-symmetry-twist defect line in 3+1D has a
non-on-site (anomalous) $Z_n\times \t Z_n$-symmetry\cite{CLW1141,CGL1314,W1313}
along the defect line.  This 1+1D anomalous symmetry makes the defect line to
be either gapless or symmetry breaking.\cite{CLW1141} This result generalizes
the one in \Ref{W1447}.  This 1+1D anomalous symmetry can be viewed as the
symmetry fractionalization on the strings.

The 1+1D anomalous symmetry also appear on the edge of 2+1D SPT state.  The
1+1D anomalous symmetry on the  $Z_n$-symmetry-twist defect line happen to be
the same 1+1D anomalous symmetry on the edge of a 2+1D $Z_n\times \t Z_n$-SPT
state characterized by the SPT invariant $ Z^\text{top}=\ee^{\ii \frac{2\pi}{n}
\int_{M^{2+1}} k_1 a^{\t Z_n} \cup \Bs_n a^{\t Z_n} -k_2 a^{Z_n} \cup \Bs_n
a^{\t Z_n} } $.

Point-like and string-like symmetry-twist defects are extrinsic defects in the
SPT states.  The above results indicates that extrinsic defects in the SPT
states can carry fractional quantum numbers or anomalous symmetry.  We would
like to remark here we need to distinguish extrinsic defects from excitations
which are intrinsic.  The point-like or string-like excitations, by definition,
can all be trapped by potential traps of the same dimension.  For example a
point-like excitation at $\v x_0$ can be trapped by a potential $V(\v x)$,
which is non zero only near $\v x_0$.  Those point-like or string-like
excitations in SPT states do not have symmetry fractionalization.  In contrast,
extrinsic defects cannot be trapped by potentials of the same dimension.  For
example, a point-like symmetry-twist defect in 2+1D can only be trapped by a
``potential'' (a change of Hamiltonian) that is non-zero along a line, where
the point-like defect is trapped at an end of the line.

\subsection{Statistical transmutation in 3+1D}

We have constructed a 3+1D exactly soluble local bosonic model 
\begin{align}
\label{Z2b}
Z(M^{3+1}) 
= 
\sum_{{b \in C^2(M^{3+1};\Z_2) \atop \dd b\se{2} 0 }} 
\ee^{\ii \pi \int_{M^{3+1}} b\cup b }
,
\end{align}
where $b$ is a 2-cochain field (see Section \ref{disltgauge} for a definition
of cochain field) and $\cup$ the cup product of cochains.  The model has an
emergent fermion, and its low energy effective theory is a EF $Z_2$-gauge
theory.  Such kind of EF $Z_2$-gauge theory has been constructed in terms of
strings in 3+1D \cite{LW0316,LW0510}. Here we give a construction in terms of
membranes (see Section \ref{ZnbSec}) \cite{YG150805689,YG14102594}.  

As a corollary of the above construction, we find a statistical transmutation
in 3+1D lattice $M^{3+1}$ (expressed in terms of partition function):
\begin{align}
Z(M^{3+1}) 
= 
\sum_{{b \in C^2(M^{3+1};\Z_2) \atop \dd b\se{2}*j }} 
\ee^{\ii \pi \int_{M^{3+1}} b\cup b }
,
\end{align}
where $j$ is a cycle corresponding to the world-line of a bosonic
scalar particle, and $*j$ is the 3-cocycle corresponding to the Poincar\'e dual
of $j$.  The term $\pi \int_{M^{3+1}} b\cup b$ changes the statistics of the
particle from bosonic to fermionic.  This is similar to the statistical
transmutation in 2+1D by Chern-Simons term.  Note that the condition  $\dd
b\se{2}*j $ means $\dd b=*j \text{ mod }2$ which can be enforced using energy
penalty $\ee^{-U\int_{M^{3+1}} |\dd b - *j|^2}$.

Does the transmuted fermion a spin-up-spin-down doublet? To address this issue,
we like to mention that the term $\pi \int_{M^{3+1}} b\cup b$ is compatible
with time-reversal symmetry. If the total model has a time-reversal symmetry,
then the particle dressed by the $b$ field, \ie the fermion, will be a
time-reversal singlet, which corresponds to a scalar fermion.  However, this
behavior can be adjusted by changing the topological term to $\pi
\int_{M^{3+1}} (\dd \t g\cup\dd \t g +b)\cup b$, where $\t g_i$ is a
$\Z_2$-0-cochain field which is a pseudo scalar.  So the new statistical
transmutation is given by
\begin{align}
Z(M^{3+1}) 
&= \hskip -1em
\sum_{
{\t g \in C^0(M^{3+1};\Z_2) } \atop
{b \in C^2(M^{3+1};\Z_2),  \dd b\se{2} *j }} \hskip -1em
\ee^{\ii \pi \int_{M^{3+1}} [\dd \t g\cup \dd \t g +b]\cup b }
.
\end{align}
The second type of statistical transmutation can still changes the statistics
of the particle from bosonic to fermionic, but now the fermion, dressed by $b$
and $\t g$ fields,  will be a Kramer doublet which corresponds to a spin-1/2
fermion (see Section \ref{proppoint}). 

\subsection{2+1D time-reversal symmetric topological orders}

\def\arraystretch{1.4} \setlength\tabcolsep{3pt}
\begin{table*}[t]
\caption{
The 3+1D time-reversal ($T$) symmetric $Z_2$-gauge theories emerged from
lattice bosonic models $Z_{k_1k_2k_3k_4k_5k_6}$ in \eqn{Z2T4d1}.  Each row
corresponds to a root family which contains a few ($N_\text{dis}$) $T$-symmetric
topological orders labeled by $k_5,k_6=0,1$.  Topological orders in the same
root family differ only by $Z_2^T$-SPT states.  $d_i$'s and $s_i$'s are the
quantum dimensions and spins of point-like excitations.  A quantum dimension
$d=2$ means that the excitation has 2 internal degrees of freedom.  $2_-$ means
that the 2 internal degrees of freedom form a Kramer doublet with $T^2=-1$.
The fourth column is the volume-independent partition function on space-time
$M^4$, where $\rw_i$ is the $i^\text{th}$ Stiefel-Whitney class.  } 
\label{Tsymm4}
 \centering
 \begin{tabular}{ |c|c|c|l|c|c| }
 \hline
$k_1k_2k_3k_4k_5k_6$  & $(d_1,d_2)$ &  $(s_1,s_2)$ &  $Z^\text{top}=\frac{|H^1(M^4;\Z_2)|\ee^{ \ii  \pi \int_{M^4} k_5 \rw_1^4 }}{|H^0(M^4;\Z_2)|\ee^{ \ii  \pi \int_{M^4} k_6 \rw_2^2 }}\times$ & $N_\text{dis}$ & As gauged SPT state\\
 \hline
$0000k_5k_6$  & $(1,1)$ & $(0,0)$ & $1$  & 4 & Bosonic $Z_2\times Z_2^T$ trivial state \\
$0100*k_6$  & $(1,1)$ & $(0,0)$ &  $\del(\rw_1^3)$ & 2  & Bosonic $Z_2\times Z_2^T$-SPT state   \\
$1000k_5k_6$  & $(1,1)$ & $(0,0)$ &  $\del(\rw_3)$ & 4  &   Bosonic $Z_2\times Z_2^T$-SPT state  \\
$1100*k_6$  & $(1,1)$ & $(0,0)$ & $\del(\rw_3+\rw_1^3)$ & 2 & Bosonic $Z_2\times Z_2^T$-SPT state  \\
\hline
$0001k_5*$  & $(1,2_-)$ & $(0,\frac12)$ &  $\del(\rw_2)$  & 2 & Free fermion $Z_4^T$-SPT state  \\
$0101**$  & $(1,2_-)$ & $(0,\frac12)$ &  $\del(\rw_1^3)\del(\rw_2)$ & 1 & ?  \\
%\dgrn{$1001k_5*$}  & \dgrn{$(1,2_-)$} & \dgrn{$(0,\frac12)$} & \dgrn{$E_{\R P^2}=\infty$} & \dgrn{$\del(\rw_3)\del(\rw_2)=\del(\rw_2)$}  & \dgrn{2} &  \\
%\dgrn{$1101**$}  & \dgrn{$(1,2_-)$} & \dgrn{$(0,\frac12)$} & \dgrn{$E_{\R P^2}=\infty$} & \dgrn{$\del(\rw_3+\rw_1^3)\del(\rw_2)=\del(\rw_1^3)\del(\rw_2)$} & \dgrn{1} & \\
\hline
$0010*k_6$  & $(1,2_-)$ & $(0,0)$ &  $\del(\rw_1^2)$  & 2 &  Bosonic $Z_4^T$-SPT state \\
\dgrn{$0110*k_6$}  & \dgrn{$(1,2_-)$} & \dgrn{$(0,0)$} & \dgrn{$\del(\rw_1^3)\del(\rw_1^2)=\del(\rw_1^2) $} & \dgrn{2} &  \\
$1010*k_6$  & $(1,2_-)$ & $(0,0)$ &  $\del(\rw_3)\del(\rw_1^2)$ & 2 & Bosonic $Z_4^T$-SPT state  \\
\dgrn{$1110*k_6$}  & \dgrn{$(1,2_-)$} & \dgrn{$(0,0)$} & \dgrn{$\del(\rw_3+\rw_1^3)\del(\rw_1^2)=\del(\rw_3)\del(\rw_1^2)$} & \dgrn{2} & \\
\hline
$0011**$  & $(1,1)$ & $(0,\frac12)$ &  $\del(\rw_1^2+\rw_2)$  & 1 & Fermionic $Z_2^f\times Z_2^T$ trivial state \\
\dgrn{$0111**$}  & \dgrn{$(1,1)$} & \dgrn{$(0,\frac12)$} & \dgrn{$\del(\rw_1^3)\del(\rw_1^2+\rw_2)=\del(\rw_1^2+\rw_2)$} & \dgrn{1} &  \\
%\dgrn{$1011(k_5k_6)$}  & \dgrn{$(1,1)$} & \dgrn{$(0,\frac12)$} & \dgrn{$\del(\rw_3)\del(\rw_1^2+\rw_2)=\del(\rw_1^2+\rw_2)$} & \dgrn{1} & \\
%\dgrn{$1111(k_5k_6)$}  & \dgrn{$(1,1)$} & \dgrn{$(0,\frac12)$} & \dgrn{$\del(\rw_3+\rw_1^3)\del(\rw_1^2+\rw_2)=\del(\rw_1^2+\rw_2)$} & \dgrn{1} & \\
 \hline
 \end{tabular}
\end{table*}

We have constructed $2^3=8$ time-reversal symmetric local bosonic models in
2+1D (see \eqn{Z2Z4T}).  
\begin{align}
\label{Z2Z4T1}
&\ \ \ \
 Z_{k_0k_1k_2;\text{t}\Z_2\text{a}T}(M^3) 
\nonumber\\
&=\sum_{ \{\t g^{\Z_2}_i, a^{\Z_2}_{ij},b^{\Z_2}_{ijk}\}}   
\ee^{ \ii  \pi \int_{M^3} b^{\Z_2}\cup (\dd a^{\Z_2}-k_0 \Bs_2 \dd \t g^{\Z_2})} \times
\\
&
\ \ \ \ \ \ \ \ \ \ \ \
\ \ \ \ \ \ \ \ 
\ee^{ \ii  \pi \int_{M^3} k_1 a^{\Z_2}\cup  a^{\Z_2}\cup  a^{\Z_2} 
+k_2 \dd \t g^{\Z_2} \cup \dd \t g^{\Z_2} \cup a^{\Z_2}
} ,
\nonumber 
\end{align}
where $\t g^{\Z_2}_i, a^{\Z_2}_{ij},b^{\Z_2}_{ijk}$ are $\Z_2$-valued
0-cochain, 1-cochain, and 2-cochain fields (see Section \ref{disltgauge}), and
$k_{0,1,2}=0,1$.  Also the time-reversal symmetry is described by group $Z_2^T$
with $T^2=1$, whose action is given by $(\t g^{\Z_2}_i,
a^{\Z_2}_{ij},b^{\Z_2}_{ijk})\to (\mod(\t g^{\Z_2}_i+1,2),
a^{\Z_2}_{ij},b^{\Z_2}_{ijk}) $ plus the complex conjugation.  (The above model
also has an additional $Z_2'$ symmetry generated by $(\t g^{\Z_2}_i,
a^{\Z_2}_{ij},b^{\Z_2}_{ijk})\to (\mod(\t g^{\Z_2}_i+1,2),
a^{\Z_2}_{ij},b^{\Z_2}_{ijk}) $ without the complex conjugation.) We see that
$\t g_i^{\Z_2}$ is a pseudo scalar field. 
%(\ie $\t g_i^{\Z_2}$ is multivalued on
%non-orientable space-time such that $\dd \t g = \rw_1$ up to coboundaries, which
%is denoted as $\dd \t g\se{\dd} \rw_1$).  
The above eight models realize five
types of time-reversal SET orders.

The four constructed models (labeled by $k_00k_2$) reduce to the
$Z_2$-topological order described by UT $Z_2$ gauge theory after we break the
time-reversal symmetry (see top three rows in Table \ref{Tsymm3}).  But three
of them have identical topological orders. Thus the four models only give us
two types of time-reversal symmetric $Z_2$ gauge theories.\cite{WS1334} They
correspond to two types of time-reversal symmetric $Z_2$ gauge theories .
Those four models are obtained by gauging the $Z_2$ subgroup in two of the four
$Z_2\times Z_2^T$ SPT states and by gauging the $Z_2$ subgroup of $Z_4^T$
SPT states ($Z_4^T$ has $T^2=-1$).

There is another type of time-reversal symmetric $Z_2$ gauge theory where the
time-reversal transformation exchange the $Z_2$-charge and
$Z_2$-vortex.\cite{HL160607816} Such a theory is missing from the table.

The other three of five constructed time-reversal SET orders correspond to
three types of time-reversal symmetric double-semion theories
\cite{FNS0428,LW0510} (see bottom three rows in Table \ref{Tsymm3}).  Those
theories are obtained by gauging the $Z_2$ subgroup in two of the four
$Z_2\times Z_2^T$ SPT states.
% and by gauging the $Z_2$ subgroup in one of two
%$Z_4^T$ SPT states ($Z_4^T$ has $T^2=-1$).  
Two of four constructed models (labeled by $k_01k_2$) have identical
topological orders. They give us three types of time-reversal symmetric
double-semion theories.  

It the interesting to note that one of the time-reversal symmetric double-semion
topological order (the last row in Table \ref{Tsymm3}) contain four types of
point-like excitations: (1) a trivial type which is a time-reversal singlet;
(2) a bosonic Kramer doublet (denoted by quantum dimension, \ie internal
degrees of freedom, $d=2_-$); (3) a $T^2=1$ time-reversal doublet formed by two
semions with spin $\frac14$ and $\frac 34$ (denoted by quantum dimension
$d=2_+$); (4) a $T^2=-1$ Kramer doublet formed by two semions with spin
$\frac14$ and $\frac 34$ (denoted by quantum dimension $d=2_-$).

\subsection{3+1D time-reversal symmetric $Z_2$ gauge theories}

We also have constructed $2^6=64$ local bosonic models in 3+1D which can
realize $20$ types of simplest topological orders with time-reversal symmetry
(see the black rows in Table \ref{Tsymm4}).  Those topological orders are
simplest since they have only one type of non-trivial point-like topological
excitations and one type of non-trivial string-like topological excitations.
The point-like topological excitations in those 3+1D SET orders can be Kramer
doublet (which corresponds to the fractionalization\cite{W0213,YFQ1070} of
time-reversal symmetry) and can be fermionic.  If we break the time-reversal
symmetry, 16 of the 20 SET orders reduce to the 3+1D $Z_2$-topological order
described by the UT $Z_2$-gauge theory, and other 4 of the 20 SET orders reduce
to the 3+1D topological order described by the EF $Z_2$-gauge theory.

Those 64 bosonic models are given by (see \eqn{Z2T4d}):
\begin{align}
\label{Z2T4d1}
&\ \ \ \
 Z_{k_1k_2k_3k_4k_5k_6}(M^4) 
\\
&=
\hskip -1em 
\sum_{ \{\t g_i^{\Z_2}, a_{ij}^{\Z_2}, b^{\Z_2}_{ijk} \}  }   
\hskip -2em 
\ee^{\ii \pi \int_{M^4} a^{\Z_2} \cup [\dd b^{\Z_2} + k_1
 a^{\Z_2} \cup a^{\Z_2} \cup a^{\Z_2} 
+(k_1+k_2)\dd \t g\cup \dd \t g\cup \dd \t g]} 
\nonumber \\
&
\ee^{ \ii  
\pi \int_{M^4} [k_4  b^{\Z_2} + (k_3+k_4) \dd \t g\cup \dd \t g]\cup b^{\Z_2} + k_5 
\dd \t g \cup
\dd \t g \cup
\dd \t g \cup
\dd \t g 
+k_6 \rw_2 \cup \rw_2
} 
\nonumber 
,
\end{align}
where $k_I=0,1$,
%$\chi(M^4)$ is the Euler number,
$b^{\Z_2}$ is a $\Z_2$-2-cocycle field, $a^{\Z_2}$ a $\Z_2$-1-cocycle field, and
$\t g_i$ a pseudo scalar field which changes under the time-reversal
transformation $\t g_i \to \mod(\t g_i+1,2)$.  The above local bosonic models
have a time-reversal symmetry: the action amplitude is invariant under the
combined transformation of $\t g_i \to \mod(\t g_i+1,2)$ and complex
conjugation.  The models also have a $Z_2'$ symmetry: the action amplitude is
invariant under $\t g_i \to \mod(\t g_i+1,2)$ (without the complex
conjugation).

But the above model 
%contains explicit coupling between the boson fields and the Stiefel-Whitney
%class which is not allowed for local bosonic model.  It turns out that 
is exactly soluble only when $k_1k_4=0$.
%, the one of the above models is a local bosonic model .without any coupling
%to the Stiefel-Whitney class.  
Those 48 exactly soluble models produce the rows in Table \ref{Tsymm4}.  The
models described by the green rows in Table \ref{Tsymm4} produce topological
orders that are identical to some black rows.  Those identities come from the
relations between the Stiefel-Whitney classes on 4-dimensional space-time (see
\eqn{wrel4}):
\begin{align}
& \rw_1\cup \rw_2=0,\ \ \rw_1\cup \rw_3=0,
\nonumber\\
& \rw_1\cup\rw_1\cup\rw_1\cup\rw_1+\rw_2\cup \rw_2+\rw_4=0.
\end{align}
We see that $\rw_1\cup\rw_1+\rw_2=0$ implies 
$\rw_1\cup\rw_1\cup\rw_1+\rw_2\cup \rw_1=\rw_1\cup\rw_1\cup\rw_1=0$.  Thus
$\del(\rw_1^3)\del(\rw_1^2+\rw_2)=\del(\rw_1^2+\rw_2)$, which implies that the
first and the second rows in the fourth block in Table \ref{Tsymm4} have the
same partition function and thus correspond to the same theory. 
%Similarly, $\Sq^1(\rw_1^2+\rw_2)=\Sq^1(\rw_2)=\rw_3+\rw_1\rw_2=\rw_3$ (see
%Appendix \ref{Rswc}).  Thus $\rw_1^2+\rw_2=0$ implies $\rw_3=0$, and
%$\del(\rw_3)\del(\rw_1^2+\rw_2)=\del(\rw_1^2+\rw_2)$. The first and the third
%rows in the fourth block in Table \ref{Tsymm4} correspond to the same theory.

We note that the four types of 3+1D $Z_2^T$-SPT states \cite{CGL1314,VS1306}
can be labeled by $k_5,k_6=0,1$ and are characterized by the SPT invariant
$Z(M^4)= \ee^{ \ii \pi \int_{M^4} k_5 \rw_1^4 + k_6 \rw_2^2 }$.  The
$Z_2^T$-SPT state $(k_5k_6)=(10)$ is the one described by group-cohomolgy
$\cH^4(Z_2^T;(\R/\Z)_T)$,\cite{CGL1314} and has a time-reversal symmetric
boundary described by an anomalous $Z_2$ gauge theory where the $Z_2$-charge
$e$ and the $Z_2$-vortex $m$ are both Kramer doublet, while the $e$ and $m$
bound state $\veps$ is a time-reversal singlet fermion.\cite{VS1306} The
$Z_2^T$-SPT state $(k_5k_6)=(01)$ is beyond $\cH^4(Z_2^T;(\R/\Z)_T)$, and has a
time-reversal symmetric boundary described by an anomalous $Z_2$ gauge theory
where $e$, $m$, and $\eps$ are all fermions.

The model with the same $k_1k_2k_3k_4$ but different $k_5k_6$ only differ by
stacking those four $Z_2^T$-SPT states.  We call two time-reversal SET order
that differ only by stacking of $Z_2^T$-SPT states as to have the same root,
since those SETs have identical bulk point-like and string-like excitations.
We find that the 20 SET orders belong to 9 root families. This is because
stacking the four $Z_2^T$-SPT states does not always produce four distinct
time-reversal SET phases, since the partition function may vanish on space-time
with non-trivial $\rw_1\cup\rw_1\cup\rw_1\cup\rw_1$, $\rw_2\cup \rw_2$, or
$\rw_1\cup\rw_1\cup\rw_1\cup\rw_1 + \rw_2\cup \rw_2$.  The number
$N_\text{dis}$ of distinct time-reversal SET phases in each root family is
given in Table \ref{Tsymm4}.  The 9 root families correspond to 9 types of 3+1D
time-reversal symmetric $Z_2$-gauge theories.

From the Table \ref{Tsymm4} and from the discussions in Section \ref{strasymm},
we also see the physical meaning of each topological term labeled by
$k_1k_2k_3k_4$:
\begin{enumerate}
\item
 $k_4=1$  makes the point-like
excitations to be fermions.
\item
 $k_4+k_3=1$ mod 2 makes the point-like
excitations to be Kramer doublet.
\item
 $k_1+k_2=1$ mod 2 makes the string-like excitations to carry 
%degrees of freedom that respect 
an anomalous $Z_2'$ symmetry that appear on the boundary of a 2+1D $Z_2'$-SPT
state.  Such an anomalous (non-on-site) $Z_2'$ symmetry is given by $U'=
\prod_I \si^x_I \prod_I \si^z_I
\frac{1+\si^z_I+\si^z_{I+1}-\si^z_I\si^z_{I+1}}{2}$, where $\si^z_i=(-)^{\t
g_i}$ and $\frac{1+\si^z_I+\si^z_{I+1}-\si^z_I\si^z_{I+1}}{2}
=CZ(\si^z_I,\si^z_{I+1})$ is the controlled-$Z$ gate acting on the two qubits
$\si_I$ and $\si_{I+1}$.
\end{enumerate}
Certainly, when $k_1+k_2=0$ mod 2, the string will not have  anomalous
symmetry, and are in general gapped and symmetric.

There are also many other ways to realize time-reversal symmetric $Z_2$ gauge
theories.  For example, one can use non-linear $\si$-model field theory to
realize many of the above time-reversal SET's with bosonic point-like
excitations.\cite{X13078131} More generally, one may start with $Z_2\times
Z_2^T$ bosonic SPT states.  There are 8 such states since $\cH^4(Z_2\times
Z_2^T;(\RZ)_T)=\Z_2^{\oplus 3}$.\cite{CGL1314} Gauging the $Z_2$ symmetry give
us 8 time-reversal symmetric $Z_2$ topological orders.  But some of them only
differ by a $Z_2^T$ SPT state.  We only obtain 4 root states (\ie 4
time-reversal symmetric $Z_2$-gauge theories), that correspond to the first
four rows in the Table \ref{Tsymm4}.  We can also start with $Z_4^T$ bosonic
SPT states.  There are 2 such states since $\cH^4(Z_4^T;(\RZ)_T)=\Z_2$.  After
gauging the unitary $Z_2$ subgroup of $Z_4^T$, we obtain two time-reversal
symmetric $Z_2$ gauge theories (see the two black rows in the third block in
Table \ref{Tsymm4}).  Those two root states have a property that stacking with
the $(k_5k_6)=(10)$ $Z_2^T$-SPT state gives us the same root states back.  The
other two root states (the green rows) are identical to the two black rows in
the third block.

The second block in the Table \ref{Tsymm4} contains two root states. The first
one can be obtained by gauging $Z_4^T$ fermionic SPT states, which
is also known as the $T^2=-1$ fermionic topological
superconductor.\cite{K0986,RSF0957}  There are at least 16 $Z_4^T$ fermionic
SPT states labeled by $\nu=0,1,\cdots,15$.\cite{WS14011142,MV14063032,YC14090168}
Gauging the fermion-parity $Z_2^f$ subgroup in $Z_4^T$ fermionic SPT states
will produce several time-reversal symmetric topological orders that contain
Kramer-doublet fermions.  The string-like excitations (\ie the $Z_2^f$ vortex
lines or the $Z_2^f$-symmetry-twist defect line) in those topological orders
must be gapless unless $\nu=$ even, if the time-reversal symmetry is not
broken.\cite{YC14046256} In comparison, the strings in the three time-reversal
symmetric topological orders described by the two rows in the second block do
not carry any anomalous time-reversal symmetry.  In other words, the
excitations on the strings can be gapped even if we do not break the time
reversal symmetry.  
%Thus the time-reversal symmetric topological orders in the
%second block may come from gauging the $Z_2^f$ symmetry of the $Z_4^T$
%fermionic topological superconductor.  

All the states in the second block have a property that stacking with the
$(k_5k_6)=(01)$ bosonic $Z_2^T$-SPT state (characterized by the SPT invariant
$\ee^{ \ii  \pi \int_{M^4} k_5 \rw_2\cup\rw_2 }$) does not change their $Z_2^T$
SET orders.  Similarly,  the $Z_4^T$ fermionic topological superconductors also
have the property that stacking with the $(k_5k_6)=(01)$ $Z_2^T$-SPT state does
not change the SPT order.  For example, the $\nu=0$ $Z_4^T$ fermionic
topological superconductor has a boundary with two types of quasiparticles
$\{1,c\}$, where $1$ is the trivial type and $c$ is a Kramer-doublet fermion.
The $(k_5k_6)=(01)$ $Z_2^T$-SPT state has boundary with four types of
quasiparticles $\{1,f_1,f_2,\veps\}$, where $f_1$ and $f_2$ are Kramer-doublet
fermions and $\veps$ is a time-reversal singlet fermion.  Also $f_{1,2}$ and
$\veps$ have $\pi$-mutual statistics among them.  The stacking of the two
states have a boundary with quasiparticles $\{1,f_1,f_2,\veps\}\times \{1,c\}$.
We may condense the time-reversal singlet boson $f_2 c$.  Then, the new
boundary state with have quasiparticles $\{1,c\}$.  The quasiparticle $f_2$ is
also not confined, but it is equivalent to $c$, since the two only differ by a
condensed boson.  Thus the stacking of the $\nu=0$ state and the
$(k_5k_6)=(01)$ state can have the same boundary as the $\nu=0$ state. The
stacking of $(k_5k_6)=(01)$ state does not change the SPT order in $Z_4^T$
fermionic topological superconductor.

The two states that correspond to the first row in the second block differ by
stacking with the $(k_5k_6)=(10)$ $Z_2^T$-SPT state characterized by the SPT
invariant $\ee^{ \ii  \pi \int_{M^4} k_5 \rw_1\cup\rw_1\cup\rw_1\cup\rw_1}$.
For  $Z_4^T$ fermionic topological superconductors, stacking with the
$(k_5k_6)=(10)$ state will shift $\nu$ by
8.\cite{FV13055851,WS14011142,YC14046256} This suggests that the two states are
the $\nu=0$ and the $\nu=8$ $Z_4^T$ fermionic topological superconductor with
gauged $Z_2^f$ symmetry.\cite{FV13055851,WS14011142,YC14046256} 

On the other hand, for the time-reversal symmetric topological order described
by the second row in the second block, stacking with any $Z_2^T$-SPT states
does not change its $Z_2^T$ SET order. It is not clear if the  $Z_2^T$ SET
order can be viewed as the $Z_2^f$-gauged $\nu=\pm 4$ $Z_4^T$ fermionic
topological superconductor or not.

\subsection{Vanishing of the volume-independent partition function}

We have calculated many volume-independent partition functions, and find they
vanish some times.  In general, a  partition function may have a form
\begin{align}
 Z(M^d) = \ee^{ -c_d L^d -c_{d-1} L^{d-1} -\cdots - c_0 L^0 - c_{-1}L^{-1}
-\cdots},
\end{align}
where $L$ is the linear size of $M^d$.  If the ground state does not contain
point-like, string-like, \etc defects, then $c_1=c_2=\cdots=c_{d-1}=0$.  In
this case,
\begin{align}
 Z^\text{top}(M^d) \equiv \lim_{L\to\infty} \frac{Z(M^d)}{\ee^{ -c_d L^d}} = \ee^{ - c_0 L^0}
\end{align}
is the  volume-independent partition function.  When the calculated
volume-independent partition function vanishes, it does not mean the partition
function to vanish nor the theory to be anomalous. It just means that $c_i>0$,
for some $0<i<d$. This implies that the given space-time topology $M^d$ induces
point-like, string-like, \etc topological excitations.  

We have calculated volume-independent partition functions for many constructed
systems and for many space-time manifolds (see Table \ref{Tsymm3}, 
\ref{Tsymm4}, and \ref{topinv}). From those results, we conjecture that:\\
\emph{A local bosonic model with emergent fermion always has
vanishing volume-independent partition function $Z^\text{top}(M^d)=0$ if the
orientable $M^d$ is not spin.  }\\
In the presence of time-reversal symmetry:\\
\emph{(1) A local bosonic model with emergent Kramer doublet fermions always has
vanishing volume-independent partition function $Z^\text{top}(M^d)=0$ if $M^d$
is not pin$^+$ (\ie $\rw_2\neq 0$).\\
(2)  A local bosonic model with emergent time-reversal singlet
fermions always has vanishing volume-independent partition function
$Z^\text{top}(M^d)=0$ if $M^d$ is not pin$^-$ (\ie $\rw_2+\rw_1^2\neq 0$). \\
(3)  A local bosonic model with emergent Kramer doublet bosons
always has vanishing volume-independent partition function
$Z^\text{top}(M^d)=0$ if $\rw_1^2\neq 0$ on $M^d$. 
}\\
(See Appendix \ref{spinstructure} for a brief introduction of spin, pin$^+$,
and pin$^-$ manifolds.) Those properties has been
used to develop cobordism theory for fermionic SPT states \cite{KTT1429}.

In the rest of this paper, we will present detailed constructions and
calculation. 

\section{A generic construction of exactly soluble bosonic lattice models on
space-time lattice}

In this section, we are going to introduce a general way to construct exactly
soluble local bosonic models.  Those models are written in terms of path
integral on space-time lattice.  Those models are also designed to have
topologically ordered ground states.  In other words, those models have
emergent topological field theory at low energies. 

First, we will briefly review the related mathematics. Then we will construct
models that realize some well known topological orders, such as those described
by discrete gauge theories, and by Dijkgraaf-Witten theories.
After that we will construct models that realize more general topological
orders whose low energy effective theories are beyond Dijkgraaf-Witten
theories. We will also compute the volume-independent partition functions for
those constructed models on several choices of space-time manifolds.  The
results are summarized in Table \ref{topinv}.

\subsection{Space-time complex, cochains, and cocycles} \label{disltgauge}

Our local bosonic models will be defined on a space-time lattice.  A space-time
lattice is a triangulation of the $d$-dimensional space-time, which is denoted
as $M^d_\text{latt}$.  We will also call the triangulation $M^d_\text{latt}$ as
a space-time complex.  A cell in the complex is called a simplex.  We will use
$i,j,\cdots$ to label vertices of the space-time complex.  The links of the
complex (the 1-simplices) will be labeled by $(i,j),(j,k),\cdots$.  Similarly,
the triangles of the complex  (the 2-simplices)  will be labeled by
$(i,j,k),(j,k,l),\cdots$.

A cochain $f_n$ is an assignment of values in $\M$ to each $n$-simplex, for
example a value $f_{n;i,j,\cdots,k}\in \M$ for $n$-simplex $(i,j,\cdots,k)$.  So
\emph{a cochain $f_n$ can be viewed as a bosonic field on the space-time
lattice}. In this paper, we will use such cochain bosonic field to construct
our models.

In this paper, we will assume $\M$ to be a ring which support addition
and multiplication operations, as well as scaling by an integer:
\begin{align}
	 x+y &= z,\ \ \ \ x*y=z, \ \ \ \ mx=y,
	\nonumber\\
	x,y,z & \in \M,\ \ \ m \in \Z.
\end{align}
We see that $\M$ can also be viewed a $\Z$-module (\ie a vector space with
integer coefficient) that also allows a multiplication operation.  In this
paper we will view $\M$ as a $\Z$-module.  The direct sum of two modules
$\M_1\oplus \M_2$ (as vector spaces) is equal to the direct product of the two
modules (as sets):
\begin{align}
 \M_1\oplus \M_2 \stackrel{\text{as set}}{=} \M_1\times \M_2
\end{align}

\begin{figure}[tb]
\begin{center}
\includegraphics[scale=0.5]{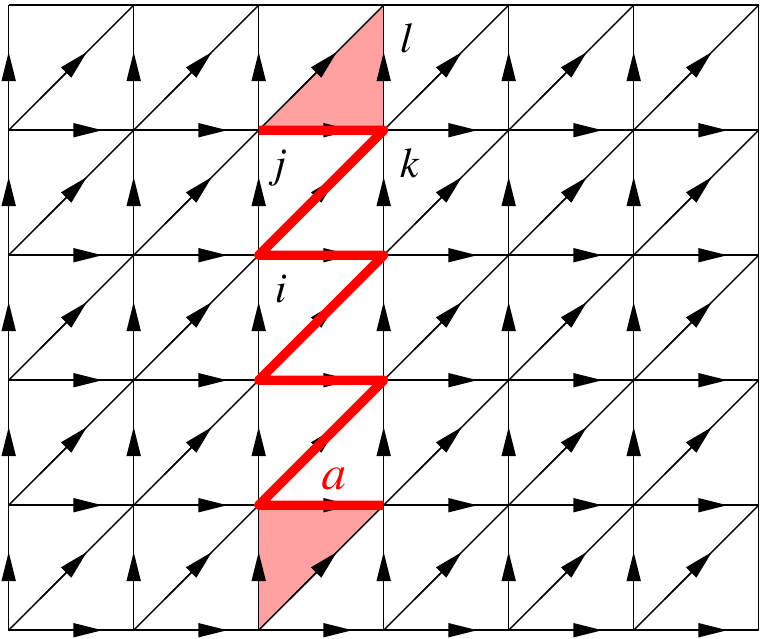} \end{center}
%2
\caption{ (Color online)
A 1-cochain $a$ has a value $1$ on the red links: $ a_{ik}=a_{jk}= 1$ and a value $0$ on other links: $ a_{ij}=a_{kl}=0 $.
$\dd a$ is non-zero on the shaded triangles: $(\dd a)_{jkl} = a_{jk} +
a_{kl} - a_{jl}$.
For such 1-cohain, we also have $a\cup a=0$.
So when viewed as a $\Z_2$-valued cochain, $\Bs_2 a \neq a\cup a$ mod 2.
}
\label{dcochain}
\end{figure}

We like to remark that a simplex $(i,j,\cdots,k)$ can have two different
orientations. We can use $(i,j,\cdots,k)$ and $(j,i,\cdots,k)=-(i,j,\cdots,k)$
to denote the same simplex with opposite orientations.  The value
$f_{n;i,j,\cdots,k}$ assigned to the simplex with opposite  orientations should
differ by a sign: $f_{n;i,j,\cdots,k}=-f_{n;j,i,\cdots,k}$.  So to be more
precise $f_n$ is a linear map $f_n: n\text{-simplex} \to \M$. We can denote the
linear map as $\<f_n, n\text{-simplex}\>$, or
\begin{align}
 \<f_n, (i,j,\cdots,k)\> = f_{n;i,j,\cdots,k} \in \M.
\end{align}
More generally, a \emph{cochain} $f_n$ is a linear map
of $n$-chains:
\begin{align}
	f_n:  n\text{-chains} \to \M,
\end{align}
or (see Fig. \ref{dcochain})
\begin{align}
 \<f_n, n\text{-chain}\> \in \M,
\end{align}
where a \emph{chain} is a composition of simplices. For example, a 2-chain can
be a 2-simplex: $(i,j,k)$, a sum of two 2-simplices: $(i,j,k)+(j,k,l)$, a more
general composition of 2-simplices: $(i,j,k)-2(j,k,l)$, \etc.  The map $f_n$ is
linear respect to such a composition.  For example, if a chain is $m$ copies of
a simplex, then its assigned value will be $m$ times that of the simplex.
$m=-1$ correspond to an opposite orientation.  

The total space-time lattice $M^d_\text{latt}$ correspond to a $d$-chain.  We
will use the same $M^d_\text{latt}$ to denote it.  Viewing $f_d$ as a linear
map of $d$-chains, we can define an ``integral'' over $M^d_\text{latt}$:
\begin{align}
 \int_{M^d_\text{latt}} f_d \equiv \<f_d,M^d_\text{latt}\>.
\end{align}

In this paper, we usually take $\M$ to be integer $\Z$ or mod $n$ integer
$\Z_n=\{0,1,\cdots,n-1\}$.  So not only the field $f_{n;i,j,\cdots,k}$ is
defined on a discrete space-time lattice, even the value of the field is
discrete.  We will use $C^n(M^d_\text{latt};\M)$ to denote the set of all
$n$-cochains on $M^d_\text{latt}$.  $C^n(M^d_\text{latt};\M)$ can also be
viewed as a set all $\M$-values fields (or paths) on  $M^d_\text{latt}$.  Note
that $C^n(M^d_\text{latt};\M)$ is an abelian group under the $+$-operation.

\begin{figure}[tb]
\begin{center}
\includegraphics[scale=0.5]{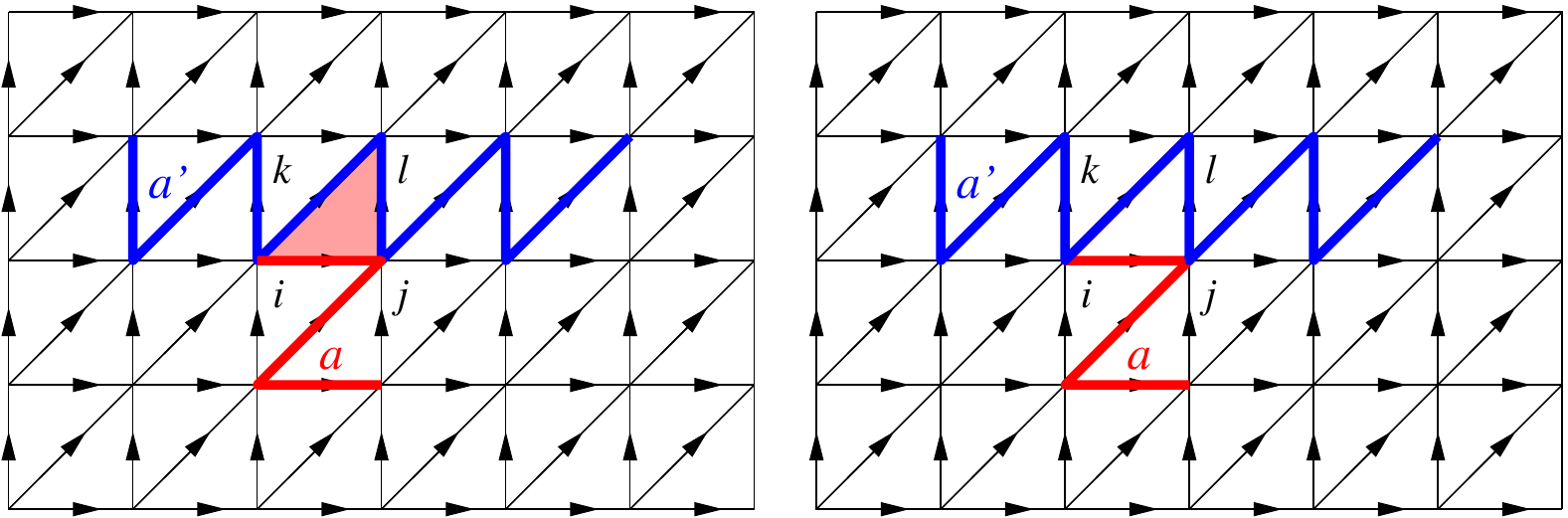} \end{center}
%3
\caption{ (Color online)
A 1-cochain $a$ has a value $1$ on the red links, Another
1-cochain $a'$ has a value $1$ on the blue links.
On the left, $a\cup a'$ is non-zero on the shade triangles:
$(a\cup a')_{ijl}=a_{ij}a'_{jl}=1$, while
on the right, $a'\cup a$ is zero.
Thus $a\cup a'+a'\cup a$ is not a coboundary.
}
\label{cupcom}
\end{figure}

We can define a derivative operator $\dd$ acting on an $n$-cochain $f_n$, which
give us an $n+1$-cochain (see Fig. \ref{dcochain}):
\begin{align}
&\ \ \ \ \<\dd f_n, (i_0i_1i_2\cdots i_{n+1})\>
\nonumber\\
&=\sum_{m=0}^{n+1} (-)^m
\<f_n, (i_0i_1i_2\cdots\hat i_m\cdots i_{n+1})\>
\end{align}
where $i_0i_1i_2\cdots \hat i_m \cdots i_{n+1}$ is the sequence
$i_0 i_1 i_2 \cdots i_{n+1}$ with $i_m$ removed, and
$i_0, i_1,i_2 \cdots i_{n+1}$ are the ordered vertices of the $(n+1)$-simplex
$(i_0 i_1 i_2 \cdots i_{n+1})$.

A cochain $f_n \in C^n(M^d_\text{latt};\M)$ is called a \emph{cocycle} if $\dd
f_n=0$.   The set of cocycles is denoted as $Z^n(M^d_\text{latt};\M)$.  A
cochain $f_n$ is called a \emph{coboundary} if there exist a cochain $f_{n-1}$
such that $\dd f_{n-1}=f_n$.  The set of coboundaries is denoted as
$B^n(M^d_\text{latt};\M)$.  Both $Z^n(M^d_\text{latt};\M)$ and
$B^n(M^d_\text{latt};\M)$ are abelian groups as well.  Since $\dd^2=0$, a
coboundary is always a cocycle: $B^n(M^d_\text{latt};\M) \subset
Z^n(M^d_\text{latt};\M)$.  We may view two  cocycles differ by a coboundary as
equivalent.  The equivalence classes of cocycles, $[f_n]$, form the so called
cohomology group denoted as
\begin{align}
H^n(M^d_\text{latt};\M)=   Z^n(M^d_\text{latt};\M)/ B^n(M^d_\text{latt};\M),
\end{align}
$H^n(M^d_\text{latt};\M)$, as a group quotient of $Z^n(M^d_\text{latt};\M)$ by
$B^n(M^d_\text{latt};\M)$, is also an abelian group.

\def\arraystretch{1.4} 
\setlength\tabcolsep{2pt}
\begin{table*}[t]
\caption{Volume independent partition function $Z^\text{top}(M^4)$ for the
constructed local bosonic models, on closed 4-dimensional space-time manifolds
The space-time $M^4$ considered here satisfy $\chi(M^4)=P_1(M^4)=0$, which
makes $Z^\text{top}(M^4)$ to be a topological invariant \cite{KW1458}.  The
topological invariants listed below are also the ground state degeneracy on the
corresponding spatial manifold $M^3_\text{space}$.  Here $L^3(p)$ is the
3-dimensional lens space and $F^4=(S^1\times S^3)\# (S^1\times S^3)\#\C P^2\#
\overline{\C P}^2$.  $F^4$ is not spin.  The different models are labeled by
$k_I$ which all have a range $k_I=0,1,\cdots,n-1$. } \label{topinv}
 \centering
 \begin{tabular}{ |c|c|c|c|c|c| }
 \hline
Models $\backslash\ M^4$:  & $T^4 $ & $T^2\times S^2$ & $S^1\times L^3(p) $  & $F^4$ & Low energy effective theory \\
\hline
$Z_{\Z_n\text{a}}^\text{top}(M^4)$  & $n^3$ & $n$ & $\<n,p\>$ & $n$ & \parbox{1.8in}{UT $Z_n$ gauge theory}\\
\hline
$Z_{k;b^2\Z_n}^\text{top}(M^4)$  & $\<2k,n\>^3$ & $\<2k,n\>$ & $\<2k,n,p\>$  & 
\parbox{1.5in}{$\<2k,n\>$ if $\frac{2kn}{\<2k,n\>^2}=$ even\\
0 if $\frac{2k_2n}{\<2k_2,n\>^2}=$ odd}
& \parbox{1.8in}{$Z_{\<2k,n\>}$ gauge theory with fermions iff $\frac{2kn}{\<2k,n\>^2}=$ odd}\\
 \hline
$Z_{k_1k_2;aa'\Bs a'\Z_n}^\text{top}(M^4)$  &  $n^6$ &  $n^2$ & 
\parbox{1.1in}{
$ \<n,p\>\<n,p,k_1,k_2\> $\\
if $p$ has no repeated prime factors.
}
& $n^2$ & \parbox{1.8in}{$Z_n\times Z_n$ Dijkgraaf-Witten theory}\\
\hline
$Z_{k_1k_2;b\Bs a\text{-}bb\Z_n}^\text{top}(M^4)$  & $n^3\<2k_2,n\>^3$ & $n\<2k_2,n\>$ & $\<n,p\>\<2k_2,n, p,  \frac{k_1p}{\<n,p\>}\>$ & 
\parbox{1.5in}{$n\<2k_2,n\>$ if $\frac{2k_2n}{\<2k_2,n\>^2}=$ even\\
0 if $\frac{2k_2n}{\<2k_2,n\>^2}=$ odd}
& \parbox{1.8in}{$Z_{\frac{n\<2k_2,n\>}{\<k_1,2k_2,n\>} } \times Z_{\<k_1,2k_2,n\>}$ gauge theory with fermions\\ iff $\frac{2k_2n}{\<2k_2,n\>^2}=$ odd}\\
\hline
 \end{tabular}
\end{table*}

From two cochains $f_m$  and $h_n$, we can construct a third cochain
$p_{m+n}$ via the cup product (see Fig. \ref{cupcom}):
\begin{align}
p_{m+n} &= f_m \cup h_n ,
\nonumber\\
\<p_{m+n}, (i_0 \cdots i_{m+n})\> 
&= 
\<f_m, (i_0 i_1 \cdots i_m)\> \times
\nonumber\\
&\ \ \ \ 
\<h_n,(i_m i_{m+1} \cdots i_{m+n}) \>
\end{align}
The cup product has the following property (see Fig. \ref{cupcom}):
\begin{align}
\label{cupprop}
 \dd(f_m \cup h_n) &= 
(\dd h_n) \cup f_m 
+ (-)^n h_n \cup (\dd f_m) 
\end{align}
We see that $f_m \cup h_n$ is a cocycle if both $f_m$ and $h_n$ are 
cocycles.  If both $f_m$ and $h_n$ are  cocycles, then $f_m \cup h_n$ is
a coboundary if one of $f_m$ and $h_n$ is a coboundary.
So the cup product is also an operation on cohomology groups
$\cup: H^m(M^d;\M)\times H^n(M^d;\M) \to  H^{m+n}(M^d;\M)$.
When both $f_m$ and $h_n$ are cocycles, we also have
\begin{align}
\label{cupprop1}
 f_m \cup h_n &= (-)^{mn} h_n \cup f_m + \text{coboundary}.
\end{align}

In the rest of this paper, we abbreviate the cup product $a\cup b$ as $ab$ by
dropping $\cup$.  Also, we will use $\Z_n=\{0,1,\cdots,n-1\}$ and
$Z_n=\{1,\ee^{\ii \frac{2\pi}{n}},\ee^{\ii 2 \frac{2\pi}{n}},\cdots,\ee^{\ii
(n-1) \frac{2\pi}{n}} \}$ to denote the same abelian group.  In $\Z_n$, the
group multiplication is mod-$n$ ``+'' and in $Z_n$, the group multiplication is ``$*$''.

\subsection{$\Z_n$-1-cocycle model and emergent $Z_n$ gauge theory}
\label{Znamdl}

\subsubsection{Model construction}

Using the above mathematical formalism, let us construct a local bosonic model
on a space-time lattice $M^{d+1}_\text{Latt}$, where the local degrees of freedom
live on the links.  The possible values on each link are
$a^{\Z_n}_{ij}=0,1,\cdots,n-1 \in \Z_n$.  

The action amplitude $\ee^{-S_\text{cell}}$ for a ${d+1}$-simplex $(ij\cdots l)$ is
a complex function of $a^{\Z_n}_{ij}$: $\ee^{ - L_{ij\cdots
l}(\{a^{\Z_n}_{ij}\})}$.  The total action amplitude $\ee^{-S}$ for a
configuration (or a path) is given by
\begin{align}
\label{lattS}
\ee^{-S(\{a^{\Z_n}_{ij}\}) }=
\prod_{(ij\cdots l)} \ee^{- L_{ij\cdots l}(\{a^{\Z_n}_{ij}\})}
\end{align}
where $\prod_{(ij\cdots l)}$ is the product over all the ${d+1}$-simplices $(ijkl)$.  
Our local bosonic model is defined by the following imaginary-time path integral
(or partition function)
\begin{align}
 Z_{\Z_n\text{a}} 
&=\sum_{ \{a^{\Z_n}_{ij}\} } \ee^{-S(\{a^{\Z_n}_{ij}\}) }
\nonumber\\
&=\sum_{ \{a^{\Z_n}_{ij}\} }   \ee^{ - \sum_{(ij\cdots l)} L_{ij\cdots l}(\{a^{\Z_n}_{ij}\})}
\end{align}
where $\sum_{ \{a^{\Z_n}_{ij}\} }$ is a sum over all paths (\ie the
path integral).

We may view $a^{\Z_n}_{ij}$ as $\Z_n$-valued 1-cochain on the space-time
complex $M^3_\text{latt}$:
\begin{align}
 a^{\Z_n}_{ij} = \<a^{\Z_n}, (ij)\>,\ \ \ a^{\Z_n} \in C^1(M^3_\text{latt}, \Z_n)
\end{align}
The Lagrangian $L_{ij\cdots l}(\{a^{\Z_n}_{ij}\})$ will produce an emergent low energy
$Z_2$-gauge theory (\ie have a $Z_2$ topological order) if we choose it to be
\begin{align}
 L_{ij\cdots l}(\{a^{\Z_n}_{ij}\}) &=  +\infty, &\text{ if } &(\dd a^{\Z_n}) \neq 0 \text{ on } (i j \cdots  l),
\nonumber\\
 L_{ij\cdots l}(\{a^{\Z_n}_{ij}\}) &= 0, &\text{ if } &(\dd a^{\Z_n})= 0 \text{ on } (i j \cdots  l).
\end{align}
So the action amplitude  $\ee^{ - L_{ij\cdots l}(\{a^{\Z_n}_{ij}\})} $ is
non-zero only when $a^{\Z_n}$ is a cocycle, and the non-zero value is always
$1$.  In other words, our local bosonic model is described by an action
$S(a^{\Z_n})=0$ when $a^{\Z_n}$ is a cocycle, and $S(a^{\Z_n})=+\infty$ when
$a^{\Z_n}$ is not a cocycle.  We see that the configurations described by
non-cocycles cost an infinity energy.  We will call the local bosonic model
described by the above $L_{ij\cdots l}$ as a $\Z_n$-1-cocycle model.

\subsubsection{Topological partition functions}

The partition function $Z_{\Z_n\text{a}}(M^{d+1}_\text{latt})$ of the
$\Z_n$-1-cocycle model can be calculated exactly, which is given by the number
of 1-cocycles $|Z^1(M^{d+1}_\text{Latt};\Z_n)|$, where $|S|$ denotes that number
elements in set $S$.  The number of 1-cochains is given by
$|H^1(M^{d+1}_\text{Latt};\Z_n)|$ times the number of 0-cochains whose derivatives
is non-zero.  The number of 0-cochains whose derivatives is non-zero is the
number of 0-cochains, $|C^0(M^{d+1}_\text{Latt};\Z_n)|$,  divide by
$|H^0(M^{d+1}_\text{Latt};\Z_n)|$.  Since $|C^0(M^{d+1}_\text{Latt};\Z_n)|=2^{N_v}$,
where $N_v$ is the number of vertices (the ``volume'' of space-time), we find
that the partition function is
\begin{align}
Z_{\Z_n\text{a}} (M^{d+1}_\text{Latt}) 
&= |Z^1(M^{d+1}_\text{Latt};\Z_n)|
\nonumber\\
&= |H^1(M^{d+1}_\text{Latt};\Z_n)| 
\frac{|C^0(M^{d+1}_\text{Latt};\Z_n)|}{|H^0(M^{d+1}_\text{Latt};\Z_n)|}
\nonumber\\
&= 2^{N_v}
 \frac{|H^1(M^{d+1}_\text{Latt};\Z_n)|}{|H^0(M^{d+1}_\text{Latt};\Z_n)|} .
\end{align}
According to \Ref{KW1458}, the topological information is given by the
volume-independent part of partition function, which is obtained by taking the
limit $N_v \to 0$:
\begin{align}
Z_{\Z_n\text{a}}^\text{top} (M^{d+1}) 
&=
 \frac{|H^1(M^{d+1};\Z_n)|}{|H^0(M^{d+1};\Z_n)|} ,
\end{align}
The volume-independent of partition function can be a topological invariant
\cite{KW1458} if the Euler number and the  Pontryagin number vanish:
$\chi(M^{d+1})=P(M^{d+1})=0$. Such topological invariant
characterizes the topological order realized by the model.  Since the $Z_2$
gauge theory will produce the same volume-independent partition function
$Z_{\Z_n\text{a}}^\text{top} (M^{d+1})$ in large system size and low energy limit,
this allows us to determine that the $\Z_n$-1-cocycle model realizes the
\emph{$Z_2$-topological order} \cite{RS9173,W9164} -- the topological order
described  UT $Z_n$-gauge theory.  

In 2+1D and for $n=2$, the $Z_2$-topological order has two bosonic topological
quasiparticles, $Z_2$-charge $e$ and $Z_2$-vortex $m$, and one fermionic
topological quasiparticles $f$ which is a bound state of $e$ and $m$.  In
higher dimensions, the $Z_n$-topological order has $n$ types of bosonic
point-like excitations -- the $Z_n$-charge $q=0,1,\cdots, n-1$.  It also has
$n$ types of $(d-2)$-dimensional brane-like excitations -- the $Z_n$-flux
$m=0,\frac{2\pi}{n},\cdots, \frac{2\pi(n-1)}{n}$.

We note that the volume-independent partition function on space-time $S^1\times
S^1\times S^{d-1} = T^2\times S^{d-1}$ is given by
\begin{align}
Z_{\Z_n\text{a}}^\text{top} (T^2\times S^{d-1}) 
&= n ,
\end{align}
Since the volume-independent partition function on $S^1\times M^{d}$ equal to
the ground state degeneracy (GSD) on space $M^{d}$:
\begin{align}
 \text{GSD}(M^{d}) = Z^\text{top}(S^1\times M^{d}),
\end{align}
we find that the GSD of our $\Z_n$-1-cocycle model on space $S^1\times S^{d-1}$
is given by GSD$(S^1\times S^{d-1}) = n$.

\begin{figure}[tb]
\begin{center}
\includegraphics[scale=0.8]{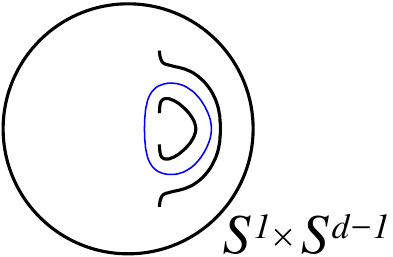} \end{center}
%4
\caption{ (Color online)
A particle-hole tunneling process is a process where we create a particle-hole
pair, move the particle around a non-contractible loop, and then annihilate the
particle and the hole.  The GSD on a $d$-dimensional space $S^1\times S^{d-1}$
is generated by the particle-hole tunneling process described by the blue loop.
Thus, each degenerate ground state correspond to a type of particle, and
GSD$(S^1\times S^{d-1}) =$ number of types of point-like excitations.
Similarly, GSD$(S^{k}\times S^{d-k}) =$ number of types of $(k-1)$-dimensional
excitations.
}
\label{parttype}
\end{figure}

It turns out that, \frmbox{for any topological order, GSD$(S^1\times S^{d-1})$
is always equal to the number of types of point-like topological excitations.}
Such a result can be understood by the particle-hole tunneling
process in Fig. \ref{parttype}.  Such a  particle tunneling process changes one
ground state to another degenerate one, and relate the number types of
point-like topological excitations to GSD$(S^1\times S^{d-1})$.  It is also
true that, \frmbox{for any topological order, GSD$(S^1\times S^{d-1})$ is
always equal to the number of types of $(d-2)$-dimensional brane-like
topological excitations \cite{WW1454}.} (The notion of types of topological
excitations, in particular high dimensional  topological excitations was
discussed in \Ref{KW1458}. It is very tricky to define the types of  high
dimensional  topological excitations.) This can be understood by a similar
brane tunneling process around $S^{d-1}$. 

In general, \frmbox{in $d$-dimensional space, the number of types of
$k-1$-dimensional brane-like excitations is equal to the number of types of
$d-k-1$-dimensional brane-like excitations, and they both equal to
GSD$(S^k\times S^{d-k})$.}

\subsubsection{Boundary effective theory}

Using the cocycle model, we can also easily study the properties of the
boundary.  Consider a space-time $M^{d+1}$ whose boundary is $N^d = \prt M^d$.
What is the low energy effective theory of our $\Z_n$-1-cocycle model on the
boundary $N^d$?  To be more concrete, what is the partition function for the
boundary effective theory?  Here, we propose that the partition function for
the boundary effective theory is simply given by
\begin{align}
\label{Zbndr1}
 Z_{\Z_n\text{a}}^\text{bndr}(N^d) = Z_{\Z_n\text{a}}(M^{d+1}) .
\end{align}
However, the above definition has a problem:
the same $N^d$ can be viewed as boundary of different space-time manifolds
$N^d=\prt M^{d+1} = \prt \t M^{d+1} $. In general
\begin{align}
  Z_{\Z_n\text{a}}(M^{d+1}) \neq Z_{\Z_n\text{a}}(\t M^{d+1})
\end{align}
so the above definition of $Z_{\Z_n\text{a}}^\text{bndr}(N^d)$ is not self
consistent.

In order for the definition to be self consistent, we require that
\begin{align}
  Z_{\Z_n\text{a}}(M^{d+1}) = Z_{\Z_n\text{a}}(\t M^{d+1})
\end{align}
for all $M^{d+1}$ and $\t M^{d+1}$ with $\prt M^{d+1} = \prt \t M^{d+1}$.  This
implies that the bulk model on $M^{d+1}$ has no topological order.  So the
boundary effective theory is well defined by itself iff the bulk theory on
$M^{d+1}$ has no topological order.  This is exactly the
gravitational-anomaly-free condition discussed in \Ref{W1313,KW1458,K1467}.

Since the bulk $\Z_n$-1-cocycle model has a non-trivial topological order, the
boundary effective theory is anomalous. This implies that the boundary
effective partition function $Z_{\Z_n\text{a}}^\text{bndr}(N^d)$ not only
depends on $N^d$, it also depends on how $N^d$ is extended to one higher
dimension, \ie depend on $M^{d+1}$.  The definition \eqn{Zbndr1} correctly
reflects such anomaly effect, and thus is a proper definition.
However, to stress the dependence on the extension, we rewrite
 \eqn{Zbndr1} as
\begin{align}
\label{Zbndr}
 Z_{\Z_n\text{a}}^\text{bndr}(N^d, M^{d+1}) = Z_{\Z_n\text{a}}(M^{d+1}) .
\end{align}

Even though the boundary partition function depend on the bulk extension, it is
still very useful in determine boundary low energy properties, such as if the
boundary gapped or not.  Let us first choose $N^d=S^d$ and choose its extension
to be $M^{d+1}=B^{d+1}$, where $B^{d+1}$ is a $d+1$-dimensional ball.  We find
the boundary partition function to be
\begin{align}
 Z_{\Z_n\text{a}}^\text{bndr}(S^d,B^{d+1})
&= \frac 1n n^{N_v^\text{bndr}} n^{N_v^\text{blk}} .
\end{align}
where $N_v^\text{bndr}$ is the number of vertices on the boundary $S^d$ and
$N_v^\text{blk}$ is the number of vertices \emph{inside} the ball $B^{d+1}$.
The partition function only depend on the ``volume'' of the
boundary and does not depend on the shape of the
boundary. This implies that the boundary theory is gapped.

Next, let us choose $N^d=S^1_t\times S^{d-1}$, where we use $S^1_t$ to
represent the closed time direction. We choose its extension to be
$M^{d+1}=S^1_t\times B^{d}$.  We find the boundary partition function to be
\begin{align}
 Z_{\Z_n\text{a}}^\text{bndr}(S^1_t\times S^{d-1} ,S^1_t\times
B^{d})
= n^{N_v^\text{bndr}} n^{N_v^\text{blk}} .
\end{align}
We see that the volume-independent boundary partition function
is
\begin{align}
 Z_{\Z_n\text{a}}^\text{bndr,top}(S^1_t\times S^{d-1} ,S^1_t\times
B^{d})
&= 1.
\end{align}
This implies that the gapped boundary has no ground state degeneracy (for the
boundary $S^{d-1}$). For example there is no symmetry breaking.

To see if the boundary carries an anomalous topological order, let us choose
$N^d=S^1_t\times S^{k+1}\times S^{d-2-k}$
and choose its extension to be $M^d=S^1_t\times S^{k+1}\times B^{d-1-k}$.
Since the tunneling process of $k$-dimensional brane-like topological
excitations around $S^{k+1}$ on the boundary corresponds to a non-contractible
loop in the bulk $S^1_t\times S^{k+1}\times B^{d-1-k}$, the tunneling process will generate a map between
different degenerate ground states.  
In contrast, the brane tunnel process 
around $S^{d-2-k}$ on the boundary corresponds to a
contractible ``loop'' in the bulk $S^1_t\times S^{k+1}\times B^{d-1-k}$ and does not generate non-trivial map
between degenerate ground states.  
Therefore, similar to the
bulk case, $Z_{\Z_n\text{a}}^\text{bndr}(S^1_t\times S^{k+1}\times S^{d-2-k},
S^1_t\times S^{k+1}\times B^{d-1-k})$ can tell us the number of types of
$k$-dimensional brane-like topological excitations on the boundary.

For our $\Z_n$-1-cocycle model, we found that the volume-independent partition
function to be
\begin{align}
&\ \ \ \
Z_{\Z_n\text{a}}^\text{bndr,top}(S^1_t\times S^{k+1}\times S^{d-2-k},
S^1_t\times S^{k+1}\times B^{d-1-k})
\nonumber\\
&=
\frac{|H^1(S^1_t\times S^{k+1}\times B^{d-1-k};\Z_n)|}{|H^0(S^1_t\times S^{k+1}\times B^{d-1-k};\Z_n)|} =
\begin{cases}
 1, & k>0\\
 n, & k=0\\ 
\end{cases}
\end{align}
Thus the boundary theory contains $n$ types of point-like excitations, and
no non-trivial brane-like excitations of dimensions greater then 0.

The $n$ types of point-like topological excitations on the boundary contain a
trivial type and $n-1$ non-trivial type.  When $n>1$, the existence  of
non-trivial topological excitations on the boundary implies that the boundary
carries a non-trivial topological order (which is anomalous).
This agrees with the previous known result \cite{KK1251,KW1458}.

A given bulk model can have many types of boundaries.  For our $\Z_n$-1-cocycle
model, the bulk contain $n$ types of point-like topological excitations and $n$
types of $(d-2)$-dimensional brane-like topological excitations.  One type of the
boundary is formed by the brane condensation.  Such a boundary has $n$ types of
point-like topological excitations only.  Another type of boundary is formed by
the particle condensation.  Such a boundary has $n$ types of $(d-2)$-dimensional
brane-like topological excitations only.  We see that our boundary of
$\Z_n$-1-cocycle model is the first type induced by the condensation of branes.
We will call such boundary as ``free boundary'' since the 1-cocycle field has
a free boundary condition on the boundary.

To realize the second type of the boundary, we need to use the fixed boundary
condition by setting the 1-cocycle field to be $a^{\Z_n}_{ij}=0$ on the
boundary.  Again $Z_{\Z_n\text{a}}^\text{bndr}(S^1_t\times S^{k+1}\times
S^{d-2-k}, S^1_t\times S^{k+1}\times B^{d-1-k})$ can tell us the number of
types of $k$-dimensional brane-like topological excitations on the boundary.
To compute such partition function, we notice that when $k<d-2$, the 1-cocycle
$a^{\Z_n}$ can be written as $a^{\Z_n}=\dd g^{\Z_n}$ is a $\Z_n$-valued
0-cochain which vanishes on the boundary.  The correspondence between
$a^{\Z_n}_{ij}$ and $g^{\Z_n}_i$ is one-to-one.  This is because even when
$k=0$, $\oint_{S^{k+1}} a^{\Z_n}=0$ since we have fixed $a^{\Z_n}=0$ on the
boundary.  Thus
\begin{align}
&\ \ \ \ Z_{\Z_n\text{a}}^\text{bndr}(S^1_t\times S^{k+1}\times
S^{d-2-k}, S^1_t\times S^{k+1}\times B^{d-1-k}) 
\nonumber\\
&= n^{N_v^\text{blk}}.
\end{align}
The volume-independent partition function is
\begin{align}
&\ \ \ \ Z_{\Z_n\text{a}}^\text{bndr,top}(S^1_t\times S^{k+1}\times
S^{d-2-k}, S^1_t\times S^{k+1}\times B^{d-1-k})
\nonumber\\
&=1,  \ \ \ \text{ for } k<d-2.
\end{align}
Thus, there is no non-trivial $k$-dimensional brane-like excitations on the
boundary for $k< d-2$.  There is no  non-trivial point-like excitations on the
boundary which is the $k=0$ case included above.
When $k=d-2$, $S^{d-2-k}=S^0$ is a set of two points.
In this case, the boundary contains two disconnected pieces.
We may set the 0-cochain field $g^{\Z_n}=0$ on one piece. But we need to set
$g^{\Z_n}=$ const. on the other piece.
We find that
\begin{align}
&\ \ \ \ Z_{\Z_n\text{a}}^\text{bndr}(S^1_t\times S^{d-1}\times
S^{0}, S^1_t\times S^{d-1}\times B^{1}) 
\nonumber\\
&= n n^{N_v^\text{blk}}.
\end{align}
Or the volume-independent one
\begin{align}
&\ \ \ \ Z_{\Z_n\text{a}}^\text{bndr,top}(S^1_t\times S^{d-1}\times
S^{0}, S^1_t\times S^{d-1}\times B^{1}) 
\nonumber\\
&= n .
\end{align}
There is $n$ types of $(d-2)$-dimensional brane-like excitations on the
boundary.  The $a^{\Z_n}=0$ boundary gives us the second type of boundary
formed by condensing the point-like excitations.

\subsection{Twisted 2+1D $\Z_n$-1-cocycle model and emergent
Dijkgraaf-Witten theory}

\subsubsection{Model construction}

To construct another local bosonic model that realize a different topological
order, we may choose $L_{ijkl}$ to be
\begin{align}
 L_{ijkl} &=  +\infty, &\text{ if } &(\dd a^{\Z_n}) \neq 0,
\nonumber\\
 L_{ijkl} &= -\ii k\frac{2\pi}{n} (a^{\Z_n}\Bs_n a^{\Z_n})(i,j,k,l), &\text{ if } &(\dd a^{\Z_n})= 0.
\end{align}
Here we have use Bockstein homomorphism  $\Bs_n: H^m(M^d;\Z_n) \to
H^{m+1}(M^d;\Z_n)$,
\begin{align}
\label{Bsn}
\Bs_n x&\se{n} \frac1n \dd x , 
\nonumber\\
x &\in  H^m(M^d;\Z_n),\ \
\Bs_n x \in  H^{m+1}(M^d;\Z_n)
.
\end{align}
To understand the Bockstein homomorphism, we note that $x$ in the above is a
cocycle with $\Z_n$. If we view it as a cochain with integer coefficient $\Z$,
then $\dd x$ is an cochain whose values are always multiples of $n$. Thus
$\frac1n \dd x$ is a valid cochain with integer coefficient. In fact, it is
$(m+1)$-cocycle with integer coefficient.  After a mod $n$ reduction, $\frac1n
\dd x$ mod $n$ becomes a  $(m+1)$-cocycle with $\Z_n$ coefficient.   This is
why $\Bs_n$ is a map from $H^m(M^d;\Z_n)$ to $H^{m+1}(M^d;\Z_n)$.
Therefore $\Bs_n a^{\Z_n}$ is a 2-cocycle
and $a^{\Z_n}\Bs_n a^{\Z_n}$ is a 3-cocycle. Here, we use such a 3-cocycle
to construct the action $L_{ijkl}$

The total action amplitude $\ee^{-S(\{a^{\Z_n}_{ij}\}) }$ is given by
\begin{align}
 \ee^{-S(\{a^{\Z_n}_{ij}\}) } = \ee^{\ii k\frac{2\pi}{n} \int_{M^3_\text{Latt}}
a^{\Z_n}\Bs_n a^{\Z_n} } 
\end{align}
for $\dd a^{\Z_n}=0$, and $\ee^{-S(\{a^{\Z_n}_{ij}\}) }=0$ for $\dd a^{\Z_n}
\neq 0$.  The partition function is given by
\begin{align}
\label{ZaBa}
 Z_{k;a\Bs a\Z_n} (M^3_\text{Latt}) = \sum_{\{a^{\Z_n}_{ij}\}, \dd a^{\Z_n}=0} \ee^{\ii k\frac{2\pi}{n} \int_{M^3_\text{Latt}} a^{\Z_n} \Bs_n a^{\Z_n}}.
\end{align}
Such a partition function defines the twisted 2+1D $\Z_n$-1-cocycle model.

The volume-independent part of partition function \eqn{ZaBa} is given by
\begin{align}
 Z_{k;a\Bs a\Z_n}^\text{top} (M^3)
= \frac{\sum_{a^{\Z_n}\in H^1(M^3;\Z_n)}
  \ee^{\ii k\frac{2\pi}{n} \int_{M^3} a^{\Z_n}\Bs_n a^{\Z_n}}}{|H^0(M^3;\Z_n)|}
\end{align}
Since the Euler number on odd-dimensional closed manifolds vanishes, the above
volume-independent partition function is a topological invariant.

\subsubsection{Topological partition functions}
\label{ZnDW3Ztop}

In this section, we are going to calculate some topological invariants.
On $M^3=S^3$, $S^1\times S^2$, or $T^3=S^1\times S^1\times S^1$, 
$\Bs_n a^{\Z_n}=0$ and the topological term  $ k\frac{2\pi}{n} \int_{M^3}
a^{\Z_n}\Bs_n a^{\Z_n}$ vanishes. We find
\begin{align}
 Z_{k;a\Bs a\Z_n}^\text{top} (S^3) &= \frac 1n ,
\nonumber\\
 Z_{k;a\Bs a\Z_n}^\text{top} (S^1\times S^2) &= 1 ,
\nonumber\\
 Z_{k;a\Bs a\Z_n}^\text{top} (T^3) &= n^2 .
\end{align}
From  $Z_{1;a\Bs a\Z_n}^\text{top} (M^2\times S^1) $, we can determine
the ground state degeneracy (GSD) on $M^2$:
\begin{align}
 Z_{1;a\Bs a\Z_n}^\text{top} (M^2\times S^1)=\text{GSD}_{1;a\Bs a\Z_n}(M^2).
\end{align}
Using
\begin{align}
Z_{1;a\Bs a\Z_n}^\text{top} (S^2\times S^1) &= 1,
\nonumber\\
Z_{1;a\Bs a\Z_n}^\text{top} (T^2\times S^1) &= n^2,
\end{align}
we find that the GSD on a
sphere $S^2$ is $1$ and the GSD on a torus $T^2=S^1\times S^1$ is $n^2$.

To obtain the  topological invariant that see the  topological term,
we put the system on the lens space $L^3(p)$ (see Appendix \ref{Lpq}).
We find from
\begin{align}
\label{LqpZ}
 H_1(L^3(p), \Z) &= \Z_{p},  
\nonumber\\
 H_2(L^3(p), \Z) &= 0, 
\nonumber\\
 H_3(L^3(p), \Z) &= \Z .
\end{align}
that (using \eqn{ucfH})
\begin{align}
\label{LpqH}
 H^1(L^3(p), \Z_n) &= \Z_{\<p,n\>}=\{a\},  
\nonumber\\
 H^2(L^3(p), \Z_n) &= \Z_{\<p,n\>}=\{b\},  
\nonumber\\
 H^3(L^3(p), \Z_n) &= \Z_n=\{c\},  
\end{align}
where we have also listed the generators $\{a,b,c\}$.  Here, $\<l,m\>$ is the
greatest common divisor of $l$ and $m$, and $\<0,m\>\equiv m$. In Appendix
\ref{Lpq}, we have computed the cohomology ring $H^*(L^3(p), \Z_n)$ (see
\eqn{ringLpq}):
\begin{align}
\label{Lpqcup}
a^2 = \frac{n^2p(p-1)}{2\<p,n\>^2} b, \ \ \ \ 
ab= \frac{n}{\<p,n\>} c ,\ \ \ \
b^2=ac=0.
\end{align}
We have also computed the Bockstein homomorphism
\begin{align}
\label{LpqB}
\Bs_n a=\frac{p}{\<p,n\>}b.
\end{align}
We can parametrize $a^{\Z_n}$ as
\begin{align}
 a^{\Z_n} = \al a, \ \ \ \al \in \Z_{\<n,p\>}.
\end{align}
and find that
\begin{align}
 Z_{k;a\Bs a\Z_n}^\text{top} (L^3(p))
= \frac 1n\sum_{\al=0}^{\<n,p\>-1}
  \ee^{\ii 2\pi
\al^2 \frac{kp}{ \<n,p\>^2} } .
\end{align}
We find the above topological invariant is identical to the topological
invariant of 2+1D $Z_n$ Dijkgraaf-Witten theory on lens space $L^3(p)$ for any
$p$ (see \eqn{ZDWLp}). In fact, one can show that the $\Z_n$-1-cocycle model
realize the 2+1D $Z_n$ Dijkgraaf-Witten theory \cite{DW9093} (see discussions
below \eqn{Lngco}).  In other words, the above topological invariant is the
topological invariant of a $Z_n$-gauge theory twisted by a quantized
topological term \cite{HW1267} $ k\frac{2\pi}{n} \int_{M^3} a^{\Z_n}\Bs_n
a^{\Z_n}$. The quantized topological term correspond to a group-cocycle in
$\cH^3(Z_n,\RZ)=\Z_n$.  It is the simplest Dijkgraaf-Witten theory.  Such a
Dijkgraaf-Witten theory can be obtained by gauging the $Z_n$-symmetry of a
$Z_n$-SPT state \cite{LG1209}.  When $k=0$, our model realizes the $Z_n$
topological order described by UT $Z_n$ gauge theory.  For $(n,k)=(2,1)$,
our model realizes the double-semion topological order
\cite{FNS0428,LW0510,LG1209}.

We like to remark that the twisted 2+1D $\Z_n$-1-cocycle model and
Dijkgraaf-Witten theory are different.  Dijkgraaf-Witten theory is a gauge
theory where two $a^{\Z_n}_{ij}$ configurations differ by a $Z_n$ gauge
transformation are regard as the same  configuration.  In other words, two
$a^{\Z_n}_{ij}$ configurations differ by a coboundary are regard as the same
configuration.  Thus the Dijkgraaf-Witten $Z_n$-gauge theory may be called
twisted $\Z_n$-1-cohomology model.  In our twisted 2+1D $\Z_n$-1-cocycle model,
different $a^{\Z_n}_{ij}$ configurations are always different with no gauge
redundancy.  So the cocycle model is not a gauge theory but a local bosonic
system. However, the cocycle model has an emergent gauge theory at low energies
which is described by Dijkgraaf-Witten theory.

\subsection{Twisted 3+1D $\Z_n\oplus \Z_n$-1-cocycle model and emergent
Dijkgraaf-Witten theory}

\subsubsection{Model construction}

In this section, we like to design a 3+1D local bosonic model that realizes the
Dijkgraaf-Witten twisted gauge theory at low energies.  Since
$\cH^4(Z_n,\RZ)=0$, there is no $Z_n$ Dijkgraaf-Witten gauge theory in 3+1D.
So here we try to realize the  $Z_n\times Z_n$ Dijkgraaf-Witten gauge theory.
Such theory exists since $\cH^4(Z_n\times Z_n,\RZ)=\Z_n\oplus \Z_n$.

To realize  $Z_n\times Z_n$ gauge theory, we construct a $\Z_n\oplus
\Z_n$-1-cocycle theory on 3+1D space-time lattice.  The local degrees of
freedom of the model correspond to two $1$-cochains $a_1^{\Z_n}, a_2^{\Z_n}\in
C^2(M^4_\text{Latt};\Z_n)$ (\ie the local degrees of freedom are described by
$\Z_n\oplus \Z_n$ on each 1-simplex).  The partition function on a oriented
space-time $M^4_\text{Latt}$ is given by \cite{WGW1489,YG150805689}
\begin{align}
&\ \ \ \ Z_{k_1k_2;aa'\Bs a'\Z_n} (M^4_\text{Latt}) =
\\
& \sum_{\{a^{\Z_n}_{I;ij}\}, \dd a_I^{\Z_n}=0} 
\ee^{\ii \frac{2\pi}{n} \int_{M^4_\text{Latt}} 
k_1 a_1^{\Z_n}a_2^{\Z_n} \Bs_n a_2^{\Z_n}
+k_2 a_2^{\Z_n}a_1^{\Z_n} \Bs_n a_1^{\Z_n}
},
\nonumber 
\end{align}
where $k_1,k_2=0,1,\cdots,n-1$,
We have assumed that the configuration with $\dd a_I^{\Z_n} \neq 0$, $I=1,2$,
have infinite energy and do not contribute to the partition function.  The term
$ \frac 1n\int_{M^4_\text{Latt}} k_1 a_1^{\Z_n}a_2^{\Z_n} \Bs_n a_2^{\Z_n} +k_2
a_2^{\Z_n}a_1^{\Z_n} \Bs_n a_1^{\Z_n} $ corresponds to a cocycle $(k_1,k_2) \in
\cH^4(Z_n\times Z_n,\RZ)=\Z_n\oplus\Z_n$.  

There are other possible choices of the action amplitude, such as
\begin{align}
 \ee^{\ii k\frac{2\pi}{n} \int_{M^4_\text{Latt}} \Bs_n a_1^{\Z_n} \Bs_n a_2^{\Z_n}}
\end{align}
But
\begin{align}
\int_{M^4_\text{Latt}} \Bs_n a_1^{\Z_n} \Bs_n a_2^{\Z_n} &=  \int_{M^4_\text{Latt}} \frac1n \dd  a_1^{\Z_n} \frac1n \dd  a_2^{\Z_n} 
\nonumber\\
&= 0 ,
\end{align}
if $M^4_\text{Latt}$ is orientable.
So such a term always vanishes.  Yet another possible choice is
$\int_{M^4_\text{Latt}} a_1^{\Z_n} (
a_2^{\Z_n})^3$ 
But when $n=2$, it is the same as
$\int_{M^4_\text{Latt}} a_1^{\Z_n}a_2^{\Z_n} \Bs_n a_2^{\Z_n}$, and when $n=$
odd, it vanishes.  So here we do not discuss it further.

\subsubsection{Topological partition functions}

When $k_1,k_2=0$, the partition function is given by the square of the number
of 1-cocycles, $|Z^1(M^4_\text{Latt};\Z_n)|^2$.
$|Z^1(M^4_\text{Latt};\Z_n)|$
is $|H^1(M^4_\text{Latt};\Z_n)|$ times the
number of 0-cochains whose derivatives is non-zero.  The number of 0-cochains
whose derivatives is non-zero is the number of 0-cochains
($|C^0(M^4_\text{Latt};\Z_n)|=n^{N_v}$) divide by $|H^0(M^4_\text{Latt};\Z_n)|$.
Thus the partition function is
\begin{align}
&\ \ \ \ Z_{0,0;aa'\Bs a'\Z_n} (M^4_\text{Latt}) =  |Z^1(M^4_\text{Latt};\Z_n)|^2
\nonumber\\
&= |H^1(M^4_\text{Latt};\Z_n)|^2 
\frac{|C^0(M^4_\text{Latt};\Z_n)|^2}{|H^0(M^4_\text{Latt};\Z_n)|^2} 
\nonumber\\
&= n^{N_v}
 \frac{|H^1(M^4_\text{Latt};\Z_n)|^2}{|H^0(M^4_\text{Latt};\Z_n)|^2} .
\end{align}
The volume-independent topological partition function is given by
\begin{align}
Z_{0,0;aa'\Bs a'\Z_n} (M^4_\text{Latt})
= 
\frac{|H^1(M^4_\text{Latt};\Z_n)|^2}{|H^0(M^4_\text{Latt};\Z_n)|^2}
\end{align}

When $k_1,k_2\neq 0$,
the volume-independent topological partition function is given by
\begin{align}
&\ \ \ \
Z_{k_1k_2;aa'\Bs a'\Z_n} (M^4_\text{Latt}) =
\\ & 
\sum_{a_I^{\Z_n} \in H^1(M^4;\Z_n)}
\frac{
\ee^{\ii \frac{2\pi}{n} \int_{M^4_\text{Latt}} 
k_1 a_1^{\Z_n}a_2^{\Z_n} \Bs_n a_2^{\Z_n}
+k_2 a_2^{\Z_n}a_1^{\Z_n} \Bs_n a_1^{\Z_n}}}
 {|H^0(M^4;\Z_n)|^2} ,
\nonumber 
\end{align}
where $|H^1(M^4_\text{Latt};\Z_n)|^2$ is replaced by the summation of phase
factors.

Now, let us compute $Z_{k_1k_2;aa'\Bs a'\Z_n} (M^4)$ on several $M^4$.  On
$M^4=S^1\times S^1\times S^1\times S^1 =T^4$ or $M^4=S^2\times S^1\times S^1
=S^2\times T^2$, $\Bs_n a_I^{\Z_n}=0$. Thus $Z_{k_1k_2;aa'\Bs a'\Z_n}
(M^4)=Z_{0,0;aa'\Bs a'\Z_n} (M^4)$ on those manifolds.
Using
\begin{align}
 H^1(T^4;\Z_n)=4\Z_n, \ \ \ \
 H^1(T^2\times S^2;\Z_n)=\Z_n^{\oplus 2},
\end{align}
we find that (see Table \ref{topinv})
\begin{align}
 Z_{k_1k_2;aa'\Bs a'\Z_n}(T^4) &= n^6,
\nonumber\\
 Z_{k_1k_2;aa'\Bs a'\Z_n}(S^2\times T^2) & =n^2.
\end{align}

On $M^4=S^1\times L^3(p)$, from
\begin{align}
 H_1(L^3(p), \Z) = \Z_{p},  
\
 H_2(L^3(p), \Z) = 0, 
\
 H_3(L^3(p), \Z) = \Z .
\end{align}
we find that (using \eqn{kunnH})
\begin{align}
 H_1(S^1\times L^3(p), \Z) &= \Z\oplus\Z_p,  
\nonumber\\
 H_2(S^1\times L^3(p), \Z) &= \Z_p,  
\nonumber\\
 H_3(S^1\times L^3(p), \Z) &= \Z,
\nonumber\\
 H_4(S^1\times L^3(p), \Z) &= \Z.
\end{align}
This allows us to obtain (using \eqn{ucfH})
\begin{align}
\label{S1LpqH}
 H^1(S^1\times L^3(p), \Z_n) &= \Z_n\oplus\Z_{\<p,n\>}=\{a_1,a\},  
\nonumber\\
 H^2(S^1\times L^3(p), \Z_n) &= \Z_{\<p,n\>}\oplus\Z_{\<p,n\>}=\{a_1a,b\},  
\nonumber\\
 H^3(S^1\times L^3(p), \Z_n) &= \Z_n\oplus \Z_{\<p,n\>}=\{c,a_1b\},  
\nonumber\\
 H^4(S^1\times L^3(p), \Z_n) &= \Z_n =\{a_1c\}.
\end{align}
where we have also listed the generators, where $a_1$ comes from $S^1$ and
$a,b,c$ from $L^3(p)$.  Here, $\<l,m\>$ is the greatest common divisor of $l$
and $m$, and $\<0,m\>\equiv m$

In Appendix \ref{Lpq}, we have computed 
the cohomology ring $H^*(S^1\times L^3(p), \Z_n)$
 (see \eqn{ringLpq}):
\begin{align}
\label{S1Lpqcup}
a_1^2=0,\ \  a^2 = \frac{n^2p(p-1)}{2\<p,n\>^2} b,  \ \ 
ab= \frac{n}{\<p,n\>} c ,\ \ 
b^2=ac=0.
\end{align}
We have also computed the Bockstein homomorphism
\begin{align}
\label{S1LpqB}
\Bs_n a=\frac{p}{\<p,n\>}b,\ \ \ \ \ \Bs_na_1=0. 
\end{align}

We see that for $\<n,p\>=1$, $a=b=0$, and thus $\Bs_n a_I^{\Z_n}=0$. Therefore
$\int_{M^4_\text{Latt}} k_1 a_1^{\Z_n}a_2^{\Z_n} \Bs_n a_2^{\Z_n} +k_2
a_2^{\Z_n}a_1^{\Z_n} \Bs_n a_1^{\Z_n}=0$.  So $Z_{k_1k_2;aa\bt
a\Z_n}(S^1\times L^3(p))=1$.

For $\<n,p\>\neq 1$,  
we can parametrize $a_I^{\Z_n}$ as
\begin{align}
 a_I^{\Z_n} = \al_I a_1 +\t \al_I a, \ \ \
\al_I \in \Z_n,\ 
\t \al_I \in \Z_{\<n,p\>}.
\end{align}
Using \eqn{S1LpqH}, \eqn{S1Lpqcup}, and \eqn{S1LpqB}, we find that
\begin{align}
\label{ZZnZna}
&
Z_{k_1k_2;aa'\Bs a'\Z_n} (S^1\times L^3(p)) 
 =\frac{1}{n^2}
\sum_{\al_{1,2}\in \Z_n,\t \al_{1,2}\in Z_{\<n,p\>}}
\nonumber\\
&\ \ \ \ee^{\ii \frac{2\pi p}{\<p,n\>^2}[ k_1 (\al_1 \t \al_2^2-\t \al_1\al_2\t \al_2) 
+k_2 (\al_2 \t \al_1^2 -\t \al_2\al_1\t \al_1)]  }
\nonumber\\
&\ \ \ = s^2
\sum_{\t \al_{1,2}=0}^{m-1}
\del_m(k_1\t \al_2^2-k_2\t\al_1\t\al_2)\times
\nonumber\\
&
\ \ \ \ \ \ \ \ \ \ 
\ \ \ \ \ \ \ \ \ \ 
\del_m(k_2\t \al_1^2-k_1\t\al_1\t\al_2),
\nonumber\\
& s= \< \frac{p}{\< n,p\>},\<n,p\> \>,\ \ \ m=\<n,p\>/s.
\end{align}
When $p$ has no repeated prime factor, the above sum has a simple expression:
\begin{align}
Z_{k_1k_2;aa'\Bs a'\Z_n} (S^1\times L^3(p)) & =
\<n,p\>\<n,p,k_1,k_2\> 
\nonumber\\
&=s^2 m\<m,k_1,k_2\>
\end{align}

On $M^4=F^4 \equiv (S^1\times S^3)\# (S^1\times S^3)\#\C P^2\# \overline{\C
P}^2 $, we note that the cup product of $1$-cocycles are always zero (see
Appendix \ref{F4ring}).  Thus $Z_{k_1k_2;aa'\Bs a'\Z_n} (F^4)=Z_{0,0;aa'\bt
a'\Z_n} (F^4)=n^2$. 

\subsubsection{Dimension reduction}
\label{DW4Dred}

Last, let us consider $M^4=M^3\times S^1$, where $M^4$ and $M^3$ are assumed to
be closed manifolds.  We write $a_I^{\Z_n}$ as
\begin{align}
 a_I^{\Z_n}=a_{I,M^3}^{\Z_n} + a_{I,S^1}^{\Z_n},
\end{align}
where $a_{I,M^3}^{\Z_n}$ lives on $M^3$ and $ a_{I,S^1}^{\Z_n}$ on $S^1$.  We
also fix $\oint_{S^1} a_{I,S^1}^{\Z_n}=\al_I \in \Z$.  The partition function
now has a form
\begin{align}
&\ \ \ \
Z_{k_1k_2;aa'\Bs a'\Z_n} (M^3\times S^1,\al_1,\al_2) =
\frac{1 } {|H^0(M^3;\Z_n)|^2} \times
\\ & 
\sum_{a_{I,M^3}^{\Z_n} \in H^1(M^3;\Z_n)}
\ee^{ \ii \frac{2\pi}{n} \int_{M^3} 
(k_1\al_2 - k_2\al_1)a_{1,M^3}^{\Z_n} \Bs_n a_{2,M^3}^{\Z_n}
}\times
\nonumber\\
&\ \ \ \ \ \ \ \ \ \ \ \ 
\ee^{ \ii \frac{2\pi}{n} \int_{M^3} 
k_1\al_1 a_{2,M^3}^{\Z_n} \Bs_n a_{2,M^3}^{\Z_n}
+k_2\al_2 a_{1,M^3}^{\Z_n} \Bs_n a_{1,M^3}^{\Z_n}
} .
\nonumber 
\end{align}

In fact, $\al_I$ in the above happen to label the different sectors. We find
the topological theory in each sector from the partition function
$Z_{k_1k_2;aa'\Bs a'\Z_n} (M^3\times S^1,\al_1,\al_2)$. As we can see that
they are 2+1D Dijkgraaf-Witten theories.

Since the Dijkgraaf-Witten theories can be viewed as gauged SPT states
\cite{LG1209}, the dimension reduction of the Dijkgraaf-Witten theories implies
a similar  the dimension reduction of SPT states: If we compact a 3+1D
$Z^{(1)}_n\times Z^{(2)}_n$-SPT state to 2+1D via a circle $S^1$, and add a
symmetry twist around $S^1$ described by  $\ee^{\ii 2\pi \al_1/n}$ for the
$Z^{(1)}_n$ and $\ee^{\ii 2\pi \al_2/n}$ for the $Z^{(2)}_n$, then the
resulting 2+1D SPT state is a stacking of a $Z^{(1)}_n$-SPT state labeled by
$k_2\al_2 \in \cH^3(Z^{(1)}_n,\RZ)=\Z_n$, a $Z^{(2)}_n$-SPT state labeled by
$k_1\al_1 \in \cH^3(Z^{(2)}_n,\RZ)=\Z_n$, and a $Z^{(1)}_n\times
Z^{(2)}_n$-SPT state labeled by $k_1\al_2 - k_2\al_1 \in \cH^3(Z^{(1)}_n\times
Z^{(2)}_n,\RZ)$ \cite{WGW1489,W1477}.

This implies that the symmetry-twist defect line (twisted by $\ee^{\ii 2\pi
\al_1/n}$ for the $Z^{(1)}_n$ and $\ee^{\ii 2\pi \al_2/n}$ for the $Z^{(2)}_n$)
in the  3+1D $Z^{(1)}_n\times Z^{(2)}_n$-SPT state (labeled by $(k_1,k_2)\in
\cH^4(Z^{(1)}_n\times Z^{(2)}_n,\RZ)=\Z_n\oplus\Z_n$), will carry gapless
1+1D excitations along the symmetry-twist defect line described by the boundary
of the $k_2\al_2$th $Z^{(1)}_n$-SPT state, the $k_1\al_1$th $Z^{(2)}_n$-SPT
state, and  the $(k_1\al_2 - k_2\al_1 )$th $Z^{(1)}_n\times Z^{(2)}_n$-SPT
state, provided that the $Z^{(1)}_n\times Z^{(2)}_n$ symmetry is not broken.
This result generalized the one in \Ref{W1447}.

\subsection{Twisted 3+1D $\Z_n$-2-cocycle model 
and emergence of fermions
}
\label{ZnbSec}

\subsubsection{Model construction}

In this section, we will study $\Z_n$-2-cocycle theory on 3+1D space-time
lattice. The local degrees of freedom of the model correspond to $2$-cochains
$b^{\Z_n} \in C^2(M^4_\text{Latt};\Z_n)$ (\ie the local degrees of freedom are
described by $\Z_n$ on each 2-simplex).  The partition function is given by,
for $k=0,1,\cdots,n-1$ \cite{YG14102594},
\begin{align}
\label{tZnbMdl}
 Z_{k;b^2\Z_n} (M^4_\text{Latt}) = \sum_{\{b^{\Z_n}_{ij}\}, \dd b^{\Z_n}=0} 
\ee^{\ii k\frac{2\pi}{n} \int_{M^4_\text{Latt}} (b^{\Z_n})^2},
\end{align}
(\ie the configuration with $\dd b^{\Z_n} \neq 0$ have infinite energy.) Note
that the source (or ``charge'') of the 2-cocycle field $b$ is a $\Z_n$ string.
When $k=0$, it describes a $\Z_n$-2-cocycle theory.  When $k\neq 0$, it
describes a twisted $\Z_n$-2-cocycle theory.

When $k=0$, the partition function is given by the number of 2-cocycles
$|Z^2(M^4_\text{Latt};\Z_n)|$, which is $|H^2(M^4_\text{Latt};\Z_n)|$ times the
number of 1-cochains whose derivatives is non-zero.  The number of 1-cochains
whose derivatives is non-zero is the number of 1-cochains
($|C^1(M^4_\text{Latt};\Z_n)|=n^{N_e}$) divide by $|H^1(M^4_\text{Latt};\Z_n)|$
and by the number of number of 0-cochains whose derivatives is non-zero.  The
number of 0-cochains whose derivatives is non-zero is the number of 0-cochains
($|C^0(M^4_\text{Latt};\Z_n)|=n^{N_v}$) divide by
$|H^0(M^4_\text{Latt};\Z_n)|$.  Thus the partition function is
\begin{align}
&\ \ \ \ Z_{0;b^2\Z_n} (M^4_\text{Latt}) =  |Z^2(M^4_\text{Latt};\Z_n)|
\nonumber\\
&= |H^2(M^4_\text{Latt};\Z_n)| 
\frac{|C^1(M^4_\text{Latt};\Z_n)|}{|H^1(M^4_\text{Latt};\Z_n)|} \frac{|H^0(M^4_\text{Latt};\Z_n)|}{|C^0(M^4_\text{Latt};\Z_n)|}
\nonumber\\
&= n^{N_e-N_v}
 \frac{|H^2(M^4_\text{Latt};\Z_n)||H^0(M^4_\text{Latt};\Z_n)|}{|H^1(M^4_\text{Latt};\Z_n)|} .
\end{align}
The volume-independent topological partition function is given by
\begin{align}
 Z_{0;b^2\Z_n}^\text{top} (M^4) = 
 \frac{|H^2(M^4;\Z_n)||H^0(M^4;\Z_n)|}{|H^1(M^4;\Z_n)|}
\end{align}
When $k\neq 0$,
The volume-independent topological partition function is given by
\begin{align}
&\ \ \ \
 Z_{k;b^2\Z_n}^\text{top} (M^4) 
\\ & = 
 \frac{|H^0(M^4;\Z_n)|}{|H^1(M^4;\Z_n)|} 
\sum_{b^{\Z_n} \in H^2(M^4;\Z_n)}
\ee^{\ii k\frac{2\pi}{n} \int_{M^4} (b^{\Z_n})^2 }
\nonumber 
\end{align}
where $\sum_{b^{\Z_n} \in H^2(M^4;\Z_n)} \ee^{\ii k\frac{2\pi}{n} \int_{M^4}
(b^{\Z_n})^2 }$ replaces $|H^2(M^4;\Z_n)|$.

\subsubsection{Topological partition functions}

Now, let us compute topological invariants (see Table \ref{topinv}).  On
$M^4=T^4$, the cohomology ring $H^*(T^4;\Z_n)$ is generated by $a_I,\
I=1,2,3,4$, where $a_I \in H^1(T^4;\Z_n)=4\Z_n$.  Using the cohomology ring
\eqn{ringT4} in Appendix \ref{crings}, we can parametrize $b^{\Z_n}$ as
\begin{align}
 b^{\Z_n} =\bt_{IJ} a_Ia_J,\ \ \ \bt_{IJ}=-\bt_{JI} 
\in \Z_n.
\end{align}
Thus
\begin{align}
&\ \ \ \
 Z_{k;b^2\Z_n}^\text{top} (T^4) 
\\ & = 
 \frac{1}{n^3} 
\sum_{\bt_{IJ} \in \Z_n}
\ee^{\ii k\frac{2\pi}{n} 
(
\bt_{12}\bt_{34} -\bt_{13}\bt_{24}+ \bt_{14}\bt_{23}
)
}
\nonumber 
\end{align}
Using $\sum_{\bt_1,\bt_2\in \Z_n} \ee^{\ii k\frac{2\pi}{n} 2\bt_1\bt_2}=
\<2k,n\> n$,
we find that
\begin{align}
 Z_{k;b^2\Z_n}^\text{top} (T^4)=\<2k,n\>^3.
\end{align}

On $M^4=S^2\times T^2$, the cohomology ring $H^*(T^2\times S^2;\Z_n)$ is
generated by $a_I,\ I=1,2$ and $b$, where 
$a_I \in H^1(T^2\times S^2;\Z_n)=\Z_n^{\oplus 2}$ and
$b \in H^2(T^2\times S^2;\Z_n)=\Z_n^{\oplus 2}$.
Using the cohomology ring \eqn{ringS2T2} in Appendix \ref{crings}, we can
parametrize $b^{\Z_n}$ as
\begin{align}
 b^{\Z_n} =\bt_1 a_1a_2 +\bt_2 b , \ \ \ \bt_1,\bt_2 \in \Z_n.
\end{align}
Thus
\begin{align}
\label{ZnbT2S2}
&\ \ \ \
 Z_{k;b^2\Z_n}^\text{top} (S^2\times T^2) 
\\ & = 
 \frac{1}{n} 
\sum_{\bt_1,\bt_2 \in \Z_n}
\ee^{\ii k\frac{2\pi}{n} 2\bt_1\bt_2}
 =\<2k,n\>.
\nonumber 
\end{align}

On $M^4=S^1\times L^3(p)$, we need to use the cohomology ring $H^*(S^1\times
L^3(p);\Z_n)$ as described in \eqn{S1LpqH}, \eqn{S1Lpqcup}, and \eqn{S1LpqB}.
For $\<n,p\>=1$,
$ Z_{k;b^2\Z_n}^\text{top} (S^1\times L^3(p) )=1$.
For $\<n,p\>\neq 1$,
we can parametrize $b^{\Z_n}$ as
\begin{align}
 b^{\Z_n} =\bt_1 aa_1 +\bt_2 b , \ \ \ \bt_1,\bt_2 \in \Z_{\<n,p\>}.
\end{align}
Using $aa_1b=\frac n{\<n,p\>} a_1c$ and $(aa_1)^2=b^2=0$,
we find that 
\begin{align}
&\ \ \ \
 Z_{k;b^2\Z_n}^\text{top} (S^1\times L^3(p)) 
\\ & = 
 \frac{1}{\<n,p\>} 
\sum_{\bt_1,\bt_2 =0}^{\<n,p\>-1}
\ee^{\ii 2k\frac{2\pi}{\<n,p\>} \bt_1\bt_2}
 =\<2k,n,p\>.
\nonumber 
\end{align}

On $M^4=F^4$, we need to use the cohomology ring $H^*(F^4;\Z_n)$ as described in
Appendix \ref{F4ring}.  We can parametrize $b^{\Z_n}$ as
\begin{align}
 b^{\Z_n} =\bt_1 b_1 +\bt_2 b_2 , \ \ \ \bt_1,\bt_2 \in \Z_n,
\end{align}
where $b_1,b_2$ are generators of $H^2(F^4;\Z_n)$.
Using $b_1^2=-b_2^2=v$ and $b_1b_2=0$,
we find that 
\begin{align} 
\label{ZZnbF4}
 Z_{k;b^2\Z_n}^\text{top} (F^4) 
 & = 
 \frac{1}{n} 
\sum_{\bt_1,\bt_2 =0}^{n-1}
\ee^{\ii k\frac{2\pi}{n} (\bt_1^2-\bt_2^2)}
\\
 &= 
\begin{cases}
\<2k,n\>, &\text{ if } \frac{2kn}{\<2k,n\>^2}= \text{ even};\\
0,        &\text{ if } \frac{2kn}{\<2k,n\>^2}= \text{ odd}.\\
\end{cases}
\nonumber 
\end{align}
The above results are summarized in Table \ref{topinv}.

\subsubsection{Point-like and string-like topological excitations}

When $k\neq 0$, the twisted 3+1D $\Z_n$-2-cocycle theory realizes a topological
order that is not described by $Z_n$-gauge theory, nor by the
group-cocycle-twisted Dijkgraaf-Witten theory, since the group-cohomology
$\cH^4(Z_n,\RZ)=0$.  Here, we will show that \frmbox{the 3+1D twisted
$\Z_n$-2-cocycle theory realizes a 3+1D $Z_{\<2k,n\>}$-gauge theory.  The
$Z_{\<2k,n\>}$-gauge theory is a EF $Z_{\<2k,n\>}$-gauge theory if
$2kn/\<2k,n\>^2 = $ odd, and it is  a UT $Z_{\<2k,n\>}$-gauge theory if
$2kn/\<2k,n\>^2 = $ even. } 

The reduction form $Z_n$ to $Z_{\<2k,n\>}$ by the twist can be seen from the
GSD of the model.  The GSD on $S^1\times S^2$ count the number of types of
point-like topological excitations, and the number of types of string-like
topological excitations.  From \eqn{ZnbT2S2}, we see that
twisted $\Z_n$-2-cocycle model gives rise to a
topological order with $\<2k,n\>$ types of  point-like topological excitations
and $\<2k,n\>$ types of string-like topological excitations.

It is interesting to see that the twisted model describes an invertible
topological order when $\<2k,n\>=1$. Since all 3+1D invertible topological
orders are trivial topological orders, thus \frmbox{the twisted $\Z_n$-2-cocycle
model describes an trivial product state when $\<2k,n\>=1$.}

Naively, the twisted $\Z_n$-2-cocycle model should have $n$ types of
point-like topological excitations and $n$ types of  string-like topological
excitations. But actually, there are only  $\<2k,n\>$ types of point-like
topological excitations and $\<2k,n\>$ types of  string-like topological
excitations. Other excitations are confined.

To understand the unconfined topological excitations in level-$k$
$\Z_n$-2-cocycle model, we note that we can view $b^{\Z_n}$ as the field
strength 2-form of a $U(1)$ gauge theory
\begin{align}
2\pi b^{\Z_n} = f,
\end{align}
where the $2\pi$ factor comes from the different quantization convention
$\int_{M^2_\text{closed}}  b^{\Z_n} =$ integer and $\int_{M^2_\text{closed}}  f
= 2\pi \times$ integer.  In this case, the point-like topological
excitations correspond to the monoples in the $U(1)$ gauge theory.  Such a
$U(1)$ gauge theory is described by the partition function
\begin{align}
&\ \ \ \ Z_{k,U(1)}(M^4) 
\\
&= 
\int D[a]
\ee^{\ii  \frac{\Th}{8\pi^2} \int_{M^4} ff  +\cdots }
,
\nonumber 
\end{align}
where $\Th=\frac{4\pi k}{n}$, and $\cdots$ represents additional interactions.
Without the additional interactions, the particle like excitation in the $U(1)$
gauge theory are labeled by two integers $(q,m)$ where $m=M$ is the magnetic
charge.  The $U(1)$ charge of $(q,m)$ is given by $Q_{q,m} = q+
\frac{\Th}{2\pi} m$.  The statistics of particle $(q,m)$ is determined by
$\ee^{\ii \th}=(-)^{mq}$, where $\ee^{\ii \th}=1$ correspond to boson and
$\ee^{\ii \th}=-1$ correspond to fermion.  Let us express the statistics in
terms of physical quantities $(Q,M)$: $\ee^{\ii
\th}=(-)^{MQ-\frac{\Th}{2\pi}M^2}$.  We see that when $\Th=0$ or $\Th=2\pi$,
both $Q$ and $M$ are integers, but the statistics of particles with charge
$(Q,M)$ are different for $\Th=0$ and $\Th=2\pi$.  Thus changing $\Th$ by
$2\pi$ will lead to a different $U(1)$ gauge theory.  Changing $\Th$ by $4\pi$
will give us the same $U(1)$ gauge theory. This is consistent with the mod $n$
periodicity of $k$.

For $\Th=\frac{4\pi k}{n}$, we note that the $(q,m)=(-2k, n)$ particle has a
vanishing $U(1)$ charge and is a boson.  We can use the additional interactions
to condense such a dyon \cite{YW1427}.  Such a condensation will make the
$U(1)$ gauge theory to be our $\Z_n$-2-cocycle theory.  This is because a
change of $\int_{M^2_\text{closed}} b^{\Z_n}$ by $n$ is a trivial change, which
means a change of $\int_{M^2_\text{closed}} f$ by $2\pi n$ should also be a
trivial change in the $U(1)$ gauge theory.  This is achieved by condensing $n$
unit of magnetic charge that is carried by $(q,m)=(-2k, n)$ particle.

Since the condensing particles have a non-zero magnetic charge, in the
condensed phase, all the particles with non-zero $U(1)$ charge, $Q_{q,m}\neq
0$, are confined.  Thus the unconfined point-like topological excitations
are given by $(q,m)= l(\frac{-2k}{\<2k,n\>}, \frac{n}{\<2k,n\>})$, with
$l=0,1,\cdots, \<2k,n\>-1$.  We see that the GSD on $S^1\times S^2$ corresponds
to the number of types of point-like topological excitations.  We also note
that when $2kn/\<2k,n\>^2 = $ odd (such as $k=1,\ n=2$), some
point-like topological excitations are fermions.  When $2kn/\<2k,n\>^2 = $
even, all point-like topological excitations are bosons.

In the condensed phase, the electric flux-lines are quantized as
$\int_{M^2_\text{closed}} \dd \v S \cdot \v E = \frac{1}{n}\times m$, $m \in
\Z$.  They are the string-like topological excitations.  Moving a point-like
excitation labeled by $l$ around a string-like excitation labeled by $m$ give
rise a phase $\ee^{\ii \frac{lm}{\<2k,n\>}}$.  So the string labeled by $m$ and
$m+\<2k,n\>$ are indistinguishable.  This suggests the we have $\<2k,n\>$ type
of string-like excitations.

We find that the point-like and string-like topological excitations in the
level-$k$ $\Z_n$-2-cocycle model are very similar to those in
$Z_{\<2k,n\>}$-gauge theory, except that the odd $Z_{\<2k,n\>}$ charges are
fermions when $2kn/\<2k,n\>^2 = $ odd.  The emergence of fermions  is supported
by the vanishing of volume-independent partition function on a non-spin
manifold $F^4=(S^1\times S^3)\# (S^1\times S^3)\#\C P^2\# \overline{\C P}^2$
(see \eqn{ZZnbF4}), which happens exactly at $2kn/\<2k,n\>^2 = $ odd.

It was first pointed out in the
string-net theory \cite{LW0510} that a 3+1D gauge theory can be twisted which
makes some gauge charge described by ``odd'' representations to be fermionic.
But when we use cocycles in $\cH^4(G,\RZ)$ to twist a $G$-gauge theory
\cite{DW9093}, the point-like topological excitations are always boson
\cite{WL14121781}.  Thus the level-$k$ $\Z_n$-2-cocycle model is a different
realization of the twist discussed in the 3+1D string-net theory.

\subsubsection{Including excitations in the path integral}

We know that the point-like excitations are described by the world-lines
$M^1_\text{WL}$ in space-time.  A world-line $M^1_\text{WL}$ can be viewed as a
$\Z_n$-valued 1-cycle, which is dual to a 3-coboundary $C_\text{WL}^{\Z_n}$.
In the twisted $\Z_n$-2-cocycle model, such a point-like excitation is
described by the 2-cochain field $b^{\Z_n}$ that satisfies $\dd b^{\Z_n} = \t p
C_\text{WL}^{\Z_n}$, where $\t p$ is the charge of the point-like excitation.
The world-sheet can be viewed as $\Z_n$-valued 2-cycles $M^2_\text{WS}$ in the
space-time lattice.  Therefore, in the presence of point-like topological
excitations described by $C_\text{WL}^{\Z_n}$ and string-like topological
excitations described by $M^2_\text{WS}$, the partition function becomes
\begin{align}
&\ \ \ \ Z_{k;b^2\Z_n}(M^4_\text{Latt};\t pC_\text{WL}^{\Z_n},sM^2_\text{WS}) 
\\
&= 
\sum_{{\{b^{\Z_n}_{ijk}\} , \dd b^{\Z_n}=\t p C_\text{WL}^{\Z_n}}} 
\ee^{\ii k \frac{2\pi}{n} \int_{M^4_\text{Latt}} (b^{\Z_n})^2 +\ii s \frac{2\pi}{n} \int_{M^2_\text{WS}} b^{\Z_n} }
,
\nonumber 
\end{align}
where $s$ is the charge of the string-like excitation.

We first solve $\dd b^{\Z_n}=\t pC_\text{WL}^{\Z_n}$ mod $n$ as
\begin{align}
 b^{\Z_n} \se{n} \t p b^{\Z_n}_\text{WL} + b^{\Z_n}_0 + \dd a^{\Z_n} ,
\end{align}
where $b^{\Z_n}_\text{WL}$ is a fixed 2-cochain field that satisfies $\dd
b^{\Z_n}_\text{WL} = C_\text{WL}^{\Z_n}$ and $b^{\Z_n}_0 \in H^2(M^4;\Z_n)$.
We can rewrite the partition function as
\begin{align}
&Z_{k;b^2\Z_n}(M^4;\t pC_\text{WL}^{\Z_n},sM^2_\text{WS}) 
\propto
\ee^{\ii  \frac{2\pi}{n} \int_{M^4} k \t p^2 (b^{\Z_n}_\text{WL})^2}
\\
&
\ee^{ \ii  \frac{2\pi}{n} \int_{M^2_\text{WS}} s \t p b^{\Z_n}_\text{WL} }
\hskip -1.5em
\sum_{\{a^{\Z_n}_{ij}\},b^{\Z_n}_0\in H^2(M^4;\Z_n)} 
\hskip -2.0em
\ee^{\ii k \frac{2\pi}{n} \int_{M^4} 
2 \t p(b^{\Z_n}_0+\dd a^{\Z_n}) b^{\Z_n}_\text{WL} 
}
\nonumber\\
& = 
\ee^{\ii \frac{2\pi k \t p^2 }{n} \int_{M^4} (b^{\Z_n}_\text{WL})^2}
\ee^{\ii \frac{ 2\pi s\t p }{n} \int_{M^2_\text{WS}} b^{\Z_n}_\text{WL}} \times
\nonumber\\
&
\sum_{b^{\Z_n}_0\in H^2(M^4;\Z_n)} \hskip -2.5em
\ee^{\ii k\frac{ 2\pi }{n} \int_{M^4}  
b^{\Z_n}_0 (2\t p b^{\Z_n}_\text{WL}+b^{\Z_n}_0)
}
\sum_{\{a^{\Z_n}_{ij}\}} \hskip -.5em
\ee^{\ii 2k\t p\frac{ 2\pi }{n} \int_{M^4}  a^{\Z_n}C_\text{WL}^{\Z_n} }
.
\nonumber
\end{align}

Let $D^3_\text{WS}$ be the extension of $M^2_\text{WS}$, \ie $\prt
D^3_\text{WS} = M^2_\text{WS}$. Then we can rewrite $ \int_{M^2_\text{WS}}
b^{\Z_n}_\text{WL} =  \int_{D^3_\text{WS}} \dd b^{\Z_n}_\text{WL} =
\int_{D^3_\text{WS}} C_\text{WL}^{\Z_n}$.  In fact $\int_{D^3_\text{WS}}
C_\text{WL}^{\Z_n} =\text{Int}(D^3_\text{WS}, M^1_\text{WL})$, is the
intersection number between $D^3_\text{WS}$ and $M^1_\text{WL}$, which in turn
is the linking number between $M^2_\text{WS}$ and $M^1_\text{WL}$:
$\text{Lnk}(M^2_\text{WS},M^1_\text{WL})$.

Using the Poincar\'e duality we can also rewrite $\int_{M^4}
a^{\Z_n}C_\text{WL}^{\Z_n} $ as $\int_{M^1_\text{WL}} a^{\Z_n}$.  Then
\begin{align}
 \sum_{\{a^{\Z_n}_{ij}\}} 
\ee^{\ii 2k\t p\frac{ 2\pi }{n} \int_{M^4}  
  a^{\Z_n}C_\text{WL}^{\Z_n}
}=
\sum_{\{a^{\Z_n}_{ij}\}} 
\ee^{\ii 2k\t p\frac{ 2\pi }{n} \int_{M^1_\text{WL}} a^{\Z_n}  
}
 \neq 0 
\end{align}
only when $[2k\t p]_n=0$, \ie when $\t p$ is
quantized as $\t p =p \frac{n}{\<2k,n\>}$, $p\in \Z_{\<2k,n\>}$.  If $\t p$ is
not quantized as the above, the correspond point-like excitation is confined.

Thus the above partition function for unconfined excitations can be
rewritten as
\begin{align}
&\ \ \ \
Z_{k;b^2\Z_n}(M^4;\frac{p n}{\<2k,n\>}C_\text{WL}^{\Z_n},sM^2_\text{WS}) 
\\
&\propto 
\ee^{ \ii sp\frac{2\pi}{\<2k,n\>} \text{Lnk}(M^2_\text{WS},M^1_\text{WL})}
\ee^{\ii \pi p^2\frac{2 nk}{\<2k,n\>^2} \int_{M^4} (b^{\Z_n}_\text{WL})^2} 
\times
\nonumber\\
&\ \ \ \
\sum_{b^{\Z_n}_0 \in H^2(M^4;\Z_n)} \hskip -1em
\ee^{\ii k\frac{ 2\pi }{n} \int_{M^4}  b^{\Z_n}_0 ( \frac{2pn}{\<2k,n\>}b^{\Z_n}_\text{WL}+b^{\Z_n}_0)}
.
\nonumber 
\end{align}

The above expression tells us the braiding statistics of point-like excitations
and string-like excitations.  Let assume $ H^2(M^4;\Z_n)=0$.  In this case
$\int_{M^4} b^{\Z_n}_\text{WL}b^{\Z_n}_\text{WL} $ is an integer, and
$p^2\frac{2 nk}{\<2k,n\>^2}$ is also an integer.  Thus $\ee^{\ii \pi p^2\frac{2
nk}{\<2k,n\>^2} \int_{M^4} b^{\Z_n}_\text{WL}b^{\Z_n}_\text{WL}}$ is always 1
when $p^2\frac{2 nk}{\<2k,n\>^2}=$ even and $\ee^{\ii \pi p^2\frac{2
nk}{\<2k,n\>^2} \int_{M^4} b^{\Z_n}_\text{WL}b^{\Z_n}_\text{WL}} $ can be $-1$
when $p^2\frac{2 nk}{\<2k,n\>^2}=$ odd.  This factor determines the statistics
of the point-like excitations since $b^{\Z_n}_\text{WL}$ is determined by the
particle world-line $M^1_\text{WL}$.  Comparing with the results obtained in
the last section, we find that \emph{when the factor  $\ee^{\ii \pi p^2\frac{2
nk}{\<2k,n\>^2} \int_{M^4} b^{\Z_n}_\text{WL}b^{\Z_n}_\text{WL}}$ can be $-1$
(depending one the braiding of the world-line $M^1_\text{WL}$), then the
correspond particle is a fermion.}  This means that when $p^2\frac{2
nk}{\<2k,n\>^2}=$ odd, the charge $p$ particle is a fermion.  

The factor $\ee^{ \ii sp\frac{2\pi}{\<2k,n\>}
\text{Lnk}(M^2_\text{WS},M^1_\text{WL})} $ determines the mutual statistics
(\ie the Aharonov-Bohm phase) between point-like excitations and string-like
excitations. We see that it is the usual mutual statistics of
$Z_{\<2k,n\>}$-gauge theory.  We also see that there is no non-trivial braiding
statistics between string-like excitations.  This confirms our result in the
last section that \frmbox{the $\Z_n$-2-cocyle model produces a low energy
effective $Z_{\<2k,n\>}$-gauge theory.  It is a UT $Z_{\<2k,n\>}$-gauge
theory if $\frac{2 nk}{\<2k,n\>^2}=$ even, and a EF $Z_{\<2k,n\>}$-gauge
theory  if $\frac{2 nk}{\<2k,n\>^2}=$ odd.}

The term $\sum_{b^{\Z_n}_0 \in H^2(M^4;\Z_n)} 
\ee^{\ii k\frac{ 2\pi }{n} \int_{M^4}  b^{\Z_n}_0 ( \frac{2pn}{\<2k,n\>}b^{\Z_n}_\text{WL}+b^{\Z_n}_0)}
$
tells us when the partition function will vanishes in the presence of
emergent fermions, \ie when $\frac{2 nk}{\<2k,n\>^2}=$ odd.  Let us assume
there is no world-line and  $\frac{2 nk}{\<2k,n\>^2}=$ odd. In this case
the above factor becomes
$\sum_{b^{\Z_n}_0\in H^2(M^4;\Z_n)} 
\ee^{\ii k \frac{2 \pi }{n} \int_{M^4} 
(b^{\Z_n}_0)^2}$.
We note that $\frac{2 nk}{\<2k,n\>^2}=$ odd implies that $k$ and $\frac n2$ are
both odd integers.
Since $n$ is even and $\frac n2$ is odd, we have
$\Z_n =\Z_{n/2}\oplus\Z_2$. Therefore, 
$b^{\Z_n}_0$ can be expressed as
\begin{align}
b^{\Z_n}_0 = 2 b^{\Z_{n/2}}_0 +  \frac n2 b^{\Z_2}_0.
\end{align}
We obtain
\begin{align}
&\ \ \ \
\sum_{b^{\Z_n}_0\in H^2(M^4;\Z_n)} \hskip -1.5em
\ee^{\ii k \frac{2 \pi }{n} \int_{M^4} 
(b^{\Z_n}_0)^2}
\nonumber\\
&= 
\sum_{b^{\Z_{n/2}}_0\in H^2(M^4;\Z_{n/2})}
\ee^{\ii k \frac{2 \pi }{n} \int_{M^4} 
4(b^{\Z_{n/2}}_0)^2 
}\times
\nonumber\\
&\ \ \ \
\sum_{b^{\Z_2}_0\in H^2(M^4;\Z_2)} 
\ee^{\ii k \frac{2 \pi }{n} \int_{M^4} 
(\frac n2)^2  (b^{\Z_2}_0)^2
}
\nonumber\\
&= 
\sum_{b^{\Z_{n/2}}_0\in H^2(M^4;\Z_{n/2})}
\ee^{\ii 2k \frac{2 \pi }{n/2} \int_{M^4} 
(b^{\Z_{n/2}}_0)^2 
}\times
\nonumber\\
&\ \ \ \
\sum_{b^{\Z_2}_0\in H^2(M^4;\Z_2)} 
\ee^{\ii \pi \int_{M^4} 
(b^{\Z_2}_0)^2
}
\end{align}
The factor $\sum_{b^{\Z_2}_0\in H^2(M^4;\Z_2)} 
\ee^{\ii \pi \int_{M^4} 
(b^{\Z_2}_0)^2
}
$ can be rewritten as
\begin{align}
 \sum_{b^{\Z_2}_0\in H^2(M^4;\Z_2)} \hskip -1.5em
\ee^{\ii \pi \int_{M^4} 
(b^{\Z_2}_0)^2
}
=
\sum_{b^{\Z_2}_0\in H^2(M^4;\Z_2)} \hskip -1.5em
\ee^{\ii \pi \int_{M^4} 
\rw_2 b^{\Z_2}_0
}
\end{align}
since $M^4$ is orientable.  Now we see that \frmbox{$Z_{k;b^2\Z_n}(M^4)=0$ when
$\rw_2\neq 0$ (\ie when the orientable $M^4$ is not spin), if there is an
emergence of fermions.}

\subsection{3+1D twisted $\Z_n{\rm a}\Z_n{\rm b}$ model}

\subsubsection{Model construction}

In this section, we are going to construct a local bosonic model on space-time
lattice $M^4_\text{Latt}$.  Our model is a mixture of $\Z_n$-1-cocycle model
and $\Z_n$-2-cocycle model. The local degrees of freedom of our model are
$\Z_n$ indices $a^{\Z_n}_{ij}$ on the links and $b^{\Z_n}_{ijk}$ on the
triangles.  We view $a^{\Z_n}_{ij}$ as a 1-cochain in
$C^1(M^4_\text{Latt};\Z_n)$ and $b^{\Z_n}_{ijk}$ as a 2-cochain in
$C^2(M^4_\text{Latt};\Z_n)$.  

Using the Bockstein homomorphism for $\Z_n$, $\Bs_n: H^m(M^d;\Z_n) \to
H^{m+1}(M^d;\Z_n)$, the partition function of our model is defined as
\begin{align}
&\ \ \ \ Z_{k_1k_2;b\Bs a\text{-}bb\Z_n}(M^4_\text{Latt}) 
\\
&= 
\sum_{\{a^{\Z_n}_{ij}\}\atop \dd a^{\Z_n}=0} \ \
\sum_{\{b^{\Z_n}_{ijk}\}\atop \dd b^{\Z_n}=0} 
\ee^{\ii  \frac{2\pi}{n} \int_{M^4_\text{Latt}} 
k_1 b^{\Z_n}\Bs_n a^{\Z_n}+ 
k_2 b^{\Z_n}  b^{\Z_n}
}
.
\nonumber 
\end{align}
The volume-independent topological partition function is given by
\begin{align}
\label{ZtopZ2Z2}
&\ \ \ \
 Z_{k_1k_2;b\Bs a\text{-}bb\Z_n}^\text{top} (M^4) 
\nonumber\\
& = 
\sum_{
a^{\Z_n} \in H^1(M^4;\Z_n) \atop
b^{\Z_n} \in H^2(M^4;\Z_n)
}
\frac{
\ee^{\ii \frac{2\pi}{n} \int_{M^4}  
k_1 b^{\Z_n}\Bs_n a^{\Z_n}
+k_2 b^{\Z_n}b^{\Z_n}
}
} {|H^1(M^4;\Z_n)|} 
\end{align}

\subsubsection{Topological partition functions}

On $M^4=T^4$ or $M^4 =S^2\times T^2$, $\Bs_n a^{\Z_n}=0$. Thus the partition
function is a product of the partition function of the $\Z_n$a model in Section
\ref{Znamdl} and the partition function of the $\Z_n$b model in Section \ref{ZnbSec}.  We
find that (see Table \ref{topinv})
\begin{align}
  Z_{k_1k_2;b\Bs a\text{-}bb\Z_n}^\text{top} (T^4) &= n^3\<2k_2,n\>^3,
\nonumber\\
  Z_{k_1k_2;b\Bs a\text{-}bb\Z_n}^\text{top} (S^2\times T^2) & =n\<2k_2,n\>.
\end{align}

On $M^4=S^1\times L^3(p)$, for $\<n,p\>=1$, we find that $ \int_{M^4}
b^{\Z_n}\Bs_n a^{\Z_n}= \int_{M^4} b^{\Z_n} b^{\Z_n} =0$, since $H^2(S^1\times
L^3(p);\Z_n)=0$.  So $Z_{k_1k_2;b\Bs a\text{-}bb\Z_n}^\text{top}(S^1\times
L^3(p))=1$.
For $\<n,p\>\neq 1$, 
we can parametrize $a^{\Z_n},\ b^{\Z_n}$ as
\begin{align}
 a^{\Z_n} &= \al_1 a_1 + \al_2 a, \ \ \ \al_1 \in \Z_n,\ \al_2 \in \Z_{\<n,p\>}, 
\nonumber\\
 b^{\Z_n} &= \bt_1  a_1 a +\bt_2 b, \ \ \ \bt_1,\bt_2  \in \Z_{\<n,p\>} .
\end{align}
Using $\Bs_n a=\frac{p}{\<n,p\>}b$, $\Bs_na_1=0$, $a_1ab=
\frac{n}{\<n,p\>}ca_1$, and $b^2=(a_1a)^2=0$ (see \eqn{S1LpqH}, \eqn{S1Lpqcup}, and
\eqn{S1LpqB}), we find that
\begin{align}
&\ \ \ \
Z_{k_1k_2;b\Bs a\text{-}bb\Z_n}^\text{top} (S^1\times L^3(p)) 
\\ & =
\sum_{\al_1\in \Z_n; \al_2,\bt_1,\bt_2\in Z_{\<n,p\>}}
\frac{\ee^{\ii \frac{2\pi}{\<n,p\>} (k_1 \frac{p}{\<n,p\>}\al_2 \bt_1 + 2k_2\bt_1\bt_2 ) }}
 {n\<n,p\>} 
\nonumber\\
&=\sum_{\al_2,\bt_2\in Z_{\<n,p\>}}
\frac{\del_{\<n,p\>}(k_1 \frac{p}{\<n,p\>}\al_2  + 2k_2\bt_2)}{\<n,p\>}
\nonumber\\
&= \<n,p\>\<2k_2, k_1 \frac{p}{\<n,p\>},\<n,p\>\>.
\nonumber 
\end{align}

On $M^4=F^4$, we note that the Bockstein homomorphism
$\Bs_n$ maps all 1-cocycles to 0.
Thus the partition
function is a product of the partition function of the $\Z_n$a model in Section
\ref{Znamdl} and the partition function of the $\Z_n$b model in Section \ref{ZnbSec}: 
\begin{align}
  Z_{k_1k_2;b\Bs a\text{-}bb\Z_n}^\text{top} (F^4)
 &= 
\begin{cases}
n\<2k_2,n\>, &\text{if } \frac{2k_2n}{\<2k_2,n\>^2}= \text{even}\\
0,        &\text{if } \frac{2k_2n}{\<2k_2,n\>^2}= \text{odd}\\
\end{cases}
\end{align}

\subsubsection{Point-like and string-like topological excitations}

Here we are going to study more physical properties of the
$\Z_n\text{a}\Z_n\text{b}$ model.  The GSD of our model on space $M^3$ is given
by $\text{GSD}_{k_1k_2;b\Bs a\text{-}bb\Z_n}(M^3)=Z_{k_1k_2;b\Bs
a\text{-}bb\Z_2}^\text{top} (S^1\times M^3)$.  If we choose $M^3=S^1\times
S^2$, we find that 
\begin{align}
\text{GSD}_{k_1k_2;b\Bs a\text{-}bb\Z_n}(S^1\times S^2) =n\<2k_2,n\>.
\end{align}
The GSD on $S^1\times S^2$ implies that there are $n\<2k_2,n\>$ types of
point-like and string-like excitations regardless the value of $k_1$.  This
result is unexpected, since one may guess the number of types of point-like and
string-like excitations are $n^2$. The reduction is due to confinement as will
be explained below.

Again, we will view $b^{\Z_n}$ as the field strength 2-form of a $U(1)$ gauge
theory
\begin{align}
2\pi b^{\Z_n} = f,
\end{align}
We will also  view $\Bs_n a^{\Z_n}$ as the field strength 2-form of another
 $U(1)$ gauge theory
\begin{align}
2\pi \Bs_n a^{\Z_n} = f'.
\end{align}
So the twisted 3+1D $b\Bs a\text{-}bb\Z_n$ model can be viewed as
$U(1)\times U'(1)$ gauge theory with some proper condensations.  The $U(1)\times
U'(1)$ gauge theory has a form
\begin{align}
&\ \ \ \ Z_{1,U^2(1)}(M^4) 
\\
&= 
\int D[a]D[a']
\ee^{
\ii  \frac{\Th_1}{4\pi^2} \int_{M^4} ff'  
\ii  \frac{\Th_2}{8\pi^2} \int_{M^4} ff  
+\cdots }
,
\nonumber 
\end{align}
with $\Th_1=k_1\frac{2\pi}{n}$ and $\Th_2=k_2\frac{4\pi}{n}$.

Let us consider a more general $U^\ka(1)$ model
\begin{align}
Z= \int \prod_I D[a_I]
\ee^{
\ii  \frac{2\pi}{8\pi^2} \int_{M^4} f_I\La_{IJ}f_J  
+\cdots }
\end{align}
where $\La_{IJ}$ is a symmetric rational matrix.
On the boundary, the action amplitude becomes
\begin{align}
\ee^{
\ii  \frac{1}{4\pi} \int_{\prt M^4} \La_{IJ}a_I\dd a_J  
+\cdots }
\end{align}
We see that $2\pi$ flux of $a_J$ carries $a_I$-charge $Q_I=\La_{IJ}$.

Before the condensation, the point-like excitations are labeled by
$(q,m,q',m')$.  The magnetic charges for the two $U(1)$ gauge fields are $M=m$
and $M'=m'$.  Using the above result with
\begin{align}
 \La=
{\scriptsize  \begin{pmatrix}
 0 & \frac{k_1}{n} \\
 \frac{k_1}{n} & \frac{2k_2}{n} \\
\end{pmatrix}},
\end{align}
we see that the electric charges for the two $U(1)$ gauge fields are $Q=q+\frac
{k_1}{n} m'$ and $Q'=q'+\frac{k_1}{n} m+\frac{2k_2}{n} m'$. The statistics of
the $(q,m,q',m')$-excitation is $\ee^{\ii \th}=(-)^{qm+q'm'}$.

Next, we condense $(q,m,q',m')=(-k_1,0,-2k_2,n)$ excitations that have
$Q=Q'=0$.  Since $(M,M')=(0,n)$ for such excitations, it breaks the second
$U'(1)$ to $Z_n$ (in the dual picture).  We also condense
$(q,m,q',m')=(n,0,0,0)$ particles with $(Q,Q',M,M')=(n,0,0,0)$.  It breaks the
first $U(1)$ to $Z_n$.  The unconfined particles must have $M=Q'=0$, \ie
$q'=m=0$.  Thus the unconfined particles are generated by
$(q,m,q',m')=(1,0,0,0)$ with $(Q,M,Q',M')=(1,0,0,0)$ and
$(q,m,q',m')=(0,0,-\frac{2k_2}{\<2k_2,n\>},\frac{n}{\<2k_2,n\>})$ with
$(Q,M,Q',M')=(\frac{k_1}{\<2k_2,n\>},0,0,\frac{n}{\<2k_2,n\>})$.  We see that the point-like
excitations are labeled by $(p,p')$ (a bound state of $p$
type-$(q,m,q',m')=(1,0,0,0)$ and $p'$
type-$(q,m,q',m')=(0,0,-\frac{2k_2}{\<2k_2,n\>},\frac{n}{\<2k_2,n\>})$
excitations).  Two particles that differ by a condensing particle are regarded
as equivalent. Thus $(p,p')$ labels have the following equivalent relation
\begin{align}
(p+n,p') \sim (p,p') \sim (p-k_1, p'+\<2k_2,n\>).
\end{align}
So there are $n\<2k_2,n\>$ distinct types of point-like excitations.  The
type-$(q,m,q',m')=(1,0,0,0)$ excitation is a boson.  The
type-$(q,m,q',m')=(0,0,-\frac{2k_2}{\<2k_2,n\>},\frac{n}{\<2k_2,n\>})$
excitation has a statistics $(-)^{\frac{2k_2 n}{\<2k_2,n\>^2}}$.  

We note that the point-like excitations are labeled by the integer points
$(p,p')$ in a 2-dimensional unit cell with basis vectors $(n,0)$ and $(-k_1,
\<2k_2,n\>)$.  We put the two basis vectors together to form a matrix
${\scriptsize  \begin{pmatrix} n & 0\\ -k_1 & \<2k_2,n\>\\ \end{pmatrix}} $. The fusion of the
point-like excitations is described by an abelian group $G{\scriptsize  \begin{pmatrix} n &
0\\ -k_1 & \<2k_2,n\>\\ \end{pmatrix}}$ characterized by the matrix.  In
general, the fusion rule of the point-like excitations is not given by
$Z_n\times Z_{\<2k_2,n\>}$.

The string-like excitations are generated by the $2\pi/n$ magnetic flux-line of
the first $U(1)$ and the $1/n$-unit electric flux-line of the second $U'(1)$.
So the generic string-like excitations are labeled by $(s,s')$.  Two strings
that can join are regarded as equivalent \cite{KW1458}.  Note 
we can attach a $(q,m,q',m')$ excitation to change string $(s,s')$ 
to an equivalent one, which
generates the following  equivalence relation
\begin{align}
 (s,s') \sim (s+nm,s'+nq'+k_1m+2k_2m')
\end{align}
The above can be rewritten as
\begin{align}
(s+n,s'+k_1) \sim (s,s') \sim (s, s'+\<2k_2,n\>).
\end{align}
We see that there are $n\<2k_2,n\>$ distinct types of string-like excitations.

The fusion of the string-like excitations is described by an abelian group
$G{\scriptsize  \begin{pmatrix} n & k_1\\ 0 & \<2k_2,n\>\\ \end{pmatrix}}$.
It turns out that the
fusion of the point-like excitations and the fusion of
the string-like excitations are described by the same abelian group
\begin{align}
G{\scriptsize  \begin{pmatrix} n & 0\\ -k_1 & \<2k_2,n\>\\ \end{pmatrix}}
=
 G{\scriptsize  \begin{pmatrix} n & k_1\\ 0 & \<2k_2,n\>\\ \end{pmatrix} }.
\end{align}
In general, two integer  matrices $M_1$ and $M_2$ describe the same abelian
group if $ M_2=WM_1U$ where $U,W$ are invertible integer matrices.  In this
case, we say $M_1\sim M_2$.  
Let ${\scriptsize  \begin{pmatrix} m_1 & 0\\ 0 & m_2\\ \end{pmatrix}}$ be the
Smith normal form of ${\scriptsize  \begin{pmatrix} n & 0\\ k_1 & \<2k_2,n\>\\
\end{pmatrix}}$, \ie 
\begin{align}
W {\scriptsize  \begin{pmatrix} n & 0\\ k_1 & \<2k_2,n\>\\
\end{pmatrix}} U = {\scriptsize  \begin{pmatrix} m_1 & 0\\ 0 & m_2\\ \end{pmatrix}}.
\end{align}
This implies that
\begin{align}
U_T {\scriptsize  \begin{pmatrix} n & k_1\\ 0 & \<2k_2,n\>\\
\end{pmatrix}} W^T = {\scriptsize  \begin{pmatrix} m_1 & 0\\ 0 & m_2\\ \end{pmatrix}}.
\end{align}
We see that
\begin{align}
{\scriptsize  \begin{pmatrix} n & k_1\\ 0 & \<2k_2,n\>\\ \end{pmatrix} }
\sim
{\scriptsize   \begin{pmatrix} n & 0\\ k_1 & \<2k_2,n\>\\ \end{pmatrix} } 
\sim
{\scriptsize  \begin{pmatrix} n & 0\\ -k_1 & \<2k_2,n\>\\ \end{pmatrix} }.
\end{align}
Via direct numerical calculation, we find that
\begin{align}
 G{\scriptsize  \begin{pmatrix} n & k_1\\ 0 & \<2k_2,n\>\\ \end{pmatrix} }
=Z_{\frac{n\<n,2k_2\>}{\<n,k_1,2k_2\>} } \times Z_{\<n,k_1,2k_2\>}
\end{align}

The mutual braiding phase between 
a type-$(p,p')$ point-like excitation and
a type-$(s,s')$ string-like excitation is given by
\begin{align}
\th=2\pi\Big(\frac{ps}{n}- \frac{p's'}{\<2k_2,n\>}+ \frac{p's k_1}{n\<2k_2,n\>}\Big).
\end{align}

Since, both point-like excitations and string-like excitations have a
fusion described by $Z_{\frac{n\<n,2k_2\>}{\<n,k_1,2k_2\>} } \times
Z_{\<n,k_1,2k_2\>}$, we will call the corresponding theory a
$Z_{\frac{n\<n,2k_2\>}{\<n,k_1,2k_2\>} } \times Z_{\<n,k_1,2k_2\>}$
fusion theory.  When $\frac{2k_2 n}{\<2k_2,n\>^2}=$ odd, some point-like
excitations are fermions.

\subsubsection{Including excitations in the path integral}

In the $\Z_n$a$\Z_n$b model, there are two kinds of point-like excitations
described by the world-lines $M^1_\text{WL}$ and $N^1_\text{WL}$, which are
$\Z$-valued 1-cycles.  Let 3-coboundary $C_\text{WL}^{\Z}$ be the
Poincar\'e dual of $M^1_\text{WL}$.  Then the point-like excitation that
corresponds to $M^1_\text{WL}$ is described the 2-cochain field $b^{\Z_n}$ that
satisfies 
\begin{align}
\dd b^{\Z_n} \se{n} \t p_1 C_\text{WL}^{\Z} ,
\end{align}
where $\t p_1$ is the charge of the point-like excitation.

The $\Z_n$a$\Z_n$b model also contains two kind of string-like excitations
described by the world-sheets $M^2_\text{WS}$ and $N^2_\text{WS}$ in
space-time.  The world-sheet  $N^2_\text{WS}$ can be viewed as a $\Z$-valued
2-cycle, which is dual to a $\Z$-valued 2-coboundary $B^{\Z}_\text{WS}$.
Such a string-like excitation is described the 1-cochain field $a^{\Z_n}$ that
satisfies 
\begin{align}
\dd a^{\Z_n} \se{n}  s_2 B^{\Z}_\text{WS}  , 
\end{align}
where $ s_2$ is the charge of
the string-like excitation.  Therefore, in the presence of point-like
topological excitations described by $C_\text{WL}^{\Z},\ N^1_\text{WL}$ and
string-like topological excitations described by $M^2_\text{WS},\
B^{\Z}_\text{WS}$, the partition function becomes
\begin{align}
&Z_{k_1k_2;b\Bs a\text{-}bb\Z_n}(M^4,
\t p_1 C_\text{WL}^{\Z_n}, p_2 N^1_\text{WL},
 s_1M^2_\text{WS}, s_2 B^{\Z}_\text{WS}) =
\nonumber \\
&
\sum_{\{a^{\Z_n}_{ij}\}\atop \dd a^{\Z_n}\se{n} s_2 B^{\Z}} \ \
\sum_{\{b^{\Z_n}_{ijk}\}\atop \dd b^{\Z_n}\se{n}\t p_1 C_\text{WL}^{\Z}} 
\ee^{\ii  \frac{2\pi}{n} \int_{M^4} 
k_1 b^{\Z_n}\Bs_n a^{\Z_n}+ 
k_2 (b^{\Z_n})^2 
}
\times
\nonumber\\
&
\ \ \ \ \ \ \
\ \ \ \ \ \ \
\ \ \ \ \ \ \
\ \ \
\ee^{\ii \frac{2\pi}{n} (
  p_2 \int_{ N^1_\text{WL}}  a^{\Z_n}
+ s_1 \int_{ M^2_\text{WS}}  b^{\Z_n})
}  .
\end{align}
where $ p_2$ is the charge of the point-like excitation, and $ s_1$ the charge
of the string-like excitation.  However, the above  partition function is not
well defined.  It is well defined only when $b^{\Z_n}$ and $a^{\Z_n}$ are
cocycles.  When $b^{\Z_n}$ and $a^{\Z_n}$ are not cocycles, the partition
function is not invariant under the shift $b^{\Z_n} \to b^{\Z_n}+ n \t b^{Z_n}$
and/or $a^{\Z_n} \to a^{\Z_n}+ n \t a^{Z_n}$.

So to remove such ambiguity,
we write $b^{\Z_n}$ and $a^{\Z_n}$ as
\begin{align}
 b^{\Z_n} & = \t p_1 b^{\Z}_\text{WL} + b^{\Z_n}_0+\dd \t a^{\Z_n}, 
\nonumber\\
 a^{\Z_n} & =  s_2 a^{\Z}_\text{WS} + a^{\Z_n}_0+\dd g^{\Z_n}. 
\end{align}
Here $b^{\Z}_\text{WL}$ is a fixed $\Z$-valued 2-cochain field that satisfies 
\begin{align}
\dd b^{\Z}_\text{WL} = C_\text{WL}^{\Z} ,
\end{align}
and $ b^{\Z_n}_0 \in H^2(M^4;\Z_n)$.  Also $a^{\Z}_\text{WS}$ is a fixed
$\Z$-valued 1-cochain fields that satisfies 
\begin{align}
\dd a^{\Z}_\text{WS} = B^{\Z}_\text{WS}.  
\end{align}
and $ a^{\Z_n}_0 \in H^1(M^4;\Z)$.  $ a^{\Z_n}_0,\ g^{\Z_n},\  b^{\Z_n}_0,\ \t
a^{\Z_n}$ are $\Z_n$-valued.  The partition function on orientable $M^4$ is
defined by summing over those $\Z_n$-valued fields:
\begin{align}
&Z_{k_1k_2;b\Bs a\text{-}bb\Z_n}(M^4,
\t p_1 C_\text{WL}^{\Z_n}, p_2 N^1_\text{WL},
 s_1M^2_\text{WS}, s_2 B^{\Z}_\text{WS}) =
\nonumber \\
&
\sum_{\{\t a^{\Z_n}_{ij}, g^{\Z_n}_i\}} 
\sum_{a^{\Z_n}_0 \in  H^1(M^4;\Z_n) \atop b^{\Z_n}_0 \in  H^2(M^4;\Z_n)} 
\ee^{\ii  \frac{2\pi}{n} \int_{M^4} 
k_2 (\t p_1 b^{\Z}_\text{WL} + b^{\Z_n}_0+\dd \t a^{\Z_n})^2 
}\times
\nonumber\\
&\ \ \ \
\ee^{\ii  \frac{2\pi}{n} \int_{M^4} 
k_1 (\t p_1 b^{\Z}_\text{WL} + b^{\Z_n}_0) \Bs_n (s_2 a^{\Z}_\text{WS} + a^{\Z_n}_0)
}
\times
\nonumber\\
&
\ \ \ \
\ee^{\ii \frac{2\pi}{n} [
  p_2 \int_{ N^1_\text{WL}}  (s_2 a^{\Z}_\text{WS} + a^{\Z_n}_0)
+ s_1 \int_{ M^2_\text{WS}}  (\t p_1 b^{\Z}_\text{WL} + b^{\Z_n}_0)]
}  .
\end{align}
We note that
\begin{align}
&\ \ \ \
\ee^{\ii  \frac{2\pi}{n} \int_{M^4} k_2 (\t p_1 b^{\Z}_\text{WL} + b^{\Z_n}_0+\dd \t a^{\Z_n})^2 }
\nonumber\\
&=
\ee^{\ii  \frac{2\pi}{n} \int_{M^4} k_2 \t p_1^2 (b^{\Z}_\text{WL})^2 }
\ee^{\ii  \frac{2\pi}{n} \int_{M^4} k_2 (b^{\Z}_0)^2 }
\ee^{\ii  \frac{2\pi}{n} \int_{M^4} 2 k_2 \t p_1 b^{\Z}_\text{WL} \dd \t a^{\Z_n} }
\nonumber\\
&=
\ee^{\ii  \frac{2\pi}{n} \int_{M^4} k_2 \t p_1^2 (b^{\Z}_\text{WL})^2 }
\ee^{\ii  \frac{2\pi}{n} \int_{M^4} k_2 (b^{\Z}_0)^2 }
\ee^{-\ii  \frac{2\pi}{n} \int_{M^4} 2 k_2 \t p_1 C^{\Z}_\text{WL} \t a^{\Z_n} }
\nonumber\\
&=
\ee^{\ii  \frac{2\pi}{n} \int_{M^4} k_2 \t p_1^2 (b^{\Z}_\text{WL})^2 }
\ee^{\ii  \frac{2\pi}{n} \int_{M^4} k_2 (b^{\Z}_0)^2 }
\ee^{-\ii  \frac{2\pi}{n} \int_{M^1_\text{WL}} 2 k_2 \t p_1 \t a^{\Z_n} }
\end{align}
Also
\begin{align}
&\ \ \ \
\ee^{\ii  \frac{2\pi}{n} \int_{M^4} k_1 (\t p_1 b^{\Z}_\text{WL} + b^{\Z_n}_0) \Bs_n (s_2 a^{\Z}_\text{WS} + a^{\Z_n}_0) }
\nonumber\\
&=
\ee^{\ii  \frac{2\pi }{n^2} \int_{M^4} k_1 \t p_1 s_2 b^{\Z}_\text{WL}  B^{\Z}_\text{WS}}
\ee^{\ii  \frac{2\pi}{n^2} \int_{M^4} k_1 s_2 b^{\Z_n}_0 B^{\Z}_\text{WS}}
\times
\nonumber\\
&\ \ \ \
\ee^{\ii  \frac{2\pi}{n^2} \int_{M^4} k_1 \t p_1  C^{\Z}_\text{WL} a^{\Z_n}_0 }
\ee^{\ii  \frac{2\pi}{n} \int_{M^4} k_1 b^{\Z_n}_0 \Bs_n a^{\Z_n}_0 }
\nonumber\\
&=
\ee^{\ii  \frac{2\pi }{n^2} \int_{M^4} k_1 \t p_1 s_2 b^{\Z}_\text{WL}  B^{\Z}_\text{WS}}
\ee^{\ii  \frac{2\pi}{n^2} \int_{N^2_\text{WS}} k_1 s_2 b^{\Z_n}_0 }
\times
\nonumber\\
&\ \ \ \
\ee^{\ii  \frac{2\pi}{n^2} \int_{M^1_\text{WL}} k_1 \t p_1  a^{\Z_n}_0 }
\ee^{\ii  \frac{2\pi}{n} \int_{M^4} k_1 b^{\Z_n}_0 \Bs_n a^{\Z_n}_0 }
\end{align}
We can rewrite the partition function as
\begin{align}
&\ \ \ \
Z_{k_1k_2;b\Bs a\text{-}bb\Z_n}(M^4,
\t p_1 C_\text{WL}^{\Z_n}, p_2 N^1_\text{WL},
 s_1M^2_\text{WS}, s_2 B^{\Z}_\text{WS}) 
\nonumber\\
&  
\propto
\ee^{\ii  \frac{2\pi}{n} \int_{M^4} k_2 \t p_1^2 (b^{\Z}_\text{WL})^2 }
\ee^{\ii  \frac{2\pi }{n^2} \int_{M^4} k_1 \t p_1 s_2 b^{\Z}_\text{WL}  B^{\Z}_\text{WS}}
\times
\nonumber\\
&\ \ \ \
\ee^{\ii \frac{2\pi}{n} [
  p_2 \int_{ N^1_\text{WL}}  s_2 a^{\Z}_\text{WS} 
+ s_1 \int_{ M^2_\text{WS}}  \t p_1 b^{\Z}_\text{WL} ]
}  \times
\nonumber\\
&\ \ \ \
\sum_{a^{\Z_n}_0 \in  H^1(M^4;\Z_n) \atop b^{\Z_n}_0 \in  H^2(M^4;\Z_n)} 
\ee^{\ii  \frac{2\pi}{n} \int_{M^4} k_2 (b^{\Z}_0)^2 }
\ee^{\ii  \frac{2\pi}{n} \int_{M^4} k_1 b^{\Z_n}_0 \Bs_n a^{\Z_n}_0 }
\times
\nonumber\\
&\ \ \ \ 
\ee^{\ii  \frac{2\pi}{n^2} \int_{N^2_\text{WS}} k_1 s_2 b^{\Z_n}_0 }
\ee^{\ii  \frac{2\pi}{n^2} \int_{M^1_\text{WL}} k_1 \t p_1  a^{\Z_n}_0 }
\times
\nonumber\\
&\ \ \ \
\ee^{\ii \frac{2\pi}{n} [
  p_2 \int_{ N^1_\text{WL}}   a^{\Z_n}_0
+ s_1 \int_{ M^2_\text{WS}}   b^{\Z_n}_0]
}  \hskip -1em
\sum_{\{\t a^{\Z_n}_{ij}, g^{\Z_n}_i\}} \hskip -1em
\ee^{-\ii  \frac{2\pi}{n} \int_{M^1_\text{WL}} 2 k_2 \t p_1 \t a^{\Z_n} }
\end{align}

Using the Poincar\'e duality, we can
rewrite $\int_{M^4} b^{\Z_n}_\text{WL} B^{\Z_n}_\text{WS}$ as
$\int_{N^2_\text{WS}} b^{\Z_n}_\text{WL}$.  
Let $D^3_\text{WS}$ be the extension of $N^2_\text{WS}$, \ie $\prt
D^3_\text{WS} = N^2_\text{WS}$. Then we can rewrite $ \int_{N^2_\text{WS}}
b^{\Z_n}_\text{WL} =  \int_{D^3_\text{WS}} \dd b^{\Z_n}_\text{WL} = 
\int_{D^3_\text{WS}} C_\text{WL}^{\Z_n}$.  In fact $\int_{D^3_\text{WS}}
C_\text{WL}^{\Z_n} =\text{Int}(D^3_\text{WS}, M^1_\text{WL})$, is the
intersection number between $D^3_\text{WS}$ and $M^1_\text{WL}$ which is the
linking number between $N^2_\text{WS}$ and $M^1_\text{WL}$:
$\text{Lnk}(N^2_\text{WS},M^1_\text{WL})$.  

Also $ \sum_{\{\t a^{\Z_n}_{ij}\}} \ee^{\ii \frac{ 2\pi}{n}
\int_{M^1_\text{WL}} 2k_2\t p_1 \t a^{\Z_n}} \neq 0 $ only when $[2k_2\t p_1]_n=0$, or when $\t p_1$ is quantized as $\t p_1 =p_1
\frac{n}{\<2k_2,n\>}$, $p_1\in \Z_{\<2k,n\>}$.  If $\t p_1$ is not quantized as
the above, the corresponding point-like excitation is confined.

Thus the above partition function for unconfined like excitations can be
rewritten as
\begin{align}
& Z_{k_1k_2;b\Bs a\text{-}bb\Z_n}(M^4,
\frac{np_1 C_\text{WL}^{\Z_n}}{\<2k_2,n\>}, p_2 N^1_\text{WL},
 s_1M^2_\text{WS}, s_2 B^{\Z_n}_\text{WS}) 
\nonumber\\
& =
\ee^{\ii \pi \frac{2 nk_2 p_1^2}{\<2k_2,n\>^2} \int_{M^4} 
b^{\Z_n}_\text{WL}b^{\Z_n}_\text{WL}} 
\ee^{ \ii\frac{2\pi  s_1p_1}{\<2k_2,n\>} \text{Lnk}(M^2_\text{WS},M^1_\text{WL})}
\nonumber\\
&\ \ \ \ 
\ee^{ \ii \frac{2\pi s_2p_1k_1}{n\<2k_2,n\>} \text{Lnk}(N^2_\text{WS},M^1_\text{WL})}
\ee^{ \ii \frac{2\pi s_2p_2}{n} \text{Lnk}(N^2_\text{WS},N^1_\text{WL})}
\times
\nonumber\\
&\ \ \ \
\sum_{a^{\Z_n}_0 \in  H^1(M^4;\Z_n) \atop b^{\Z_n}_0 \in  H^2(M^4;\Z_n)} 
\ee^{\ii  \frac{2\pi}{n} \int_{M^4} k_2 (b^{\Z}_0)^2 }
\ee^{\ii  \frac{2\pi}{n} \int_{M^4} k_1 b^{\Z_n}_0 \Bs_n a^{\Z_n}_0 }
\times
\nonumber\\
&\ \ \ \ 
\ee^{\ii  \frac{2\pi k_1 s_2}{n^2} \int_{N^2_\text{WS}}  b^{\Z_n}_0 }
\ee^{\ii  \frac{2\pi p_1k_1}{n\<2k_2,n\>} \int_{M^1_\text{WL}}  a^{\Z_n}_0 }
\times
\nonumber\\
&\ \ \ \
\ee^{\ii \frac{2\pi}{n} [
  p_2 \int_{ N^1_\text{WL}}   a^{\Z_n}_0
+ s_1 \int_{ M^2_\text{WS}}   b^{\Z_n}_0]
} 
.
\end{align}

The factors
\begin{align}
&\ \ \ \ \ee^{\ii \pi \frac{2 nk_2 p_1^2}{\<2k_2,n\>^2} \int_{M^4} 
b^{\Z_n}_\text{WL}b^{\Z_n}_\text{WL}} 
\ee^{ \ii\frac{2\pi  s_1p_1}{\<2k_2,n\>} \text{Lnk}(M^2_\text{WS},M^1_\text{WL})}
\nonumber\\
&\ \ \ \ 
\ee^{ \ii \frac{2\pi s_2p_1k_1}{n\<2k_2,n\>} \text{Lnk}(N^2_\text{WS},M^1_\text{WL})}
\ee^{ \ii \frac{2\pi s_2p_2}{n} \text{Lnk}(N^2_\text{WS},N^1_\text{WL})}
\end{align}
in the above expression determines the braiding statistics of point-like
excitations and string-like excitations.  We see that \emph{there is no
non-trivial braiding for string-like excitations}.  But \emph{there are
non-trivial mutual statistics (\ie the Aharonov-Bohm phase) between point-like
excitations and string-like excitations}.  Also \emph{when $\frac{2
nk_2}{\<2k_2,n\>^2}=$ odd, the theory contain fermions}.

\section{Comparison between the 3+1D $\Z_n$-2-cocycle model and 3+1D
$Z_n$-1-cocycle model}
\label{comp}

There is a well known duality between the 3+1D $Z_n$-1-cocycle theory (with
emergent $Z_n$-gauge theory)  and the above 3+1D $\Z_n$-2-cocycle theory with
$k=0$.  In the following, we will compare the two theories in detail.  We find
that the two theories are equivalent, if they are viewed as pure topological
theory without any symmetry.  So both 3+1D $Z_n$-1-cocycle theory and 3+1D
$\Z_n$-2-cocycle theory realize the same topological order described by
UT $Z_n$-gauge theory.  However, if we view the two theories as topological theory
with time-reversal symmetry or parity symmetry, then the two theories are not
equivalent.  In other words, the two models realize the same topological
orders, but different symmetry enriched topological orders (with time-reversal
symmetry or parity symmetry).

\subsection{Duality}

To see the above mentioned  duality, let us describe
the lattice Hamiltonian of the two theories. We consider a 3D cubic lattice
whose sites are labeled by $i$.  To obtain a  $\Z_n$-1-cocycle theory, we put a
$\Z_n$ degrees of freedom $a_{ij}^{\Z_n}=0,1,\cdots,n-1=-a_{ji}^{\Z_n}$ on each
nearest neighbor links $(ij)$.  Let $U_{ij}=\ee^{\ii \frac{2\pi}{n}
a_{ij}^{\Z_n}}$ and $V_{ij}$ is an operator that raise $a_{ij}^{\Z_n}$ by one:
$V_{ij}|a_{ij}^{\Z_n}=m\>=|a_{ij}^{\Z_n}=m+1\>$.  Noting that the
$\Z_n$-1-cocycle theory is a theory of closed $\Z_n$-loops at low energy, we
find that the lattice  Hamiltonian for the $\Z_n$-1-cocycle theory will be
\begin{align}
\label{HZng}
 H_{\Z_n\text{a}} 
&=-\sum_i (Q_i + Q_i^\dag) - \sum_{(ijkl)} ( B_{ijkl} +B_{ijkl}^\dag) ,
\nonumber\\
Q_i &= 
\prod_{j\text{ next to } i} U_{ij},
\nonumber\\
B_{ijkl} &= V_{ij} V_{jk} V_{kl} V_{li}, 
\end{align}
where $\sum_i$ sum over all sites and $\sum_{(ijkl)}$ sum over all squares
$(ijkl)$.  The $- (Q_i + Q_i^\dag)$ terms enforce the closed-loop
condition and the $-  ( B_{ijkl} +B_{ijkl}^\dag)$ terms are the
loop hopping and/or loop creation/annihilation terms.

To obtain a  $\Z_n$-2-cocycle theory, we put a $\Z_n$ degrees of freedom
$b_{ijkl}^{\Z_n}=0,1,\cdots,n-1=-b_{lkji}^{\Z_n}$ on each square $(ijkl)$.  But
this is equivalent to  put a $\Z_n$ degrees of freedom
$a_{IJ}^{\Z_n}=0,1,\cdots,n-1=-a_{JI}^{\Z_n}$ on each link $(IJ)$ of the dual
lattice.  The dual lattice of a cubic lattice is also a cubic lattice.  The
$\Z_n$-2-cocycle theory is a theory of closed $\Z_n$-membranes at low energy.
Thus the lattice  Hamiltonian for the $\Z_n$-2-cocycle theory with $k=0$ is
\begin{align}
 H_{0;b^2\Z_n} 
&= - \sum_I (Q_I+Q_I^\dag) - \sum_{(IJKL)} ( B_{IJKL} +B_{IJKL}^\dag) ,
\nonumber\\
Q_I &= 
 \prod_{J\text{ next to } I} V_{IJ},
\nonumber\\
B_{IJKL} &= U_{IJ} U_{JK} U_{KL} U_{LI}. 
\end{align}
The $- ( B_{IJKL} +B_{IJKL}^\dag)$ terms enforces the closed-membrane condition
and the $- (Q_I+Q_I^\dag )$ are the membrane hopping and/or membrane
creation/annihilation terms.  The two Hamiltonians  $H_{\Z_n\text{a}}$ and
$H_{0;b^2\Z_n}$ are equivalent under a local unitary
transformation that exchanges $U$ and $V$.  This implies that the two theories
are really equivalent.

\subsection{Topological invariants for orientable space-time}

To compare the two theory at Lagrangian level, we note that the
volume-independent topological partition function for 3+1D $\Z_n$-1-cocycle
theory is given by
\begin{align}
 Z_{Z_n\text{a}}^\text{top} (M^4) = 
 \frac{|H^1(M^4;\Z_n)|}{|H^0(M^4;\Z_n)|}.
\end{align}
while the
volume-independent topological partition function for 3+1D $\Z_n$-2-cocycle
theory (with $k_1=k_2=0$) is given by
\begin{align}
Z_{00;b\Bs a\text{-}bb\Z_n}^\text{top} (M^4) =
\frac{|H^0(M^4;\Z_n)||H^2(M^4;\Z_n)|}{|H^1(M^4;\Z_n)|}
\end{align}
So their ratio is given by
\begin{align}
\frac{Z_{00;b\Bs a\text{-}bb\Z_n}^\text{top} (M^4)}{ Z_{Z_n\text{a}}^\text{top} (M^4)}
=
\frac{|H^0(M^4;\Z_n)|^2|H^2(M^4;\Z_n)|}{|H^1(M^4;\Z_n)|^2}
\end{align}
In Appendix \ref{Hfactor}, we will show that for
orientable close space-time $M^4$,
\begin{align}
\frac{Z_{00;b\Bs a\text{-}bb\Z_n}^\text{top} (M^4)}{ Z_{Z_n\text{a}}^\text{top} (M^4)}
=n^{\chi(M^4)},
\end{align}
where $\chi(M^4)$ is the Euler number.  The volume-independent topological
partition functions of the two models are different, which may lead one to
conclude that the $\Z_n$-1-cocycle model and the $\Z_n$-2-cocycle model realize
different topological orders.  However, in \Ref{KW1458}, it was conjectured
that two 3+1D topological partition functions $ Z_1^\text{top} (M^4)$ and $
Z_2^\text{top} (M^4)$ describe the same L-type topological orders iff their
ratio has a form
\begin{align}
\frac{Z_1^\text{top} (M^4)}{Z_2^\text{top} (M^4)} = \rho^{\chi(M^4)}
 \la^{P_1(M^4)}.
\end{align}
where  $P_1(M^4)$ is the Pontryagin number of $M^4$.  Therefore, the above
result implies that the $\Z_n$-1-cocycle model and the $\Z_n$-2-cocycle model
realize the same topological order.

\subsection{Ground state degeneracy for non-orientable spaces}

Now we turn to study the  ground state degeneracy of the two models.  To
calculate the GSD on closed space manifold $M^3$, we compute the
volume-independent partition function on $M^3\times S^1$ space-time:
\begin{align}
\text{GSD}(M^3) = Z^\text{top} (M^3\times S^1).
\end{align}
We see that the  ground state degeneracy of the two models are the same on
orientable spaces $M^3$ since their partition functions are the same on
orientable space-times $M^3\times S^1$.

However, for non orientable space $M^3$, the GSDs of the two models can be
different. For example, let us assume the space to be $M^3=S^1\times KB$, where
$KB$ is the Klein bottle. We note that
\begin{align}
 H_2(KB;\Z)&= 0,\ H_1(KB;\Z)= \Z\oplus \Z_2,\  H_0(KB;\Z)= \Z
\end{align}
and
\begin{align}
 H_2(S^1\times KB;\Z)&= H_2(KB;\Z)\oplus H_1(KB;\Z),
\nonumber\\ 
&= \Z\oplus \Z_2;
\nonumber\\
 H_1(S^1\times KB;\Z)&= H_1(KB;\Z)\oplus \Z.
\nonumber\\ 
&= \Z\oplus \Z_2 \oplus \Z;
\end{align}
Then using the universal coefficient theorem \eqn{ucfH}, we find that
\begin{align}
 H^2(S^1\times KB;\Z_n)&= \Z_n \oplus \Z_{\<n,2\>}^{\oplus 2};
\nonumber\\
 H^1(S^1\times KB;\Z) &= \Z_n^{\oplus 2} \Z_{\<n,2\>}.
\end{align}
Thus
\begin{align}
  \text{GSD}_{0;b^2\Z_n}(S^1\times KB) &= n \<n,2\>^2,
\nonumber\\ 
  \text{GSD}_{Z_n\text{a}}(S^1\times KB) &= n^2 \<n,2\>.
\end{align}
When $n>2$, the GSDs of the two theories are different.  Since the difference
only appear in non-orientable manifolds, \frmbox{the $\Z_n$-2-cocycle model and
the $\Z_n$-1-cocyle model realize two different time-reversal symmetry enriched
topological orders.} This is consistent with the fact that the two theories
realize the same topological order if we ignore the time-reversal symmetry.

Both topological orders has point-like excitations labeled by $i\in \Z_n$
and string-like excitations labeled by $s\in \Z_n$.  But they transform
differently under time-reversal.  For the $\Z_n$-1-cocyle theory $(i,s) \to
(i,-s)$ under time reversal.  For the $\Z_n$-2-cocycle theory $(i,s) \to
(-i,s)$ under time reversal.  Both the $\Z_n$-1-cocyle theory and the
$\Z_n$-2-cocycle theory are described by the same Hamiltonian \eqn{HZng}.  But
the time-reversal symmetry are realized differently.  In the $\Z_n$-1-cocyle
theory, we assume $|a_{ij}^{\Z_n}\>$, the eigenstates of $U_{ij}$, are invariant
under time-reversal. Thus $(U_{ij},V_{ij})\to (U^\dag_{ij},V_{ij})$ under time
reversal.  In the $\Z_n$-2-cocycle theory, we assume that the eigenstates of
$V_{ij}$ are invariant under time-reversal. Thus $(U_{ij},V_{ij})\to
(U_{ij},V^\dag_{ij})$ under time reversal.

\section{Non-abelian cocycle models}

So far, we have constructed many local bosonic models -- the cocycle models.
But in those construction, the local degrees of freedom are always described by
an abelian group, such as $\Z_n$.  In this section, we will use group-cocycles
in group cohomology theory (see Appendix \ref{gcoh}) to generalize the cocycle
models so that the local degrees of freedom are described by a non-abelian
group $G$.  To use group-cocycles to construct the cocycle models, we need to
map the group-cocycles in group cohomology theory to topological-cocycles in
topological cohomology theory.  To obtain such a map, we need to first introduce
the branching structure in space-time lattice.

\subsection{The branching structure of space-time lattice}

\begin{figure}[tb]
\begin{center}
\includegraphics[scale=0.6]{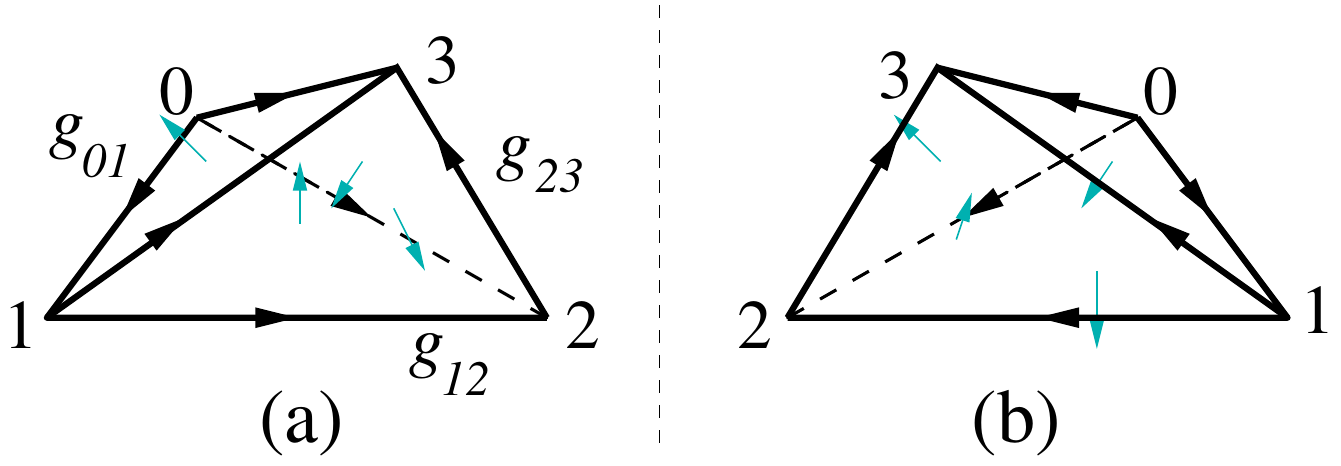} \end{center}
%6
\caption{ (Color online) Two branched simplices with opposite orientations.
(a) A branched simplex with positive orientation and (b) a branched simplex
with negative orientation.  }
\label{mir}
\end{figure}

In order to define a generic lattice theory on the space-time complex
$M^d_\text{latt}$ using group cocycles, it is important to give the vertices of
each simplex a local order.  A nice local scheme to order  the vertices is
given by a branching structure.\cite{C0527,CGL1314,CGL1204} A branching
structure is a choice of orientation of each link in the $d$-dimensional
complex so that there is no oriented loop on any triangle (see Fig. \ref{mir}).

The branching structure induces a \emph{local order} of the vertices on each
simplex.  The first vertex of a simplex is the vertex with no incoming links,
and the second vertex is the vertex with only one incoming link, \etc.  So the
simplex in  Fig. \ref{mir}a has the following vertex ordering: $0,1,2,3$.

The branching structure also gives the simplex (and its sub simplices) an
canonical orientation.  Fig. \ref{mir} illustrates two $3$-simplices with
opposite canonical orientations compare the 3-dimension space in which they are
embedded.  The blue arrows indicate that canonical orientations of the
$2$-simplices.  The black arrows indicate that canonical orientations of the
$1$-simplices.

\subsection{Group-vertex models that realize $G$-SPT orders}
\label{Gver}

\Ref{CLW1141,CGL1314,CGL1204} have constructed exactly soluble local
bosonic models using homogeneous group cocycles (see Appendix \ref{gcoh}) of
group $G$ to realize $G$-SPT orders.  Those models are actually cocycle models
on space-time lattice.  In this section, we will review those results using the
cocycle notation introduced above.

The local degrees of freedom of our model are now group elements living on the
vertices of the orientable space-time lattice $M^d_\text{Latt}$: $g_i\in G$.
Let $\nu_n(g_0,\cdots,g_n)$ be a homogeneous group $n$-cocycle:
$\nu_n(g_0,\cdots,g_n) \in \cH^n(G,\RZ)$.  From $\nu_n$, we can construct a
topological $n$-cocycle $\t \nu_n$ on  $M^d_\text{Latt}$:
\begin{align}
 \t \nu_n(i_0,i_1,\cdots,i_n)=\nu_n(g_{i_0},g_{i_1},\cdots,g_{i_n})
\end{align}
where $(i_0,i_1,\cdots,i_n)$ is an $n$-simplex with the canonical orientation
and the vertex ordering $i_0<i_1\cdots < i_n$.  Below, we will drop the $\sim$
and denote $ \t \nu_n(i_0,i_1,\cdots,i_n)$ as
$\nu_n(g_{i_0},g_{i_1},\cdots,g_{i_n})$.

Using such mapping, we can construct a group-vertex model
on orientable space-time $M^d_\text{Latt}$:
\begin{align}
 Z_{\nu_d}(M^d_\text{Latt}) 
&=\sum_{ \{g_i\} }   
\ee^{ \ii  2\pi \int_{M^d_\text{Latt}} 
\nu_d(\{g_i\})
} . 
\end{align}
Since $\nu_d(\{fg_i\})=\nu_d(\{g_i\}),\ f \in G$, the group-vertex model has a
global on-site $G$-symmetry.  Since $\ee^{ \ii  2\pi \int_{M^d} \nu_d(\{g_i\})
} =1$ on any closed orientable manifold $M^d$. 
We find that the constructed model is
gapped. We also see that
\begin{align}
 Z_{\nu_d}(M^d_\text{Latt}) &= |G|^{N_v} .
\end{align}
So the volume-independent partition function $Z_{\nu_d}^\text{top}(M^d) =1$,
for all closed orientable manifolds $M^d$, which implies that the model does
not have any topological order regardless the choice of the group cocycle
$\nu_d$.  $Z_{\nu_d}^\text{top}(M^d) =1$ also implies that the group-vortex
model does not break the $G$ symmetry (as one can see from the ground state
degeneracy on closed orientable space manifold $M^{d-1}_\text{space}$:
$\text{GSD}_{\nu_d}^\text{top}(
M^{d-1}_\text{space})=Z_{\nu_d}^\text{top}(S^1\times M^{d-1}_\text{space})
=1$).

But $Z_{\nu_d}^\text{top}(M^d) =1$ also means that  volume-independent
partition function fails to detect SPT orders. In fact, we do not even know
weather the lattice models with different $\nu_d$'s belong to different SPT
phases, if we just look at $Z_{\nu_d}^\text{top}(M^d)$.

To detect SPT order via the partition function \cite{HW1339,W1447,K1459,K1467},
we need to add the symmetry twist \cite{LG1209} in space-time. A symmetry twist
is described by $a_{ij}\in G$ on each link (\ie 1-simplex), that satisfy
\begin{align}
 a_{ij}=a_{ji}^{-1},\ \ \
a_{ij}a_{jk}a_{ki}=1.
\end{align}
Such a $a_{ij}$ configuration define a so called ``flat $G$-connection''
on space-time $M^d$.
In the presence of symmetry twist, the partition function becomes
 \begin{align}
 Z_{\nu_d}(M^d_\text{Latt},a_{ij}) 
&=\sum_{ \{g_i\} }   
\ee^{ \ii  2\pi \int_{M^d_\text{Latt}} 
\nu_d^\text{g}(\{g_i\},\{a_{ij}\})
} ,
\end{align}
where
\begin{align}
\nu_d^\text{g}&(\{g_i\},\{a_{ij}\}) \equiv
\nu_d^\text{g}(g_{i_0},g_{i_1},\cdots,g_{i_d}; a_{i_0i_1},a_{i_1i_2},\cdots)
\nonumber\\
&\equiv \nu_d( 
g_{i_0}, 
a_{i_0i_1} g_{i_1},
a_{i_0i_1} a_{i_1i_2} g_{i_2}, \cdots).
\end{align}
Clearly the partition function $Z_{\nu_d}(M^d_\text{Latt},a_{ij})$
is invariant under the gauge transformation
\begin{align}
 g_i \to f_i g_i, &\ \ \ \ \ a_{ij}\to f_ia_{ij}f_j^{-1};
\nonumber\\
\nu_d^\text{g}(\{f_ig_i\},\{f_ia_{ij}f_j^{-1}\}) &= \nu_d^\text{g}(\{g_i\},\{a_{ij}\});
\nonumber\\
Z_{\nu_d}(M^d_\text{Latt},a_{ij}) &= Z_{\nu_d}(M^d_\text{Latt},f_ia_{ij}f_j^{-1}).
\end{align}
So the partition function $Z_{\nu_d}(M^d_\text{Latt},a_{ij})$ only depend on
the gauge equivalent class of the flat connection $a_{ij}$.  

The volume-independent  partition functions
$Z_{\nu_d}^\text{top}(M^d_\text{Latt},a_{ij})$ are the so called SPT invariants
that suppose to fully characterize the SPT order \cite{W1447,HW1339,W1477,K1459,K1467}.  Using a gauge transformation
to change $g_i \to 1$, we find the SPT invariant to be given by
\begin{align}
 Z_{\nu_d}^\text{top}(M^d_\text{Latt},a_{ij})
&=\ee^{ \ii  2\pi \int_{M^d_\text{Latt}} 
\nu_d^\text{g}(\{g_i=1\},\{a_{ij}\})
}
\nonumber\\
&=
\ee^{ \ii  2\pi \int_{M^d_\text{Latt}} 
\om_d(\{a_{ij}\})
}
\end{align}
where $\om_d$ is the inhomogeneous group-cocycle that corresponds to the
homogeneous group-cocycle $\nu_d$ (see \eqn{homoinhomo}).
The above expression allows us to compute the SPT invariant.

In the following, we will list some the SPT invariants for some simple
SPT states:
\begin{enumerate}
\item
The $Z_n$-SPT states in 2+1D are classified by $\cH^3(Z_n;\R/\Z)=\Z_n$.  For a
$Z_n$-SPT state labeled by $k\in \Z_n$, its SPT invariant is
\begin{align}
Z_k^\text{top}(M^3,a^{\Z_n})=\ee^{ \ii  k\frac{2\pi}{n} \int_{M^3}
a^{\Z_n}\Bs_n a^{\Z_n}}.
\end{align}
\item
The $Z_n\times \t Z_n$-SPT states in 3+1D are classified by $\cH^4(Z_n\times \t Z_n;\R/\Z)=\Z_n^{\oplus 2}$.  For a
$Z_n\times \t Z_n$-SPT state labeled by $i(k_1,k_2)\in \Z_n^{\oplus 2}$, its SPT invariant is
\begin{align}
&\ \ \ \
Z_{k_1,k_2}^\text{top}(M^4,a^{\Z_n},\t a^{\Z_n})
\nonumber\\
&=
\ee^{ \ii  \frac{2\pi}{n} \int_{M^4} 
k_1 a^{\Z_n} \t a^{\Z_n}\Bs_n \t a^{\Z_n}
+k_2 \t a^{\Z_n}  a^{\Z_n}\Bs_n  a^{\Z_n}
}.
\end{align}
\end{enumerate}

\subsection{Group-vertex models that realize $Z_2^T$-SPT orders}
\label{GverT}

To construct a local bosonic model that realize the time-reversal $Z_2^T$ SPT
order, we consider a $\Z_2$-group-vertex model: $g_i \in \Z_2=\{0,1\}$.
The $\Z_2$-group-vertex model on orientable space-time $M^d_\text{Latt}$ is
given by
\begin{align}
 Z_{\nu_d}(M^d_\text{Latt}) 
&=\sum_{ \{g_i\} }   
\ee^{ \ii  2\pi \int_{M^d_\text{Latt}} 
\nu_d(\{g_i\})
} ,
\end{align}
where the homogeneous $\Z_2$-group cocycle $\nu_d(\{g_i\}) \in
\cH^d(\Z_2,(\RZ)_{\Z_2})$ satisfies 
\begin{align}
 \nu_d(\{g_i+1\}) = -\nu_d(\{g_i\}) \text{ mod } 1.
\end{align}
The extra ``$-$'' sign implies that the $\Z_2$ group has a non-trivial action
on $\RZ$ which is indicated by the subscript $\Z_2$ in $(\RZ)_{\Z_2}$.
For example, in 1+1D,
\begin{align}
 \nu_2(g_0,g_1,g_2) =\frac12 [g_1-g_0]_2 [g_2-g_1]_2
\end{align}

Since the $\Z_2$ action correspond to the time-reversal (or orientation
reversal) transformation, to obtain partition function 
with the symmetry twist, we need to put the system on
non-orientable space-time and
to introduce a $\Z_2$ valued 1-cocycle $a_{ij}$
to describe orientation reversal:
\begin{align}
 Z_{\nu_d}(M^d_\text{Latt}) 
&=\sum_{ \{g_i\} }   
\ee^{ \ii  2\pi \int_{M^d_\text{Latt}} 
\nu_d^\text{g}(\{g_i\},\{a_{ij}\}))
} .
\end{align}
where
\begin{align}
\nu_d^\text{g}&(\{g_i\},\{a_{ij}\}) \equiv
\nu_d^\text{g}(g_{i_0},g_{i_1},\cdots,g_{i_d}; a_{i_0i_1},a_{i_1i_2},\cdots)
\nonumber\\
&\equiv \nu_d( 
g_{i_0}, 
a_{i_0i_1}+g_{i_1},
a_{i_0i_1}+a_{i_1i_2}+g_{i_2}, \cdots).
\end{align}

\begin{figure}[tb]
\begin{center}
\includegraphics[scale=0.6]{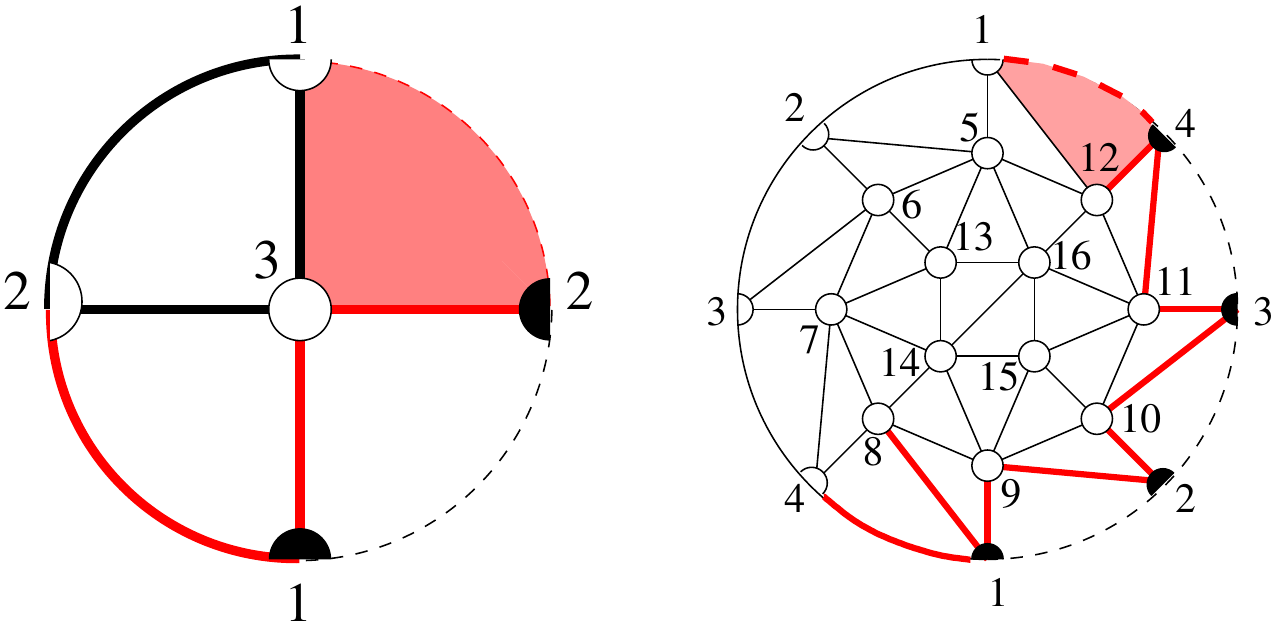}
%7
\end{center}
\caption{ (Color online) 
Two triangulations of $\R P^2$ where the opposite points on the boundary are
identified.  One triangulation has 3 vertices and the other has 16 vertices.
The open dots represent $\t g_i = 0$ and the filled dots represent $\t g_i = 1$
at the vertices.  $\t g_i$ is multivalued since it takes different values on
the same vertex, such as vertex-1 and vertex-2.  The black links represent
$a_{ij}=(\dd \t g)_{ij}=\t g_i-\t g_j=(\rw_1)_{ij}=0$ and the red links
represent $a_{ij}=(\dd \t g)_{ij}=\t g_i-\t g_j=(\rw_1)_{ij}=1$ mod 2.  The
unshaded triangles represent $(\dd \t g\dd \t g)_{ijk} = (\rw_1^2)_{ijk}=\Bs_2 \dd \t g=\Bs_2 \rw_1=0$ and
the shaded triangle represents   $(\dd \t g\dd \t g)_{ijk} = (\rw_1^2)_{ijk}=\Bs_2 \dd \t g=\Bs_2 \rw_1=1$.
(By definition, $(\dd \t g \dd \t g)_{ijk}=[(\t g_i-\t g_j)(\t g_j-\t g_k)]_2$ and
$(\Bs_2 \rw_1)_{ijk}=[\frac {(\rw_1)_{ij}+(\rw_1)_{jk}-(\rw_1)_{ik}}{2}]_2 $ where $i,j,k$ are ordered as
$i<j<k$.)  We see that $\int_{\R P^2} \rw_1^2 = \int_{\R P^2} \dd \t g\dd \t g
=1$.
}
\label{RP2}
\end{figure}

Here $a_{ij}$ is the $\Z_2$ flat connection that describe the orientation of
the manifold (see Fig. \ref{RP2}). In other words, if the orientation does not
change around a loop $C$, then $\sum_{(ij)\in C} a_{ij}=\oint_C a =0$;  if the
orientation changes around a loop $C$, then $\sum_{(ij)\in C} a_{ij}=\oint_C
a=1$ (see Fig. \ref{RP2}).   The above definition implies that $a_{ij}$ is a
$\Z_2$ valued 1-cocycle $a \in C^1(M^d;\Z_2)$.  In fact $a=\rw_1$.  

We can use a multivalued $\Z_2$ gauge transformation to make $a_{ij}=0$, which
changes the single-valued $g_i$ to multivalued $\t g_i$.  If the orientation
changes around a loop $C$, $\t g$ will have to take different values on the
same vertex somewhere on $C$ (see Fig. \ref{RP2}). We see that to realize
$Z_2^T$-SPT order, the local bosonic degrees of freedom must couple to
space-time orientation.  In other words, $(-)^{\t g_i}$ is a pseudo scalar,
which changes sign under time-reversal and parity transformations.  In this
paper, we will also refer $\t g$ as a pseudo scalar field.  Thus if we view $\t
g_i$ as a $\Z_2$-valued 0-cochain, we have (see Fig. \ref{RP2}) 
\begin{align}
\label{w1dg}
 a=\rw_1 = \dd \t g.
\end{align}
In terms of such multivalued $\t g_i$, the partition function can be written as
\begin{align}
\label{Znud}
 Z_{\nu_d}(M^d_\text{Latt}) 
&=\sum_{ \{\t g_i\} }   
\ee^{ \ii  2\pi \int_{M^d_\text{Latt}} 
\nu_d(\{\t g_i\}))
} .
\end{align}

The $Z_2^T$ SPT invariant is given by the corresponding
inhomogeneous cocycle $\om_d$:
\begin{align}
 Z_{\nu_d}^\text{top}(M^d_\text{Latt})
&=\ee^{ \ii  2\pi \int_{M^d_\text{Latt}} 
\nu_d^\text{g}(\{g_i=1\},\{a_{ij}\})
}
\nonumber\\
&=
\ee^{ \ii  2\pi \int_{M^d_\text{Latt}} 
\om_d(\{a_{ij}\})
} .
\end{align}
We can express  $\om_d(\{a_{ij}\})$ in terms of $a_{ij}$ (see Fig. \ref{RP2}):
\begin{align}
 \om_d (\{a_{ij}\}) =  \begin{cases}
 \frac 12 a^d &\text{ if } d = \text{even},\\
 0 &\text{ if } d = \text{odd}.\\
\end{cases}
\end{align}
Thus, the $Z_2^T$ SPT invariant is given by
\begin{align}
 Z_{\nu_d}^\text{top}(M^d)
&=\ee^{ \ii  \pi \int_{M^d} 
\rw_1^d
}.
\end{align}
From $\rw_1^d=\Sq^1(\rw_1^{d-1}) = (d-1) \rw_1^d $, we see that
$\rw_1^d=0$ mod 2 automatically, when $d=$ odd.  So the above expression for
the $Z_2^T$ SPT invariant is valid for both $d=$ even and $d=$ odd.

Last, we like to mention that, using multivalued $\t g_i$, we can also 
express  the non-trivial homogeneous cocycle $\nu_d(\{\t g_{i}\})$ as (see Fig.
\ref{RP2})
\begin{align}
 \nu_d (\{\t g_{i}\}) =  \begin{cases}
 \frac 12 (\dd \t g)^d &\text{ if } d = \text{even},\\
 0 &\text{ if } d = \text{odd} ,\\
\end{cases}
\end{align}
since $a=\dd \t g$.
This allows us to rewrite \eqn{Znud} as (see Fig. \ref{RP2})
\begin{align}
 Z_{\nu_d}(M^d_\text{Latt}) 
&=\sum_{ \{\t g_i\} }   
\ee^{ \ii  \pi \int_{M^d_\text{Latt}} 
(\dd \t g)^d
} ,
\end{align}
for even $d$.

\subsection{Group-link model and emergent Dijkgraaf-Witten gauge theory}

Now let us construct local bosonic models -- group-link models, whose
topological orders are described by Dijkgraaf-Witten gauge theory.  The local
degrees of freedom of the group-link model are group elements living on the
links of the space-time lattice $M^d_\text{Latt}$: $a_{ij}\in G$ that satisfies
$a_{ij}=G_{ji}^{-1}$.  Then, using the inhomogeneous group-cocycle
$\om_d(\{a_{ij}\})$, we can construct a group-link model
\cite{DW9093,HW1267,WW1454,WH14093216}
\begin{align}
&\ \ \ \ Z_{G,\om_d}(M^d_\text{Latt}) 
\\
&=\sum_{ \{a_{ij}\}\atop a_{ij}a_{jk}a_{ki}=1}   
\ee^{ \ii  2\pi \int_{M^d_\text{Latt}} 
\om_d(\{a_{ij}\}) - U \sum_{(ijk)} |a_{ij}a_{jk}a_{ki}-1|
},
\nonumber 
\end{align}
where $\sum_{(ijk)}$ sums over all 3-simplices, and $U \to +\infty $.

Note that the above model is a local bosonic model, not the Dijkgraaf-Witten
gauge theory.  The  Dijkgraaf-Witten gauge theory is defined by
\begin{align}
&\ \ \ \ Z_{G,\om_d,\text{DW}}(M^d_\text{Latt}) 
\\
&=\sum_{ [\{a_{ij}\}]\atop a_{ij}a_{jk}a_{ki}=1 }  
\ee^{ \ii  2\pi \int_{M^d_\text{Latt}} 
\om_d(\{a_{ij}\}) - U \sum_{(ijk)}  |a_{ij}a_{jk}a_{ki}-1|
} , 
\nonumber 
\end{align}
where the summation $\sum_{ [\{a_{ij}\}]}$ is over the gauge equivalent class,
$[\{a_{ij}\}]$, of the configurations, $\{a_{ij}\}$.  In contrast, the
summation  $\sum_{ \{a_{ij}\}}$ in the group-link model is over all the
configurations, $\{a_{ij}\}$ (without the gauge reduction).  However, the
volume-independent partition function of the two models are the same
\begin{align}
 Z_{G,\om_d,\text{DW}}^\text{top}(M^d_\text{Latt})=Z_{G,\om_d}^\text{top}(M^d_\text{Latt})
\end{align}
So the two models have the same emergent topological order.

\begin{figure}[tb]
\begin{center}
\includegraphics[scale=1.0]{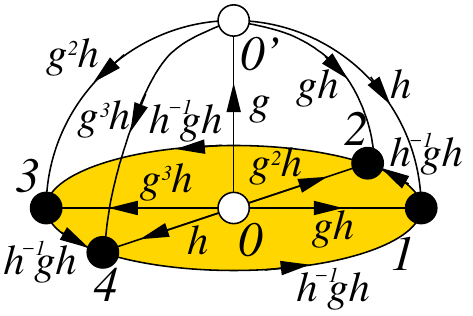}
%8
\end{center}
\caption{ (Color online) 
The lens space, $L^3(p)$, is obtain by identifying the bottom and the top disk
after a $2\pi/p$ rotation, \eg link-$(01)$ and link-$(0'2)$ are identified,
link-$(02)$ and link-$(0'3)$ are identified, link-$(12)$ and link-$(23)$ are
identified, \etc.
}
\label{lensCW1}
\end{figure}

As an example, let us compute the topological invariant for 2+1D lens space
$L^3(p)$, using the explicit  CW-complex decomposition in Fig. \ref{lensCW1}:
\begin{align}
\label{ZDWLp}
 Z_{G,\om_3}^\text{top}(L^3(p))
& =
\frac{1}{|G|^2} \sum_{g,h\in G\atop g^p=1} 
 \ee^{\ii 2\pi \sum_{m=0}^{p-1} \om_3(g,g^mh,h^{-1}gh) } .
\end{align}
For $G=\Z_n$, $\om_3 \in \cH^3(\Z_n,\RZ)=\Z_n$ is labeled by $k\in \Z_n$:
\begin{align}
\label{Lngco}
 \om_3(g_1,g_2,g_3)=\frac{k}{n^2} g_1(g_2+g_3-[g_2+g_3]_n).
\end{align}
We find that
\begin{align}
 Z_{\Z_n,k}^\text{top}(L^3(p)) =
 Z_{k;a\Bs a\Z_n}^\text{top}(L^3(p)) .
\end{align}
In fact, the topological term in the $Z_n$ Dijkgraaf-Witten theory and the
topological term in the $\Z_n$-1-cocycle model are directly related
\begin{align}
 2\pi \int_{M^3} \om_3(\{a^{\Z_n}\})
 =k\frac {2\pi} n \int_{M^3} a^{\Z_n} \Bs_n a^{\Z_n} ,
\end{align}
as one can see from \eqn{Lngco} and the explicit expression of $a^{\Z_n} \Bs_n a^{\Z_n}$:  
\begin{align}
 \<a^{\Z_n} \Bs_n  a^{\Z_n}, &(ijkl)\>
 =a^{\Z_n}_{ij} \<\Bs_n a^{\Z_n},(jkl)\>;
\nonumber\\
\<\Bs_n a^{\Z_n},(jkl)\> &=
\frac 1n (a^{\Z_n}_{jk}+a^{\Z_n}_{kl} - a^{\Z_n}_{jl})
\\
&=
\frac 1n (a^{\Z_n}_{jk}+a^{\Z_n}_{kl} - [a^{\Z_n}_{jk}+a^{\Z_n}_{kl}]_n).
\nonumber 
\end{align}
Therefore, the $\Z_n$-1-cocycle model
realizes the $Z_n$ Dijkgraaf-Witten theory.

\subsection{Symmetric topological orders described by gauge theories}

We can also construct local bosonic models (called mixed group-vertex
group-link models) that will produce topological orders described by a
$G_\text{gauge}$-gauge theory that also have a symmetry $G_\text{symm}$.  In
the mixed model, the local degrees of freedom of are group elements $g_{i} \in
G_\text{symm}$ living on the links group elements $a_{ij} \in G_\text{gauge}$
living on the links of the space-time lattice $M^d_\text{Latt}$.  Then, using
the homogeneous group-cocycle $\nu_n(\{g_{i}\}) \in
\cH^n(G_\text{symm},\RZ)$, and the inhomogeneous group-cocycle
$\om_{d-n}(\{a_{ij}\}) \in \cH^{d-n}(G_\text{gauge},\RZ)$, we can construct
the mixed model
\begin{align}
&Z_{\nu_n\om_{d-n}}(M^d_\text{Latt}) 
\\
&= \sum_{ \{g_i,a_{ij}\}, a_{ij}a_{jk}a_{ki}=1}   
\ee^{ \ii  2\pi \int_{M^d_\text{Latt}} 
\nu_n(\{g_{i}\}) \om_{d-n}(\{a_{ij}\}) 
},
\nonumber 
\end{align}

We can also  construct a more general mixed model using  inhomogeneous
group-cocycle $\om_d \in \cH^d(G_\text{symm}\times G_\text{gauge},\RZ)$:
\begin{align}
&Z_{\nu_n\om_{d-n}}(M^d_\text{Latt}) 
\\
&= \sum_{ \{g_i,a_{ij}\}, a_{ij}a_{jk}a_{ki}=1}   
\ee^{ \ii  2\pi \int_{M^d_\text{Latt}} 
\om_d(\{ (g_i^{-1}g_j, a_{ij})\}) 
},
\nonumber 
\end{align}
where $(g_i,a_{ij})$ is the group element of $G_\text{symm}\times
G_\text{gauge}$.  

We can construct an even more general mixed model using
inhomogeneous group-cocycle $\om_d \in \cH^d(G_\text{PSG},\RZ)$ \cite{HW1267}:
\begin{align}
&Z_{\om_d;G_\text{PSG}}(M^d_\text{Latt}) 
\\
&= \sum_{ \{g_i,a_{ij}\}, a^\text{PSG}_{ij}a^\text{PSG}_{jk}a^\text{PSG}_{ki}=1}   
\ee^{ \ii  2\pi \int_{M^d_\text{Latt}} 
\om_d(\{ (g_i^{-1}g_j, a_{ij})\}) 
},
\nonumber 
\end{align}
where $G_\text{PSG}$ is a group that contains $G_\text{gauge}$ as a normal
subgroup such that $G_\text{PSG}/G_\text{gauge}=G_\text{symm}$, and
$a_{ij}^\text{PSG}=(g_i^{-1}g_j,a_{ij})$ is the group element of $G_\text{PSG}$
\cite{W0213}.  In other words, $G_\text{PSG}$ is an extension of
$G_\text{symm}$ by $G_\text{gauge}$, which is also described by the following
short exact sequence
\begin{align}
 1\to G_\text{gauge} \to G_\text{PSG} \to G_\text{symm} \to 1.
\end{align}
In this case, as discussed in \Ref{W0213}, a gauge charge does not transform as
a representation of $G_\text{gauge}$, but rather, transforms as a
representation of $G_\text{PSG}$. Under the symmetry transformation, the gauge
charge transforms according to $G_\text{PSG}$ (which is called the projective
symmetry group).  In fact, $G_\text{PSG}$ describes the so called ``symmetry
fractionalization''.

If there is a symmetry twist described by $a^\text{symm}_{ij} \in
G_\text{symm}$ on the links, then the partition function will be
\begin{align}
\label{ZPSG}
&Z_{\om_d;G_\text{PSG}}(M^d_\text{Latt},a^\text{symm}_{ij}) 
\\
&= \sum_{ \{g_i,a_{ij}\}, a_{ij}a_{jk}a_{ki}=1}   
\ee^{ \ii  2\pi \int_{M^d_\text{Latt}} 
\om_d(\{ (g_i^{-1}a^\text{symm}_{ij}g_j, a_{ij})\}) 
}.
\nonumber 
\end{align}
The above construction also applies to the situation where $G_\text{symm}$
contains time-reversal symmetry. In that case, $a^\text{symm}_{ij}$ will
contain contribution from the change of the orientations of the manifold, and
$\om_d \in \cH^d(G_\text{PSG},(\RZ)_T)$ where time-reversal $T\in
G_\text{PSG}$ will have an sign-changing action on $\RZ$.

If we include $\Z_n$-2-cochain field $b^{\Z_n}$,  we can construct new general
local boson models with emergent symmetric topological order, such as
\cite{KT1321}
\begin{align}
&\ \ \ \
Z_{b^{\Z_n}\om_{d-2};G_\text{PSG}}(M^d_\text{Latt},a^\text{symm}_{ij})  =
\\
& \sum_{ \{g_i,a_{ij},b^{\Z_n}_{ijk}\}, \dd b^{\Z_n}=0 \atop a_{ij}a_{jk}a_{ki}=1}   
\ee^{ \ii  2\pi \int_{M^d_\text{Latt}} 
b^{\Z_n}\om_{d-2}(\{ (g_i^{-1}a^\text{symm}_{ij}g_j, a_{ij})\}) 
},
\nonumber 
\end{align}
where we have assumed that $n \om_{d-2} =0$.
This model has an emergent $Z_n\times G_\text{gauge}$-gauge theory with
$G_\text{symm}$ symmetry.  When, $G_\text{gauge}=1$, the $Z_n$ charge may carry
a projective representation of $G_\text{symm}$.  When, $G_\text{symm}=1$, the
$Z_n$ charge may carry a projective representation of $G_\text{gauge}$.  In
general, the  $Z_n$ charge may carry projective representation of
$G_\text{PSG}$ (\ie with mixed fractionalized symmetry $G_\text{symm}$ charge
and gauge $G_\text{gauge}$ charge).

\section{Time-reversal symmetric topological orders}

In this section, we are going construct exactly soluble local bosonic models
that have time-reversal symmetry and emergent time-reversal symmetric
topological orders.  The time-reversal symmetry $T$ is described by the
symmetry group $Z_2^T$, which means $T^2=1$.  We will first construct 2+1D
models and then 3+1D models.  All the  3+1D models realize time-reversal
symmetric $Z_2$-gauge theories at low energies.

\subsection{2+1D time-reversal symmetric $\Z_2$-1-cocycle models}
\label{Z2aT3}

\subsubsection{Model construction}

We start with the $\Z_2$-1-cocycle models which produce time-reversal-symmetry
enriched $Z_2$ topological orders and double-semion topological orders in 2+1D.
The partition function has a form
\begin{align}
 Z_{\Z_2\text{a}T}(M^3_\text{Latt}) 
=\sum_{ \{a^{\Z_2}_{ij}\}, \dd a^{\Z_2}=0 }   
\ee^{ \ii \pi \int_{M^3_\text{Latt}} 
W(a^{\Z_2},\rw_m)
}
\end{align}
The possible topological terms $W(a^{\Z_2},\rw_m)$ are mixture of 1-cocycle
$a^{\Z_2}$ and Stiefel-Whitney classes $\rw_m$.  Here $W(a^{\Z_2},\rw_m)$ has
its value in $\Z_2$.  Thus $\ee^{ \ii \pi \int_{M^3_\text{Latt}}
W(a^{\Z_2},\rw_m) } =\pm 1$ and there is time-reversal symmetry in our model.
Also since $W(a^{\Z_2},\rw_m) \in C^3(M^3_\text{Latt};\Z_2)$, $\ee^{ \ii \pi
\int_{M^3_\text{Latt}} W(a^{\Z_2},\rw_m) } $ is well define even for
non-orientable manifold $M^3_\text{non}$ where $H_3(M^3_\text{non};\Z)=0$ but
$H_3(M^3_\text{non};\Z_2)=\Z_2$. We also note that for non-orientable manifold,
$M^3_\text{non}$ itself is a chain with boundary(\ie $M^3_\text{non}$ is not a
cycle).  Therefore $\int_{M^3_\text{non}} \dd b \neq 0$, for a 2-cochain $b$. 

The possible topological terms are given by
the combinations of the following six 3-cocycles:
\begin{align}
& \rw_1^3, && \rw_1 \rw_2, && \rw_3,
\nonumber\\
& (a^{\Z_2})^3, && \rw_1 (a^{\Z_2})^2, && \rw_1^2 a^{\Z_2}.
\end{align}
From Appendix \ref{aSWrel3}, we find many relations between Stiefel-Whitney
and the $\Z_2$-1-cocycle:
\begin{align}
\rw_1^2  &=\rw_2,\ \ \rw_1\rw_2=\rw_3=0,
\nonumber\\
\rw_1(a^{\Z_2})^2& =\Sq^1((a^{\Z_2})^2)=2(a^{\Z_2})^3=0.
\end{align}
So the most general time-reversal symmetric $\Z_2$-1-cocycle model that couples
to Stiefel-Whitney classes is given by
\begin{align}
&\ \ \ \
 Z_{k_1k_2;\text{t}\Z_2\text{a}T}(M^3_\text{Latt}) 
\nonumber\\
&=\sum_{ \{a^{\Z_2}_{ij}\}, \dd a^{\Z_2}=0 }   
\ee^{ \ii  \pi \int_{M^3_\text{Latt}} 
 k_1 a^{\Z_2}\Bs_2a^{\Z_2}
+k_2 \rw_1^2  a^{\Z_2}
}
\end{align}
where $k_1,k_2 \in \Z_2$, and we have used $(a^{\Z_2})^3 =a^{\Z_2}\Bs_2a^{\Z_2}
$.  
%This represents a classification of time-reversal symmetric $Z_2$ gauge
%theory, where $k_1,k_2$ represent the group-cocycle twists to the $Z_2$ gauge
%theory.  This agrees with the group cohomology result $\cH^3(Z_2\times
%Z_2^T,\RZ)=\Z_2\oplus \Z_2$ \cite{CGL1314,CGL1204,K1459}.  

We like to remark that the Stiefel-Whitney class $\rw_1$ in the above
path integral can be induced by a local degrees of freedom -- 
a pseudo-scalar $\t g_i$ introduced in Section \ref{GverT}.
Using $\rw_1=\dd \t g_i - \dd g_i$, where $ g_i$ is $\Z_2$-single-valued
0-cochain, we can rewrite the above path integral as
(the $g_i$ dependence disappears)
\begin{align}
\label{tZ2aT3}
&\ \ \ \
 Z_{k_1k_2;\text{t}\Z_2\text{a}T}(M^3_\text{Latt}) 
\nonumber\\
&=\sum_{ \{\t g_i, a^{\Z_2}_{ij}\}, \dd a^{\Z_2}=0 }   
\ee^{ \ii  \pi \int_{M^3_\text{Latt}} 
 k_1 a^{\Z_2}\Bs_2a^{\Z_2}
+k_2 \Bs_2\dd \t g  a^{\Z_2}
}
\end{align}
which is a pure local bosonic model. 
%More detailed discussions can be found in Section \ref{remark}.

The above four local bosonic models with different values of $k_1,k_2$ give
rise to four different time-reversal-symmetry enriched topological orders.  If
we break the time-reversal symmetry, the  above local bosonic model will only
give rise to two different topological orders labeled by $k_1$: the $Z_2$
topological order (\ie the $Z_2$ gauge theory) for $k_1=0$ and the
double-semion topological order for $k_1=1$ .

\subsubsection{Topological partition functions}
\label{Z2aTpf}

Next, we will compute the volume independent partition function, which is given
by
\begin{align}
\label{Z2aTtop}
&\ \ \ \
 Z_{k_1k_2;\text{t}\Z_2\text{a}T}^\text{top}(M^3) 
\nonumber\\
&=\frac 12 \sum_{a^{\Z_2}\in H^1(M^3;\Z_2)}   
\ee^{ \ii  \pi \int_{M^3} 
 k_1 a^{\Z_2}\Bs_2a^{\Z_2}
+k_2 \rw_1^2  a^{\Z_2}
}
\end{align}
On $M^3=S^1\times \Si_g$, $\int_{M^3} 
 k_1 a^{\Z_2}\Bs_2a^{\Z_2}
+k_2 \rw_1^2  a^{\Z_2}=0$. Thus
\begin{align}
  Z_{k_1k_2;\text{t}\Z_2\text{a}T}^\text{top}(S^1\times \Si_g)=2^{2g}
\end{align}

On $M^3=S^1\times \Si^\text{non}_g$, 
we note that the cohomology ring $H^*(S^1\times \Si^\text{non}_g;\Z_2)$
has a basis
\begin{align}
 H^*(S^1\times \Si^\text{non}_g;\Z_2)=
\{ a_0, a_i|_{i=1,\cdots,g}, a_0a_i,b,a_0b 
\}
\end{align}
with $a_0,a_i \in H^1(S^1\times \Si^\text{non}_g;\Z_2)$ and $b \in
H^2(S^1\times \Si^\text{non}_g;\Z_2)$, which have the following cup product:
\begin{align}
 a_i^2=b,\ \ \ a_0^2=a_ib=0.
\end{align}
The Stiefel-Whitney classes are given by 
\begin{align}
\rw_1=\sum_{i=1}^g a_i, \ \ \ \ \ \ 
\rw_2=\rw_1^2=[g]_2 b,
\end{align}
and the Bockstein homomorphism is given by
\begin{align}
 \Bs_2 a_i =(a_i)^2=b,\ \ \
 \Bs_2 a_0 = 0.
\end{align}
Expand
\begin{align}
 a^{\Z_2}=\sum_{\mu=0}^g\al_\mu a_\mu,
\end{align}
we find that
\begin{align}
&\ \ \ \
 Z_{k_1k_2;\text{t}\Z_2\text{a}T}^\text{top}(S^1\times \Si^\text{non}_g) 
\nonumber\\
&=\frac 12 \sum_{a_\mu=0,1}   
\ee^{ \ii  \pi ( 
 k_1 \al_0 \sum_{i=1}^g \al_i 
+k_2 g \al_0)
}
\nonumber\\
&=
\sum_{a_i=0,1} \del_2(k_1 \sum_{i=1}^g \al_i +k_2 g) 
\nonumber\\
&=
(1-k_1) [k_2g+1]_2 2^g + k_1 2^{g-1}
\end{align}
The results are summarized in  Table \ref{Tsymm3}.

We like to remark that $Z_2\times Z_2^T$-SPT states are classified by
$\cH^3(Z_2\times Z_2^T; (\R/\Z)_T)=\Z_2^{\oplus 2}$.  For a $Z_2\times
Z_2^T$-SPT state labeled by $(k_1,k_2)\in \Z_2^{\oplus 2}$, its SPT invariant
is given by $Z^\text{top}(M^3,a^{\Z_2})=\ee^{ \ii  \pi \int_{M^3} k_1
a^{\Z_2}\Bs_2a^{\Z_2} +k_2 \rw_1^2  a^{\Z_2} }$, where $a^{\Z_2}$ describes the
$Z_2$-symmetry twist on $M^3$.  Such SPT invariant happen to be the phase
factor in \eqn{Z2aTtop}, and the in summation in \eqn{Z2aTtop} happen to be the
summation of all possible $Z_2$-symmetry twists.  The implies that the
topological orders produced by the 2+1D $\Z_2$-1-cocycle model can be regarded
as the $Z_2$-gauged $Z_2\times Z_2^T$-SPT states.

\subsubsection{Properties of excitations}
\label{Z2aT3pex}

When $k_1=0$, the 2+1D $\Z_2$-1-cocycle model has an emergent $Z_2$ topological
order described by a low energy $Z_2$-gauge theory. It has four types of
point-like excitations: $1$, $e$, $m$, $\veps=em$, where $\veps$ is a fermion
and others are bosons.  When $k_1=1$, the cocycle model has an emergent
double-semion topological order. It has four types of point-like excitations:
$1$, $e$, $m$, $\veps$, where $e$ is an semion with spin $\frac 14$, and
$\veps$ an semion  with spin $-\frac 14$. $1$ and $e$ are bosons, and they
carry $Z_2$-charge 0 and 1 respectively. 

To obtain more properties of the excitations in those $T$-symmetric topological
orders, let us consider dimension reduction.  In general, when we reduce a
stable phase $\cC^d$ in $d$-dimension to lower dimension $d'$ via a
compactification $M^d \to M^{d'}\times N^{d-d'}$, the resulting lower
dimensional phase on $M^{d'}$ may correspond to several stable phases
$\cC^{d'}_i$ with accidental degenerate energy.\cite{MW1514} We denote such
dimension reduction as
\begin{align}
 \cC^d = \bigoplus_i \cC^{d'}_i ,
\end{align}
and refer $\cC^{d'}_i$'s as different sectors.  The different sector arise from
different field configurations on  $N^{d-d'}$.  We like to ask: what are
effective theories for those $d'$-dimensional systems in each sector? 

To apply the above general picture to our case, let us assume the space-time to
be $M^3=M^2\times S^1$ and $S^1$ is a small circle. We can view the 2+1D
$\Z_2$-1-cocycle models as a 1+1D local bosonic systems.  Then what is the
effective theory for such 1+1D systems?

To answer the above question, we can write $a^{\Z_2}$ as
$a^{\Z_2}=a^{\Z_2}_{M^2}+a^{\Z_2}_{S^1}$, where $a^{\Z_2}_{M^2}$ are low energy
degrees of freedom only live on $M^2$ (\ie constant in the $S^1$ direction),
and $a^{\Z_2}_{S^1}$ are high energy degrees of freedom only live on $S^1$ (\ie
constant in the $M^2$ directions).  The different  field configurations on
$S^1$ are labeled by $\al = \int_S^1 a^{\Z_2}_{S^1} \in \Z_2$.
So the different sectors are also  labeled by $\al=0,1$.
The partition function on $M^2\times S^1$
becomes
\begin{align}
\label{1Deff}
&\ \ \ \
 Z_{k_1k_2;\text{t}\Z_2\text{a}T}(M^2\times S^1) 
\nonumber\\
&=\sum_{ \{a^{\Z_2}_{ij}\}, \dd a^{\Z_2}=0 }   
\ee^{ \ii  \pi \int_{M^2\times S^1} 
 k_1 a^{\Z_2}\Bs_2a^{\Z_2}
+k_2 \Bs_2\dd \t g  a^{\Z_2}
}
\nonumber\\
&=
\sum_{ \{a^{\Z_2}_{ij}\}, \dd a^{\Z_2}_{M^2}=0 }   
\ee^{ \ii  \pi  \al \int_{M^2} 
 k_1 \Bs_2 a^{\Z_2}_{M^2}
+k_2 \Bs_2\dd \t g  
}
\end{align}

\begin{figure}[tb]
\begin{center}
\includegraphics[scale=0.5]{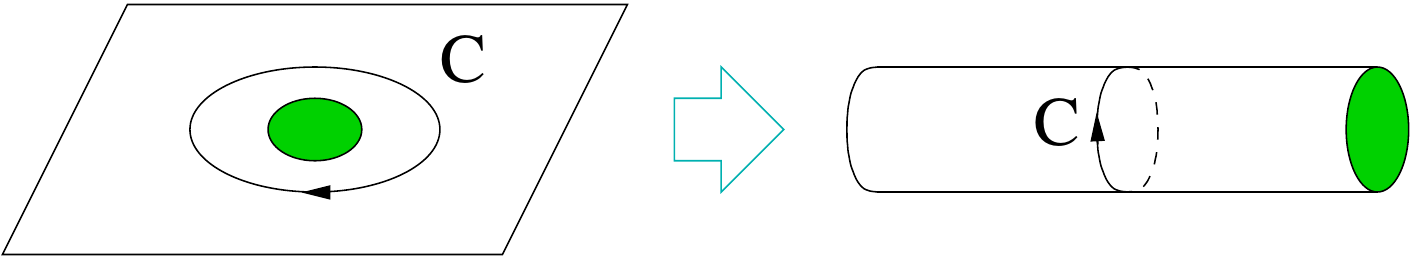} \end{center}
%9
\caption{ (Color online) 
In a dimension reduction from 2D space to 1D space (a cylinder),
a hole in the 2D space becomes an end of the 1D space.
The $Z_2$-vortex with $\int_C a^{Z_2}=1$ in 2D space becomes
the $\int_C a^{Z_2}=1$ sector in the 1D space.
}
\label{2D1D}
\end{figure}

We see that in the sector $\al=0$, the resulting 1+1D $Z_2^T$ SPT order is
trivial.  In contrast, in the sector $\al=1$, the resulting 1+1D $Z_2^T$-SPT
order is non-trivial.  Usually, in 1+1D, the gauge field $a^{\Z_2}_{M^2}$
fluctuate strongly. Here, we want to treat the 1+1D system as reduced from the
2+1D system as shown in Fig. \ref{2D1D}. In this case, we can assume the  gauge
field $a^{\Z_2}_{M^2}$ to fluctuate weakly, and treat $a^{\Z_2}_{M^2}$ as a
background probe field.  Therefore, we can view the 1+1D system as a system
with $Z_2\times Z_2^T$ symmetry.  Then form the 1+1D effective theory
\eqn{1Deff} which can be viewed as an SPT invariant \cite{W1447}, we see that
in the sector $\al=1$ is described by a $Z_2\times Z_2^T$-SPT state labeled by
$(k_1,k_2)$, which agrees with the group cohomology result $\cH^2(Z_2\times
Z_2^T,\RZ_T)=\Z_2^{\oplus 2}$.

If $(k_1,k_2)=(0,1)$, the 1+1D SPT state is a pure $Z_2^T$-SPT state as
indicated by the term $\ee^{ \ii  \pi \int_{M^2} k_2 \Bs_2\dd \t g  }$.  Such SPT
state has Kramer doublet at the chain end.  In fact, the chain end has to
sector with $Z_2$-charge 0 and with $Z_2$-charge 1.  Both sectors are  Kramer
doublets.  We may view the 1+1D system with a chain end as a 2+1D system with a
hole as described in Fig. \ref{2D1D}.  Thus the $\al=1$ sector, correspond to a
$\pi$-flux in 2+1D. We see that a $\pi$-flux carries a Kramer doublet
regardless if it carries addition $Z_2$-charge or not.  Similarly, the $\al=0$
sector gives rise to trivial 1+1D SPT state, and thus a $\pi$-flux carries a
time-reversal singlet regardless if it carries addition $Z_2$-charge or not.
To summarize, \frmbox{the 2+1D $\Z_2$-1-cocycle model labeled by
$(k_1,k_2)=(0,1)$ has four types of point-like excitations $1$, $e$, $m$,
$\veps=em$. The excitations $m,\ \veps$ carry $\pi$-flux, while the excitations $e,\ \veps$
carry a $Z_2$-gauge 1. The excitations $m, \ \veps$ are Kramer doublets and the
excitation $\veps$ is a fermion (see Table \ref{Tsymm3}).}

The time-reversal singlet has a quantum dimension $d=1$ and the Kramer doublet
has a quantum dimension $d=2$.  (Quantum dimension is the dimension of the
Hilbert space for the internal degrees of freedom carried by a particle.)  Thus
the four types of particles have the following quantum dimensions
$(d_1,d_e,d_m,d_f)=(1,1,2_-,2_-)$, where the subscript $-$ indicates the Kramer
doublet.  A particle can also carry spin $s$, which is defined mod 1.  A boson
has spin 0 mod 1 and a fermion has spin $\frac12$ mod 1.  Thus, the four types
of particles have the following spins  $(s_1,s_e,s_m,s_f)=(0,0,0,\frac12)$ (see
Table \ref{Tsymm3}).
 
If $(k_1,k_2)=(1,0)$, the cocycle model has four excitations: $1$, $e$, $m$,
$\veps$. $1$ and $e$ transform as time-reversal singlet.  $m$ and $\veps$
transform into each other and form a time-reversal doublet.  Since $m$ and
$\veps$ are always degenerate with time-reversal symmetry, we view them as a
single type of excitation with quantum dimension $2$.  Thus \frmbox{the 2+1D
$\Z_2$-1-cocycle model labeled by $(k_1,k_2)=(1,0)$ has three types of
point-like excitations with quantum dimensions $(d_i)=(1,1,2)$ and spins
$(s_i)=(0,0,[\frac14,\frac 34])$.}

Under the dimension reduction, the 1+1D state in $\al=1$ sector is a $Z_2\times
Z_2^T$-SPT state described by the SPT invariant $\ee^{ \ii  \pi \int_{M^2}
\Bs_2a^{\Z_2}_{M^2}  }$.  The chain end for such a $Z_2\times Z_2^T$-SPT is a
doublet with fraction $Z_2$-charge $\pm \frac12$.  Under the time reversal, the
$+\frac12$ and $-\frac12$ $Z_2$-charge states get exchanged and $T^2=1$.  Thus,
the $\pi$-flux in 2+1D ground state will carries a doublet of $\pm \frac12$
$Z_2$-charges.  There are 2 types of $0$-flux excitations with $0$ and $1$
$Z_2$-charges.  Those two types of excitations are time-reversal singlet.  Thus
we denote that quantum dimensions for those excitations as $(d_i)=(1,1,2_+)$,
where subscript $+$ indicates $T^2=1$ (see Table \ref{Tsymm3}).

If $(k_1,k_2)=(1,1)$, under the dimension reduction, the 1+1D state in $\al=1$
sector is a $Z_2\times Z_2^T$-SPT state described by the SPT invariant $\ee^{
\ii  \pi \int_{M^2} \Bs_2a^{\Z_2}_{M^2} +\Bs_2\dd \t g }$.  The chain end for
such a $Z_2\times Z_2^T$-SPT states may contain four degenerate states formed
by a doublet with fraction $Z_2$-charge $\pm \frac12$ and a Kramer doublet.
The time-reversal transformation is described by
\begin{align}
 T &= {\footnotesize\begin{pmatrix}
 0 & 0 & 0 & -1 \\
 0 & 0 & 1 & 0 \\
 0 & -1 & 0 & 0 \\
 1 & 0 & 0 & 0 \\
\end{pmatrix}} K = \si^1\otimes \ii\si^2 K, \ \
T^2=-1,
\end{align}
where $K$ is the anti-unitary transformation, and $\si^{1,2,3}$ are the Pauli
matrices.  
The $Z_2$ symmetry is generated by
\begin{align}
 Q={\footnotesize\begin{pmatrix}
 \ii & 0 & 0 & 0 \\
 0 & \ii & 0 & 0 \\
 0 & 0 & -\ii & 0 \\
 0 & 0 & 0 & -\ii \\
\end{pmatrix}}  = \ii \si^3\otimes \si^0 , \ \
Q^2=-1.
\end{align}
However, the four states can be split by a time-reversal and $Z_2$ symmetric
perturbation
\begin{align}
 \del H = \Del {\footnotesize \begin{pmatrix}
 1 & 0 & 0 & 0 \\
 0 & -1 & 0 & 0 \\
 0 & 0 & -1 & 0 \\
 0 & 0 & 0 & 1 \\
\end{pmatrix} }=\Del \si^3\otimes \si^3 .
\end{align}
Thus the chain end in general has a doublet with fractional $Z_2$-charge $\pm
\frac12$ which is also a $T^2=-1$ Kramer doublet at the same time.  As a
result, the $\pi$-flux in 2+1D ground state will carries  a Kramer doublet with
fractional $Z_2$-charge $\pm \frac12$.  We stress that there is no
time-reversal symmetric perturbation that can give rise to $T^2=1$  doublet.
To summarize, \frmbox{the 2+1D $\Z_2$-1-cocycle model labeled by
$(k_1,k_2)=(1,1)$ has three types of point-like excitations with quantum
dimensions $(d_1,d_e,d_s)=(1,1,2_-)$ and spins
$(s_1,s_e,s_s)=(0,0,[\frac14,\frac 34])$, where subscript ``$-$'' indicates
$T^2=-1$ (see Table \ref{Tsymm3}).}

\subsection{2+1D time-reversal symmetric $Z_4^T$-group cohomology models}

\subsubsection{Model construction}

Using the group cocycles, we can construct more local bosonic models that can
produce time-reversal symmetric 2+1D (twisted) $Z_2$-gauge theories at low
energy (see \eqn{ZPSG}).  In this section, we will discuss those models.

We put $\Z_2$ degrees of freedom on both vertices and links: $\t g_i \in \Z_2$
and $a_{ij}^{\Z_2}\in \Z_2$.  Note that $\t g_i$ is a pseudo scalar as
discussed in Section \ref{GverT} (see Fig. \ref{RP2}).  Using
\begin{align}
\label{Z2Z4Z2}
 1\to \Z_2 \to \Z_4 \to \Z_2\to 1,
\end{align}
we can construct a $\Z_4$-1-cocycle field
\begin{align}
 a_{ij}^{\Z_4} 
= 2 a_{ij}^{\Z_2} + (\dd \t g)_{ij}.
\end{align}
Notice that
$\cH^3(\Z_4,(\RZ)_{\Z_4}) = 0$.  Thus there is no group cocycle term in the
action amplitude.  We obtain the following time-reversal symmetric model
\begin{align}
 Z_{Z_4^T}(M^3) 
&=\sum_{ \{a_{ij}^{\Z_2}, \t g_i\}, \dd a^{\Z_4}=0 \atop
a^{\Z_4}=2 a^{\Z_2} + \dd \t g 
 }   
1
\end{align}
The condition $\dd a^{\Z_4}=0$ becomes (when we view the cochains as $\Z$
valued)
\begin{align}
 &\dd a^{\Z_4} = 2\dd a^{\Z_2} +\dd (\dd \t g) \se{4} 0 
\nonumber\\
\to\ \  & \dd a^{\Z_2} +\Bs_2 \dd \t g \se{2} 0 .
\end{align}
We can rewrite the above
partition function as
\begin{align}
 Z_{Z_4^T}(M^3) 
&=\sum_{ \{a_{ij}^{\Z_2}, \t g_i\}, \dd a^{\Z_2}\se{2} \Bs_2 \dd \t g
 }   
1 .
\end{align}
We see that such a
model is different from the model \eqn{tZ2aT3} with $k_{1,2}=0$.  The condition
$\dd a^{\Z_2}=\Bs_2\dd \t g $ encode the non-trivial group extension
\eq{Z2Z4Z2}.

Due to the relation $\Bs_2\dd \t g = \Bs_2\rw_1\se{2} \rw_1^2$,
$Z_{Z_4^T}(M^3)\neq 0$ only when $\rw_1^2=0$ as $\Z_2$-valued cohomology class.
Thus we introduce
\begin{align}
\bar \del_m(c) = \begin{cases}
0 & \text{ if } c \neq \dd b  \text{ mod } m,\\
1 & \text{ if } c = \dd b  \text{ mod } m.\\
\end{cases}
\end{align}
So $Z_{Z_4^T}(M^3)$ contains a factor $\bar\del_2(\Bs_2 \rw_1)$.  Furthermore,
on space-time $M^3$ with $\Bs_2 \rw_1=0$, we have $\dd a^{\Z_2}\se{2} 0$. In
this case, we can combine the $Z_2$-1-cocycle model and the $Z_4^T$ group
cohomology model together:
\begin{align}
\label{Z2Z4T}
&\ \ \ \
 Z_{k_0k_1k_2;\text{t}\Z_2\text{a}T}(M^3) 
\\
&=
\hskip -1em
\sum_{ \{\t g_i, a^{\Z_2}_{ij}\}, \dd a^{\Z_2}\se{2} k_0\Bs_2 \dd\t g}   
\hskip -3em
\ee^{ \ii  \pi \int_{M^3} 
 k_1 (a^{\Z_2})^3
+k_2 a^{\Z_2} \Bs_2 \dd\t g  
} 
\nonumber\\
&=
\hskip -1em
\sum_{ \{\t g_i, a^{\Z_2}_{ij}, \t a^{\Z_2}_{ij}\} }   
\hskip -1em
\ee^{ \ii  \pi \int_{M^3} k_1 (a^{\Z_2})^3  +\t a^{\Z_2}  \dd a^{\Z_2} +k_0 \t a^{\Z_2} \Bs_2 \dd\t g  
 +k_2 a^{\Z_2} \Bs_2 \dd\t g  } 
.
\nonumber 
\end{align}
When $k_0=0$, the above model reduces to the $Z_2$-1-cocycle model
\eqn{tZ2aT3}. When $k_0=1$ and $k_1=k_2=0$, the above becomes the  $Z_4^T$
group cohomology model.  The volume-independent partition function is given by
\begin{align}
\label{Z4Tmdl3}
&
Z_{k_0k_1k_2;\text{t}\Z_2\text{a}T}^\text{top}(M^3)
= \frac{\bar\del_2(k_0\Bs_2 \rw_1)} {|H^0(M^3;\Z_2)|} \times
\nonumber\\
&
\sum_{ a^{\Z_2} \in  H^1(M^3;\Z_2)}   
\ee^{ \ii  \pi \int_{M^3} 
 k_1 (a^{\Z_2})^3
+k_2 \rw_1^2  a^{\Z_2}
} 
.
\end{align}
In the above, we have assumed that when $k_0\Bs_2 \rw_1$ is a coboundary, we
will choose such a coboundary to be zero.  We note that
$Z_{k_0k_1k_2;\text{t}\Z_2\text{a}T}^\text{top}(M^3)$ is simply given by
$Z_{k_1k_2;\text{t}\Z_2\text{a}T}^\text{top}(M^3)$ (see Section \ref{Z2aTpf})
with an extra $\bar\del_2(k_0\Bs_2\rw_1)$ term.

When $k_0=0$, the above model becomes the one studied in Section \ref{Z2aT3},
and the topological order that it produces can be viewed as a gauged $Z_2\times
Z_2^T$-SPT state.  

%When $k_0=0$, the above model produces a topological order
%that can be viewed as a gauged $Z_4^T$-SPT state (which is classified by
%$\cH^3(Z_4^T,(\R/\Z)_T)=\Z_2$).  To write down the $Z_4^T$-SPT invariant, we
%note that the $Z_4^T$ symmetry twist is given by $a^{\Z_4}$ that satisfies
%\begin{align}
%\label{aZ4w1}
% \dd a^{\Z_4} =0, \ \ \ \ a^{\Z_4} = \rw_1 \text{ mod } 2.
%\end{align}
%Thus we can write $a^{\Z_4}$ as $a^{\Z_4} = 2 a^{\Z_2}+\rw_1$.
%In terms of $a^{\Z_2}$ and $\rw_1$, the $Z_4^T$-SPT invariant
%is given by
%\begin{align}
%Z_{k_1}^\text{top}(M^3,a^{\Z_4}) = 
%\bar\del_2(\Bs_2 \rw_1) \ee^{ \ii  \pi \int_{M^3} k_1 (a^{\Z_2})^3 },
%\end{align}
%where $k_1 \in \cH^3(Z_4^T,(\R/\Z)_T)=\Z_2$.  We note that when $\Bs_2 \rw_1
%\neq 0$, the condition \eqn{aZ4w1} cannot be satisfied, and thus
%$Z_{k_1}^\text{top}(M^3,a^{\Z_4})=0$.
%

\subsubsection{Properties of excitations}
\label{Z4T3pex}

When $k_0=1$, the non-trivial group extension makes the time-reversal
transformation $T$ to have a property that $T^2$ is a $Z_2$-gauge
transformation.  So $T^2=-1$ for a non-trivial $Z_2$-charge. In other words,
the $e$ particle with $Z_2$-charge 1 carries a Kramer doublet. $e$ is also a
boson, since if we break the time-reversal symmetry, the above model give rise
to the $Z_2$ or double-semion topological orders, where in both cases, the
$Z_2$-charge is a boson.  We also note that when $k_0=1$, $k_2=0,1$ gives rise
to the same model.

When $(k_0,k_1,k_2)=(1,0,*)$, the dimension reduction $M^3\to M^2\times S^1$
does not produce non-trivial $Z_2\times Z_2^T$ SPT state in 1+1D, thus the
$Z_2$ vortex $m$ in 2+1D is a time-reversal singlet and is a boson.  The bound
state of a $Z_2$ charge and a $Z_2$ vortex is a fermion that carries a Kramer
doublet. The results are summarized in Table \ref{Tsymm3}.

When $(k_0,k_1,k_2)=(1,1,0)$, the dimension reduction $M^3\to M^2\times S^1$
produce a non-trivial $Z_2\times Z_2^T$ SPT state in 1+1D, thus the $Z_2$
vortex $m$. In fact the $Z_2$ vortex $m$ is a $T^2=1$ time-reversal doublet
that carries $Z_2$-gauge-charge $\pm 1/2$ (the same as discussed in Section
\ref{Z2aT3pex} for the  $(k_0,k_1,k_2)=(0,1,0)$ case).  The  $Z_2$-gauge-charge
$\pm 1/2$ doublet is formed by a semion with spin $s=1/4$ and a conjugate
semion with spin $s=3/4$.  The  bound state of a $Z_2$  charge and a $Z_2$
vortex is $\veps$ which also form a time-reversal doublet.  But $\veps$ is a
$T^2=-1$ Kramer doublet that carries $Z_2$-gauge-charge $\pm 1/2$.  To
summarize, \frmbox{the 2+1D $\Z_4^T$ group-cocycle model labeled by
$(k_0,k_1,k_2)=(1,1,0)$ has four types of point-like excitations with quantum
dimensions $(d_1,d_e,d_m,d_\veps)=(1,2_-,2_+,2_-)$ and spins
$(s_1,s_e,s_m,s_\veps)=(0,0,[\frac14,\frac 34],,[\frac14,\frac 34])$ (see Table
\ref{Tsymm3}).}

For $(k_0,k_1,k_2)=(1,1,1)$ the results is the same as that for
$(k_0,k_1,k_2)=(1,1,0)$, except that the properties of $m$ and $\veps$ are
exchanged. This is why $(k_0,k_1,k_2)=(1,1,0)$ and $(k_0,k_1,k_2)=(1,1,1)$
correspond to the same time-reversal SET order.

\subsubsection{Including excitations in the path integral}

Now, let us include the excitations in the partition function \eqn{Z2Z4T}.  Let
$M^1_e$ be the $\Z_2$-valued 1-cycle that correspond to the world-line of the
$Z_2$-charge $e$: $M^1_e \in Z_1(M^4;\Z_2)$.  Let $M^1_m$ be the $\Z_2$-valued
1-cycle that correspond to the world-line of the $Z_2$-vortex $m$: $M^1_m \in
Z_1(M^4;\Z_2)$.  The Poincar\'e dual of $M^1_e$ is a $\Z_2$-valued 2-cocycle
$B^{\Z_2}_{e}$ and the Poincar\'e dual of $M^1_m$ is a $\Z_2$-valued
2-cocycle $B^{\Z_2}_{m}$: $B^{\Z_2}_{e}\in Z^2(M^4;\Z_2)$ and
$B^{\Z_2}_{m}\in Z^2(M^4;\Z_2)$.  The partition function with excitations
is given by
\begin{align}
\label{Z2Z4Tex}
&\ \ \ \
 Z_{k_0k_1k_2;\text{t}\Z_2\text{a}T}(M^3) 
\\
&=
\hskip -1em
\sum_{ \{\t g_i, a^{\Z_2}_{ij}\}, \dd a^{\Z_2} \se{2}
k_0\Bs_2 \dd\t g +B^{\Z_2}_{m} }   
\hskip -4.5em
\ee^{ \ii  \pi \int_{M^3} 
 k_1 (a^{\Z_2})^3
+k_2  a^{\Z_2}\Bs_2 \dd\t g  
} 
\ee^{ \ii  \pi \int_{M^1_e} a^{\Z_2} } 
\nonumber\\
&=
\hskip -1em
\sum_{ \{\t g_i, a^{\Z_2}_{ij},\t a^{\Z_2}_{ij}\}}   
\hskip -1em
\ee^{ \ii  \pi \int_{M^3}  \t a^{\Z_2}(\dd a^{\Z_2} + k_0\Bs_2 \dd\t g +B^{\Z_2}_{m}) } 
\nonumber\\
&\ \ \ \
\ee^{ \ii  \pi \int_{M^3} k_1 (a^{\Z_2})^3 +k_2  a^{\Z_2}\Bs_2 \dd\t g  } 
\ee^{ \ii  \pi \int_{M^1_{e}} a^{\Z_2} } 
\nonumber\\
&=
%\hskip -1em
\sum_{ \{\t g_i, a^{\Z_2}_{ij},\t a^{\Z_2}_{ij}\}}   
\hskip -1em
\ee^{ \ii  \pi \int_{M^3}  \t a^{\Z_2}\dd a^{\Z_2}} 
\ee^{ \ii  \pi \int_{M^3} k_1 (a^{\Z_2})^3 } 
\nonumber\\
&\ \ \ \
\ee^{\ii \pi \int_{M^3}\t a^{\Z_2} (k_0\Bs_2 \dd\t g +B^{\Z_2}_m) +a^{\Z_2} (k_2\Bs_2 \dd\t g +B^{\Z_2}_e) } 
.
\nonumber 
\end{align}
%Since $\Bs_2 =\frac12 \dd$, when $a^{\Z_2}$ is not a cocycle, the above
%expression is not invariant under $a^{\Z_2} \to a^{\Z_2} + 2 \t a$. So it is
%not quite well defined.  In the following, we try to define the above partition
%function for the case when $a^{\Z_2}$ is not a cocycle.

Let us change the variables to
\begin{align}
a^{\Z_2}\se{2}a^{\Z_2}_m +a^{\Z_2}_0, \ \ \ \ \ 
\t a^{\Z_2}\se{2}a^{\Z_2}_e +\t a^{\Z_2}_0, 
\end{align}
where 
%$h^{\Z_2}\in C^0(M^4;\Z_2)$, 
$a^{\Z_2}_0, \t a^{\Z_2}_0 \in C^1(M^2;\Z_2)$, 
and $a^{\Z_2}_m,\ a^{\Z_2}_e$  are fixed
$\Z_2$-valued 1-cochains satisfying 
\begin{align}
 \dd a^{\Z_2}_m  \se{2} B^{\Z_2}_{m} + k_0\Bs_2 \rw_1 , \ \
 \dd a^{\Z_2}_e  \se{2} B^{\Z_2}_{e} + k_2\Bs_2 \rw_1 .
\end{align}
(Here we have assumed that $B^{\Z_2}_{m} + k_0\Bs_2 \rw_1$ and $B^{\Z_2}_{e} +
k_2\Bs_2 \rw_1$ are coboundaries.) Now we can rewrite the partition function as
\begin{align}
&\ \ \ \
 Z_{k_0k_1k_2;\text{t}\Z_2\text{a}T}(M^3) 
\\
&=
%\hskip -1em
\sum_{ \t g, a^{\Z_2}_{0},\t a^{\Z_2}_{0}}   
\hskip -1em
\ee^{ \ii  \pi \int_{M^3}  a^{\Z_2}_e\dd a^{\Z_2}_m+\t a^{\Z_2}_0\dd a^{\Z_2}_0} 
\ee^{ \ii  \pi \int_{M^3} k_1 (a^{\Z_2}_m+a^{\Z_2}_0)^3 } 
\nonumber\\
&\ \ \ \
\ee^{\ii \pi \int_{M^3} a^{\Z_2}_e (k_0\Bs_2 \dd\t g +B^{\Z_2}_m) +a^{\Z_2}_m (k_2\Bs_2 \dd\t g +B^{\Z_2}_e) } 
\nonumber\\
&=
%\hskip -1em
\sum_{ \t g, \dd a^{\Z_2}_{0}=0}   
\hskip -1em
\ee^{ \ii  \pi \int_{M^3}  a^{\Z_2}_e\dd a^{\Z_2}_m} 
\ee^{ \ii  \pi \int_{M^3} k_1 (a^{\Z_2}_m+a^{\Z_2}_0)^3 } 
\nonumber\\
&\ \ \ \
\ee^{\ii \pi \int_{M^3} a^{\Z_2}_e (k_0\Bs_2 \dd\t g +B^{\Z_2}_m) +a^{\Z_2}_m (k_2\Bs_2 \dd\t g +B^{\Z_2}_e) } 
.
\nonumber 
\end{align}
Since $a^{\Z_2}_0$ becomes a cocycle, we can further simplify the factor 
$ \ee^{ \ii  \pi \int_{M^3} k_1 (a^{\Z_2}_m+a^{\Z_2}_0)^3 }$
using
\eqn{cupprop1}
\begin{align}
&\ \ \ \
 \ee^{ \ii  \pi \int_{M^3} k_1 (a^{\Z_2}_m+a^{\Z_2}_0)^3 }
\nonumber\\
& =\ee^{ \ii  \pi \int_{M^3} k_1 [(a^{\Z_2}_m)^3+ (a^{\Z_2}_0)^3 
+(a^{\Z_2}_m)^2 a^{\Z_2}_0+a^{\Z_2}_m (a^{\Z_2}_0)^2
] }.
\end{align}
The partition function now becomes
\begin{align}
&\ \ \ \
 Z_{k_0k_1k_2;\text{t}\Z_2\text{a}T}(M^3) 
%\bar\del_2(M^1_m + k_0M^1_\rw) 
%\bar\del_2(M^1_e + k_2M^1_\rw) 
%\times
 \\
&=
\ee^{ \ii  \pi \int_{M^3}  a^{\Z_2}_m B^{\Z_2}_e} 
%\hskip -1em
\sum_{ \t g, \dd a^{\Z_2}_{0}=0}   
\hskip -1em
\ee^{\ii \pi \int_{M^3} k_0 a^{\Z_2}_e \Bs_2 \dd\t g  + k_2 a^{\Z_2}_m \Bs_2 \dd\t g  } \times
\nonumber\\
&\ \ \ \ \
\ee^{ \ii  \pi \int_{M^3} k_1 [(a^{\Z_2}_m)^3+ (a^{\Z_2}_0)^3 
+(a^{\Z_2}_m)^2 a^{\Z_2}_0+a^{\Z_2}_m (a^{\Z_2}_0)^2
] }
.
\nonumber 
\end{align}

The above partition function can be expressed in terms of linking numbers.
Consider $\int_{M^3} B^{\Z}_e a^{\Z_2}_m = \int_{M^1_e} a^{\Z_2}_m  $.  If
$M^1_e$ is a boundary $M^1_e=\prt D^2_e$, then we can relate the above to
the intersection number and the linking number:
\begin{align}
&\ \ \ \
\int_{M^1_e} a^{\Z_2}_m  
= \int_{D^2_e} \dd a^{\Z_2}_m
= \int_{D^2_e} B^{\Z_2}_{m} + k_0\rw_1^2
\\
&= \text{Int}(D^2_e, M^1_m+k_0 M^1_\rw)
= \text{Lnk}(M^1_e, M^1_m+k_0 M^1_\rw).
\nonumber 
\end{align}
where $M^1_\rw$ is the $\Z_2$-valued 1-cycle which is the Poincar\'e dual of
$\Bs_2 \rw_1$. Here $\text{Int}(D^2_e, M^1_m)$ is the intersection number
between $D^2_e$ and $M^1_m$, and $\text{Lnk}(M^1_e, M^1_m)$ the linking number
between $M^1_e$ and $M^1_m$.  The linking number satisfies 
\begin{align}
\text{Lnk}(M^1_e, M^1_m)=\text{Lnk}(M^1_m, M^1_e).  
\end{align}
Using the  linking number, we can rewrite the partition function as
\begin{align}
\label{Z2Z4TexL}
&\ \ \ \
 Z_{k_0k_1k_2;\text{t}\Z_2\text{a}T}(M^3)
\\
&\propto 
\ee^{\ii \pi \int_{M^3} k_1 (a^{\Z_2}_m)^3} 
\ee^{\ii \pi \text{Lnk}(k_2 M^1_\rw + M^1_e, M^1_m) } 
\times
\nonumber\\
&\ \ \ \
\sum_{ \t g, \dd a^{\Z_2}_{0}=0}   
\hskip -1em
\ee^{\ii \pi \int_{M^3} k_0 a^{\Z_2}_e \Bs_2 \dd\t g  + k_2 a^{\Z_2}_m \Bs_2 \dd\t g  } \times
\nonumber\\
&
\ \ \ \ \
\ \ \ \ \
\ \ \ 
\ee^{ \ii  \pi \int_{M^3} k_1 [(a^{\Z_2}_0)^3 
+(a^{\Z_2}_m)^2 a^{\Z_2}_0+a^{\Z_2}_m (a^{\Z_2}_0)^2
] }
.
\nonumber 
\end{align}
We like to stress that the above path integral has a time-reversal symmetry: it
is invariant under a combined transformation $\t g_i \to [\t g_i +1]_2$,
$a^{\Z_2}_{0,ij} \to a^{\Z_2}_{0,ij}$, and complex conjugation.

The physical properties of excitations can be obtained from the above effective
theory.  Let us first assume $k_1=0$, and rewrite the partition function as
\begin{align}
&\ \ \ \
 Z_{k_0k_1k_2;\text{t}\Z_2\text{a}T}(M^3)
\\
&\propto 
\bar\del_2(M^1_m + k_0M^1_\rw) 
\bar\del_2(M^1_e + k_2M^1_\rw) 
\ee^{\ii \pi \text{Lnk}(k_2 M^1_\rw + M^1_e, M^1_m) } 
\nonumber\\
&\ \ \ \
\sum_{ \t g, \dd a^{\Z_2}_{0}=0}   
\hskip -1em
\ee^{\ii \pi \int_{M^3} k_0 a^{\Z_2}_e \Bs_2 \dd\t g  + k_2 a^{\Z_2}_m \Bs_2 \dd\t g  } 
%\nonumber\\
%&=
%\bar\del_2(M^1_m + k_0M^1_\rw) 
%\bar\del_2(M^1_e + k_2M^1_\rw) \times
%\nonumber\\
%&\ \ \ \
%\ee^{\ii \pi \int_{M^3} k_1 (a^{\Z_2}_m)^3} 
%\ee^{\ii \pi \text{Lnk}(k_2 M^1_\rw + M^1_e, k_0M^1_\rw) }
%\nonumber\\
%&\ \ \ \
%\sum_{ \t g, \dd a^{\Z_2}_{0}=0}   
%\hskip -1em
%\ee^{\ii \pi \int_{M^3} k_0 a^{\Z_2}_e \Bs_2 \dd\t g  + k_2 a^{\Z_2}_m \Bs_2 \dd\t g  } \times
%\nonumber\\
%&\ \ \ \ \
%\ee^{ \ii  \pi \int_{M^3} k_1 [(a^{\Z_2}_m)^3+ (a^{\Z_2}_0)^3 
%+(a^{\Z_2}_m)^2 a^{\Z_2}_0+a^{\Z_2}_m (a^{\Z_2}_0)^2
%] }
,
\nonumber 
\end{align}
where we have restored the two $\del$-functions.  For simplicity, we will also
assume $\rw_1^2=0$, and choose $a^{\Z_2}_e$ be the Poincar\'e dual of $D^2_e$
and $a^{\Z_2}_m$ be the Poincar\'e dual of $D^2_m$.  Here $D^2_e$ and $D^2_m$
are the disks bonded by the world-lines $M^1_e$ and $M^1_m$.
The dynamical part of the partition function can be written as
\begin{align}
&\ \ \ \
\sum_{ \t g}   
\ee^{\ii \pi \int_{M^3} k_0 a^{\Z_2}_e \Bs_2 \dd\t g  + k_2 a^{\Z_2}_m \Bs_2 \dd\t g  } 
\nonumber\\
&=
\sum_{ \t g}   
\ee^{\ii \pi \int_{D^2_e} k_0 \Bs_2 \dd\t g} \ee^{\ii \pi \int_{D^2_m} k_2 \Bs_2 \dd\t g  } 
\nonumber\\
&\propto
\ee^{\ii \pi \int_{D^2_e} k_0 \Bs_2\rw_1 } 
\ee^{\ii \pi \int_{D^2_m} k_2 \Bs_2\rw_1  } 
.
\nonumber 
\end{align}
From the above, we see that, when $k_0=1$, there is $Z_2^T$-SPT state described
by the SPT invariant $\ee^{\ii \pi \int_{D^2_e} \Bs_2\rw_1 }$ on $D^2_e$.  In
this case, the boundary of $D^2_e$, \ie the $e$ particle described by the
world-line $M^1_e=\prt D^2_e$, will carry a Kramer doublet.  This agrees with
the result in Section \ref{Z4T3pex}.  Similarly,  when $k_2=1$, there is
$Z_2^T$-SPT state described by the SPT invariant $\ee^{\ii \pi \int_{D^2_m}
\Bs_2\rw_1 }$ on $D^2_m$, and the $m$ particle will carry a Kramer doublet.

The term $\ee^{\ii \pi \text{Lnk}(M^1_e, M^1_m) }$ tell us that the $e$ and $m$
have a mutual $\pi$ statistics between them.  The absence of self-linking terms,
$\ee^{\ii \th \text{Lnk}(M^1_e, M^1_e) }$ and $\ee^{\ii \th \text{Lnk}(M^1_m,
M^1_m) }$, implies that the $e$ and $m$ are bosons.  We also see that the
emergence of Kramer-doublet bosons cause the partition function to vanish on
the space-time with $\rw_1^2\neq 0$.  From the form of $\bar\del_2(M^1_m +
k_0M^1_\rw) \bar\del_2(M^1_e + k_2M^1_\rw)$, we see that space-time with
$\rw_1^2\neq 0$ will nuclear a $m$ particle (or more precisely, a
non-contractible world-line of the $m$) if the bosonic $e$ particle is a Kramer
doublet.  Similarly,  space-time with $\rw_1^2\neq 0$ will nuclear a $e$
particle if the bosonic $m$ particle is a Kramer doublet.  In other words,
\frmbox{if there is an emergent bosonic Kramer doublet, then a space-time with
$\rw_1^2\neq 0$ will create a world-line of a particle  that has a mutual $\pi$
statistics with the bosonic Kramer doublet.  The world-line is equal to the
Poincar\'e dual of $\rw_1^2$.} Those results are summarized by the top three
rows in Table \ref{Tsymm3}.

Next, we consider the case of $k_1=1$.
The partition function now reads
\begin{align}
&\ \ \ \
 Z_{k_0k_1k_2;\text{t}\Z_2\text{a}T}(M^3)
\\
&\propto 
\bar\del_2(M^1_m + k_0M^1_\rw) 
\ee^{\ii \pi \int_{M^3} k_1 (a^{\Z_2}_m)^3} 
\ee^{\ii \pi \text{Lnk}(k_2 M^1_\rw + M^1_e, M^1_m) } 
\times
\nonumber\\
&\ \ \ \
\sum_{ \t g, \dd a^{\Z_2}_{0}=0}   
\hskip -1em
\ee^{\ii \pi \int_{M^3} k_0 a^{\Z_2}_e \Bs_2 \dd\t g  + k_2 a^{\Z_2}_m \Bs_2 \dd\t g  } \times
\nonumber\\
&
\ \ \ \ \
\ \ \ \ \
\ \ \ 
\ee^{ \ii  \pi \int_{M^3} (a^{\Z_2}_0)^3 
+(a^{\Z_2}_m)^2 a^{\Z_2}_0+a^{\Z_2}_m (a^{\Z_2}_0)^2
}
.
\nonumber 
\end{align}
Note that we only have one $\del$-function in this case.  The above result for
the $e$ particle is not changed: the $e$ is still a boson, which carries Kramer
doublet if $k_0=1$ and time-reversal singlet if $k_0=0$.

But the result for the $m$ particle is changed.  The effective theory on
$D^2_m$ now becomes
\begin{align}
\sum_{ \t g, a^{\Z_2}_0}   
\ee^{\ii \pi \int_{D^2_m} k_2 \Bs_2 \dd\t g  + (a^{\Z_2}_0)^2} . 
\end{align}
If we treat the emergent $Z_2$ gauge symmetry as a $Z_2$ symmetry, then the
above can be viewed as a $Z_2\times Z_2^T$-SPT  state on $D^2_m$. The SPT
state is characterized by SPT invariant $\ee^{\ii \pi \int_{D^2_e} k_2
\Bs_2\rw_1+ (a^{Z_2})^2}$ where $a^{Z_2}$ is the symmetry twist of
$Z_2$.  As discussed in Section \ref{Z2aT3pex}, when $k_2=0$, the
$m$-particle will carry $\pm 1/2$ $Z_2$-gauge-charge, which form a $T^2=1$
time-reversal doublet (labeled by $2_+$).  When $k_2=1$, the $m$-particle will
carry $\pm 1/2$ $Z_2$-gauge-charge, which form a $T^2=-1$ Kramer doublet
(labeled by $2_-$). The above applies for both $k_0=0,1$ cases.

For the bond state of $e$ and $m$, the $\veps$ particle, the $Z_2\times
Z_2^T$-SPT  state on the corresponding $D^2_\veps$ is described by 
\begin{align}
\sum_{ \t g, a^{\Z_2}_0}   
\ee^{\ii \pi \int_{D^2_m} (k_0+k_2) \Bs_2 \dd\t g  + k_1 (a^{\Z_2}_0)^2} . 
\end{align}
We see that the $\veps$ is always a $\pm 1/2$ $Z_2$-gauge-charge doublet.  It
is a $T^2=-1$ Kramer doublet ($2_-$) if $(k_0+k_2)=1$ and a $T^2=1$
time-reversal doublet ($2_+$) if $(k_0+k_2)=0$.

The statistics of the $m$ particle is no longer bosonic due the self-braiding
term $\ee^{\ii \pi \int_{M^3} k_1 (a^{\Z_2}_m)^3}$ (which can be viewed as the
triple self-intersection of $D^2_m$).  We note that $\ee^{\ii \pi \int_{M^3}
k_1 (a^{\Z_2}_m)^3}=\pm 1$ respects the time-reversal symmetry.  But one expect
$m$ to a semion described by the self-linking term $\ee^{\ii \frac{\pi}{2}
\text{Lnk}(M^1_m, M^1_m) }$.  In fact, the above self-linking term breaks the
time-reversal symmetry, and does not describe the statics of $m$ which in our
case is a particle the respect the time-reversal symmetry.  In other words, due
to the time reversal symmetry, $m$ is not a semion.

In fact, $m$ is a $T^2=-1$ Kramer doublet or a $T^2=1$ time-reversal doublet
formed by a semion (with spin $s=1/4$) and a conjugate semion  (with spin
$s=3/4$).  The statistics of such a time-reversal symmetric doublet is not
described by  the self-linking term $\ee^{\ii \frac{\pi}{2} \text{Lnk}(M^1_m,
M^1_m) }$ or  the self-linking term $\ee^{-\ii \frac{\pi}{2} \text{Lnk}(M^1_m,
M^1_m) }$.  Our calculation suggest that the statistics of the time-reversal
symmetric doublet is described by $\ee^{\ii \pi \int_{M^3} k_1 (a^{\Z_2}_m)^3}$
-- the triple self-intersection of $D^2_m$.  Those results are summarized by
the bottum three rows in Table \ref{Tsymm3}.

\subsection{3+1D time-reversal symmetric model}

\subsubsection{Model construction}

In this section, we are going to study a class of 3+1D time-reversal symmetric
local bosonic models, that can produce the simplest time-reversal symmetric
topological orders. 
%The motivation of this class of models is given in Appendix
%\ref{}.  Here we will just study the properties of the models.
The 3+1D time-reversal symmetric local bosonic models contain
$\Z_2$-multi-valued 0-cochain field $\t g_i$, $\Z_2$-valued 1-cochain field
$a^{\Z_2}_{ij}$, and $\Z_2$-valued 2-cochain field $b^{\Z_2}_{ijk}$.  Its path
integral is given by
\begin{align}
\label{Z2T4d}
&
Z_{k_1k_2k_3k_4k_5k_6}(M^4)  
\\
&=
\hskip -1em 
\sum_{ \{\t g_i^{\Z_2}, a_{ij}^{\Z_2}, b^{\Z_2}_{ijk} \}  }  \hskip -1.5em 
%\ee^{-U \int_{M^4} |[\dd b^{Z_2}]_2|^2 +|[\dd a^{Z_2}]_2|^2}
\ee^{\ii \pi \int_{M^4} b^{\Z_2}\dd a^{\Z_2}}
\ee^{\ii \pi \int_{M^4}  (k_3+k_4) b^{\Z_2}\Bs_2\dd \t g +k_4 (b^{\Z_2})^2} 
\times
\nonumber  \\
&
\ee^{ \ii  \pi \int_{M^4}   k_1 (a^{\Z_2})^4  + (k_2+k_1)a^{\Z_2}  (\dd \t g)^3   }
 \ee^{ \ii  \pi \int_{M^4} k_5 (\dd \t g )^4 +k_6 (\rw_2)^2 } 
.
\nonumber 
\end{align}
%We will take large $U$ limit and thus the term $\ee^{-U \int_{M^4} |[\dd
%b^{Z_2}]_2|^2 +|[\dd a^{Z_2}]_2|^2}$ enforces $\dd a^{Z_2} \se{2} 0$ and $\dd
%b^{Z_2} \se{2} 0$ conditions.  
The 0-cocycle field $\t g_i$ is a pseudo scalar
as introduced in Section \ref{GverT}.  It satisfies $\dd \t g_i =\rw_1 +\dd g$,
where $g_i$ is a $\Z_2$-single-valued 0-cochain field.  Thus $\Bs_2 \dd \t g_i
= \Bs_2 \rw_1$.  The above path integral defines the system for both closed and
open space-time manifold $M^4$.  But in the following, we will assume $M^4$ to
be closed.  The index $k_I=0,1$. So there are $2^6=64$ different models.

We note that the above path integral has the time-reversal symmetry $Z_2^T$,
\ie invariant under the combined transformation of $\t g_i \to [\t g_i +1]_2$
and complex conjugation. (Under the transformation $\t g_i \to \t g_i' =[\t g_i
+1]_2$, $\dd \t g_i  = -\dd \t g_i'$.) This is a designed property.  However,
the  path integral also has an extra $Z'_2$ symmetry: $\t g_i \to [\t
g_i +1]_2$ (without the complex conjugation).  

Let us also include the excitations in the path integral.  We know that the
point-like excitations are described by the world-lines in space-time.  A world
line $M^1_\text{WL}$ can be viewed as a $\Z_2$-valued 1-cycle, which is
Poincar\'e dual to a $\Z_2$-valued 3-cochain $C_\text{WL}^{\Z_2}$.  The
string-like excitations are described by the world-sheet in space-time, which
can be viewed as $\Z_2$-valued 2-cycles $M^2_\text{WS}$ in the space-time
lattice, whose Poincar\'e dual is a $\Z_2$-valued  2-cocycle
$B_\text{WS}^{\Z_2}$.  

Just like the $Z_2$-gauge theory, we can include those excitations in path
integral \eqn{Z2T4d}, by adding the $Z_2$-charge coupling term $\ee^{\ii \pi
\int_{M^1_\text{WL}} a^{\Z_2}}$ and the $Z_2$-flux coupling term $\ee^{\ii \pi
\int_{M^2_\text{WS}} b^{\Z_2}}$. Due to the Poincar\'e duality,
\begin{align}
 \ee^{\ii \pi \int_{M^1_\text{WL}} a^{\Z_2}} 
&=\ee^{\ii \pi \int_{M^4} C^{\Z_2}_\text{WL} a^{\Z_2}},
\nonumber\\
 \ee^{\ii \pi \int_{M^2_\text{WS}} b^{\Z_2}} 
&=\ee^{\ii \pi \int_{M^4} B^{\Z_2}_\text{WS} b^{\Z_2}}.
\nonumber\\
\end{align}
Thus, in the presence of point-like topological
excitations described by $C_\text{WL}^{\Z_2}$ and string-like topological
excitations described by $B^2_\text{WS}$, the partition
function \eqn{Z2T4d} becomes
\begin{align}
&\ \ \ \
 Z_{k_1k_2k_3k_4k_5k_6}(M^4) 
\\
&= 
\sum_{ \t g,a^{\Z_2}, b^{\Z_2} }  
\hskip -1em
\ee^{\ii \pi \int_{M^4} b^{\Z_2}(\dd a^{Z_2}+k_4 b^{\Z_2}+(k_3+k_4) \Bs_2\dd\t g+ B_\text{WS}^{\Z_2})}
\times
\nonumber\\
&
\ee^{ \ii  \pi \int_{M^4} k_1    (a^{\Z_2})^4 +[(k_2+k_1) (\dd\t g)^3+ C^{\Z_2}_\text{WL}]  a^{\Z_2} } 
\ee^{ \ii  \pi \int_{M^4} k_5 (\dd\t g)^4 +k_6 \rw_2^2 }
.
\nonumber 
\end{align}

\subsubsection{Partition function}

To understand the physical properties of those 64 models, we like to compute
the corresponding partition functions on closed space-time $M^4$.  However,
unlike other models constructed in this paper, the above models are not exactly
soluble.  They are exactly soluble only in the cases $k_1=0$ or $k_4=0$.  So we
will calculate the partition functions for those two cases.

When $k_4=0$, the action is linear in $b^{\Z_2}$,  and we can integrate out
$b^{\Z_2}$ first, which lead to a constraint 
\begin{align}
 \dd a^{\Z_2} \se{2} k_3 \Bs_2\dd\t g+ B_\text{WS}^{\Z_2}
=k_3 \Bs_2\dd\t g_0+ B_\text{WS}^{\Z_2}.
\end{align}
where $\t g_0$ is a fixed $\Z_2$-multi-valued 0-cochain such that
\begin{align}
 \t g - \t g_0 \se{2} g
\end{align}
is a $\Z_2$-single-valued 0-cochain $g$.
We see that the partition function is zero when $k_3\Bs_2\dd \t g_0+
B_\text{WS}^{\Z_2}$ is not a coboundary.  Thus the partition function contain a
factor $\bar\del_2(k_3\Bs_2\dd \t g_0+ B_\text{WS}^{\Z_2})$.  We may solve the $\dd
a^{\Z_2} \se{2} k_3 \Bs_2\dd\t g+ B_\text{WS}^{\Z_2}$ constraint via the
following ansatz
\begin{align}
 a^{\Z_2} \se{2} a^{\Z_2}_\text{WS} + a^{\Z_2}_0 
\end{align}
where  $a^{\Z_2}_\text{WS}$ is a $\Z_2$-valued 1-cochain that satisfies
\begin{align}
 \dd a^{\Z_2}_\text{WS} \se{2} k_3\Bs_2 \dd \t g+ B_\text{WS}^{\Z_2},
\end{align}
and $a^{\Z_2}_0 $ is a $\Z_2$-valued 1-cocycle field $a^{\Z_2}_0 \in
Z^1(M^4;\Z_2)$.  The partition function now becomes
\begin{align}
\label{Z2Tk40dg}
&\ \ \ \
 Z_{k_1k_2k_30k_5k_6}(M^4) = \bar\del_2(k_3\Bs_2 \dd\t g + B_\text{WS}^{\Z_2})
\\
&
\sum_{ \t g,\dd a^{\Z_2}_0\se{2}0}  
\hskip -1em
\ee^{ \ii  \pi \int_{M^4} k_1    (a^{\Z_2}_\text{WS})^4 +[(k_2+k_1) (\dd\t g)^3+ C^{\Z_2}_\text{WL}]  a^{\Z_2}_\text{WS} } 
\times
\nonumber\\
&
\ee^{ \ii  \pi \int_{M^4}  k_1    (a^{\Z_2}_0)^4 +[(k_2+k_1) (\dd\t g)^3+ C^{\Z_2}_\text{WL}]  a^{\Z_2}_0 } 
\ee^{ \ii  \pi \int_{M^4} k_5 (\dd\t g)^4 +k_6 \rw_2^2 }
.
\nonumber 
\end{align}
Since $a^{\Z_2}_0$ is a cocycle, we can replace $\dd \t g$ by $\dd \t g_0$ in
the last line above, and use many relations between $a^{\Z_2}_0$ and
Stiefel-Whitney classes, such as (see Appendix \ref{aSWrel4} where $\rw_1$ is
replaced by $\dd \t g_0$)
\begin{align}
\label{warel4}
& (\dd \t g_0)^2(a^{\Z_2}_0)^2\se{2,\dd}(\dd \t g_0)^3a^{\Z_2}_0,
\  
\rw_2 (a^{\Z_2}_0)^2\se{2,\dd} \rw_3 a^{\Z_2}_0, 
\\
&  
(a^{\Z_2}_0)^4 \se{2,\dd}\dd \t g_0 (a^{\Z_2}_0)^3, 
\ 
[a^{\Z_2}_0)^2 + (\dd \t g_0)^2  +\rw_2](a^{\Z_2}_0)^2\se{2,\dd}0
,
\nonumber 
\end{align}
to simplify the last line.  Note that those relations are valid only
when $a^{\Z_2}_0$ is a cocycle and when $M^4$ is closed.  Therefore, we can
rewrite the above partition function on closed $M^4$ as
\begin{align}
\label{Z2Tk40}
&\ \ \ \
 Z_{k_1k_2k_30k_5k_6}(M^4) = \bar\del_2(k_3\Bs_2 \dd \t g_0+ B_\text{WS}^{\Z_2})
\nonumber \\
&
\sum_{ \t g,\dd a^{\Z_2}_0\se{2}0}  
\hskip -1em
\ee^{ \ii  \pi \int_{M^4} k_1    (a^{\Z_2}_\text{WS})^4 +[(k_2+k_1) (\dd\t g)^3+ C^{\Z_2}_\text{WL}]  a^{\Z_2}_\text{WS} } 
\times
\nonumber\\
&
\ee^{ \ii  \pi \int_{M^4}  (k_1   \rw_3 + k_2 (\dd \t g_0)^3+ C^{\Z_2}_\text{WL})  a^{\Z_2}_0 } 
\ee^{ \ii  \pi \int_{M^4} k_5 (\dd\t g)^4 +k_6 \rw_2^2 }
\nonumber\\
&= 
\bar\del_2(k_3\Bs_2 \dd \t g_0+ B_\text{WS}^{\Z_2})
\bar\del_2(k_1   \rw_3 + k_2 (\dd \t g_0)^3+ C^{\Z_2}_\text{WL})
\nonumber \\
&\ \ \ \
\ee^{ \ii  \pi \int_{M^4} k_5 (\dd \t g_0)^4 +k_6 \rw_2^2 }\times
\\
&\ \ \ \
\sum_{ \t g}  
\ee^{ \ii  \pi \int_{M^4} k_1    (a^{\Z_2}_\text{WS})^4 +[(k_2+k_1) (\dd\t g)^3+ C^{\Z_2}_\text{WL}]  a^{\Z_2}_\text{WS} } 
.
\nonumber 
\end{align}
We note that $x^2 \se{2} \Bs_2 x$ for any $Z_2$-valued 1-cocycle $x$,  and $\dd
\t g= \dd \t g_0+\dd g$. Thus
\begin{align}
&
 \ee^{ \ii  \pi \int_{M^4} (k_2+k_1) (\dd\t g)^3  a^{\Z_2}_\text{WS} } 
 =
\ee^{ \ii  \pi \int_{M^4} (k_2+k_1) \dd\t g \Bs_2 \dd \t g_0  a^{\Z_2}_\text{WS} } 
\\
&=
\ee^{ \ii  \pi \int_{M^4} (k_2+k_1) \dd \t g_0 \Bs_2 \dd \t g_0  a^{\Z_2}_\text{WS} } 
\ee^{ \ii  \pi \int_{M^4} (k_2+k_1) \dd g \Bs_2 \dd \t g_0  a^{\Z_2}_\text{WS} } 
\nonumber\\
&=
\ee^{ \ii  \pi \int_{M^4} (k_2+k_1) (\dd \t g_0)^3   a^{\Z_2}_\text{WS} } 
\ee^{ \ii  \pi \int_{M^4} (k_2+k_1) g \Bs_2 \dd \t g_0  ( k_3\Bs_2 \dd \t g_0+ B_\text{WS}^{\Z_2}) } 
\nonumber 
\end{align}
Therefore, the volume independent partition function is given by
\begin{align}
&
 Z_{k_1k_2k_30k_5k_6}^\text{top}(M^4) 
= 
\frac{|H^0(M^4;\Z_2)| |H^2(M^4;\Z_2)|}{|H^1(M^4;\Z_2)|}
\times
\nonumber\\
&
\ee^{ \ii  \pi \int_{M^4} k_5 (\dd \t g_0)^4 +k_6 \rw_2^2 
+ k_1    (a^{\Z_2}_\text{WS})^4 +[(k_2+k_1) (\dd \t g_0)^3+ C^{\Z_2}_\text{WL}]  a^{\Z_2}_\text{WS} } 
\times
\nonumber \\
&\ \
\bar\del_2(k_3\Bs_2 \dd \t g_0+ B_\text{WS}^{\Z_2})
\bar\del_2(k_1   \rw_3 + k_2 (\dd \t g_0)^3+ C^{\Z_2}_\text{WL})
\times
\nonumber\\
&\ \
\del_2[(k_2+k_1) \Bs_2 \dd \t g_0  ( k_3\Bs_2 \dd \t g_0+ B_\text{WS}^{\Z_2})]
.
\nonumber 
\end{align}
We note that $\t g_0$ is multivalued only on $\prt M^4$ (which is a non-zero
even cycle when $ M^4$ is not orientable).  So $\Bs_2 \dd \t g_0$ is non-zero
only near the ``boundary''  $\prt M^4$ (see Fig. \ref{RP2}).  Therefore,
$\del_2[(k_2+k_1) \Bs_2 \dd \t g_0 ( k_3\Bs_2 \dd \t g_0+ B_\text{WS}^{\Z_2})]$
is a boundary term.

When $k_1=0$, the action is linear in $a^{\Z_2}$. In this case, we can
integrate out $a^{\Z_2}$ first, which lead to a constraint 
\begin{align}
 \dd b^{\Z_2} = k_2 (\dd\t g)^3+ C^{\Z_2}_\text{WL}.
\end{align}
So the partition function is zero when $k_2 (\dd \t g_0)^3+ C^{\Z_2}_\text{WL}$
is not a coboundary.  Thus the partition function contains a factor
$\bar\del_2(k_2 (\dd \t g_0)^3+ C^{\Z_2}_\text{WL})$.  We may solve the $\dd
b^{\Z_2} \se{2} k_2 (\dd \t g)^3+ C^{\Z_2}_\text{WL}$ constraint via the
following ansatz
\begin{align}
 b^{\Z_2} \se{2} b^{\Z_2}_\text{WL} + b^{\Z_2}_0 
\end{align}
where  $b^{\Z_2}_\text{WL}$ is a $\Z_2$-valued 2-cochain that satisfies
\begin{align}
 \dd b^{\Z_2}_\text{WL} \se{2} k_2 (\dd \t g)^3+ C^{\Z_2}_\text{WL},
\end{align}
and $b^{\Z_2}_0 $ is a $\Z_2$-valued 2-cocycle field $b^{\Z_2}_0 \in
Z^2(M^4;\Z_2)$.  The partition function now becomes
\begin{align}
\label{Z2Tk10dg}
&\ \ \ \
 Z_{0k_2k_3k_4k_5k_6}(M^4) = \bar\del_2[k_2 (\dd \t g_0)^3+ C^{\Z_2}_\text{WL}]
\times
\\
&\ \ \ \
\sum_{ \t g,\dd b^{\Z_2}_0\se{2}0}  
\hskip -1em
\ee^{\ii \pi \int_{M^4} b^{\Z_2}_\text{WL}[ k_4 b^{\Z_2}_\text{WL}+(k_3+k_4) \Bs_2\dd\t g+ B_\text{WS}^{\Z_2}]}
\times
\nonumber\\
&\ \ \ \
\ee^{\ii \pi \int_{M^4}b^{\Z_2}_0[ k_4b^{\Z_2}_0+ (k_3+k_4) \Bs_2\dd\t g+ B_\text{WS}^{\Z_2}]}
\ee^{ \ii  \pi \int_{M^4} k_5 (\dd \t g_0)^4 +k_6 \rw_2^2 }
.
\nonumber 
\end{align}
Since $b^{\Z_2}_0$ is a cocycle, we can
replace $\dd \t g$ by $\dd \t g_0$ in the last line above,
and use many relations between $b^{\Z_2}_0$ and
Stiefel-Whitney classes, such as (see Appendix \ref{bSWrel4})
\begin{align}
\dd \t g_0 \Bs_2 b^{\Z_2}=0,\ \ \ \
(b^{\Z_2})^2+[(\dd \t g_0)^2+\rw_2]b^{\Z_2}  =  0,
\end{align}
to simplify the last line.  Therefore, we can rewrite the above partition
function on closed $M^4$ as
\begin{align}
\label{Z2Tk10}
&\ \ \ \
 Z_{0k_2k_3k_4k_5k_6}(M^4) = \bar\del_2(k_2 (\dd \t g_0)^3+ C^{\Z_2}_\text{WL})
\times
\\
&\ \ \ \
\sum_{ \t g,\dd b^{\Z_2}_0\se{2}0}  
\hskip -1em
\ee^{\ii \pi \int_{M^4}  b^{\Z_2}_\text{WL}[k_4b^{\Z_2}_\text{WL} + (k_3+k_4) \Bs_2\dd\t g+ B_\text{WS}^{\Z_2}]}
\times
\nonumber\\
&\ \ \ \
\ee^{\ii \pi \int_{M^4}b^{\Z_2}_0[ k_4\rw_2 +k_3 \Bs_2\dd \t g_0+ B_\text{WS}^{\Z_2}]}
\ee^{ \ii  \pi \int_{M^4} k_5 (\dd \t g_0)^4 +k_6 \rw_2^2 }
\nonumber\\
 &= 
\bar\del_2(k_2 (\dd \t g_0)^3+ C^{\Z_2}_\text{WL})
\bar\del_2(k_4\rw_2 +k_3 \Bs_2\dd \t g_0+ B_\text{WS}^{\Z_2})
\times
\nonumber\\
&\ \ \ \
\ee^{ \ii  \pi \int_{M^4} k_5 (\dd \t g_0)^4 +k_6 \rw_2^2 }
\times
\nonumber \\
&\ \ \ \
\sum_{ \t g}  
\ee^{\ii \pi \int_{M^4}  b^{\Z_2}_\text{WL}[k_4b^{\Z_2}_\text{WL} + (k_3+k_4) \Bs_2\dd\t g+ B_\text{WS}^{\Z_2}]}
.
\nonumber 
\end{align}
The above partition function can be simplified further.  Let $\bar
b^{\Z_2}_\text{WL}$ be a fixed 2-cocycle that satisfies
\begin{align}
 \dd \bar b^{\Z_2}_\text{WL} \se{2} k_2 (\dd \t g_0)^3+ C^{\Z_2}_\text{WL}.
\end{align}
and let
\begin{align}
 b^{\Z_2}_\text{WL} \se{2}  \bar b^{\Z_2}_\text{WL} +  b^{\Z_2}_1
\end{align}
In this case, $b^{\Z_2}_1$ satisfies
\begin{align}
\dd b^{\Z_2}_1 & \se{2} k_2 [(\dd\t g)^3-(\dd \t g_0)^3]
\se{2} k_2 [\dd\t g\Bs_2\dd\t g -\dd \t g_0\Bs_2 \dd \t g_0]
\nonumber\\
&
= k_2\dd (g\Bs_2 \dd \t g_0)
\end{align}
where we have used $\dd \t g=\dd \t g_0+\dd g$ and $x^2 \se{2} \Bs_2 x$ for any
$Z_2$-valued 1-cocycle $x$.  So $b^{\Z_2}_\text{WL}$ is given by 
\begin{align} 
b^{\Z_2}_\text{WL} \se{2} k_2 g \Bs_2\dd \t g_0 +  \bar b^{\Z_2}_\text{WL}. 
\end{align} 
Now, we can rewrite the partition function \eqn{Z2Tk10dg} in the following form
(using the relations obtained in Appendix \ref{bSWrel4})
\begin{align}
&\ \ \ \
 Z_{0k_2k_3k_4k_5k_6}(M^4) 
\\
 &= 
\bar\del_2[k_2 (\dd \t g_0)^3+ C^{\Z_2}_\text{WL}]
\bar\del_2(k_4\rw_2 +k_3 \Bs_2\dd \t g_0+ B_\text{WS}^{\Z_2})
\times
\nonumber\\
&
\ee^{ \ii  \pi \int_{M^4} k_5 (\dd \t g_0)^4 +k_6 \rw_2^2 }
\ee^{\ii \pi \int_{M^4} \bar b^{\Z_2}_\text{WL} [k_4\bar b^{\Z_2}_\text{WL}+ (k_3+k_4) \Bs_2\dd \t g_0+ B_\text{WS}^{\Z_2}]}
\times
\nonumber \\
&
\sum_{g}  
\ee^{\ii \pi \int_{M^4} k_2 g  \Bs_2 \dd \t g_0[k_4g  \Bs_2 \dd \t g_0+ (k_3+k_4) \Bs_2\dd \t g_0+ B_\text{WS}^{\Z_2}]}
.
\nonumber 
\end{align}
Using the fact (see \eqn{cupprop1}) $ g  \Bs_2 \dd \t g_0 g   \se{2,\dd}  g^2
\Bs_2 \dd \t g_0 = g  \Bs_2 \dd \t g_0$, we can simplify
\begin{align}
&\ \ \ \
 \sum_g 
\ee^{\ii \pi \int_{M^4} 
k_2 g  \Bs_2 \dd \t g_0[k_4g  \Bs_2 \dd \t g_0+ (k_3+k_4) \Bs_2\dd \t g_0+ B_\text{WS}^{\Z_2}]}
\nonumber\\
&=
\sum_g \ee^{\ii \pi \int_{M^4} k_2g  \Bs_2 \dd \t g_0[k_4\Bs_2 \dd \t g_0 +   (k_3+k_4) \Bs_2\dd \t g_0+ B_\text{WS}^{\Z_2}]}
\nonumber\\
& =\del_2[k_2 \Bs_2 \dd \t g_0(k_3\Bs_2 \dd \t g_0 +   B_\text{WS}^{\Z_2})]
\end{align}
%So we may choose the world-sheet to make 
%\begin{align}
% k_3\Bs_2 \dd \t g_0 +   B_\text{WS}^{\Z_2}=0.
%\end{align}
Thus, the volume independent partition function is given by
\begin{align}
\label{Z2Ttopk10}
&
 Z_{0k_2k_3k_4k_5k_6}^\text{top}(M^4) 
= 
\frac{|H^0(M^4;\Z_2)| |H^2(M^4;\Z_2)|}{|H^1(M^4;\Z_2)|}
\times
\nonumber\\
&\ \ \ \
\ee^{ \ii  \pi \int_{M^4} k_5 (\dd \t g_0)^4 +k_6 \rw_2^2 
+ \bar b^{\Z_2}_\text{WL} [k_4\bar b^{\Z_2}_\text{WL}+ (k_3+k_4) \Bs_2\dd \t g_0+ B_\text{WS}^{\Z_2}]}
\times
\nonumber \\
&\ \ \ \
\bar\del_2[k_2 (\dd \t g_0)^3+ C^{\Z_2}_\text{WL}]
\bar\del_2(k_4\rw_2 +k_3 \Bs_2\dd \t g_0+ B_\text{WS}^{\Z_2}) \times
\nonumber\\
&\ \ \ \
\del_2[k_2 \Bs_2 \dd \t g_0(k_3\Bs_2 \dd \t g_0 +   B_\text{WS}^{\Z_2})]
.
\nonumber 
\end{align}

\subsubsection{Physical properties of ground states}

Using the above partition functions, we can obtain many physical properties of
ground states by setting $B^{\Z_2}_\text{WS}=C^{\Z_2}_\text{WL}=0$.  For
simplicity, we will also assume that $\rw_i=0$ on $M^4$, so that we can choose
$a^{\Z_2}_\text{WS}=\bar b^{\Z_2}_\text{WL}=0$.  

First, we see that the
partition functions for different $k_I$'s do not depend the shape or the
metrics of space-time manifold $M^4$. So the ground states of the 48 models
with $k_1k_4=0$ are all gapped.  The  partition functions also do not depend on
the triangulation of the space-time. So the ground states are all gapped
liquids.\cite{ZW1490,SM1403} If we choose space-time to be $M^4=S^1\times S^3$
where $\rw_1=\rw_2=\rw_3=0$, we find the volume-independent partition functions
to be $Z_{k_1k_2k_3k_4k_5k_6}^\text{top}(S^1\times S^3)=1$.  This means that
the ground state degeneracies on $S^3$ for the 48 models (with $k_1k_4=0$) are
all equal to 1, and there is no spontaneous symmetry breaking of $Z_2^T$ or
$Z_2'$.

The  volume-independent partition functions are not equal to 1 for other closed
space-times with vanishing Euler number and  Pontryagin number. For example on
$M^4=T^2\times S^2$ where $\rw_1=\rw_2=\rw_3=0$, $Z_{k_1k_2k_3k_4k_5k_6}^\text{top}(T^2\times S^2)=2$.
Thus, those 48 models all realize non-trivial 3+1D topological orders in their
ground states. The 2-fold ground state degeneracy on space $S^1\times S^2$
tells us that the topological orders are simple since they all have only one
non-trivial point-like topological excitation and one non-trivial string-like
topological excitation.  In fact the emergent topological orders are $Z_2$
topological orders described by UT or EF $Z_2$ gauge theories with
$a^{\Z_2}$ as the $Z_2$-gauge field.  Because the ground states also have
symmetries, we may view those topological orders as $Z_2^T$-SET orders or as
$Z_2'\times Z_2^T$-SET orders.

We remark that the action amplitude $\ee^{ \ii  \pi \int_{M^4} k_5 (\dd\t g)^4
+k_6 \rw_2^2}=\ee^{ \ii  \pi \int_{M^4} k_5 \rw_1^4 +k_6 \rw_2^2}$ is the SPT
invariant for the $Z_2^T$ SPT states.  So different $k_5,k_6$ correspond to
stacking with different $Z_2^T$-SPT states.

\subsubsection{Properties of point-like excitations}
\label{proppoint}

First, if we break the time-reversal symmetry (\ie only put the system on
orientable space-time with $\rw_1=0$), then our models with $k_1=0$ reduce to
the $\Z_n$-2-cocycle model \eqn{tZnbMdl} with $n=2$ and $k=k_4$.  So when
$k_1=0$, the point-like topological excitation in our model is a fermion if
$k_4=1$, and a boson if $k_4=0$ (see Table \ref{Tsymm4} where a fermion is
indicated by spin $s_2=\frac12$ and a boson by spin $s_2=0$).

When $k_4=0$ (and without time-reversal symmetry), our model reduce to the
UT $Z_2$-gauge theory (note that $(a^{\Z_2}_0)^4=(a^{\Z_2}_0)^3\rw_1$, and
$(a^{\Z_2}_0)^4=0$ when $\rw_1=0$). So the  point-like topological excitation
in our model is a boson if $k_4=0$, even when $k_1\neq 0$.

In the presence of time-reversal symmetry $Z_2^T$ with $T^2=1$, the point-like
topological excitation may carry fractionalized time-reversal symmetry with
$T^2=-1$, \ie it may carry Kramer doublet.  In fact, in this section, we will
consider both time-reversal symmetry and the extra $Z_2'$ symmetry $\t g_i \to
[\t g_i+1]_2$ of our models. So the total symmetry group is $Z_2'\times Z_2^T$.
In this case, the multivalueness of $\t g_i$ is not only due the orientation
twist around a loop, it is also due to the $Z_2'$-symmetry twist around a loop.
Thus 
\begin{align}
 \dd \t g=\rw_1+a^{\prime\Z_2}
\end{align}
where $a^{\prime\Z_2}$ is the 1-cocycle that describes the $Z_2'$-symmetry
twist in space-time.\cite{LG1209,HW1339,W1447,SCR1325}

To see the time-reversal and  $Z_2'$ symmetry properties of the point-like
topological excitation, we first consider the $k_4=0$ case and start with the
path integral \eqn{Z2Tk40dg}.  We like to stress that in our calculation to
obtain  \eqn{Z2Tk40dg}, we did not use the relation $\dd \t g=\rw_1$ which is
not valid in the presence of $Z_2'$-symmetry twist, which is necessary to
consider $Z_2'$ symmetry.  The only term that involves the world-line of the
point-like topological excitation is $\ee^{\ii \pi \int_{M^4}
C^{\Z_2}_\text{WL} (a^{\Z_2}_\text{WS}}+a_0^{\Z_2})$, which can be expanded as
\begin{align}
 \ee^{\ii \pi \int_{M^4} C^{\Z_2}_\text{WL} (a^{\Z_2}_\text{WS}+a^{\Z_2}_0)}
&=
\ee^{\ii \pi \int_{M^1_\text{WL}} a^{\Z_2}_\text{WS}+a^{\Z_2}_0}
=
\ee^{\ii \pi \int_{D^2_\text{WL}} \dd a^{\Z_2}_\text{WS}}
\nonumber\\
&=
\ee^{\ii \pi \int_{D^2_\text{WL}} k_3 \Bs_2\dd\t g+ B_\text{WS}^{\Z_2}},
\end{align}
where $D^2_\text{WL}$ is the 2-dimensional submanifold whose boundary is
the world-line $\prt D^2_\text{WL}=M^1_\text{WL}$.
The term $\ee^{\ii \pi \int_{M^1_\text{WL}} a^{\Z_2}_0}$ indicates that the
point-like excitation carries an unit of $Z_2$-gauge charge.

The term $\ee^{\ii \pi \int_{D^2_\text{WL}} B_\text{WS}^{\Z_2}} =\ee^{\ii \pi
\text{Lnk}(M^1_\text{WL},M^2_\text{WS})}$ is determined by the linking number $
\text{Lnk}(M^1_\text{WL},M^2_\text{WS})$ between the world-line $M^1_\text{WL}$
of the point-like excitation and the world-sheet $M^2_\text{WS}$ of the
string-like excitation.  It describes the $\pi$ phase change as a point-like
excitation goes around the string-like excitation.

The term $\ee^{\ii \pi \int_{D^2_\text{WL}} k_3 \Bs_2\dd\t g}$ 
give rise to a $Z_2'\times Z_2^T$-SPT invariant
\begin{align}
 \ee^{\ii \pi \int_{D^2_\text{WL}} k_3 \Bs_2\dd\t g} \to
 \ee^{\ii \pi \int_{D^2_\text{WL}} k_3 (\Bs_2\rw_1 + \Bs_2a^{\prime\Z_2})} 
\end{align}
which describes a 1+1D $Z_2'\times Z_2^T$-SPT state on $ D^2_\text{WL}$ when
$k_3=1$.  Due to the term $\Bs_2\rw_1$ in the SPT invariant, the boundary of
the 1+1D $Z_2^T$-SPT state carries a Kramer doublet. Thus 
\frmbox{the world-line, \ie
the point-like excitation, carries a Kramer doublet if $k_3=1$ and carries a
time-reversal singlet if $k_3=0$}
 (see Table \ref{Tsymm4} where a Kramer doublet
is indicated by quantum dimension $d_2=2_-$ and a time-reversal singlet by
quantum dimension $d_2=1$).  Due to the term $\Bs_2a^{\prime\Z_2}$, the Kramer
doublet on the point-like excitation is formed by $Z_2'$-charge $\pm \frac12$
states. So \frmbox{the  Kramer doublet also carries fractional $Z_2'$-charge
$\pm \frac12$.}

We next consider the $k_1=0$ case and start with the path integral
\eqn{Z2Tk10dg}.  Again, in our calculation to obtain  \eqn{Z2Tk10dg}, we did
not use the relation $\dd \t g=\rw_1$. The only term that involves the world
line of the point-like topological excitation is 
\begin{align}
\ee^{\ii \pi \int_{M^4}  b^{\Z_2}_\text{WL}[k_4b^{\Z_2}_\text{WL} + (k_3+k_4) \Bs_2\dd\t g+ B_\text{WS}^{\Z_2}]}
\end{align}
Let us consider a particular world-line which is a boundary:
$M^1_\text{WL}=\prt D^2_\text{WL}$, and write $b^{\Z_2}_\text{WL}$ as
\begin{align}
 b^{\Z_2}_\text{WL}\se{2} \bar b^{\Z_2}_{\text{WL}}+b^{\prime\Z_2}_{\text{WL}}
\end{align}
where
\begin{align}
 \dd b^{\prime\Z_2}_{\text{WL}}= C^{\Z_2}_\text{WL}
\end{align}
comes from the world-line and  $\bar b^{\Z_2}_{\text{WL}}$ from the background
Stiefel-Whitney class and other world-lines.
We obtain
\begin{align}
&\ \ \ \
\ee^{\ii \pi \int_{M^4}  b^{\Z_2}_\text{WL}[k_4b^{\Z_2}_\text{WL} + (k_3+k_4) \Bs_2\dd\t g+ B_\text{WS}^{\Z_2}]}
\nonumber\\
&=
\ee^{\ii \pi \int_{M^4}  \bar b^{\Z_2}_\text{WL}[k_4\bar b^{\Z_2}_\text{WL} + (k_3+k_4) \Bs_2\dd\t g+ B_\text{WS}^{\Z_2}]}\times
\nonumber\\
&
\ \ \ \ 
\ \ \ \ 
\ee^{\ii \pi \int_{M^4}  b^{\prime\Z_2}_\text{WL}[k_4b^{\prime\Z_2}_\text{WL} + (k_3+k_4) \Bs_2\dd\t g+ B_\text{WS}^{\Z_2}]}
\end{align}
which can be viewed as the effective action amplitude on the word line.

Compare with our previous result, we see that \frmbox{the term $\ee^{\ii \pi \int_{M^4}
k_4(b^{\prime\Z_2}_\text{WL})^2}$ should describe the Fermi statistics of the
point-like excitation when $k_4=1$.}  
Using the fact that  Poincar\'e dual of $b^{\prime\Z_2}_\text{WL}$ is
$D^2_\text{WL}$, we can express
$\ee^{\ii \pi \int_{M^4}
k_4(b^{\prime\Z_2}_\text{WL})^2}$ in terms of self-intersection number
of $D^2_\text{WL}$:
\begin{align}
 \ee^{\ii \pi \int_{M^4} k_4(b^{\prime\Z_2}_\text{WL})^2}
=\ee^{\ii \pi \text{Int}(D^2_\text{WL},D^2_\text{WL})}.
\end{align}
We see that \frmbox{the Fermi statistics in 3+1D is described by the
self-intersection number of the disk whose boundary is the world-line of the
fermion.} The term $\ee^{\ii \pi \int_{M^4} b^{\prime\Z_2}_\text{WL}
B_\text{WS}^{\Z_2}}$ describes the $\pi$ phase change as a point-like
excitation goes around the string-like excitation.

Now, let us concentrate on
\begin{align}
&\ \ \ \
\ee^{\ii \pi \int_{M^4}  (k_3+k_4)b^{\prime\Z_2}_\text{WL}  \Bs_2\dd\t g}
&=
\ee^{\ii \pi \int_{D^2_\text{WL}}  (k_3+k_4)\Bs_2\dd\t g}
\end{align}
where we have used the fact that Poincar\'e dual of $b^{\prime\Z_2}_\text{WL}$
is $D^2_\text{WL}$.  As discussed before, due to such a term will make 
\frmbox{the point-like
excitation carry a Kramer doublet formed by fractional $Z_2'$-charge $\pm
\frac12$, if $k_3+k_4\se{2} 1$ and carry a time-reversal singlet with integer
$Z_2'$-charge, if $k_3+k_4\se{2} 0$.} 

\subsubsection{Properties of string-like excitations}
\label{strasymm}

To obtain physical properties of string excitations, let us consider a
dimension reduction $M^4=M^3\times S^1$ (for details see Section
\ref{Z2aT3pex}).

Let us first consider the case for $k_4=0$ and start from \eqn{Z2Tk40dg}.  We
can choose $a^{\Z_2}_\text{WS}$ to make $\int_{S^1}a^{\Z_2}_\text{WS}=0$.  The
two sectors after the reduction are labeled by $\al=\int_{S^1}
a^{\Z_2}=\int_{S^1} a^{\Z_2}_0$.  The effective theory on $M^3$ after the
dimension reduction is given by
\begin{align}
&
 Z_{k_1k_2k_3k_4k_5k_6}(M^4) 
= \bar\del_2(k_3\Bs_2 \rw_1+ B_\text{WS}^{\Z_2})
\times
\nonumber\\
&
\hskip -1em
\sum_{ \t g,  \dd a^{\Z_2}_0\se{2}0 }  
\hskip -1em
\ee^{ \ii  \pi \int_{M^3} B^{\Z_2}_\text{WL}a^{\Z_2}_\text{WS} } 
 \ee^{ \ii  \pi \int_{M^3}   B^{\Z_2}_\text{WL} a^{\Z_2}_0 
+  \al (k_2+k_1) (\dd\t g)^3 
}
,
\nonumber 
\end{align}
where $a^{\Z_2}_0$ now lives on $M^3$ and $B^{\Z_2}_\text{WL}$ is the
Poincar\'e dual of the world-line in $M^3$.

For simplicity, let us choose the world-line to make
$B^{\Z_2}_\text{WL}=0$.  The effective theory on $M^3$ now
becomes (only the dynamical part)
\begin{align}
 Z_{k_1k_2k_3k_4k_5k_6}(M^4) 
&=
 \sum_{ \t g,  \dd a^{\Z_2}_0\se{2}0 }  
\hskip -1em
\ee^{ \ii  \pi \int_{M^3}   \al (k_2+k_1) (\dd\t g)^3 }
.
\end{align}
If we view the above effective theory as a 2+1D theory with time-reversal
$Z_2^T$ symmetry that acts on $\t g_i$, 
%and $Z_2$-gauge symmetry that acts on $a^{\Z_2}_0$, 
then the above effective theory describe trivial $Z_2^T$-SPT states
since the SPT invariant
\begin{align}
&\ \ \ \
 \ee^{ \ii  \pi \int_{M^3}   \al (k_2+k_1) (\dd\t g)^3  } 
\nonumber\\
&=\ee^{ \ii  \pi \int_{M^3}   \al (k_2+k_1) \rw_1^3 } =1
\end{align}
becomes trivial in 2+1D (see Appendix \ref{aSWrel3}).  The 1+1D boundary of the
2+1D theory in the $\al=1$ sector corresponds to the $Z_2$-vortex line.  So the
above result implies that the $Z_2$-vortex line of our model just behave like
the $Z_2$-vortex line of UT $Z_2$-gauge theory regardless the values of
$k_I$.

Our model actually has a $Z_2'\times Z_2^T$ symmetry. So the
2+1D effective theory can be viewed as a model with  $Z_2'$
symmetry.  In this case, the model describes a non-trivial $Z_2'$-SPT state, when $ \al (k_2+k_1) \neq 0$.  To see this,
we note that the $Z_2'$ acts like $\t g_i\to [\t g_i+1]_2$.  So to
obtain the $Z_2'$-SPT invariant, we need to gauge the
$Z_2'$ symmetry (see Section \ref{Gver}) by replacing $\dd \t g$ by
$a^{\prime\Z_2}$:
\begin{align}
 \ee^{ \ii  \pi \int_{M^3}   \al (k_2+k_1) (\dd\t g)^3 } 
&=\ee^{ \ii  \pi \int_{M^3}   \al (k_2+k_1) 
(a^{\prime\Z_2})^3 } .
\end{align}
The above SPT invariant allows us to show our 2+1D effective theory leads to a
non-trivial $Z_2'$-SPT state, which was first studied in \Ref{CLW1141}.  Since
the 1+1D boundary of the 2+1D theory in the $\al=1$ sector corresponds to the
$Z_2$-vortex line,  so the above result implies that 
\frmbox{the $Z_2$-vortex line of our model carry non-trivial edge excitations
of $Z_2'$-SPT state described by SPT invariant $\ee^{ \ii  \pi \int_{M^3}
(k_2+k_1) (a^{\prime\Z_2})^3 }$.}

The above results about the  $Z_2$-vortex line can be obtained by directly
calculating the effective theory on the  $Z_2$-vortex line.  We start from the
theory with excitations \eqn{Z2Tk40dg}.  Let the world-sheet of the string (\ie
the $Z_2$-vortex line), $M^2_\text{WS}$, be the boundary of $D^3_\text{WS}$.
For simplicity, let us assume that $\rw_2=\rw_1=0$ and $a^{\prime\Z_2}=0$ (\ie
no $Z_2'$-symmetry twist) on $M^4$.  In this case, $a^{\Z}_\text{WS}$ can be
chosen to be the Poincar\'e dual of $D^3_\text{WS}$.

The  effective theory on the string comes from the factor $\ee^{ \ii  \pi
\int_{M^4}  (k_2+k_1)
a^{\Z_2}_\text{WS} (\dd\t g)^3}$ in \eqn{Z2Tk40dg}, which leads to the
following effective theory
\begin{align}
Z 
&=
\sum_{\t g_i}
 \ee^{ \ii  \pi \int_{D^3_\text{WS}} (k_2+k_1) (\dd\t g)^3  }.
\end{align}

If we identify $(-)^{\t g_i}$ as $\si^z_i$ of spin-1/2, then the above action
amplitude describes a 2+1D spin-1/2 model with $Z'_2\times Z_2^T$ symmetry
acting on $\t g_i$'s:
\begin{align}
Z'_2:&\  \prod_i \si^x_i ,
&
Z_2^T:&\  K\prod_i \si^x_i,
\end{align}
where $K$ is the complex conjugation.  The effective theory actually describes
a non-trivial $Z'_2\times Z_2^T$-SPT state on
$D^3_\text{WS}$.  So the effective theory on the world-sheet $M^2_\text{WS}$ is
the effective boundary theory of the $Z'_2\times Z_2^T$-SPT
state. In other worlds, the string will carry non-trivial boundary excitations
of the 2+1D $Z'_2\times Z_2^T$-SPT state.  The
non-trivialness of the excitations on the string is protected by the anomalous
symmetry on the boundary.\cite{W1313}  This can be viewed as the symmetry
fractionalization (or quantum number fractionalization) on strings.  We have
seen that on point-like excitation, the $T^2=1$ $Z_2^T$ time-reversal symmetry
can be fractionalized into $T^2=-1$ Kramer doublet.  In contrast, on strings,
the symmetry fractionalization is realized as the anomalous (\ie non-on-site)
symmetry that constrains the effective theory for degrees of freedom on the
strings.

So the key to calculate the symmetry fractionalization is to calculate the
non-on-site (\ie anomalous) symmetry on the strings.  Let us do the calculation
for the case $k_2+k_1=1$, which leads to the following effective theory
\begin{align}
Z &= 
\sum_{\t g_i}
 \ee^{ \ii  \pi \int_{D^3_\text{WS}} (\dd\t g)^3  } .
\end{align}
which describes a $Z'_2\times Z_2^T$-SPT state.
The group-cocycle that describes the
$Z'_2\times Z_2^T$-SPT phase is in fact the topological term
$\int_{D^3_\text{WS}} (\dd\t g)^3$:
\begin{align}
 \nu_3( \t g_0, \t g_1, \t g_2, \t g_3)
= -(\t g_0 -\t g_1) (\t g_1 -\t g_2) (\t g_2 -\t g_3)
\end{align}
The $Z_2\times Z_2^T$ symmetry on the string are twisted by the group-cocycle
and becomes non-on-site 
\begin{align}
Z'_2:&\ \ \ \prod_I \si^x_I \ee^{\ii \pi \nu_3(1,0,\t g_I,\t g_{I+1})},
\nonumber\\
Z_2^T:&\ \ \ K\prod_I \si^x_I  \ee^{\ii \pi \nu_3(1,0,\t g_I,\t g_{I+1})} ,
\end{align}
where 
\begin{align}
\ee^{\ii \pi \nu_3(1,0,\t g_I,\t g_J)}
&
= (-)^{\t g_I (\t g_I-\t g_J)}
= (-)^{\t g_I} (-)^{\t g_I\t g_J}
\nonumber\\
&=
\si^z_{I}\frac{1+\si^z_{I}+\si^z_{J}  -\si^z_{I}\si^z_{J}}{2}
\end{align}
%In fact, the $Z'_2$ symmetry is an emergent symmetry, which can be
%broken explicitly on the string.  So we only have the anomalous (\ie
%non-on-site) $Z_2^T$ symmetry on the strings.

The effective Hamiltonian on the string respects the anomalous $Z_2'\times Z_2^T$ symmetry,
which may take a form
\begin{align}
 H_\text{str} &= \sum_I J^z_I \si^z_I \si^z_{I+1} +
\sum_I K^x_I( \si^z_{I-1} \si^x_I \si^z_{I+1} - \si^x_I)
%\nonumber\\
%&\ \ \ \
%+\sum_I K^y_I( \si^z_{I-1} \si^y_I \si^z_{I+1} - \si^y_I).
\end{align}
%(The anomalous $Z_2'$ symmetry will require $K^y_I=0$.) 
The ground state of such Hamiltonian is gapless or spontaneously breaks the
$Z_2'$ symmetry.  So when $k_1+k_2=1$ and $k_4=0$, the strings carry non-trivial
excitations described by the above Hamiltonian with an anomalous $Z_2'\times
Z_2^T$ symmetry: $ U'=\prod_I \si^x_I
\prod_I\si^z_{I}\frac{1+\si^z_{I}+\si^z_{I+1} -\si^z_{I}\si^z_{I+1}}{2}$ and $
U_T=K\prod_I \si^x_I \prod_I\si^z_{I}\frac{1+\si^z_{I}+\si^z_{I+1}
-\si^z_{I}\si^z_{I+1}}{2}$.  

Next, let us consider the case for $k_1=0$ and start from \eqn{Z2Tk10}.
The only term that involve the word sheet is
$\ee^{\ii \pi \int_{M^4} b^{\Z_2}_\text{WL}B^{\Z_2}_\text{WS}}$, which can be 
rewritten as
\begin{align}
 \ee^{\ii \pi \int_{M^4} b^{\Z_2}_\text{WL}B^{\Z_2}_\text{WS}}
&= \ee^{\ii \pi \int_{M^2_\text{WS}} b^{\Z_2}_\text{WL}}
= \ee^{\ii \pi \int_{D^3_\text{WS}} \dd b^{\Z_2}_\text{WL}}
\nonumber\\
&
= \ee^{\ii \pi \int_{D^3_\text{WS}} k_2 (\dd \t g)^3+ C^{\Z_2}_\text{WL}}
\end{align}
Repeat the above calculation, we see that 
\frmbox{when $k_2=1$ and $k_1=0$, the strings
carry non-trivial excitations with an
anomalous $Z_2'\times Z_2^T$ symmetry: $ U'=\prod_I \si^x_I
\prod_I\si^z_{I}\frac{1+\si^z_{I}+\si^z_{I+1} -\si^z_{I}\si^z_{I+1}}{2}$ and $
U_T=K\prod_I \si^x_I \prod_I\si^z_{I}\frac{1+\si^z_{I}+\si^z_{I+1}
-\si^z_{I}\si^z_{I+1}}{2}$.}  

We like remark that potentially, the strings may carry an anomalous $Z_2\times
Z_2'\times Z_2^T$ symmetry, where $Z_2$ is associated with $a^{\Z_2}$.  From
the above calculation, we see that the  anomalous symmetry only come from the
$Z_2'$ symmetry. There is no anomalous symmetry from $Z_2$.

%From the companion term
%\begin{align}
%& \ \ \ \
%\sum_{\dd a^{\Z_2}_0 \se{2}0} \hskip -0.5em
%\ee^{ \ii  \pi \int_{M^4} k_1   a^{\Z_2}_0 \Bs_2 a^{\Z_2}_0 \dd\t g+(k_2+k_1) (\dd\t g)^3  a^{\Z_2}_0}
%%\nonumber\\
%&=
%\sum_{\dd a^{\Z_2}_0 \se{2}0} \hskip -0.5em
%\ee^{ \ii  \pi \int_{M^4} k_1   (a^{\Z_2}_0)^3 \rw_1 +(k_2+k_1) \rw_1^3  a^{\Z_2}_0}
%%\\
%&=
%\sum_{\dd a^{\Z_2}_0 \se{2}0} \hskip -0.5em
%%\ee^{ \ii  \pi \int_{M^4}  a^{\Z_2}_0 (k_1\rw_3+k_2\rw_1^3)}
%= \bar\del_2(k_1  \rw_3+k_2\rw_1^3)  ,
%\nonumber 
%\end{align}
%we see that the models with such an anomalous-$Z_2^T$ string have
%vanishing partition function on space-time with $\rw_1^3 \neq 0$ (since $k_1=0$ and $k_2=1$).

\section{Acknowledgement} 

I would like to thank Meng Cheng, Cenke Xu, and Liujun Zou for many helpful
discussions.  This research was supported by NSF Grant No.  DMR-1506475 and
NSFC 11274192.

\appendix

\section{The K\"unneth formula} \label{HBGRZ}

The K\"unneth formula is a very helpful formula that allows us to calculate the
cohomology of chain complex $X\times X'$ in terms of  the cohomology of chain
complex $X$ and chain complex $X'$.  The K\"unneth formula is expressed in
terms of the tensor-product operation $\otimes_R$ and the torsion-product
operation $\Tor_1^R$ that act on $R$-modules $\M,\M',\M''$.  Here $R$ is a ring
and a $R$-module is like a vector space over $R$ (\ie we can ``multiply'' a
``vector'' in $\M$ by an element of $R$, and two ``vectors'' in $\M$ can add.)
The tensor-product operation $\otimes_R$ has the following properties:
\begin{align}
\label{tnprd}
& \M \otimes_\Z \M' \simeq \M' \otimes_\Z \M ,
\nonumber\\
&  (\M'\oplus \M'')\otimes_R \M = (\M' \otimes_R \M)\oplus (\M'' \otimes_R \M)   ,
\nonumber\\
& \M \otimes_R (\M'\oplus \M'') = (\M \otimes_R \M')\oplus (\M \otimes_R \M'')   ;
\nonumber\\
& \Z \otimes_\Z \M \simeq \M \otimes_\Z \Z =\M ,
\nonumber\\
& \Z_n \otimes_\Z \M \simeq \M \otimes_\Z \Z_n = \M/n\M ,
%\nonumber\\
%& \Z_n \otimes_\Z \RZ \simeq \RZ \otimes_\Z \Z_n = 0,
\nonumber\\
& \Z_m \otimes_\Z \Z_n  =\Z_{\<m,n\>} ,
%\nonumber\\
%& \RZ \otimes_\Z \RZ = 0,
%\nonumber\\
%& \R \otimes_\Z \RZ = 0,
%\nonumber\\
%& \R \otimes_\Z \R = \R,
\end{align}
The torsion-product operation $\Tor_1^R$ has the following properties:
\begin{align}
\label{trprd}
& \Tor^1_R(\M , \M') \simeq \Tor^1_R (\M', \M)  ,
\nonumber\\
& \Tor^1_R(\M'\oplus \M'' , \M) = \Tor^1_R(\M', \M) \oplus \Tor^1_R(\M'',\M),
\nonumber\\
& \Tor^1_R(\M,\M'\oplus \M'') = \Tor^1_R(\M,\M')\oplus \Tor^1_R(\M,  \M'')
\nonumber\\
& \Tor^1_\Z(\Z,  \M) = \Tor^1_\Z(\M,  \Z) = 0,
\nonumber\\
& \Tor^1_\Z(\Z_n, \M) = \{m\in \M| nm=0\},
%\nonumber\\
%& \Tor^1_\Z(\Z_n, \RZ) = \Z_n,
\nonumber\\
& \Tor^1_\Z(\Z_m, \Z_n) = \Z_{\<m,n\>} ,
%\nonumber\\
%& \text{Tor}_1^\Z(U(1), U(1)) = 0 ,
\end{align}
where $\<m,n\>$ is the greatest common divisor of $m$ and $n$.  These
expressions allow us to compute the tensor-product $\otimes_R$ and  the
torsion-product $\Tor^1_R$.  We will use abbreviated $\Tor$ to denote
$\Tor^1_\Z$.

The K\"unneth formula itself is given by (see
\Ref{Spa66} page 247)
\begin{align}
\label{kunn}
&\ \ \ \ H^d(X\times X',\M\otimes_R \M')
\nonumber\\
&\simeq \Big[\oplus_{k=0}^d H^k(X,\M)\otimes_R H^{d-k}(X',\M')\Big]\oplus
\nonumber\\
&\ \ \ \ \ \
\Big[\oplus_{k=0}^{d+1}
\Tor_R^1(H^k(X,\M), H^{d-k+1}(X',\M'))\Big]  .
\end{align}
Here $R$ is a principle ideal domain and $\M,\M'$ are $R$-modules such that
$\Tor_R^1(\M, \M')=0$.  
We also require either\\
(1) $H_d(X;\Z)$ and  $H_d(X';\Z)$ are finitely generated, or\\
(2) $\M'$ and $H_d(X';\Z)$ are
finitely generated.\\
%For example, $\M'=\Z\oplus \cdots \oplus \Z\oplus \Z_n\oplus\cdots\oplus\Z_m$.
%Note that $\R$ and $\RZ$ are not finitely generated if the coefficient ring is
%$\Z$.

For more details on principal ideal domain and $R$-module, see the
corresponding Wiki articles.  Note that $\Z$ and $\R$ are principal ideal
domains, while $\RZ$ is not.  Also, $\R$ and $\RZ$ are not finitely
generate $R$-modules if $R=\Z$.  The K\"unneth formula also works for
topological cohomology where $X$ and $X'$ are treated as topological spaces.  

For homology, there is a similar
 K\"unneth formula
\begin{align}
\label{kunnH}
&\ \ \ \ H_d(X\times X';\Z)
\nonumber\\
&\simeq \Big[\oplus_{k=0}^d H_k(X;\Z)\otimes H_{d-k}(X';\Z)\Big]\oplus
\nonumber\\
&\ \ \ \ \ \
\Big[\oplus_{k=0}^{d-1}
\Tor(H_k(X;\Z), H_{d-k-1}(X';\Z))\Big]  .
\end{align}

%But, it does not work for group cohomology by treating  $H^d$ as $\cH^d$ and
%$X$ and $X'$ as groups, $X=G$ and $X'=G'$, as demonstrated by the example
%$\M=\M'=\RZ$ and $X=X'=Z_n$.  However, since $\cH^d(G;\Z)=H^d(BG;\Z)$, the
%above K\"unneth formula works for group cohomology when $\M=\M'=\Z$.  The
%above K\"unneth formula also works for group cohomology when $G,G'$ are finite
%or when $G'$ is finite and $\M'$ is finitely generate (such as $\M'$ is $\Z$
%or $\Z_n$).

%, provided that $G'$ is a finite group.  However, the above K\"unneth formula
%does not apply for Borel-group cohomology when $X'=G'$ is a continuous group,
%since in that case the set of cycles $\cH_d(G';\Z)$ is not finitely generated.

%The K\"unneth formula works for topological cohomology where $X$ and $X'$ are
%treated as topological spaces.  However, the above K\"unneth formula  does not
%apply to Boral-group cohomology where $X$ and $X'$ are treated as groups.

As the first application of K\"unneth formula, we like to use it to calculate
$H^*(X',\M)$ from $H^*(X';\Z)$,  by choosing $R=\M'=\Z$. In this case, the
condition $\Tor_R^1(\M,\M')=\Tor_\Z^1(\M, \Z)=0$ is always
satisfied. $\M$ can be $\RZ$, $\Z$, $\Z_n$ \etc. So we have
\begin{align}
\label{kunnZ}
&\ \ \ \ H^d(X\times X',\M)
\nonumber\\
&\simeq \Big[\oplus_{k=0}^d H^k(X,\M)\otimes_{\Z} H^{d-k}(X';\Z)\Big]\oplus
\nonumber\\
&\ \ \ \ \ \
\Big[\oplus_{k=0}^{d+1}
\Tor(H^k(X,{\M}), H^{d-k+1}(X';\Z))\Big]  .
\end{align}
The above is also valid for topological cohomology.

We can further choose $X$ to be the space of one point in \eqn{kunnZ},
and use
\begin{align}
H^{d}(X,\M))=
\begin{cases}
\M, & \text{ if } d=0,\\
0, & \text{ if } d>0,
\end{cases}
\end{align}
to reduce \eqn{kunnZ} to
\begin{align}
\label{ucf}
 H^d(X,\M)
&\simeq H^d(X;\Z) \otimes_{\Z} \M 
\oplus
\Tor(H^{d+1}(X;\Z),\M)  .
\end{align}
where $X'$ is renamed as $X$.  The above is a form of the universal coefficient
theorem which can be used to calculate $H^*(X,\M)$ from $H^*(X;\Z)$ and the
module $\M$.  The  universal coefficient theorem works for topological
cohomology where $X$ is a topological space.  
In fact, we also have a similar universal coefficient
theorem for homology
\begin{align}
\label{ucfH}
 H_d(X,\M)
&\simeq H_d(X;\Z) \otimes_{\Z} \M 
\oplus
\Tor(H_{d-1}(X;\Z),\M)  .
\end{align}

Using the universal
coefficient theorem, we can rewrite \eqn{kunnZ} as
\begin{align}
\label{kunnHX}
H^d(X\times X',\M) \simeq \oplus_{k=0}^d H^k[X, H^{d-k}(X',\M)] .
\end{align}
The above is also valid for topological cohomology. We note that
\begin{align}
 H^0(X,\M)=\M.
\end{align}

There is also a universal coefficient theorem between homology and cohomolgy
\begin{align}
\label{ucf1}
 H^d(X,\M)
&\simeq \Hom(H_d(X;\Z), \M) 
\oplus
\Ext(H_{d-1}(X;\Z),\M)  .
\end{align}
Here Ext operation on modules is given by
\begin{align}
\label{extprd}
& \Ext^1_R(\M'\oplus \M'' , \M) = \Ext^1_R(\M', \M) \oplus \Ext^1_R(\M'',\M),
\nonumber\\
& \Ext^1_R(\M,\M'\oplus \M'') = \Ext^1_R(\M,\M')\oplus \Ext^1_R(\M,  \M'')
\nonumber\\
& \Ext^1_\Z(\Z,  \M) = 0,
\nonumber\\
& \Ext^1_\Z(\Z_n, \M) = \M/n\M,
\nonumber\\
& \Ext^1_\Z(\Z_n, \Z) = \Z_n,
%\nonumber\\
%& \Ext^1_\Z(\Z_n, \RZ) = 0,
\nonumber\\
& \Ext^1_\Z(\Z_m, \Z_n) = \Z_{\<m,n\>} ,
%\nonumber\\
%& \text{Tor}_1^\Z(U(1), U(1)) = 0 ,
.
\end{align}
The Hom operation on modules is given by
\begin{align}
\label{homprd}
& \Hom_R(\M'\oplus \M'' , \M) = \Hom_R(\M', \M) \oplus \Hom_R(\M'',\M),
\nonumber\\
& \Hom_R(\M,\M'\oplus \M'') = \Hom_R(\M,\M')\oplus \Hom_R(\M,  \M'')
\nonumber\\
& \Hom_\Z(\Z,  \M) = \M,
\nonumber\\
& \Hom_\Z(\Z_n, \M) = \{m\in \M| nm=0\},
\nonumber\\
& \Hom_\Z(\Z_n, \Z) = 0,
%\nonumber\\
%& \Hom_\Z(\Z_n, \RZ) = \Z_n,
\nonumber\\
& \Hom_\Z(\Z_m, \Z_n) = \Z_{\<m,n\>} ,
.
\end{align}
We will use abbreviated $\Ext$ and $\Hom$ to
denote $\Ext^1_\Z$ and $\Hom_\Z$.

\section{ Poincar\'e Duality}

Poincar\'e Duality relates $H^k(M^d,R)$ and $H_{d-k}(M^d,R)$.
We note that for a closed connected $d$-dimensional space $M^d$,
$H_0(M^d;\Z)=\Z$, $H_d(M^d;\Z)=\Z$ if $M^d$ is orientable, and $H_d(M^d;\Z)=0$
if $M^d$ is non-orientable.  Similarly, $H^0(M^d;\Z)=\Z$, $H^d(M^d;\Z)=\Z$ if
$M^d$ is orientable, and $H^d(M^d;\Z)=\Z_2$ if $M^d$ is non-orientable.

\textbf{Poincar\'e Duality}: If $M$ is a closed $R$-orientable $n$-dimensional
manifold with fundamental class $[M] \in H^n(M,R)$ (here $R$ is a ring), then
the map $D: H^k (M;R) \to H_{n - k} (M; R)$  defined by $D(\al) =[M]\cap \al$
is an isomorphism for all $k$.

The cup product pairing between $H^k(M^d,R)$ and $H^{d-k}(M^d,R)$ is
non singular for closed $R$-orientable manifolds when $R$ is a field, or when
$R= \Z$ and torsion in $H^*(M; \Z)$ is factored out.  This implies that the
free part of $H^k(M^d;\Z)$ and $H^{d-k}(M^d;\Z)$ has the same dimension.

\section{The factor $\frac{|H^0(M^4;\Z_n)|^2|H^2(M^4;\Z_n)|}{|H^1(M^4;\Z_n)|^2}$}
\label{Hfactor}

To calculate the factor
$\frac{|H^0(M^4;\Z_n)|^2|H^2(M^4;\Z_n)|}{|H^1(M^4;\Z_n)|^2}$ we first use
\eqn{ucf1} to show
\begin{align}
&\ \ \ \
 H^1(M^4;\Z_n)
\\
&=\Hom(H_1(M^4;\Z);\Z_n)\oplus \Ext(H_0(M^4;\Z),Z_n)
\nonumber\\
&=\Hom(\text{f}H_1(M^4;\Z);\Z_n)\oplus \Hom(\text{t}H_1(M^4;\Z);\Z_n)
\nonumber\\
&=\Z_n^{\oplus b_1}\oplus \Hom(\text{t}H_1(M^4;\Z);\Z_n),
\nonumber \\
\end{align}
and
\begin{align}
&\ \ \ \
 H^2(M^4;\Z_n)
\\
&=\Hom(H_2(M^4;\Z);\Z_n)\oplus \Ext(H_1(M^4;\Z),Z_n)
\nonumber\\
&=\Hom(\text{f}H_2(M^4;\Z);\Z_n)\oplus \Hom(\text{t}H_2(M^4;\Z);\Z_n)\oplus
\nonumber\\
&\ \ \ \
\Ext(\text{t}H_1(M^4;\Z),Z_n),
\nonumber\\
&=\Z_n^{\oplus b_2} \oplus \Hom(\text{t}H_2(M^4;\Z);\Z_n)\oplus
\Hom(\text{t}H_1(M^4;\Z),Z_n).
\nonumber 
\end{align}
where ``f'' and ``t'' indicate the free and torsion parts of a discrete abelian
group and $b_n$ is the dimension of $\text{f}H_n(M^4;\Z)$ (\ie the
$n^\text{th}$ Betti number).
Using $H^0(M^4;\Z_n)=\Z_n^{\oplus b_0}$ , we find that
\begin{align}
&\ \ \ \
 \frac{|H^0(M^4;\Z_n)|^2|H^2(M^4;\Z_n)|}{|H^1(M^4;\Z_n)|^2}
\nonumber\\
&= n^{2b_0+b_2-2b_1}
\frac{ |\Hom(\text{t}H_2(M^4;\Z);\Z_n)| } {|  \Hom(\text{t}H_1(M^4;\Z);\Z_n) |} .
\end{align}
We note that, according to \eqn{ucf1}
\begin{align}
&\ \ \ \
 H^2(M^d;\Z)
\nonumber\\
&=\Hom(H_2(M^d;\Z),\Z)\oplus \Ext(H_1(M^d;\Z),\Z).
\end{align}
Since $\Hom(\Z_n,\Z)=0$, $\Ext(\Z_n,\Z)=\Z_n$, and $\Ext(\Z,\Z)=0$, 
we see that t$H^2(M^d;\Z)= \text{t}H_1(M^d;\Z)$. We get
\begin{align}
&\ \ \ \
 \frac{|H^0(M^4;\Z_n)|^2|H^2(M^4;\Z_n)|}{|H^1(M^4;\Z_n)|^2}
\nonumber\\
&= n^{2b_0+b_2-2b_1}
\frac{ |\Hom(\text{t}H_2(M^4;\Z);\Z_n)| } {|  \Hom(\text{t}H^2(M^4;\Z);\Z_n) |} .
\end{align}

For 4-dimensional closed orientable manifolds $b_1=b_3$, $b_0=b_4$ and
$\chi(M^4) = \sum_{n=0}^4 (-)^n b_n$ is the Euler number of $M^4$.
Using Poincar\'e duality $H^2(M^4;\Z) = H_2(M^4;\Z)$, we can show that
\begin{align}
\label{HZnchi}
 \frac{|H^0(M^4;\Z_n)|^2|H^2(M^4;\Z_n)|}{|H^1(M^4;\Z_n)|^2}
&= n^{\chi(M^4)}.
\end{align}

When $n=2$ we have a Poincar\'e duality $H^k(M^d;\Z_2) = H^{d-k}(M^d;\Z_2)$ for
any closed manifold $M^d$ regardless if $M^d$ is orientable or not (since $M^d$
is always $Z_2$-orientable).  Thus
\begin{align}
&\ \ \ \ 
\frac{|H^0(M^4;\Z_2)|^2|H^2(M^4;\Z_2)|}{|H^1(M^4;\Z_2)|^2}
\nonumber\\
&=
 \frac{|H^0(M^4;\Z_2)||H^2(M^4;\Z_2)||H^4(M^4;\Z_2)|}{|H^1(M^4;\Z_2)||H^3(M^4;\Z_2)|}.
\end{align}
According to \eqn{ucf1}
\begin{align}
&\ \ \ \
 H^k(M^d;\Z_2)
\nonumber\\
&= \Hom(H_k(M^d;\Z),\Z_2)\oplus \Ext(H_{k-1}(M^d;\Z),\Z_2)
\nonumber\\
&= \Z_2^{\oplus b_k} \oplus \Hom(\text{t}H_k(M^d;\Z),\Z_2) \oplus
\nonumber\\
&\ \ \ \ 
\Hom(\text{t}H_{k-1}(M^d;\Z),\Z_2).
\end{align}
This allows us to show
\begin{align}
\label{HZ2chi}
\frac{|H^0(M^4;\Z_2)|^2|H^2(M^4;\Z_2)|}{|H^1(M^4;\Z_2)|^2}
&= 2^{\chi(M^4)},
\end{align}
where we have used the fact that $\text{t}H_4(M^4;\Z)=0$ for both
orientable and non-orientable closed manifolds.
On the other hand, the factor
$\frac{|H^0(M^4;\Z_n)|^2|H^2(M^4;\Z_n)|}{|H^1(M^4;\Z_n)|^2}$ is in general not of
the form $\rho^{\chi(M^4)}$ for non-orientable manifolds when $n> 2$.

\section{Relations between cocycles and Stiefel-Whitney classes
on a closed manifold}
\label{Rswc}

The cocycles and the Stiefel-Whitney classes on a closed manifold many satisfy
many relations.  In this section, we will show how to generate those relations.

\subsection{Introduction to Stiefel-Whitney classes}

The Stiefel-Whitney classes $\rw_i \in H^i(M^d;\Z_2)$ is defined for an $O(n)$
vector bundle on a $d$-dimensional space with $n\to \infty$.  If the
$O(\infty)$ vector bundle on $d$-dimensional space, $M^d$, happen to be the
tangent bundle of $M^d$ direct summed with a trivial $\infty$-dimensional
vector bundle, then the corresponding Stiefel-Whitney classes are referred as
the Stiefel-Whitney classes of the manifold $M^d$.

The Stiefel-Whitney classes of manifold behave well under the connected sum of
manifolds.  Let $\rw(M)$ be the total Stiefel-Whitney class of a manifold $M$.
If we know $\rw(M)$ and $\rw(N)$,  then we can obtain $\rw(M\#N)$:
\begin{align}
\label{SWsum}
\rw(M\#N)=\rw(M)+\rw(N) -1.
\end{align}
Under the product of manifolds, we have
\begin{align}
\rw(M\times N)=\rw(M)\rw(N) .
\end{align}

The Stiefel-Whitney numbers are non-oriented cobordism invariant.  All the
Stiefel-Whitney numbers of a smooth compact manifold vanish iff the manifold is
the boundary of some smooth compact manifold.  Here the manifold can be
non-orientable.

The Stiefel-Whitney numbers and Pontryagin numbers are oriented cobordism
invariant.  All the Stiefel-Whitney numbers and Pontryagin numbers  of a smooth
compact orientable manifold vanish iff the manifold is the boundary of some
smooth compact orientable manifold.

\subsection{Relations between Stiefel-Whitney classes of the tangent bundle}

For generic $O(\infty)$ vector bundle, the Stiefel-Whitney classes are all
independent. However, the Stiefel-Whitney classes for a manifold (\ie for the
tangent bundle) are not independent and satisfy many relations.

To obtain those relations, we note that,  for any $O(\infty)$ vector bundle,
the total Stiefel-Whitney class $\rw=1+\rw_1+\rw_2+\cdots$ is related to the
total Wu class $u=1+u_1+u_2+\cdots$ through the total Steenrod square
\cite{W5008}:
\begin{align}
 \rw=Sq(u),\ \ \ Sq=1+\Sq^1+\Sq^2+ \cdots .
\end{align}
Therefore, 
$\rw_n=\sum_{i=0}^n \Sq^i (u_{n-i})$.
The Steenrod squares have the following properties:
\begin{align}
\Sq^i(x_j) &=0, \  i>j, \ \ 
\Sq^j(x_j) =x_jx_j,  \ \  \Sq^0=1,
\end{align}
for any $x_j\in H^j(M^d;\Z_2)$.
Thus
\begin{align}
u_n=\rw_n+\sum_{i=1, 2i\leq n} \Sq^i (u_{n-i}).
\end{align}
This allows us to compute $u_n$ iteratively, using Wu formula
\begin{align}
\label{WuF}
\Sq^i(\rw_j) &=0, \ \ i>j, \ \ \ \ \
\Sq^i(\rw_i) =\rw_i\rw_i, 
\\
 \Sq^i(\rw_j) &= \rw_i\rw_j+\sum_{k=1}^i 
\frac{(j-i-1+k)!}{(j-i-1)!k!}
\rw_{i-k} \rw_{j+k},\ \ i<j ,
\nonumber\\
\Sq^1(\rw_j) &= \rw_1\rw_j + (j-1) \rw_{j+1},
\nonumber 
\end{align}
and the Steenrod relation 
\begin{align}
	\Sq^n(xy)=\sum_{i=0}^n \Sq^i(x)\Sq^{n-i}(y).
\end{align}
We find
\begin{align}
u_0&=1, 
\ \ \ \ \
u_1=\rw_1, 
\ \ \	 \ \
u_2=\rw_1^2+\rw_2, 
	\nonumber\\
u_3&=\rw_1\rw_2, 
\ \ \ \ \
u_4=\rw_1^4+\rw_2^2+\rw_1\rw_3+\rw_4, 
	\\
u_5&=\rw_1^3\rw_2+\rw_1\rw_2^2+\rw_1^2\rw_3+\rw_1\rw_4, 
	\nonumber\\
u_6&=\rw_1^2\rw_2^2+\rw_1^3\rw_3+\rw_1\rw_2\rw_3+\rw_3^2+\rw_1^2\rw_4+\rw_2\rw_4, 
	\nonumber\\
u_7&=\rw_1^2\rw_2\rw_3+\rw_1\rw_3^2+\rw_1\rw_2\rw_4, 
	\nonumber\\
u_8&=\rw_1^8+\rw_2^4+\rw_1^2\rw_3^2+\rw_1^2\rw_2\rw_4+\rw_1\rw_3\rw_4+\rw_4^2
	\nonumber\\
	&\ \ \ \ 
	+\rw_1^3\rw_5 +\rw_3\rw_5+\rw_1^2\rw_6+\rw_2\rw_6+\rw_1\rw_7+\rw_8. 
	\nonumber
\end{align}
%We note that the Steenrod squares form an algebra:
%\begin{align}
% \Sq^a\Sq^b
%&=\sum_{j=0}^{[a/2]} \bpm b-j-1 \\ a-2j \\ \epm  \Sq^{a+b-j} \Sq^j, 
%\nonumber\\
%&=\sum_{j=0}^{[a/2]} \frac{(b-j-1)!}{(a-2j)!(b-a+j-1)!} \Sq^{a+b-j} \Sq^j, 
%\nonumber\\
%& 0<a<2b.
%\end{align}
%which leads to the relation $\Sq^1\Sq^1=0$ used in the last section.

If the $O(\infty)$ vector bundle on $d$-dimensional space, $M^d$, happen to be
the tangent bundle of $M^d$, then the corresponding Wu class and the
Steenrod square satisfy 
\begin{align}
\label{SqWu}
\Sq^{d-j}(x_j)=u_{d-j} x_j,  \text{ for any } x_j \in H^j(M^d;\Z_2) .
\end{align}
We can generate many relations for cocycles and Stiefel-Whitney classes on a
manifold using the above result:
\begin{enumerate}
\item If we choose $x_j$ to be a combination of Stiefel-Whitney classes, the
above will generate many relations between Stiefel-Whitney classes.
\item If we choose $x_j$ to be a combination of Stiefel-Whitney classes and cocycles, the
above will generate many relations between Stiefel-Whitney classes and cocycles.
\item Since $\Sq^i(x_j)=0$ if $i>j$, therefore $u_ix_{d-i}=0$ for any $x_{d-i}
\in H^{d-i}(M^d;\Z_2)$ if $i>d-i$.  Since $\Z_2$ is a field and according to the Poincar\'e duality, this implies that
$u_i=0$ for  $2i>d$.  
\item $\Sq^n\cdots \Sq^m(u_i)=0$ if $2i>d$.  This also gives
us relations among  Stiefel-Whitney classes.  
%\item There is another type of relation. In $4n$-dimension, the mod 2 reduction
%of Pontryagin classes $p_{i_1}p_{i_2}\cdots$, $n=i_1+i_2+\cdots$, should be
%regarded as zero. The reason is explained below the \eqn{w22}.  This lead to
%the relations for $d$-dimensional manifold
%\begin{align}
%\label{w2i2pi}
% \rw_{2i_1}^2
% \rw_{2i_2}^2\cdots =0, \text{ if } 2i_1+2i_2+\cdots = d.
%\end{align}
\end{enumerate}

%The relation between Stiefel-Whitney classes of $M^d$ can be systematically
%described by
%\begin{align}
%\langle \sum_{i+j=d-k}u_i{\rm Sq}^{j} p, [M^d]\rangle =0
%\end{align}
%where $p=p(w_1,w_2,\ldots)\in H^k(M^d;Z/2)$ and $[M^d]$ the generator of
%$H_d(M^d;Z/2)$.  Thom showed that all relations between characteristic numbers
%arise in this way. This allowed him to compute the bordism ring of non-oriented
%manifolds exactly: it is $\Z_2[m_k]|_{k \neq 2^j-1}$.  (see
%http://mathoverflow.net/questions/40539)

\subsection{Relations between Stiefel-Whitney classes and a $\Z_2$-valued
1-cocycle in 3-dimensions}
\label{aSWrel3}

On a 3-dimensional manifold, we can find
many relations between Stiefel-Whitney classes:\\
(1) $u_2=\rw_1^2+\rw_2=0$.\\
(2) $u_3=\rw_1\rw_2=0$.\\
(3) $\Sq^1(u_2)=0$. Using $\Sq^1(\rw_i)=\rw_1\rw_i+(i+1)\rw_{i+1}$, we find that
$\Sq^1(\rw_1^2+\rw_2)=\Sq^1(\rw_1)\rw_1+\rw_1\Sq^1(\rw_1)+\Sq^1(\rw_2)=\rw_1\rw_2+\rw_3=0$.\\
This gives us three relations
\begin{align}
 \rw_1^2=\rw_2,\ \ \rw_1\rw_2=\rw_3=0.
\end{align}
Let $a^{\Z_2}$ be a $\Z_2$-valued 1-cocycle. 
We can also find a relation between the
Stiefel-Whitney classes and $a^{\Z_2}$: 
\begin{align}
\rw_1(a^{\Z_2})^2=\Sq^1((a^{\Z_2})^2)=2(a^{\Z_2})^3=0.
\end{align}

There are six possible 3-cocycles
that can be constructed from the Stiefel-Whitney classes and the 1-cocycle
$a^{\Z_2}$:
\begin{align}
& (\rw_1)^3, && \rw_1 \rw_2, && \rw_3,
\nonumber\\
& (a^{\Z_2})^3, && \rw_1 (a^{\Z_2})^2, && \rw_1^2 a^{\Z_2}.
\end{align}
From the above relations, we see that only two of them are non-zero:
\begin{align}
& (a^{\Z_2})^3, && \rw_1^2 a^{\Z_2}.
\end{align}

\subsection{Relations between Stiefel-Whitney classes and a $\Z_2$-valued
1-cocycle in 4-dimensions}
\label{aSWrel4}

The relations between the  Stiefel-Whitney classes for
4-dimensional manifold can be listed:\\
(1) $u_3=\rw_1\rw_2=0$.\\
(2) $u_4=\rw_1^4+\rw_2^2+\rw_1\rw_3+\rw_4=0$.\\
(3) $\Sq^1(u_3)=0$, which implies
$\Sq^1(\rw_1\rw_2)=\Sq^1(\rw_1)\rw_2+\rw_1\Sq^1(\rw_2)=\rw_1^2\rw_2
+\rw_1^2\rw_2+\rw_1\rw_3=\rw_1\rw_3=0$,\\
%(4) $\Sq^1(\rw_2)=\rw_1\rw_2+\rw_3$
which can be summarized as
\begin{align}
\label{wrel4}
 \rw_1\rw_2=0,\ \ \rw_1\rw_3=0,\ \ \rw_1^4+\rw_2^2+\rw_4=0.
\end{align}
We also have many relations between
the  Stiefel-Whitney classes and $a^{\Z_2}$:\\
(1) $\Sq^1((a^{\Z_2})^3)=(a^{\Z_2})^4=\rw_1(a^{\Z_2})^3$.\\
(2) $\Sq^1(\rw_1^2a^{\Z_2})=\rw_1^2(a^{\Z_2})^2=\rw_1^3a^{\Z_2}$.\\
(3) $\Sq^1(\rw_2a^{\Z_2})=(\rw_1\rw_2+\rw_3)a^{\Z_2}+\rw_2(a^{\Z_2})^2
=\rw_1\rw_2a^{\Z_2}$, which implies that
$\rw_3a^{\Z_2}=\rw_2(a^{\Z_2})^2$.\\
(4) $\Sq^2((a^{\Z_2})^2)=(a^{\Z_2})^4=(\rw_1^2+\rw_2)(a^{\Z_2})^2$.\\
(5) $\Sq^2(\rw_1a^{\Z_2})=\rw_1^2(a^{\Z_2})^2=(\rw_1^2+\rw_2)\rw_1a^{\Z_2}=\rw_1^3a^{\Z_2}$, which is the same as (2).\\
To summarize
\begin{align}
\label{warel}
& \rw_1^2(a^{\Z_2})^2=\rw_1^3a^{\Z_2},
&& (a^{\Z_2})^4 =\rw_1 (a^{\Z_2})^3, 
\\
&  \rw_2 (a^{\Z_2})^2= \rw_3 a^{\Z_2}, &&
(a^{\Z_2})^4 + \rw_1^2(a^{\Z_2})^2  +\rw_2(a^{\Z_2})^2=0
.
\nonumber 
\end{align}

There are nine 4-cocycles that can be constructed from Stiefel-Whitney classes
and a 1-cocycle $a^{\Z_2}$:
\begin{align}
& (a^{\Z_2})^4, &&
 \rw_1 (a^{\Z_2})^3, &&
 \rw_1^2 (a^{\Z_2})^2,
\nonumber\\
& \rw_2 (a^{\Z_2})^2, &&
 \rw_1^3 a^{\Z_2}, &&
 \rw_3 a^{\Z_2},
\nonumber\\
& \rw_1^4, &&
  \rw_2^2, &&
  \rw_4 .
\end{align}
Only four of them are independent
\begin{align}
 \rw_1^4,\ \  \rw_2^2,\ \ \rw_3 a^{\Z_2},\ \ \rw_1^3 a^{\Z_2}.
\end{align}

\subsection{Relations between Stiefel-Whitney classes and a $\Z_2$-valued
2-cocycle in 4-dimensions}

\label{bSWrel4}

There are two relations between
the  Stiefel-Whitney classes and a $\Z_2$-valued
2-cocycle $b^{\Z_2}$:\\
(1) $\Sq^1(\rw_1b^{\Z_2})= \rw_1^2 b^{\Z_2} + \rw_1 \Bs_2b^{\Z_2} =\rw_1^2b^{\Z_2}$, which implies $\rw_1 \Bs_2 b^{\Z_2}=0$.\\
(2) $\Sq^2(b^{\Z_2})= (b^{\Z_2})^2=(\rw_1^2+\rw_2)b^{\Z_2}$.\\
There are seven 4-cocycles that can be constructed from Stiefel-Whitney classes
and a $\Z_2$-valued 2-cocycle $b^{\Z_2}$:
\begin{align}
& (b^{\Z_2})^2, &&
 \rw_1 \Bs_2 b^{\Z_2}, &&
 \rw_1^2 b^{\Z_2}, &&
 \rw_2 b^{\Z_2},
\nonumber\\
& \rw_1^4, &&
  \rw_2^2, &&
  \rw_4 .
\end{align}
So the following four  4-cocycles are independent
\begin{align}
 \rw_1^4,\ \  \rw_2^2,\ \ \rw_2 b^{\Z_2},\ \ \rw_1^2 b^{\Z_2}.
\end{align}

\section{Spin and Pin structures}
\label{spinstructure}

Stiefel-Whitney classes can determine when a manifold can have a spin
structure.  The  spin structure is defined only for orientable manifolds.  The
tangent bundle for an orientable manifold $M^d$ is a $SO(d)$ bundle.  The group
$SO(d)$ has a central extension to the group $\text{Spin}(d)$.  Note that
$\pi_1(SO(d)=\Z_2$. The group $\text{Spin}(d)$ is the double covering of the
group  $SO(d)$.  A spin structure on $M^d$ is a $\text{Spin}(d)$ bundle, such
that under the group reduction $\text{Spin}(d) \to SO(d)$, the $\text{Spin}(d)$
bundle reduces to the $SO(d)$ bundle.  Some manifolds cannot have such a
lifting from $SO(d)$ tangent bundle to the $\text{Spin}(d)$ spinor bundle.  The
manifolds that have such a lifting is called spin manifold.  A manifold is a
spin manifold iff its first and second Stiefel-Whitney class vanishes
$\rw_1=\rw_2=0$.  

For a non-orientable manifold $N^d$, the tangent bundle is a $O(d)$ bundle.
The non-connected group $O(d)$ has two nontrivial central extensions (double
covers) by $Z_2$ with different group structures, denoted by $\text{Pin}^+(d)$
and  $\text{Pin}^-(d)$.  So the $O(d)$ tangent bundle has two types of lifting
to a $\text{Pin}^+$ bundle and a $\text{Pin}^-$ bundle, which are called
$\text{Pin}^+$ structure and $\text{Pin}^-$ structure respectively.  The
manifolds with such liftings are called $\text{Pin}^+$ manifolds or
$\text{Pin}^-$ manifolds.  We see that the concept of $\text{Pin}^\pm$
structure applies to both orientable and non-orientable manifolds.  A manifold
is a  $\text{Pin}^+$ manifold iff $\rw_2=0$.  A manifold is a  $\text{Pin}^-$
manifold iff $\rw_2+\rw_1^2=0$.  If a manifold $N^d$ does admit $\text{Pin}^+$
or $\text{Pin}^-$ structures, then the set of isomorphism classes of
$\text{Pin}^+$-structures (or $\text{Pin}^-$-structures) can be labled by
elements in $H^1(N^d;\Z_2)$. For example $\R P^4$ admits two
$\text{Pin}^+$-structures and no $\text{Pin}^-$-structures since $\rw_2(\R
P^4)=0$ and $\rw_2(\R P^4)+\rw_1^2(\R P^4)\neq 0$.

From \eqn{SWsum}, we see that $M\# N$ is pin$^+$ iff both $M$ and $N$ are
pin$^+$. Similarly, $M\# N$ is pin$^-$ iff both $M$ and $N$ are pin$^-$.

\section{Cohomology rings}
\label{crings}

In this section, we list some cohomology rings $H^*(M^4;\Z_n)$, that are used
in the main text of the paper.
First, let us list a few  theorems:\\
\textbf{The cohomology ring of product space}
(see \Ref{Hat02} page 216):\\
 Let $X$ and $Y$ be arbitrary spaces.  Assume
$H^k(Y;R)$ is a free and finitely generated $R$-module for all $k$.  Then 
\begin{align}
H^*(X;R) \otimes_R H^*(Y;R) \to H^*(X\times Y;R)
\end{align}
is an isomorphism of graded rings.  (A free $R$-module is a module that has a
basis, or equivalently, one that is isomorphic to a direct sum of copies of the
ring $R$.)
\\
\textbf{The cohomology of connected sum}:
\begin{align}
 H^k(M^d\# N^d,\M) =  H^k(M^d,\M)\oplus H^k(N^d,\M),\
0<k<d.
\end{align}
\textbf{The cup product of connected sum}:
\begin{align}
 H^k(M^d\# N^d,\M)\times & H^l(M^d\# N^d,\M)\stackrel{\cup}{\to} H^{k+l}(M^d\# N^d,\M),
\nonumber\\
&  0<k,l,k+l<d :
\nonumber\\
 (a,a')\cup (b,b') &=(a\cup b,a'\cup b'),
\end{align}
where $a \in H^k(M^d,\M)$, $b \in H^l(M^d,\M)$, $a' \in H^k(N^d,\M)$, and $b'
\in H^l(N^d,\M)$.
The above also works for $k+l=d$ is we identify
\begin{align}
 (\al v_{M^d}, \bt v_{N^d}) \sim (\al+\bt)v_{M^d\# N^d}
\end{align}
where $v_{M^d}$, $v_{N^d}$, and $v_{M^d\# N^d}$ are the generators in $
H^d(M^d,\M)$, $ H^d(N^d,\M)$, and $ H^d(M^d\# N^d,\M)$.

\subsection{ $H^*(T^4, \Z_n)$}

For $M^4=S^1\times S^1 \times S^1\times S^1=T^4$, we have
\begin{align}
\label{ringT4}
 H^*(T^4, \Z_n) = \frac{\Z_n[a_1,a_2,a_3,a_4]}{(a_1^2,a_2^2,a_3^2,a_4^2)}
\end{align}
where $a_i \in H^1(T^4, \Z_n)$ generate the ring.
The Bockstein homomorphism all vanishes:
\begin{align}
 \Bs_n a_i =0 , \ \ \ \ i=1,2,3,4.
\end{align}

\subsection{ $H^*(T^2\times S^2, \Z_n)$}

For $M^4=T^2\times S^2$ (where $T^2=S^1\times S^1$), we have
\begin{align}
\label{ringS2T2}
 H^*(T^2\times S^2, \Z_n) = \frac{\Z_n[a_1,a_2,b]}{(a_1^2,a_2^2,b^2)}
\end{align}
where $a_i \in H^1(T^2\times S^2, \Z_n)$ and $b \in H^2(T^2\times S^2, \Z_n)$
generate the ring.  We also have
\begin{align}
 \Bs_n a_i =\Bs_n b =0 , \ \ \ \ i=1,2
\end{align}

\subsection{ $H^*(L^2(p);\Z_n)$}
\label{L2pring}

\begin{figure}[tb]
\begin{center}
\includegraphics[scale=0.55]{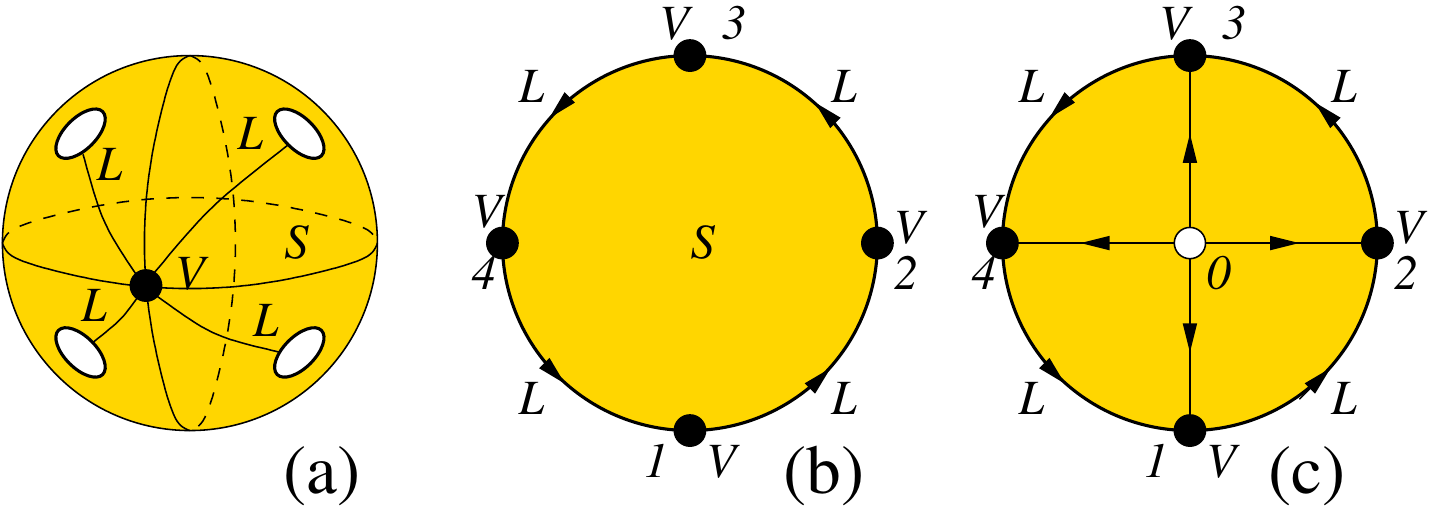}
%10
\end{center}
\caption{ (Color online) 
(a) $L^2(p)$ (with $p=4$) space. (b)
$L^2(p)$ can be described by a CW-complex with a 0-cell $V$, a 1-cell
$L$, and a 2-cell $S$.  The filled dots are identified.  The links $(12)$,
$(23)$, $(34)$, and $(41)$ are identified..  The boundary of the 2-cell $S$ is
$p$ copies of the 1-cell $L$: $\prt S = p L$ and $L$ is a cycle $\prt L=0$. 
(c) $L^2(p)$ can be described by a singular-complex with
two vertices $(0)$ and $V$, $p+1$ links $(01),\cdots, (0p)$ and $L$,
$p$ triangles $(012),\cdots,(0,p-1,p)$. The sum of the $p$ triangles
gives us $S$, the whole $L^2(p)$ space.
}
\label{L2p}
\end{figure}

$L^2(p)$ space is a 2-dimensional sphere with $p$ holes removed and with the
boundary of the $p$ holes identified (see Fig. \ref{L2p}a).
It has a CW-complex decomposition as shown in Fig. \ref{L2p}b.
Since $\prt S = p L$, $\prt L=0$,  we can compute explicitly that
\begin{align}
 H_0(L^2(p), \Z) &= \Z,  &
 H_1(L^2(p), \Z) &= \Z_p,  
\nonumber\\
 H_2(L^2(p), \Z) &= 0,  
\end{align}
\begin{align}
 H^0(L^2(p), \Z) &= \Z,  &
 H^1(L^2(p), \Z) &= 0,  
\nonumber\\
 H^2(L^2(p), \Z) &= \Z_p .
\end{align}
and
\begin{align}
 H_0(L^2(p), \Z_n) &= \Z,  \ \ \
 H_1(L^2(p), \Z_n) = \Z_{\<n,p\>} =\{L\},  
\nonumber\\
 H_2(L^2(p), \Z_n) &= \Z_{\<n,p\>} =\{\frac n{\<n,p\>}S\},  
\end{align}
\begin{align}
\label{L2pH}
 H^0(L^2(p), \Z_n) &= \Z,  \ \ \
 H^1(L^2(p), \Z_n) = \Z_{\<n,p\>}=\{a\},  
\nonumber\\
 H^2(L^2(p), \Z_n) &= \Z_{\<n,p\>} =\{b\},
\end{align}
where we have listed the generators of $H_*(L^2(p), \Z_n)$ and $H^*(L^2(p),
\Z_n)$.

Using the CW-complex of $L^2(p)$, we can compute the  Bockstein homomorphism
for $\Z_n$ coefficient.  Let $\t a \in Z^1(L^2(p);\Z)$ to be a generator of
$H^1(L^2(p);\Z_p)$, and $\t b \in C^2(L^2(p);\Z)$ to be a generator of
$H^2(L^2(p);\Z)$:
\begin{align}
\<\t a, L\> = 1, \ \ \  \<\t b, S\> = 1.
\end{align}
We see that $p=\<p\t a, L\>=\<\t a, pL\>=\<\t a, \prt S\> = \<\dd \t a, S\>$.
Thus $\dd \t a=0$ mod $p$, confirming that $\t a$ is a cocycle in
$H^1(L^3(p,q);\Z_p)$, but $\t a$ is not a cocycle in $H^1(L^3(p,q);\Z)$.  From
the above calculation, we also see that $\dd \t a=p\t b$, or $\frac1p \dd \t a
= \t b$.  Therefore, $\frac1n \frac n{\<p,n\>} \dd \t a= \frac p{\<p,n\>} \t
b$, or $\frac1n \dd (\frac n{\<p,n\>} \t a)=\Bs_n(\frac n{\<p,n\>} \t a) =
\frac p{\<p,n\>}\t b$.  We note that $\frac n{\<p,n\>} \t a$ is an integer
valued-cochain that satisfies $\dd \frac n{\<p,n\>} \t a =0$ mod $n$.  Thus
$a=\frac n{\<p,n\>} \t a$ is a cocycle and a generator in $H^1(L^3(p,q);\Z_n)$.
Also $b=\t b$ is a cocycle and a generator in
$H^2(L^3(p,q);\Z_n)$.
The Bockstein homomorphism can be written as 
\begin{align}
\label{L2pB}
\Bs_n a =\frac p{\<p,n\>} b.
\end{align}

We can calculate the cohomology ring $H^*(L^2(p);\Z_n)$, by decomposing
$L^2(p)$ into a singular-complex characterized by the
vertices $0,1,2,\cdots,p$ (see Fig. \ref{L2p}c).  Note that
$1,2,\cdots,p$  correspond to the same
vertex.
First $a,b$ (the generators of $H^1(L^2(p);\Z_n)$ and
$H^2(L^2(p);\Z_n)$) are given by 
\begin{align}
\<a,(m,m+1)\> &=\frac n{\<p,n\>},\ \ 
\nonumber \\
\<a,(0m)\>&= \frac {(m-1)n}{\<p,n\>}, 
\\
\<b,(012)\>&= \<b,(00'2)\>=1,\ \ 
\<b,\text{others}\>=0, \  \ 
\nonumber\\
m&=1,\cdots,p.
\nonumber 
\end{align}
We see that
\begin{align}
\label{aLbS}
 \<a,L\>=\frac n{\<p,n\>}, \ \ \ \
 \<b,\frac n{\<p,n\>} S\>=\frac n{\<p,n\>}
\end{align}
where $\frac n{\<p,n\>} S$ is a 2-cycle
$\prt \frac n{\<p,n\>} S = \frac {np}{\<p,n\>} L = 0$ mod $n$.

Now, we can calculate the cup product
\begin{align}
 \<a^2, (0,m,m+1)\>&=\<a, (0,m)\>\<a, (m,m+1)\>
\nonumber\\
&= \frac {(m-1)n}{\<p,n\>} \frac n{\<p,n\>}
,
\end{align}
or
\begin{align}
 \<a^2, \frac n{\<p,n\>} S\>
&=\sum_{m=1}^p 
\frac {(m-1)n}{\<p,n\>}
\frac {n^2}{\<p,n\>^2}
=\frac{n^3p(p-1)}{2\<p,n\>^3}
\nonumber \\
&\se{n} 
\begin{cases}
 \frac n2, & \text{ if } p_2>1,\  \frac {n}{2^{p_2}}=\text{ odd }
\\ 
0,  & \text{ otherwise}
\end{cases}
, 
\end{align}
where $p_2$ is the number of prime factor $2$ in $p$.
The above implies that
\begin{align}
\label{L2pcup}
a^2 
&= \frac{n^2p(p-1)}{2\<p,n\>^2} b
\nonumber\\
&=
\begin{cases}
  \frac {\<n,p\>}2 b , & \text{ if } p_2>1,\ \frac {n}{2^{p_2}}=\text{ odd }
\\ 
0, & \text{ otherwise}
\end{cases}
.
\end{align}
The ring $H^*(L^2(p);\Z_n)$ is determined by \eqn{L2pH} and \eqn{L2pcup}.

\subsection{ $H^*(L^3(p,q) \times S^1, \Z_n)$}
\label{Lpq}

We know that $S^3$ can be described by two complex numbers $z_1,z_2$ satisfying
$|z_1|^2+|z_2|^2=1$.  Let $p$ and $q$ be coprime integers. We can see that the
action $(z_1,z_2)\to (\ee^{\ii \frac{2\pi}{p}}z_1,\ee^{\ii \frac{2\pi
q}{p}}z_2)$ is a free action on $S^3$.  Quotient out such a free action, the
resulting space is the lens space $L^3(p,q)$.  We see that $L^3(2,1)=\R P^3$.
$L^3(p,q_1)$ and $L^3(p,q_2)$ are homotopically equivalent if and only if $q_1q_2=
\pm m^2$ mod $p$ for an integer $m$.

\begin{figure}[tb]
\begin{center}
\includegraphics[scale=0.8]{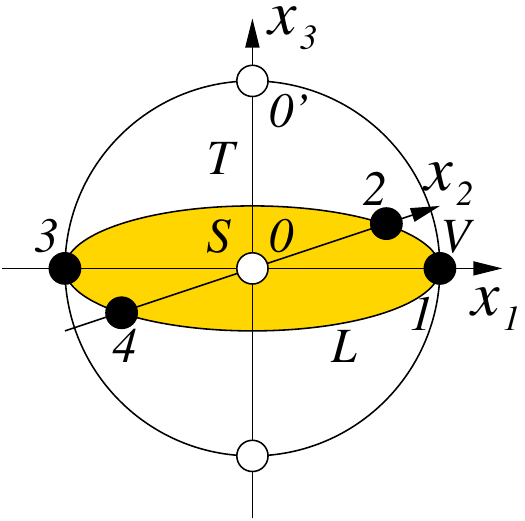}
%11
\end{center}
\caption{ (Color online) 
The $S^3$ is parametrized by $(x_1,x_2,x_3)=\frac{(\Re z_1,\Im z_1, \Re
z_2)}{1+\Im z_2}$ which is the whole $\R^3$.  The open dots are the points
$(z_1,z_2)=(0,\ee^{\ii 2\pi m/p}),\ m=0,\cdots,p-1$. The shaded disc is
$B^2_0$. The north and the south hemisphere are $B^2_{\pm 1}$.  The volume
between $B^2_0$ and  $B^2_1$ is the lens space $L^3(p,q)|_{(p,q)=(4,1)}$.  The
lens space $L^3(p,q)$ is described by a CW-complex with a 0-cell $V$, a 1-cell
$L$, a 2-cell $S$, and a 3-cell $T$.  The filled dots are identified under the
quotient map and correspond to the 0-cell $V$. The shaded disc $S$ is the
2-cell.  The boundary of the 2-cell $S$ is $p$ copies of the 1-cell $L$: $\prt
S = p L$ and $L$ is a cycle $\prt L=0$. The 3-cell $T$ is the half ball above
the shaded disc.
}
\label{lensCW}
\end{figure}

$L^3(p,q)$ is described by the CW-complex in Fig. \ref{lensCW} (for
(p,q)=(4,1)), which has a 0-cell $V$ (the 4 vertices $1,2,3,4$ are identified
and correspond to $V$), a 1-cell $L$ (the 4 links $(12),(23),(34),(41)$ are
identified and correspond to $L$), a 2-cell $S$ (which is the union of
$(012),(023),(034),(041)$), and a 3-cell $T$  (which is the union of
$(00'12),(00'23),(00'34),(00'41)$).  To describe the lens space, let us first
consider $p$ points $(z_1,z_2)=(0,\ee^{\ii 2\pi m/p}),\ m=0,\cdots,p-1$ (which
become one point after the quotient).  The 2-cell $B^2_m$ is formed by the
points $(z_1,z_2)=\cos \th (0,\ee^{\ii 2\pi m/p})+\sin \th(z_1,0),\ |z_1|=1$
(where $B^2_m$ and $B^2_{m'}$ are identified by the quotient map).  The volume
between $B^2_m$ and  $B^2_{m+1}$ is the lens space $L^3(p,q)$ which is also the
3-cell $T$.  The 0-cell is given by $(z_1,z_2)=(\ee^{\ii 2\pi m/p},0),\
m=0,\cdots,p-1$ (which become one point after the quotient).

Since $\prt S = p L$, $\prt L=\prt T=0$,  we see that
\begin{align}
 H_0(L^3(p,q), \Z) &= \Z,  &
 H_1(L^3(p,q), \Z) &= \Z_p,  
\nonumber\\
 H_2(L^3(p,q), \Z) &= 0,  &
 H_3(L^3(p,q), \Z) &= \Z ,
\end{align}
and, by Poincar\'e  duality,
\begin{align}
 H^0(L^3(p,q), \Z) &= \Z,  &
 H^1(L^3(p,q), \Z) &= 0,  
\nonumber\\
 H^2(L^3(p,q), \Z) &= \Z_p,  &
 H^3(L^3(p,q), \Z) &= \Z .
\end{align}
Then using the universal coefficient theorem \eqn{ucf1} and \eqn{ucfH}, we find that
\begin{align}
 H_0(L^3(p,q), \Z_n) &= \Z_n,  &
 H_1(L^3(p,q), \Z_n) &= \Z_{\<n,p\>},  
\nonumber\\
 H_2(L^3(p,q), \Z_n) &= \Z_{\<n,p\>},  &
 H_3(L^3(p,q), \Z_n) &= \Z_n .
\end{align}
\begin{align}
 H^0(L^3(p,q), \Z_n) &= \Z_n,  &
 H^1(L^3(p,q), \Z_n) &= \Z_{\<n,p\>},  
\nonumber\\
 H^2(L^3(p,q), \Z_n) &= \Z_{\<n,p\>},  &
 H^3(L^3(p,q), \Z_n) &= \Z_n .
\end{align}
$H_1(L^3(p,q), \Z_n)$ is generated by $L$ and
$H_2(L^3(p,q), \Z_n)$ is generated by $\frac n{\<n,p\>}S$.

The cohomology rings $H^*(L^3(p,q);\Z_p)$ are given by
(see \Ref{Hat02} page 251)
\begin{align}
&\ \ \ \  H^*(L^3(p,q);\Z_p) 
\nonumber\\
&= \{ m_0+m_1a+m_2b+m_3 ab\ |\ 
a^2= \frac p2 (\<p,2\>-1)b\},
\nonumber \\
&\ \ \ \ H^*(L^3(p,q);\Z) 
= \{ m_0+m_2b+m_3 c \}.
\end{align}
We also have $\Bs_p a=b$.

In the following, we will only consider $L^3(p,1)\equiv L^3(p)$.
We like to calculate the cohomology ring $H^*(L^3(p);\Z_n)$, by decomposing
the lens space $L^3(p)$ into a simplicial complex characterized by the
vertices $0,0',1,2,\cdots,p$ (see Fig. \ref{lensCW}).  Note that $0$ and $0'$
correspond to the same vertex and $1,2,\cdots,p$  correspond to the same
vertex.  Also note that, for example, the 2-simplices $(012)$ and $(0'23)$ are
identified.  First $a,b$ (the generators of $H^1(L^3(p);\Z_n)$ and
$H^2(L^3(p);\Z_n)$) are given by 
\begin{align}
\<a,(00')\> &= \<a,(m,m+1)\>=\frac n{\<p,n\>},\ \ 
\nonumber \\
\<a,(0m)\>&= \frac {(m-1)n}{\<p,n\>}, 
\\
\<b,(012)\>&= \<b,(00'2)\>=1,\ \ 
\<b,\text{others}\>=0, \  \ 
\nonumber\\
m&=1,\cdots,p.
\nonumber 
\end{align}
We see that
\begin{align}
 \<a,L\>=\frac n{\<p,n\>}, \ \ \ \
 \<b,\frac n{\<p,n\>} S\>=\frac n{\<p,n\>}
\end{align}
where $\frac n{\<p,n\>} S$ is a 2-cycle
$\prt \frac n{\<p,n\>} S = \frac {np}{\<p,n\>} L = 0$ mod $n$.

Now, we can calculate the cup product
\begin{align}
 \<a^2, (0,m,m+1)\>&=\<a, (0,m)\>\<a, (m,m+1)\>
\nonumber\\
&= \frac {(m-1)n}{\<p,n\>} \frac n{\<p,n\>}
,
\end{align}
or
\begin{align}
 \<a^2, \frac n{\<p,n\>} S\>
&=\sum_{m=1}^p 
\frac {(m-1)n}{\<p,n\>}
\frac {n^2}{\<p,n\>^2}
=\frac{n^3p(p-1)}{2\<p,n\>^3}
\nonumber \\
&\se{n} 
\begin{cases}
 \frac n2 , & \text{ if } p_2>1,\  \frac {n}{2^{p_2}}=\text{ odd }
\\ 
0, & \text{ otherwise}
\end{cases}
, 
\end{align}
where $p_2$ is the number of prime factor $2$ in $p$.
The above implies that
\begin{align}
a^2 
&= \frac{n^2p(p-1)}{2\<p,n\>^2} b
\nonumber\\
&=
\begin{cases}
  \frac {\<n,p\>}2 b , & \text{ if } p_2>1,\ \frac {n}{2^{p_2}}=\text{ odd }
\\ 
0, & \text{ otherwise}
\end{cases}
.
\end{align}
We also note that 
\begin{align}
\<ab,T\> &= 
-\<a,(0'0)\> \<b,(012)\> 
\nonumber\\
&= \frac n{\<p,n\>},
\end{align}
which implies that
\begin{align}
ab &= \frac {n}{\<p,n\>}c,
\end{align}
Thus, the cohomology ring $H^*(L^3(p);\Z_n)$ is given by
\begin{align}
\label{ringLpq}
 H^*(L^3(p);\Z_n) 
&= \{ \zeta+\al a+\bt b+\ga c \},
\\
\text{with }\ \ 
a^2 &=
\begin{cases}
  \frac {\<n,p\>}2 b , & \text{ if } p_2>1,\ \frac {n}{2^{p_2}}=\text{ odd }
\\ 
0, & \text{ otherwise}
\end{cases}
\nonumber \\
ab &= \frac {n}{\<p,n\>}c,
\nonumber 
\end{align}
where $\zeta,\ga \in \Z_n$ and $\al,\bt \in \Z_{\<p,n\>}$.  We also have $\Bs_n
a=\frac p{\<p,n\>} b$.

Notice that
\begin{align}
H^*(S^1;\Z_n)=\frac{\Z_n[a_1]}{(a_1^2)}
\end{align}
is a free $\Z_n$-module.  This allows us to compute the cohomology ring
$H^*(S^1\times L^3(p);\Z_n)$.

\subsection{ $H^*(F^4;\Z_n)$}
\label{F4ring}

In order for the volume-independent partition function $Z^\text{top}(M^4)$ on an
orientable space-time $M^4$ to be a topological invariant, we require the Euler
number and the Pontryagin number of $M^4$ to vanish: $\chi(M^4)=P_1(M^4)=0$.
We also like $M^4$ to be complicated enough so that its second Stiefel-Whitney
class $\rw_2$ is non-zero. How to construct such an 4-dimensional manifold?

First, let us introduce intersection form $Q_{M^4}$: $H^2(M^4;\Z)\times H^2(M^4;\Z) \to \Z$ defined by
\begin{align}
 Q_{M^4}(a,b) =\<a\cup b,[M^4]\>=\int_{M^4} ab.
\end{align}
The intersection form has the following properties
\begin{enumerate}
\item
Under connected sum, 
\begin{align}
 Q_{M^4\#N^4} = Q_{M^4} \oplus Q_{N^4}.
\end{align}
\item
Poincar\'e duality implies that the intersection form $Q_{M^4}$ is unimodular.
\item
If $M^4$ is spin, then $ Q_{M^4}(a,a) =$ even for all $a \in H^2(M^4;\Z)$.  If
$M^4$ is orientable and $ Q_{M^4}$ is even, then $M^4$ is spin.  
\item
The signature of $ Q_{M^4}$ is one third of the Pontryagin number:
$\si(M^4)=\frac13 P_1(M^4)$.  
\item
A smooth compact spin 4-manifold has a signature which
is a multiple of 16.
\item
A 4-manifold bounds a 5-manifold if and only if
it has zero signature.  
\end{enumerate}

We know that  $Q_{\C P^2}$ is 1-by-1 matrix: $Q_{\C P^2}=(1)$, while
$Q_{\overline{\C P}^2}=(-1)$.  Thus, $Q_{\C P^2\# \overline{\C
P}^2}={\scriptsize\begin{pmatrix} 1&0\\ 0& -1\\ \end{pmatrix}}$.  This means
$\C P^2\# \overline{\C P}^2$ is not spin and has a zero Pontryagin number.

The Euler number $\chi(M)$ has the following properties:
\begin{enumerate}
\item
$\chi(S^d)=1+(-)^d$.
\item
$\chi(\R P^d)=\frac{1+(-)^d}{2}$.
\item
$\chi(\C P^2)=\chi(\overline{\C P}^2)=3$.
\item
$\chi(M\times N)=\chi(M)\chi(N)$.
\item
$\chi(M^d\# N^d)=\chi(M^d)+\chi(N^d)-\chi(S^d)$.
\end{enumerate}
Using the above result, we find that
\begin{align}
F^4\equiv (S^1\times S^3)\# (S^1\times S^3)\#\C P^2\# \overline{\C P}^2 ,
\end{align}
has
\begin{align}
 Q_{F^4} &=\begin{pmatrix} 1&0\\ 0& -1\\ \end{pmatrix},\ \ \ \
\chi(F^4)=P_1(F^4)=0.
\end{align}
We see that $F^4$ is not spin.

The cohomology classes for $F^4$ are
\begin{align}
 H^1(F^4;\Z_n)&=\Z_n^{\oplus 2},\ \ 
 H^2(F^4;\Z_n)=\Z_n^{\oplus 2},\ \ 
\nonumber\\
 H^3(F^4;\Z_n)&=\Z_n^{\oplus 2},\ \ 
 H^4(F^4;\Z_n)=\Z_n .
\end{align}
Let $a_1,a_2$ be the
generators of $H^1(F^4;\Z_n)$, $b_1,b_2$ the generators of $H^2(F^4;\Z_n)$,
$c_1,c_2$ be the generators of $H^3(F^4;\Z_n)$, and $v$ be the generator of
$H^4(F^4;\Z_n)$: 
\begin{align}
 H^*(F^4;\Z_n)=\{ a_1,a_2, b_1,b_2, c_1,c_2, v \}.
\end{align}
We find that the non-zero cup products are given by
\begin{align}
 b_1^2=-b_2^2=a_1c_1=a_2c_2=v.
\end{align}
All other cup products vanish.

\subsection{ $H^*(\R P^d;\Z_2)$}
\label{RPdring}

Next, let us list some cohomology rings with $\Z_2$ coefficient for
non-orientable spaces.  The cohomology ring $H^*(\R P^d;\Z_2)$ is given by
\begin{align}
 H^*(\R P^d;\Z_2) &= \frac{\Z_2[a]}{(a^{d+1})}
\end{align}
with $a\in H^1(\R P^d;\Z_2)$.  
$\R P^d$ is non-orientable if $d$ = even.
The total Stiefel-Whitney class for $\R P^d$ is given by
\begin{align}
 \rw = (1+a)^{d+1}.
\end{align}
(see https://amathew.wordpress.com/2010/12/17/the-stiefel-whitney-classes-of-projective-space/ )
We see that for $\R P^4$, $\rw_1=a$ and $\rw_2=0$.  Thus $\R P^4$ is a pin$^+$
manifold, but not a pin$^-$ manifold.

\subsection{ $H^*(F^4_{\rm non};\Z_2)$ }
\label{F4nonring}

We note that $\R P^4$ has an intersection form $Q_{\R P^4}=(1)$ (with $\Z_2$
field), $\si(\R P^4)=1$ mod 2, and $\chi(\R P^4)=1$.  So 
\begin{align}
F^4_\text{non} \equiv  \R P^4 \# \C P^2 \# (S^1\times S^3) 
\end{align}
has  $\si(F^4_\text{non})=0$ mod 2 and $\chi(F^4_\text{non})=0$.

The cohomology classes for $F^4_\text{non}$ are
\begin{align}
 H^1(F^4_\text{non};\Z_2)&=\Z_2^{\oplus 2},\ \ 
 H^2(F^4_\text{non};\Z_2)=\Z_2^{\oplus 2},\ \ 
\nonumber\\
 H^3(F^4_\text{non};\Z_2)&=\Z_2^{\oplus 2},\ \ 
 H^4(F^4_\text{non};\Z_2)=\Z_2 .
\end{align}
Let $a^{\R P^4},a^{S^1\times S^3}$ be the generators of
$H^1(F^4_\text{non};\Z_n)$, $(a^{\R P^4})^2,b^{\C P^2}$ of
$H^2(F^4_\text{non};\Z_n)$, $(a^{\R P^4})^3,c^{S^1\times S^3}$ of
$H^3(F^4_\text{non};\Z_n)$, and $v$ the generator of
$H^4(F^4_\text{non};\Z_n)$: 
\begin{align}
 H^*(F^4_\text{non};\Z_n)=\{ (a^{\R P^4})^{m=1,2,3},a^{S^1\times S^3},b^{\C P^2},c^{S^1\times S^3},v \}.
\end{align}
We find that the non-zero cup products are given by
\begin{align}
(a^{\R P^4})^4 =
& (b^{\C P^2})^2=
a^{S^1\times S^3}c^{S^1\times S^3}=
v,
\nonumber\\
& (a^{\R P^4})^2, \ \ \
 (a^{\R P^4})^3
.
\end{align}
All other cup products vanish.  The first Stiefel-Whitney class for
$F^4_\text{non}$ is given by $ \rw_1 = a^{\R P^4}$.  Since $ \R P^4$, $ \C
P^2$, and  $S^1\times S^3$ are all pin$^+$ manifolds, their connected sum
$F^4_\text{non}$ is also a pin$^+$ manifold.  Thus the second Stiefel-Whitney
class for $F^4_\text{non}$ is  $ \rw_2=0$.  Since $\rw_2+\rw_1^2=(a^{\R
P^4})^2\neq 0$, $F^4_\text{non}$ is not a pin$^-$ manifold.

\subsection{ $H^*(K;\Z_2)$}
\label{KBring}

\begin{figure}[tb]
\begin{center}
\includegraphics[scale=1.0]{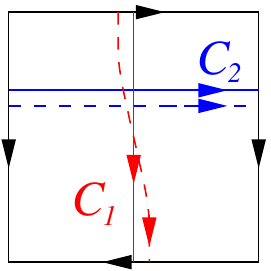}
%12
\end{center}
\caption{ (Color online) 
The Klein bottle: the top and bottom boundaries are identified with a twist and
the left and right boundaries are identified without twist.
$H_1(K;\Z_2)$ is generated by $C_1$ and $C_2$ cycles.
}
\label{KB}
\end{figure}

The Klein bottle $K$ has the following cohomology class
\begin{align}
 H^1(K;\Z_2)=\Z_2^{\oplus 2} =\{a_1,a_2\},\ \ \ \ 
 H^2(K;\Z_2)=\Z_2  =\{b \} .
\end{align}
$H^1(K;\Z_2)$ is generated by $a_1$ and $a_2$ which are the Poincar\'e dual
of $C_1$ and $C_2$ (see Fig. \ref{NgSurface}):
\begin{align}
 a_1=C_1^*, \ \ \ \ \ a_2=C_2^*  .
\end{align}
We see that $a_1a_2=b$ since $C_1$ and $C_2$ intersect once; $a_2^2=0$ since
$C_2$ does not self intersect (\ie $C_2$ and its displacement does not
intersect); $a_1^2=b$ since $C_1$ self intersects once (\ie $C_1$ and its
displacement intersect once).
Therefore $H^*(K;\Z_2)$ is determined by
\begin{align}
  a_1^2=a_1a_2=b,\ \ \ a_2^2=0.
\end{align}

\subsection{ $H^*(\Si^{\rm non}_g;\Z_2)$}
\label{Sinonring}

\begin{figure}[tb]
\begin{center}
\includegraphics[scale=0.6]{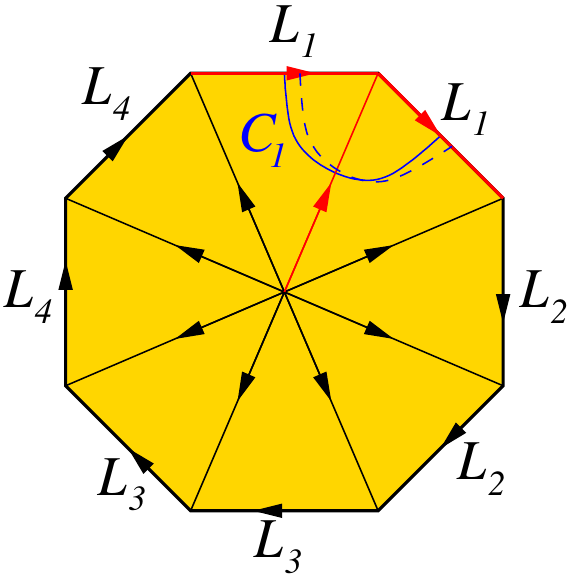}
%13
\end{center}
\caption{ (Color online) 
A non-orientable surface $\Si^\text{non}_g$ with genus $g=4$. All the corners
are identified and the edges with the same label $L_i$ are glued together along
its direction.  The Poincar\'e dual of the cycle $C_1$ is $a_1 \in
H^1(\Si^\text{non}_g;\Z_2)$: $\<a_1,\text{red link}\>=1$ and $\<a_1,\text{black
link}\>=0$.  We note that $\Si^\text{non}_1 =\R P^2$ and $\Si^\text{non}_2 =$
Klein bottle.
}
\label{NgSurface}
\end{figure}

The cohomology ring for non-orientable surface
$\Si^\text{non}_g$ (see Fig. \ref{NgSurface}), $H^*(\Si^\text{non}_g;\Z_2)$, is
given by (see \Ref{Hat02} page 208)
\begin{align}
& H^*(\Si^\text{non}_g;\Z_2) = 
\frac{\Z_2[a_i]}{(a_i^{3},  a_i^2-a_j^2, a_ia_ja_k, a_ia_j|_{i\neq j})}
\nonumber\\
 &=
\{ \zeta +\al_ia_i + \bt b\ |\ \zeta ,\al_i,\bt \in \Z_2,\  a_i^2 =\del_{ij} b\},
\end{align}
with $a_i\in H^1(\Si^\text{non}_g;\Z_2)=\Z_2^{\oplus g},\ i=1,2,\cdots,g$
and $b\in H^2(\Si^\text{non}_g;\Z_2)=\Z_2$.

To understand the above result, we note that the cycles $C_i,\ i=1,\cdots,g$,
generate $H_1(\Si^\text{non}_g;\Z_2)$ (see Fig. \ref{NgSurface} where only
$C_1$ is drawn). The Poincar\'e dual of $C_i$, $a_i=C_i^*$, generate
$H^1(\Si^\text{non}_g;\Z_2)$.  We note that the self intersection number for
$C_i$ is 1.  Thus $a_i^2=b$. $C_i$ and $C_j$ does not intersect if $i\neq j$.
Thus $a_ia_j=0$.

To calculate the Stiefel-Whitney class $\rw_i$, we note that the orientation
reverses as we go along the loop $C_i$. This implies that  $\oint_{C_i} \rw_1
=1$ mod 2.  Since $\oint_{C_i} a_j$ is the intersection number between $C_i$
and Poincar\'e dual of $a_j$ which is $C_j$, we see that $\oint_{C_i}
a_j=\del_{ij}$.  Therefore $\rw_1=\sum_{i=1}^g a_i$.  In 2-dimensions
$\rw_2=\rw_1^2= \sum_{i=1}^g a_i^2 = [g]_2 b$.  Thus $\Si^\text{non}_g$ is
a pin$^+$ manifold if $g=$ even, and it is not a pin$^+$ manifold if $g=$ odd.
$\Si^\text{non}_g$ is always a pin$^-$ manifold.

We also note that the CW-complex of $\Si^\text{non}_g$ in Fig. \ref{NgSurface}
has $V=2$ vertices, $L=3g$ links, and $T=2g$ triangles.  Thus the Euler number
$\chi(\Si^\text{non}_g)=V-L+T=2-g$.  
The top Stiefel-Whitney class is equal to
the Euler class mod 2, regardless of the $\Z$-orientability of the manifold. In
other words, every manifold is $\Z_2$-orientable. So the Euler class (with
$\Z_2$-coefficients) coincides with the top Stiefel-Whitney class.  
%(This is Corollary 11.12 in Milnor, Stasheff: Characteristic Classes).
This is another way to obtain $\rw_2= [g]_2 b$.

\section{Group cohomology theory}
\label{gcoh}

\subsection{Homogeneous group cocycle}

In this section, we will briefly introduce group cohomology.  The group
cohomology class $\cH^d(G,\M)$ is an $\Z$-model constructed from a group $G$
and a $\Z$-module $\M$ (\ie a vector space over $\Z$).   
Each elements of $G$ also induce a mapping $\M\to \M$,
which is denoted as
\begin{eqnarray}
g\cdot m = m', \ \ \ g\in G,\ m,m'\in \M.
\end{eqnarray}
The map $g\cdot$ is a group homomorphism:
\begin{eqnarray}
g\cdot (m_1+m_2)= g\cdot m_1 +g \cdot m_2.
\end{eqnarray}
The module $\M$ with such a $G$-group homomorphism, is called a $G$-module.

A homogeneous $d$-cochain
is a function $\nu_d: G^{d+1}\to \M$, that satisfies
\begin{align}
\label{scond}
\nu_d(g_0,\cdots,g_d)
=g\cdot \nu_d(gg_0,\cdots,gg_d), \ \ \ \ g,g_i \in G.
\end{align}
We denote the set of $d$-cochains as $\cC^d(G,\M)$. Clearly $\cC^d(G,\M)$ is an
abelian group.

Let us define a mapping $\dd$ (group homomorphism) from
$\cC^d(G,\M)$ to $\cC^{d+1}(G,\M)$:
\begin{align}
  (\dd \nu_d)( g_0,\cdots, g_{d+1})=
  \sum_{i=0}^{d+1} (-)^i \nu_d( g_0,\cdots, \hat g_i
  ,\cdots,g_{d+1})
\end{align}
where $g_0,\cdots, \hat g_i ,\cdots,g_{d+1}$ is the sequence $g_0,\cdots, g_i
,\cdots,g_{d+1}$ with $g_i$ removed.
One can check that $\dd^2=0$.
The homogeneous $d$-cocycles are then
the homogeneous $d$-cochains that also satisfy the cocycle condition
\begin{eqnarray}
\label{cccond}
 \dd \nu_d =0.
\end{eqnarray}
We denote the set of $d$-cocycles as
$\cZ^d(G,\M)$. Clearly $\cZ^d(G,\M)$ is an abelian subgroup of $\cC^d(G,\M)$.

Let us denote  $\cB^d(G,\M)$ as the image of the map $\dd: \cC^{d-1}(G,\M) \to
\cC^d(G,\M)$ and $\cB^0(G,\M)=\{0\}$.
The elements in $\cB^d(G,\M)$ are called $d$-coboundary.
Since $\dd^2=0$, $\cB^d(G,\M)$  is a subgroup of $\cZ^d(G,\M)$:
\begin{align}
\label{cb}
\cB^d(G,\M) =
\{\dd \nu_{d-1}| \nu_{d-1}\in \cC^{d-1}(G,\M)\} \subset \cZ^d(G,\M).
\end{align}
The group cohomology class $\cH^d(G,\M)$ is then defined as
\begin{eqnarray}
\cH^d(G,\M) =  \cZ^d(G,\M)/ \cB^d(G,\M) .
\end{eqnarray}
We note that the $\dd$ operator and the cochains $\cC^d(G,\M)$ (for all values
of $d$) form a so called cochain complex,
\begin{align}
\cdots
\stackrel{\dd}{\to}
\cC^d(G,\M)
\stackrel{\dd}{\to}
\cC^{d+1}(G,\M)
\stackrel{\dd}{\to}
\cdots
\end{align}
which is denoted as $C(G,\M)$.  So we may also write the group cohomology
$\cH^d(G,\M)$ as the standard cohomology of the cochain complex $H^d[C(G,\M)]$.

\subsection{Inhomogeneous group cocycle}
\label{inhomo}

The above definition of group cohomology class can be rewritten in terms of
inhomogeneous group cochains/cocycles.  An inhomogeneous group $d$-cochain is
a function $\om_d: G^d \to M$. All $\om_d(g_1,\cdots,g_d)$ form $\cC^d(G,\M)$.
The inhomogeneous group cochains and the homogeneous group cochains are
related as
\begin{eqnarray}
\label{homoinhomo}
\nu_d(g_0,g_1,\cdots,g_d)=
\om_d( a_{01},\cdots, a_{d-1,d}),
\end{eqnarray}
with
\begin{align}
g_0=1,\ \
g_1=g_0 a_{01}, \ \
g_2=g_1 a_{12}, \ \ \cdots \ \
g_d=g_{d-1} a_{d-1,d}.
\end{align}
Now the $\dd$ map has a form on $\om_d$:
\begin{align}
&
(\dd\om_d)(a_{01},\cdots, a_{d,d+1})=
 a_{01}\cdot \om_d( a_{12},\cdots,a_{d,d+1})
\nonumber\\
& \ \ \
+\sum_{i=1}^d (-)^i \om_d(a_{01},\cdots, a_{i-1,i}a_{i,i+1},\cdots, a_{d,d+1})
\nonumber\\
& \ \ \
+(-)^{d+1}\om_d(a_{01},\cdots,\t a_{d-1,d})
\end{align}
This allows us to define the inhomogeneous group $d$-cocycles which satisfy
$\dd \om_d=0$ and  the inhomogeneous group $d$-coboundaries which have a form
$\om_d = \dd \mu_{d-1}$.  In the following, we are going to use inhomogeneous
group cocycles to study group cohomology.  Geometrically, we may view $g_i$ as
living on the vertex $i$, while $a_{ij}$ as living on the link connecting the
two vertices $i$ to $j$.

%Hatcher 2.48: $H_n$ for mapping torus. 
%
%$H^1(M^d;\Z)= \text{Home}(\pi_1(M^d);\Z)$ that has no torsion.
%
%http://math.stackexchange.com/questions/30865/no-torsion-in-h1-cx-mathbfz/30878
%
%Serre spectral sequence to compute cohomology of fiber bundle.

\vfill
\bibliography{../../bib/wencross,../../bib/all,../../bib/publst} 

\begin{thebibliography}{93}%
\makeatletter
\providecommand \@ifxundefined [1]{%
 \@ifx{#1\undefined}
}%
\providecommand \@ifnum [1]{%
 \ifnum #1\expandafter \@firstoftwo
 \else \expandafter \@secondoftwo
 \fi
}%
\providecommand \@ifx [1]{%
 \ifx #1\expandafter \@firstoftwo
 \else \expandafter \@secondoftwo
 \fi
}%
\providecommand \natexlab [1]{#1}%
\providecommand \enquote  [1]{``#1''}%
\providecommand \bibnamefont  [1]{#1}%
\providecommand \bibfnamefont [1]{#1}%
\providecommand \citenamefont [1]{#1}%
\providecommand \href@noop [0]{\@secondoftwo}%
\providecommand \href [0]{\begingroup \@sanitize@url \@href}%
\providecommand \@href[1]{\@@startlink{#1}\@@href}%
\providecommand \@@href[1]{\endgroup#1\@@endlink}%
\providecommand \@sanitize@url [0]{\catcode `\\12\catcode `\$12\catcode
  `\&12\catcode `\#12\catcode `\^12\catcode `\_12\catcode `\%12\relax}%
\providecommand \@@startlink[1]{}%
\providecommand \@@endlink[0]{}%
\providecommand \url  [0]{\begingroup\@sanitize@url \@url }%
\providecommand \@url [1]{\endgroup\@href {#1}{\urlprefix }}%
\providecommand \urlprefix  [0]{URL }%
\providecommand \Eprint [0]{\href }%
\providecommand \doibase [0]{http://dx.doi.org/}%
\providecommand \selectlanguage [0]{\@gobble}%
\providecommand \bibinfo  [0]{\@secondoftwo}%
\providecommand \bibfield  [0]{\@secondoftwo}%
\providecommand \translation [1]{[#1]}%
\providecommand \BibitemOpen [0]{}%
\providecommand \bibitemStop [0]{}%
\providecommand \bibitemNoStop [0]{.\EOS\space}%
\providecommand \EOS [0]{\spacefactor3000\relax}%
\providecommand \BibitemShut  [1]{\csname bibitem#1\endcsname}%
\let\auto@bib@innerbib\@empty
%</preamble>
\bibitem [{\citenamefont {Landau}(1937)}]{L3726}%
  \BibitemOpen
  \bibfield  {author} {\bibinfo {author} {\bibfnamefont {L.~D.}\ \bibnamefont
  {Landau}},\ }\href@noop {} {\bibfield  {journal} {\bibinfo  {journal} {Phys.
  Z. Sowjetunion}\ }\textbf {\bibinfo {volume} {11}},\ \bibinfo {pages} {26}
  (\bibinfo {year} {1937})}\BibitemShut {NoStop}%
\bibitem [{\citenamefont {Kitaev}\ and\ \citenamefont
  {Preskill}(2006)}]{KP0604}%
  \BibitemOpen
  \bibfield  {author} {\bibinfo {author} {\bibfnamefont {A.}~\bibnamefont
  {Kitaev}}\ and\ \bibinfo {author} {\bibfnamefont {J.}~\bibnamefont
  {Preskill}},\ }\href@noop {} {\bibfield  {journal} {\bibinfo  {journal}
  {Phys. Rev. Lett.}\ }\textbf {\bibinfo {volume} {96}},\ \bibinfo {pages}
  {110404} (\bibinfo {year} {2006})}\BibitemShut {NoStop}%
\bibitem [{\citenamefont {Levin}\ and\ \citenamefont {Wen}(2006)}]{LW0605}%
  \BibitemOpen
  \bibfield  {author} {\bibinfo {author} {\bibfnamefont {M.}~\bibnamefont
  {Levin}}\ and\ \bibinfo {author} {\bibfnamefont {X.-G.}\ \bibnamefont
  {Wen}},\ }\href@noop {} {\bibfield  {journal} {\bibinfo  {journal} {Phys.
  Rev. Lett.}\ }\textbf {\bibinfo {volume} {96}},\ \bibinfo {pages} {110405}
  (\bibinfo {year} {2006})},\ \Eprint {http://arxiv.org/abs/cond-mat/0510613}
  {cond-mat/0510613} \BibitemShut {NoStop}%
\bibitem [{\citenamefont {Chen}\ \emph {et~al.}(2010)\citenamefont {Chen},
  \citenamefont {Gu},\ and\ \citenamefont {Wen}}]{CGW1038}%
  \BibitemOpen
  \bibfield  {author} {\bibinfo {author} {\bibfnamefont {X.}~\bibnamefont
  {Chen}}, \bibinfo {author} {\bibfnamefont {Z.-C.}\ \bibnamefont {Gu}}, \ and\
  \bibinfo {author} {\bibfnamefont {X.-G.}\ \bibnamefont {Wen}},\ }\href@noop
  {} {\bibfield  {journal} {\bibinfo  {journal} {Phys. Rev. B}\ }\textbf
  {\bibinfo {volume} {82}},\ \bibinfo {pages} {155138} (\bibinfo {year}
  {2010})},\ \Eprint {http://arxiv.org/abs/arXiv:1004.3835} {arXiv:1004.3835}
  \BibitemShut {NoStop}%
\bibitem [{\citenamefont {Wen}(1989)}]{Wtop}%
  \BibitemOpen
  \bibfield  {author} {\bibinfo {author} {\bibfnamefont {X.-G.}\ \bibnamefont
  {Wen}},\ }\href@noop {} {\bibfield  {journal} {\bibinfo  {journal} {Phys.
  Rev. B}\ }\textbf {\bibinfo {volume} {40}},\ \bibinfo {pages} {7387}
  (\bibinfo {year} {1989})}\BibitemShut {NoStop}%
\bibitem [{\citenamefont {Wen}(1990)}]{Wrig}%
  \BibitemOpen
  \bibfield  {author} {\bibinfo {author} {\bibfnamefont {X.-G.}\ \bibnamefont
  {Wen}},\ }\href@noop {} {\bibfield  {journal} {\bibinfo  {journal} {Int. J.
  Mod. Phys. B}\ }\textbf {\bibinfo {volume} {4}},\ \bibinfo {pages} {239}
  (\bibinfo {year} {1990})}\BibitemShut {NoStop}%
\bibitem [{\citenamefont {Keski-Vakkuri}\ and\ \citenamefont
  {Wen}(1993)}]{KW9327}%
  \BibitemOpen
  \bibfield  {author} {\bibinfo {author} {\bibfnamefont {E.}~\bibnamefont
  {Keski-Vakkuri}}\ and\ \bibinfo {author} {\bibfnamefont {X.-G.}\ \bibnamefont
  {Wen}},\ }\href@noop {} {\bibfield  {journal} {\bibinfo  {journal} {Int. J.
  Mod. Phys. B}\ }\textbf {\bibinfo {volume} {7}},\ \bibinfo {pages} {4227}
  (\bibinfo {year} {1993})}\BibitemShut {NoStop}%
\bibitem [{\citenamefont {Gu}\ and\ \citenamefont {Wen}(2009)}]{GW0931}%
  \BibitemOpen
  \bibfield  {author} {\bibinfo {author} {\bibfnamefont {Z.-C.}\ \bibnamefont
  {Gu}}\ and\ \bibinfo {author} {\bibfnamefont {X.-G.}\ \bibnamefont {Wen}},\
  }\href@noop {} {\bibfield  {journal} {\bibinfo  {journal} {Phys. Rev. B}\
  }\textbf {\bibinfo {volume} {80}},\ \bibinfo {pages} {155131} (\bibinfo
  {year} {2009})},\ \Eprint {http://arxiv.org/abs/arXiv:0903.1069}
  {arXiv:0903.1069} \BibitemShut {NoStop}%
\bibitem [{\citenamefont {Pollmann}\ \emph {et~al.}(2010)\citenamefont
  {Pollmann}, \citenamefont {Berg}, \citenamefont {Turner},\ and\ \citenamefont
  {Oshikawa}}]{PBT1039}%
  \BibitemOpen
  \bibfield  {author} {\bibinfo {author} {\bibfnamefont {F.}~\bibnamefont
  {Pollmann}}, \bibinfo {author} {\bibfnamefont {E.}~\bibnamefont {Berg}},
  \bibinfo {author} {\bibfnamefont {A.~M.}\ \bibnamefont {Turner}}, \ and\
  \bibinfo {author} {\bibfnamefont {M.}~\bibnamefont {Oshikawa}},\ }\href
  {\doibase 10.1103/PhysRevB.81.064439} {\bibfield  {journal} {\bibinfo
  {journal} {Phys. Rev. B}\ }\textbf {\bibinfo {volume} {81}},\ \bibinfo
  {pages} {064439} (\bibinfo {year} {2010})},\ \Eprint
  {http://arxiv.org/abs/arXiv:0910.1811} {arXiv:0910.1811} \BibitemShut
  {NoStop}%
\bibitem [{\citenamefont {Chen}\ \emph
  {et~al.}(2011{\natexlab{a}})\citenamefont {Chen}, \citenamefont {Liu},\ and\
  \citenamefont {Wen}}]{CLW1141}%
  \BibitemOpen
  \bibfield  {author} {\bibinfo {author} {\bibfnamefont {X.}~\bibnamefont
  {Chen}}, \bibinfo {author} {\bibfnamefont {Z.-X.}\ \bibnamefont {Liu}}, \
  and\ \bibinfo {author} {\bibfnamefont {X.-G.}\ \bibnamefont {Wen}},\
  }\href@noop {} {\bibfield  {journal} {\bibinfo  {journal} {Phys. Rev. B}\
  }\textbf {\bibinfo {volume} {84}},\ \bibinfo {pages} {235141} (\bibinfo
  {year} {2011}{\natexlab{a}})},\ \Eprint
  {http://arxiv.org/abs/arXiv:1106.4752} {arXiv:1106.4752} \BibitemShut
  {NoStop}%
\bibitem [{\citenamefont {Chen}\ \emph {et~al.}(2013)\citenamefont {Chen},
  \citenamefont {Gu}, \citenamefont {Liu},\ and\ \citenamefont
  {Wen}}]{CGL1314}%
  \BibitemOpen
  \bibfield  {author} {\bibinfo {author} {\bibfnamefont {X.}~\bibnamefont
  {Chen}}, \bibinfo {author} {\bibfnamefont {Z.-C.}\ \bibnamefont {Gu}},
  \bibinfo {author} {\bibfnamefont {Z.-X.}\ \bibnamefont {Liu}}, \ and\
  \bibinfo {author} {\bibfnamefont {X.-G.}\ \bibnamefont {Wen}},\ }\href@noop
  {} {\bibfield  {journal} {\bibinfo  {journal} {Phys. Rev. B}\ }\textbf
  {\bibinfo {volume} {87}},\ \bibinfo {pages} {155114} (\bibinfo {year}
  {2013})},\ \Eprint {http://arxiv.org/abs/arXiv:1106.4772} {arXiv:1106.4772}
  \BibitemShut {NoStop}%
\bibitem [{\citenamefont {Chen}\ \emph
  {et~al.}(2011{\natexlab{b}})\citenamefont {Chen}, \citenamefont {Gu},\ and\
  \citenamefont {Wen}}]{CGW1107}%
  \BibitemOpen
  \bibfield  {author} {\bibinfo {author} {\bibfnamefont {X.}~\bibnamefont
  {Chen}}, \bibinfo {author} {\bibfnamefont {Z.-C.}\ \bibnamefont {Gu}}, \ and\
  \bibinfo {author} {\bibfnamefont {X.-G.}\ \bibnamefont {Wen}},\ }\href@noop
  {} {\bibfield  {journal} {\bibinfo  {journal} {Phys. Rev. B}\ }\textbf
  {\bibinfo {volume} {83}},\ \bibinfo {pages} {035107} (\bibinfo {year}
  {2011}{\natexlab{b}})},\ \Eprint {http://arxiv.org/abs/arXiv:1008.3745}
  {arXiv:1008.3745} \BibitemShut {NoStop}%
\bibitem [{\citenamefont {Fidkowski}\ and\ \citenamefont
  {Kitaev}(2011)}]{FK1103}%
  \BibitemOpen
  \bibfield  {author} {\bibinfo {author} {\bibfnamefont {L.}~\bibnamefont
  {Fidkowski}}\ and\ \bibinfo {author} {\bibfnamefont {A.}~\bibnamefont
  {Kitaev}},\ }\href@noop {} {\bibfield  {journal} {\bibinfo  {journal} {Phys.
  Rev. B}\ }\textbf {\bibinfo {volume} {83}},\ \bibinfo {pages} {075103}
  (\bibinfo {year} {2011})},\ \Eprint {http://arxiv.org/abs/arXiv:1008.4138}
  {arXiv:1008.4138} \BibitemShut {NoStop}%
\bibitem [{\citenamefont {Schuch}\ \emph {et~al.}(2011)\citenamefont {Schuch},
  \citenamefont {Perez-Garcia},\ and\ \citenamefont {Cirac}}]{SPC1139}%
  \BibitemOpen
  \bibfield  {author} {\bibinfo {author} {\bibfnamefont {N.}~\bibnamefont
  {Schuch}}, \bibinfo {author} {\bibfnamefont {D.}~\bibnamefont
  {Perez-Garcia}}, \ and\ \bibinfo {author} {\bibfnamefont {I.}~\bibnamefont
  {Cirac}},\ }\href@noop {} {\bibfield  {journal} {\bibinfo  {journal} {Phys.
  Rev. B}\ }\textbf {\bibinfo {volume} {84}},\ \bibinfo {pages} {165139}
  (\bibinfo {year} {2011})},\ \Eprint {http://arxiv.org/abs/arXiv:1010.3732}
  {arXiv:1010.3732} \BibitemShut {NoStop}%
\bibitem [{\citenamefont {Chen}\ \emph
  {et~al.}(2011{\natexlab{c}})\citenamefont {Chen}, \citenamefont {Gu},\ and\
  \citenamefont {Wen}}]{CGW1128}%
  \BibitemOpen
  \bibfield  {author} {\bibinfo {author} {\bibfnamefont {X.}~\bibnamefont
  {Chen}}, \bibinfo {author} {\bibfnamefont {Z.-C.}\ \bibnamefont {Gu}}, \ and\
  \bibinfo {author} {\bibfnamefont {X.-G.}\ \bibnamefont {Wen}},\ }\href@noop
  {} {\bibfield  {journal} {\bibinfo  {journal} {Phys. Rev. B}\ }\textbf
  {\bibinfo {volume} {84}},\ \bibinfo {pages} {235128} (\bibinfo {year}
  {2011}{\natexlab{c}})},\ \Eprint {http://arxiv.org/abs/arXiv:1103.3323}
  {arXiv:1103.3323} \BibitemShut {NoStop}%
\bibitem [{\citenamefont {Zeng}\ and\ \citenamefont {Wen}(2015)}]{ZW1490}%
  \BibitemOpen
  \bibfield  {author} {\bibinfo {author} {\bibfnamefont {B.}~\bibnamefont
  {Zeng}}\ and\ \bibinfo {author} {\bibfnamefont {X.-G.}\ \bibnamefont {Wen}},\
  }\href {\doibase 10.1103/PhysRevB.91.125121} {\bibfield  {journal} {\bibinfo
  {journal} {Phys. Rev. B}\ }\textbf {\bibinfo {volume} {91}},\ \bibinfo
  {pages} {125121} (\bibinfo {year} {2015})},\ \Eprint
  {http://arxiv.org/abs/arXiv:1406.5090} {arXiv:1406.5090} \BibitemShut
  {NoStop}%
\bibitem [{\citenamefont {{Swingle}}\ and\ \citenamefont
  {{McGreevy}}(2016)}]{SM1403}%
  \BibitemOpen
  \bibfield  {author} {\bibinfo {author} {\bibfnamefont {B.}~\bibnamefont
  {{Swingle}}}\ and\ \bibinfo {author} {\bibfnamefont {J.}~\bibnamefont
  {{McGreevy}}},\ }\href {\doibase 10.1103/PhysRevB.93.045127} {\bibfield
  {journal} {\bibinfo  {journal} {Phys. Rev. B}\ }\textbf {\bibinfo {volume}
  {93}},\ \bibinfo {pages} {045127} (\bibinfo {year} {2016})},\ \Eprint
  {http://arxiv.org/abs/arXiv:1407.8203} {arXiv:1407.8203} \BibitemShut
  {NoStop}%
\bibitem [{\citenamefont {{Rowell}}\ \emph {et~al.}(2009)\citenamefont
  {{Rowell}}, \citenamefont {{Stong}},\ and\ \citenamefont {{Wang}}}]{RSW0777}%
  \BibitemOpen
  \bibfield  {author} {\bibinfo {author} {\bibfnamefont {E.}~\bibnamefont
  {{Rowell}}}, \bibinfo {author} {\bibfnamefont {R.}~\bibnamefont {{Stong}}}, \
  and\ \bibinfo {author} {\bibfnamefont {Z.}~\bibnamefont {{Wang}}},\
  }\href@noop {} {\bibfield  {journal} {\bibinfo  {journal} {Comm. Math.
  Phys.}\ }\textbf {\bibinfo {volume} {292}},\ \bibinfo {pages} {343} (\bibinfo
  {year} {2009})},\ \Eprint {http://arxiv.org/abs/arXiv:0712.1377}
  {arXiv:0712.1377} \BibitemShut {NoStop}%
\bibitem [{\citenamefont {{Wen}}(2016)}]{W150605768}%
  \BibitemOpen
  \bibfield  {author} {\bibinfo {author} {\bibfnamefont {X.-G.}\ \bibnamefont
  {{Wen}}},\ }\href {\doibase 10.1093/nsr/nwv077} {\bibfield  {journal}
  {\bibinfo  {journal} {Natl. Sci. Rev.}\ }\textbf {\bibinfo {volume} {3}},\
  \bibinfo {pages} {68} (\bibinfo {year} {2016})},\ \Eprint
  {http://arxiv.org/abs/arXiv:1506.05768} {arXiv:1506.05768} \BibitemShut
  {NoStop}%
\bibitem [{\citenamefont {{Barkeshli}}\ \emph {et~al.}(2014)\citenamefont
  {{Barkeshli}}, \citenamefont {{Bonderson}}, \citenamefont {{Cheng}},\ and\
  \citenamefont {{Wang}}}]{BBC1440}%
  \BibitemOpen
  \bibfield  {author} {\bibinfo {author} {\bibfnamefont {M.}~\bibnamefont
  {{Barkeshli}}}, \bibinfo {author} {\bibfnamefont {P.}~\bibnamefont
  {{Bonderson}}}, \bibinfo {author} {\bibfnamefont {M.}~\bibnamefont
  {{Cheng}}}, \ and\ \bibinfo {author} {\bibfnamefont {Z.}~\bibnamefont
  {{Wang}}},\ }\href@noop {} {\  (\bibinfo {year} {2014})},\ \Eprint
  {http://arxiv.org/abs/arXiv:1410.4540} {arXiv:1410.4540} \BibitemShut
  {NoStop}%
\bibitem [{\citenamefont {{Lan}}\ \emph {et~al.}(2016)\citenamefont {{Lan}},
  \citenamefont {{Kong}},\ and\ \citenamefont {{Wen}}}]{LW150704673}%
  \BibitemOpen
  \bibfield  {author} {\bibinfo {author} {\bibfnamefont {T.}~\bibnamefont
  {{Lan}}}, \bibinfo {author} {\bibfnamefont {L.}~\bibnamefont {{Kong}}}, \
  and\ \bibinfo {author} {\bibfnamefont {X.-G.}\ \bibnamefont {{Wen}}},\ }\href
  {\doibase 10.1103/PhysRevB.94.155113} {\bibfield  {journal} {\bibinfo
  {journal} {\prb}\ }\textbf {\bibinfo {volume} {94}},\ \bibinfo {pages}
  {155113} (\bibinfo {year} {2016})},\ \Eprint
  {http://arxiv.org/abs/arXiv:1507.04673} {arXiv:1507.04673} \BibitemShut
  {NoStop}%
\bibitem [{\citenamefont {Lan}\ \emph {et~al.}(2016)\citenamefont {Lan},
  \citenamefont {Kong},\ and\ \citenamefont {Wen}}]{LW160205946}%
  \BibitemOpen
  \bibfield  {author} {\bibinfo {author} {\bibfnamefont {T.}~\bibnamefont
  {Lan}}, \bibinfo {author} {\bibfnamefont {L.}~\bibnamefont {Kong}}, \ and\
  \bibinfo {author} {\bibfnamefont {X.-G.}\ \bibnamefont {Wen}},\ }\href@noop
  {} {\  (\bibinfo {year} {2016})},\ \Eprint
  {http://arxiv.org/abs/arXiv:1602.05946} {arXiv:1602.05946} \BibitemShut
  {NoStop}%
\bibitem [{\citenamefont {Haldane}(1983)}]{H8364}%
  \BibitemOpen
  \bibfield  {author} {\bibinfo {author} {\bibfnamefont {F.~D.~M.}\
  \bibnamefont {Haldane}},\ }\href@noop {} {\bibfield  {journal} {\bibinfo
  {journal} {Physics Letters A}\ }\textbf {\bibinfo {volume} {93}},\ \bibinfo
  {pages} {464} (\bibinfo {year} {1983})}\BibitemShut {NoStop}%
\bibitem [{\citenamefont {Affleck}\ \emph {et~al.}(1987)\citenamefont
  {Affleck}, \citenamefont {Kennedy}, \citenamefont {Lieb},\ and\ \citenamefont
  {Tasaki}}]{AKL8799}%
  \BibitemOpen
  \bibfield  {author} {\bibinfo {author} {\bibfnamefont {I.}~\bibnamefont
  {Affleck}}, \bibinfo {author} {\bibfnamefont {T.}~\bibnamefont {Kennedy}},
  \bibinfo {author} {\bibfnamefont {E.~H.}\ \bibnamefont {Lieb}}, \ and\
  \bibinfo {author} {\bibfnamefont {H.}~\bibnamefont {Tasaki}},\ }\href@noop {}
  {\bibfield  {journal} {\bibinfo  {journal} {Phys. Rev. Lett.}\ }\textbf
  {\bibinfo {volume} {59}},\ \bibinfo {pages} {799} (\bibinfo {year}
  {1987})}\BibitemShut {NoStop}%
\bibitem [{\citenamefont {Kane}\ and\ \citenamefont {Mele}(2005)}]{KM0502}%
  \BibitemOpen
  \bibfield  {author} {\bibinfo {author} {\bibfnamefont {C.~L.}\ \bibnamefont
  {Kane}}\ and\ \bibinfo {author} {\bibfnamefont {E.~J.}\ \bibnamefont
  {Mele}},\ }\href@noop {} {\bibfield  {journal} {\bibinfo  {journal} {Phys.
  Rev. Lett.}\ }\textbf {\bibinfo {volume} {95}},\ \bibinfo {pages} {146802}
  (\bibinfo {year} {2005})},\ \Eprint {http://arxiv.org/abs/cond-mat/0506581}
  {cond-mat/0506581} \BibitemShut {NoStop}%
\bibitem [{\citenamefont {Moore}\ and\ \citenamefont {Balents}(2007)}]{MB0706}%
  \BibitemOpen
  \bibfield  {author} {\bibinfo {author} {\bibfnamefont {J.~E.}\ \bibnamefont
  {Moore}}\ and\ \bibinfo {author} {\bibfnamefont {L.}~\bibnamefont
  {Balents}},\ }\href@noop {} {\bibfield  {journal} {\bibinfo  {journal} {Phys.
  Rev. B}\ }\textbf {\bibinfo {volume} {75}},\ \bibinfo {pages} {121306}
  (\bibinfo {year} {2007})},\ \Eprint {http://arxiv.org/abs/cond-mat/0607314}
  {cond-mat/0607314} \BibitemShut {NoStop}%
\bibitem [{\citenamefont {Roy}(2009)}]{R0922}%
  \BibitemOpen
  \bibfield  {author} {\bibinfo {author} {\bibfnamefont {R.}~\bibnamefont
  {Roy}},\ }\href {\doibase 10.1103/PhysRevB.79.195322} {\bibfield  {journal}
  {\bibinfo  {journal} {Phys. Rev. B}\ }\textbf {\bibinfo {volume} {79}},\
  \bibinfo {pages} {195322} (\bibinfo {year} {2009})},\ \Eprint
  {http://arxiv.org/abs/cond-mat/0607531} {cond-mat/0607531} \BibitemShut
  {NoStop}%
\bibitem [{\citenamefont {Fu}\ \emph {et~al.}(2007)\citenamefont {Fu},
  \citenamefont {Kane},\ and\ \citenamefont {Mele}}]{FKM0703}%
  \BibitemOpen
  \bibfield  {author} {\bibinfo {author} {\bibfnamefont {L.}~\bibnamefont
  {Fu}}, \bibinfo {author} {\bibfnamefont {C.~L.}\ \bibnamefont {Kane}}, \ and\
  \bibinfo {author} {\bibfnamefont {E.~J.}\ \bibnamefont {Mele}},\ }\href@noop
  {} {\bibfield  {journal} {\bibinfo  {journal} {Phys. Rev. Lett.}\ }\textbf
  {\bibinfo {volume} {98}},\ \bibinfo {pages} {106803} (\bibinfo {year}
  {2007})},\ \Eprint {http://arxiv.org/abs/cond-mat/0607699} {cond-mat/0607699}
  \BibitemShut {NoStop}%
\bibitem [{\citenamefont {Qi}\ \emph {et~al.}(2008)\citenamefont {Qi},
  \citenamefont {Hughes},\ and\ \citenamefont {Zhang}}]{QHZ0824}%
  \BibitemOpen
  \bibfield  {author} {\bibinfo {author} {\bibfnamefont {X.-L.}\ \bibnamefont
  {Qi}}, \bibinfo {author} {\bibfnamefont {T.}~\bibnamefont {Hughes}}, \ and\
  \bibinfo {author} {\bibfnamefont {S.-C.}\ \bibnamefont {Zhang}},\ }\href@noop
  {} {\bibfield  {journal} {\bibinfo  {journal} {Phys. Rev. B}\ }\textbf
  {\bibinfo {volume} {78}},\ \bibinfo {pages} {195424} (\bibinfo {year}
  {2008})},\ \Eprint {http://arxiv.org/abs/arXiv:0802.3537} {arXiv:0802.3537}
  \BibitemShut {NoStop}%
\bibitem [{\citenamefont {{Wang}}\ \emph {et~al.}(2014)\citenamefont {{Wang}},
  \citenamefont {{Potter}},\ and\ \citenamefont {{Senthil}}}]{WS13063238}%
  \BibitemOpen
  \bibfield  {author} {\bibinfo {author} {\bibfnamefont {C.}~\bibnamefont
  {{Wang}}}, \bibinfo {author} {\bibfnamefont {A.~C.}\ \bibnamefont
  {{Potter}}}, \ and\ \bibinfo {author} {\bibfnamefont {T.}~\bibnamefont
  {{Senthil}}},\ }\href {\doibase 10.1126/science.1243326} {\bibfield
  {journal} {\bibinfo  {journal} {Science}\ }\textbf {\bibinfo {volume}
  {343}},\ \bibinfo {pages} {629} (\bibinfo {year} {2014})},\ \Eprint
  {http://arxiv.org/abs/arXiv:1306.3238} {arXiv:1306.3238} \BibitemShut
  {NoStop}%
\bibitem [{\citenamefont {Kong}\ and\ \citenamefont {Wen}(2014)}]{KW1458}%
  \BibitemOpen
  \bibfield  {author} {\bibinfo {author} {\bibfnamefont {L.}~\bibnamefont
  {Kong}}\ and\ \bibinfo {author} {\bibfnamefont {X.-G.}\ \bibnamefont {Wen}},\
  }\href@noop {} {\  (\bibinfo {year} {2014})},\ \Eprint
  {http://arxiv.org/abs/arXiv:1405.5858} {arXiv:1405.5858} \BibitemShut
  {NoStop}%
\bibitem [{\citenamefont {{Cui}}(2016)}]{C161007628}%
  \BibitemOpen
  \bibfield  {author} {\bibinfo {author} {\bibfnamefont {S.~X.}\ \bibnamefont
  {{Cui}}},\ }\href@noop {} {\  (\bibinfo {year} {2016})},\ \Eprint
  {http://arxiv.org/abs/arXiv:1610.07628} {arXiv:1610.07628} \BibitemShut
  {NoStop}%
\bibitem [{\citenamefont {Dijkgraaf}\ and\ \citenamefont
  {Witten}(1990)}]{DW9093}%
  \BibitemOpen
  \bibfield  {author} {\bibinfo {author} {\bibfnamefont {R.}~\bibnamefont
  {Dijkgraaf}}\ and\ \bibinfo {author} {\bibfnamefont {E.}~\bibnamefont
  {Witten}},\ }\href@noop {} {\bibfield  {journal} {\bibinfo  {journal} {Comm.
  Math. Phys.}\ }\textbf {\bibinfo {volume} {129}},\ \bibinfo {pages} {393}
  (\bibinfo {year} {1990})}\BibitemShut {NoStop}%
\bibitem [{\citenamefont {Turaev}\ and\ \citenamefont {Viro}(1992)}]{TV9265}%
  \BibitemOpen
  \bibfield  {author} {\bibinfo {author} {\bibfnamefont {V.~G.}\ \bibnamefont
  {Turaev}}\ and\ \bibinfo {author} {\bibfnamefont {O.~Y.}\ \bibnamefont
  {Viro}},\ }\href@noop {} {\bibfield  {journal} {\bibinfo  {journal}
  {Topology}\ }\textbf {\bibinfo {volume} {31}},\ \bibinfo {pages} {865}
  (\bibinfo {year} {1992})}\BibitemShut {NoStop}%
\bibitem [{\citenamefont {Freedman}\ \emph {et~al.}(2004)\citenamefont
  {Freedman}, \citenamefont {Nayak}, \citenamefont {Shtengel}, \citenamefont
  {Walker},\ and\ \citenamefont {Wang}}]{FNS0428}%
  \BibitemOpen
  \bibfield  {author} {\bibinfo {author} {\bibfnamefont {M.}~\bibnamefont
  {Freedman}}, \bibinfo {author} {\bibfnamefont {C.}~\bibnamefont {Nayak}},
  \bibinfo {author} {\bibfnamefont {K.}~\bibnamefont {Shtengel}}, \bibinfo
  {author} {\bibfnamefont {K.}~\bibnamefont {Walker}}, \ and\ \bibinfo {author}
  {\bibfnamefont {Z.}~\bibnamefont {Wang}},\ }\href@noop {} {\bibfield
  {journal} {\bibinfo  {journal} {Ann. Phys. (NY)}\ }\textbf {\bibinfo {volume}
  {310}},\ \bibinfo {pages} {428} (\bibinfo {year} {2004})},\ \Eprint
  {http://arxiv.org/abs/cond-mat/0307511} {cond-mat/0307511} \BibitemShut
  {NoStop}%
\bibitem [{\citenamefont {Levin}\ and\ \citenamefont {Wen}(2005)}]{LW0510}%
  \BibitemOpen
  \bibfield  {author} {\bibinfo {author} {\bibfnamefont {M.}~\bibnamefont
  {Levin}}\ and\ \bibinfo {author} {\bibfnamefont {X.-G.}\ \bibnamefont
  {Wen}},\ }\href@noop {} {\bibfield  {journal} {\bibinfo  {journal} {Phys.
  Rev. B}\ }\textbf {\bibinfo {volume} {71}},\ \bibinfo {pages} {045110}
  (\bibinfo {year} {2005})},\ \Eprint {http://arxiv.org/abs/cond-mat/0404617}
  {cond-mat/0404617} \BibitemShut {NoStop}%
\bibitem [{\citenamefont {Gu}\ \emph {et~al.}(2015)\citenamefont {Gu},
  \citenamefont {Wang},\ and\ \citenamefont {Wen}}]{GWW1017}%
  \BibitemOpen
  \bibfield  {author} {\bibinfo {author} {\bibfnamefont {Z.-C.}\ \bibnamefont
  {Gu}}, \bibinfo {author} {\bibfnamefont {Z.}~\bibnamefont {Wang}}, \ and\
  \bibinfo {author} {\bibfnamefont {X.-G.}\ \bibnamefont {Wen}},\ }\href
  {\doibase 10.1103/PhysRevB.91.125149} {\bibfield  {journal} {\bibinfo
  {journal} {Phys. Rev. B}\ }\textbf {\bibinfo {volume} {91}},\ \bibinfo
  {pages} {125149} (\bibinfo {year} {2015})},\ \Eprint
  {http://arxiv.org/abs/arXiv:1010.1517} {arXiv:1010.1517} \BibitemShut
  {NoStop}%
\bibitem [{\citenamefont {Walker}\ and\ \citenamefont {Wang}(2011)}]{WW1132}%
  \BibitemOpen
  \bibfield  {author} {\bibinfo {author} {\bibfnamefont {K.}~\bibnamefont
  {Walker}}\ and\ \bibinfo {author} {\bibfnamefont {Z.}~\bibnamefont {Wang}},\
  }\href@noop {} {\  (\bibinfo {year} {2011})},\ \Eprint
  {http://arxiv.org/abs/arXiv:1104.2632} {arXiv:1104.2632} \BibitemShut
  {NoStop}%
\bibitem [{\citenamefont {{Williamson}}\ and\ \citenamefont
  {{Wang}}(2016)}]{WW160607144}%
  \BibitemOpen
  \bibfield  {author} {\bibinfo {author} {\bibfnamefont {D.~J.}\ \bibnamefont
  {{Williamson}}}\ and\ \bibinfo {author} {\bibfnamefont {Z.}~\bibnamefont
  {{Wang}}},\ }\href@noop {} {\  (\bibinfo {year} {2016})},\ \Eprint
  {http://arxiv.org/abs/arXiv:1606.07144} {arXiv:1606.07144} \BibitemShut
  {NoStop}%
\bibitem [{\citenamefont {{Tarantino}}\ and\ \citenamefont
  {{Fidkowski}}(2016)}]{TF160402145}%
  \BibitemOpen
  \bibfield  {author} {\bibinfo {author} {\bibfnamefont {N.}~\bibnamefont
  {{Tarantino}}}\ and\ \bibinfo {author} {\bibfnamefont {L.}~\bibnamefont
  {{Fidkowski}}},\ }\href {\doibase 10.1103/PhysRevB.94.115115} {\bibfield
  {journal} {\bibinfo  {journal} {\prb}\ }\textbf {\bibinfo {volume} {94}},\
  \bibinfo {pages} {115115} (\bibinfo {year} {2016})},\ \Eprint
  {http://arxiv.org/abs/arXiv:1604.02145} {arXiv:1604.02145} \BibitemShut
  {NoStop}%
\bibitem [{\citenamefont {{Bhardwaj}}\ \emph {et~al.}(2016)\citenamefont
  {{Bhardwaj}}, \citenamefont {{Gaiotto}},\ and\ \citenamefont
  {{Kapustin}}}]{BK160501640}%
  \BibitemOpen
  \bibfield  {author} {\bibinfo {author} {\bibfnamefont {L.}~\bibnamefont
  {{Bhardwaj}}}, \bibinfo {author} {\bibfnamefont {D.}~\bibnamefont
  {{Gaiotto}}}, \ and\ \bibinfo {author} {\bibfnamefont {A.}~\bibnamefont
  {{Kapustin}}},\ }\href@noop {} {\  (\bibinfo {year} {2016})},\ \Eprint
  {http://arxiv.org/abs/arXiv:1605.01640} {arXiv:1605.01640} \BibitemShut
  {NoStop}%
\bibitem [{\citenamefont {Lan}\ \emph {et~al.}(2017)\citenamefont {Lan},
  \citenamefont {Kong},\ and\ \citenamefont {Wen}}]{LW170404221}%
  \BibitemOpen
  \bibfield  {author} {\bibinfo {author} {\bibfnamefont {T.}~\bibnamefont
  {Lan}}, \bibinfo {author} {\bibfnamefont {L.}~\bibnamefont {Kong}}, \ and\
  \bibinfo {author} {\bibfnamefont {X.-G.}\ \bibnamefont {Wen}},\ }\href@noop
  {} {\  (\bibinfo {year} {2017})},\ \Eprint
  {http://arxiv.org/abs/arXiv:1704.04221} {arXiv:1704.04221} \BibitemShut
  {NoStop}%
\bibitem [{\citenamefont {{Kapustin}}\ and\ \citenamefont
  {{Thorngren}}(2017)}]{KT170108264}%
  \BibitemOpen
  \bibfield  {author} {\bibinfo {author} {\bibfnamefont {A.}~\bibnamefont
  {{Kapustin}}}\ and\ \bibinfo {author} {\bibfnamefont {R.}~\bibnamefont
  {{Thorngren}}},\ }\href@noop {} {\  (\bibinfo {year} {2017})},\ \Eprint
  {http://arxiv.org/abs/arXiv:1701.08264} {arXiv:1701.08264} \BibitemShut
  {NoStop}%
\bibitem [{\citenamefont {{Wang}}\ and\ \citenamefont
  {{Gu}}(2017)}]{WG170310937}%
  \BibitemOpen
  \bibfield  {author} {\bibinfo {author} {\bibfnamefont {Q.-R.}\ \bibnamefont
  {{Wang}}}\ and\ \bibinfo {author} {\bibfnamefont {Z.-C.}\ \bibnamefont
  {{Gu}}},\ }\href@noop {} {\  (\bibinfo {year} {2017})},\ \Eprint
  {http://arxiv.org/abs/arXiv:1703.10937} {arXiv:1703.10937} \BibitemShut
  {NoStop}%
\bibitem [{\citenamefont {Kapustin}\ and\ \citenamefont
  {Thorngren}(2013)}]{KT1321}%
  \BibitemOpen
  \bibfield  {author} {\bibinfo {author} {\bibfnamefont {A.}~\bibnamefont
  {Kapustin}}\ and\ \bibinfo {author} {\bibfnamefont {R.}~\bibnamefont
  {Thorngren}},\ }\href@noop {} {\  (\bibinfo {year} {2013})},\ \Eprint
  {http://arxiv.org/abs/arXiv:1309.4721} {arXiv:1309.4721} \BibitemShut
  {NoStop}%
\bibitem [{\citenamefont {{Gaiotto}}\ \emph {et~al.}(2015)\citenamefont
  {{Gaiotto}}, \citenamefont {{Kapustin}}, \citenamefont {{Seiberg}},\ and\
  \citenamefont {{Willett}}}]{GW14125148}%
  \BibitemOpen
  \bibfield  {author} {\bibinfo {author} {\bibfnamefont {D.}~\bibnamefont
  {{Gaiotto}}}, \bibinfo {author} {\bibfnamefont {A.}~\bibnamefont
  {{Kapustin}}}, \bibinfo {author} {\bibfnamefont {N.}~\bibnamefont
  {{Seiberg}}}, \ and\ \bibinfo {author} {\bibfnamefont {B.}~\bibnamefont
  {{Willett}}},\ }\href {\doibase 10.1007/JHEP02(2015)172} {\bibfield
  {journal} {\bibinfo  {journal} {Journal of High Energy Physics}\ }\textbf
  {\bibinfo {volume} {2}},\ \bibinfo {pages} {172} (\bibinfo {year} {2015})},\
  \Eprint {http://arxiv.org/abs/arXiv:1412.5148} {arXiv:1412.5148} \BibitemShut
  {NoStop}%
\bibitem [{\citenamefont {{Freed}}(2014)}]{F1478}%
  \BibitemOpen
  \bibfield  {author} {\bibinfo {author} {\bibfnamefont {D.~S.}\ \bibnamefont
  {{Freed}}},\ }\href@noop {} {\  (\bibinfo {year} {2014})},\ \Eprint
  {http://arxiv.org/abs/arXiv:1406.7278} {arXiv:1406.7278} \BibitemShut
  {NoStop}%
\bibitem [{\citenamefont {Chen}\ \emph {et~al.}(2012)\citenamefont {Chen},
  \citenamefont {Gu}, \citenamefont {Liu},\ and\ \citenamefont
  {Wen}}]{CGL1204}%
  \BibitemOpen
  \bibfield  {author} {\bibinfo {author} {\bibfnamefont {X.}~\bibnamefont
  {Chen}}, \bibinfo {author} {\bibfnamefont {Z.-C.}\ \bibnamefont {Gu}},
  \bibinfo {author} {\bibfnamefont {Z.-X.}\ \bibnamefont {Liu}}, \ and\
  \bibinfo {author} {\bibfnamefont {X.-G.}\ \bibnamefont {Wen}},\ }\href@noop
  {} {\bibfield  {journal} {\bibinfo  {journal} {Science}\ }\textbf {\bibinfo
  {volume} {338}},\ \bibinfo {pages} {1604} (\bibinfo {year} {2012})},\ \Eprint
  {http://arxiv.org/abs/arXiv:1301.0861} {arXiv:1301.0861} \BibitemShut
  {NoStop}%
\bibitem [{\citenamefont {Kapustin}(2014{\natexlab{a}})}]{K1459}%
  \BibitemOpen
  \bibfield  {author} {\bibinfo {author} {\bibfnamefont {A.}~\bibnamefont
  {Kapustin}},\ }\href@noop {} {\  (\bibinfo {year} {2014}{\natexlab{a}})},\
  \Eprint {http://arxiv.org/abs/arXiv:1404.6659} {arXiv:1404.6659} \BibitemShut
  {NoStop}%
\bibitem [{\citenamefont {Kapustin}(2014{\natexlab{b}})}]{K1467}%
  \BibitemOpen
  \bibfield  {author} {\bibinfo {author} {\bibfnamefont {A.}~\bibnamefont
  {Kapustin}},\ }\href@noop {} {\  (\bibinfo {year} {2014}{\natexlab{b}})},\
  \Eprint {http://arxiv.org/abs/arXiv:1403.1467} {arXiv:1403.1467} \BibitemShut
  {NoStop}%
\bibitem [{\citenamefont {Wen}(2015)}]{W1477}%
  \BibitemOpen
  \bibfield  {author} {\bibinfo {author} {\bibfnamefont {X.-G.}\ \bibnamefont
  {Wen}},\ }\href {\doibase 10.1103/PhysRevB.91.205101} {\bibfield  {journal}
  {\bibinfo  {journal} {Phys. Rev. B}\ }\textbf {\bibinfo {volume} {91}},\
  \bibinfo {pages} {205101} (\bibinfo {year} {2015})},\ \Eprint
  {http://arxiv.org/abs/arXiv:1410.8477} {arXiv:1410.8477} \BibitemShut
  {NoStop}%
\bibitem [{\citenamefont {{Shiozaki}}\ \emph {et~al.}(2016)\citenamefont
  {{Shiozaki}}, \citenamefont {{Shapourian}},\ and\ \citenamefont
  {{Ryu}}}]{SR160905970}%
  \BibitemOpen
  \bibfield  {author} {\bibinfo {author} {\bibfnamefont {K.}~\bibnamefont
  {{Shiozaki}}}, \bibinfo {author} {\bibfnamefont {H.}~\bibnamefont
  {{Shapourian}}}, \ and\ \bibinfo {author} {\bibfnamefont {S.}~\bibnamefont
  {{Ryu}}},\ }\href@noop {} {\  (\bibinfo {year} {2016})},\ \Eprint
  {http://arxiv.org/abs/arXiv:1609.05970} {arXiv:1609.05970} \BibitemShut
  {NoStop}%
\bibitem [{\citenamefont {Levin}\ and\ \citenamefont {Gu}(2012)}]{LG1209}%
  \BibitemOpen
  \bibfield  {author} {\bibinfo {author} {\bibfnamefont {M.}~\bibnamefont
  {Levin}}\ and\ \bibinfo {author} {\bibfnamefont {Z.-C.}\ \bibnamefont {Gu}},\
  }\href@noop {} {\bibfield  {journal} {\bibinfo  {journal} {Phys. Rev. B}\
  }\textbf {\bibinfo {volume} {86}},\ \bibinfo {pages} {115109} (\bibinfo
  {year} {2012})},\ \Eprint {http://arxiv.org/abs/arXiv:1202.3120}
  {arXiv:1202.3120} \BibitemShut {NoStop}%
\bibitem [{\citenamefont {Hung}\ and\ \citenamefont {Wen}(2014)}]{HW1339}%
  \BibitemOpen
  \bibfield  {author} {\bibinfo {author} {\bibfnamefont {L.-Y.}\ \bibnamefont
  {Hung}}\ and\ \bibinfo {author} {\bibfnamefont {X.-G.}\ \bibnamefont {Wen}},\
  }\href@noop {} {\bibfield  {journal} {\bibinfo  {journal} {Phys. Rev. B}\
  }\textbf {\bibinfo {volume} {89}},\ \bibinfo {pages} {075121} (\bibinfo
  {year} {2014})},\ \Eprint {http://arxiv.org/abs/arXiv:1311.5539}
  {arXiv:1311.5539} \BibitemShut {NoStop}%
\bibitem [{\citenamefont {Wen}(2014)}]{W1447}%
  \BibitemOpen
  \bibfield  {author} {\bibinfo {author} {\bibfnamefont {X.-G.}\ \bibnamefont
  {Wen}},\ }\href {\doibase 10.1103/PhysRevB.89.035147} {\bibfield  {journal}
  {\bibinfo  {journal} {Phys. Rev. B}\ }\textbf {\bibinfo {volume} {89}},\
  \bibinfo {pages} {035147} (\bibinfo {year} {2014})},\ \Eprint
  {http://arxiv.org/abs/arXiv:1301.7675} {arXiv:1301.7675} \BibitemShut
  {NoStop}%
\bibitem [{\citenamefont {{Sule}}\ \emph {et~al.}(2013)\citenamefont {{Sule}},
  \citenamefont {{Chen}},\ and\ \citenamefont {{Ryu}}}]{SCR1325}%
  \BibitemOpen
  \bibfield  {author} {\bibinfo {author} {\bibfnamefont {O.~M.}\ \bibnamefont
  {{Sule}}}, \bibinfo {author} {\bibfnamefont {X.}~\bibnamefont {{Chen}}}, \
  and\ \bibinfo {author} {\bibfnamefont {S.}~\bibnamefont {{Ryu}}},\ }\href
  {\doibase 10.1103/PhysRevB.88.075125} {\bibfield  {journal} {\bibinfo
  {journal} {\prb}\ }\textbf {\bibinfo {volume} {88}},\ \bibinfo {pages}
  {075125} (\bibinfo {year} {2013})},\ \Eprint
  {http://arxiv.org/abs/arXiv:1305.0700} {arXiv:1305.0700} \BibitemShut
  {NoStop}%
\bibitem [{\citenamefont {Kogut}(1979)}]{K7959}%
  \BibitemOpen
  \bibfield  {author} {\bibinfo {author} {\bibfnamefont {J.~B.}\ \bibnamefont
  {Kogut}},\ }\href {\doibase 10.1103/RevModPhys.51.659} {\bibfield  {journal}
  {\bibinfo  {journal} {Rev. Mod. Phys.}\ }\textbf {\bibinfo {volume} {51}},\
  \bibinfo {pages} {659 } (\bibinfo {year} {1979})}\BibitemShut {NoStop}%
\bibitem [{\citenamefont {{Kapustin}}\ and\ \citenamefont
  {{Seiberg}}(2014)}]{KS14010740}%
  \BibitemOpen
  \bibfield  {author} {\bibinfo {author} {\bibfnamefont {A.}~\bibnamefont
  {{Kapustin}}}\ and\ \bibinfo {author} {\bibfnamefont {N.}~\bibnamefont
  {{Seiberg}}},\ }\href {\doibase 10.1007/JHEP04(2014)001} {\bibfield
  {journal} {\bibinfo  {journal} {Journal of High Energy Physics}\ }\textbf
  {\bibinfo {volume} {4}},\ \bibinfo {pages} {1} (\bibinfo {year} {2014})},\
  \Eprint {http://arxiv.org/abs/arXiv:1401.0740} {arXiv:1401.0740} \BibitemShut
  {NoStop}%
\bibitem [{\citenamefont {Wang}\ \emph {et~al.}(2015)\citenamefont {Wang},
  \citenamefont {Gu},\ and\ \citenamefont {Wen}}]{WGW1489}%
  \BibitemOpen
  \bibfield  {author} {\bibinfo {author} {\bibfnamefont {J.}~\bibnamefont
  {Wang}}, \bibinfo {author} {\bibfnamefont {Z.-C.}\ \bibnamefont {Gu}}, \ and\
  \bibinfo {author} {\bibfnamefont {X.-G.}\ \bibnamefont {Wen}},\ }\href@noop
  {} {\bibfield  {journal} {\bibinfo  {journal} {Phys. Rev. Lett.}\ }\textbf
  {\bibinfo {volume} {114}},\ \bibinfo {pages} {031601} (\bibinfo {year}
  {2015})},\ \Eprint {http://arxiv.org/abs/arXiv:1405.7689} {arXiv:1405.7689}
  \BibitemShut {NoStop}%
\bibitem [{\citenamefont {{Ye}}\ and\ \citenamefont {{Gu}}(2015)}]{YG14102594}%
  \BibitemOpen
  \bibfield  {author} {\bibinfo {author} {\bibfnamefont {P.}~\bibnamefont
  {{Ye}}}\ and\ \bibinfo {author} {\bibfnamefont {Z.-C.}\ \bibnamefont
  {{Gu}}},\ }\href {\doibase 10.1103/PhysRevX.5.021029} {\bibfield  {journal}
  {\bibinfo  {journal} {Physical Review X}\ }\textbf {\bibinfo {volume} {5}},\
  \bibinfo {pages} {021029} (\bibinfo {year} {2015})},\ \Eprint
  {http://arxiv.org/abs/arXiv:1410.2594} {arXiv:1410.2594} \BibitemShut
  {NoStop}%
\bibitem [{\citenamefont {{Kapustin}}\ and\ \citenamefont
  {{Thorngren}}(2013)}]{KT13094721}%
  \BibitemOpen
  \bibfield  {author} {\bibinfo {author} {\bibfnamefont {A.}~\bibnamefont
  {{Kapustin}}}\ and\ \bibinfo {author} {\bibfnamefont {R.}~\bibnamefont
  {{Thorngren}}},\ }\href@noop {} {\  (\bibinfo {year} {2013})},\ \Eprint
  {http://arxiv.org/abs/arXiv:1309.4721} {arXiv:1309.4721} \BibitemShut
  {NoStop}%
\bibitem [{\citenamefont {{Gu}}\ \emph {et~al.}(2016)\citenamefont {{Gu}},
  \citenamefont {{Wang}},\ and\ \citenamefont {{Wen}}}]{GWW1568}%
  \BibitemOpen
  \bibfield  {author} {\bibinfo {author} {\bibfnamefont {Z.-C.}\ \bibnamefont
  {{Gu}}}, \bibinfo {author} {\bibfnamefont {J.~C.}\ \bibnamefont {{Wang}}}, \
  and\ \bibinfo {author} {\bibfnamefont {X.-G.}\ \bibnamefont {{Wen}}},\ }\href
  {\doibase 10.1103/PhysRevB.93.115136} {\bibfield  {journal} {\bibinfo
  {journal} {Phys. Rev. B}\ }\textbf {\bibinfo {volume} {93}},\ \bibinfo
  {pages} {115136} (\bibinfo {year} {2016})},\ \Eprint
  {http://arxiv.org/abs/arXiv:1503.01768} {arXiv:1503.01768} \BibitemShut
  {NoStop}%
\bibitem [{\citenamefont {{Ye}}\ and\ \citenamefont
  {{Gu}}(2016)}]{YG150805689}%
  \BibitemOpen
  \bibfield  {author} {\bibinfo {author} {\bibfnamefont {P.}~\bibnamefont
  {{Ye}}}\ and\ \bibinfo {author} {\bibfnamefont {Z.-C.}\ \bibnamefont
  {{Gu}}},\ }\href {\doibase 10.1103/PhysRevB.93.205157} {\bibfield  {journal}
  {\bibinfo  {journal} {\prb}\ }\textbf {\bibinfo {volume} {93}},\ \bibinfo
  {pages} {205157} (\bibinfo {year} {2016})},\ \Eprint
  {http://arxiv.org/abs/arXiv:1508.05689} {arXiv:1508.05689} \BibitemShut
  {NoStop}%
\bibitem [{\citenamefont {{Putrov}}\ \emph {et~al.}(2016)\citenamefont
  {{Putrov}}, \citenamefont {{Wang}},\ and\ \citenamefont
  {{Yau}}}]{PY161209298}%
  \BibitemOpen
  \bibfield  {author} {\bibinfo {author} {\bibfnamefont {P.}~\bibnamefont
  {{Putrov}}}, \bibinfo {author} {\bibfnamefont {J.}~\bibnamefont {{Wang}}}, \
  and\ \bibinfo {author} {\bibfnamefont {S.-T.}\ \bibnamefont {{Yau}}},\
  }\href@noop {} {\  (\bibinfo {year} {2016})},\ \Eprint
  {http://arxiv.org/abs/arXiv:1612.09298} {arXiv:1612.09298} \BibitemShut
  {NoStop}%
\bibitem [{\citenamefont {Wen}(2013)}]{W1313}%
  \BibitemOpen
  \bibfield  {author} {\bibinfo {author} {\bibfnamefont {X.-G.}\ \bibnamefont
  {Wen}},\ }\href@noop {} {\bibfield  {journal} {\bibinfo  {journal} {Phys.
  Rev. D}\ }\textbf {\bibinfo {volume} {88}},\ \bibinfo {pages} {045013}
  (\bibinfo {year} {2013})},\ \Eprint {http://arxiv.org/abs/arXiv:1303.1803}
  {arXiv:1303.1803} \BibitemShut {NoStop}%
\bibitem [{\citenamefont {Levin}\ and\ \citenamefont {Wen}(2003)}]{LW0316}%
  \BibitemOpen
  \bibfield  {author} {\bibinfo {author} {\bibfnamefont {M.}~\bibnamefont
  {Levin}}\ and\ \bibinfo {author} {\bibfnamefont {X.-G.}\ \bibnamefont
  {Wen}},\ }\href@noop {} {\bibfield  {journal} {\bibinfo  {journal} {Phys.
  Rev. B}\ }\textbf {\bibinfo {volume} {67}},\ \bibinfo {pages} {245316}
  (\bibinfo {year} {2003})},\ \Eprint {http://arxiv.org/abs/cond-mat/0302460}
  {cond-mat/0302460} \BibitemShut {NoStop}%
\bibitem [{\citenamefont {Wang}\ and\ \citenamefont {Senthil}(2013)}]{WS1334}%
  \BibitemOpen
  \bibfield  {author} {\bibinfo {author} {\bibfnamefont {C.}~\bibnamefont
  {Wang}}\ and\ \bibinfo {author} {\bibfnamefont {T.}~\bibnamefont {Senthil}},\
  }\href@noop {} {\bibfield  {journal} {\bibinfo  {journal} {Phys. Rev. B}\
  }\textbf {\bibinfo {volume} {87}},\ \bibinfo {pages} {235122} (\bibinfo
  {year} {2013})},\ \Eprint {http://arxiv.org/abs/arXiv:1302.6234}
  {arXiv:1302.6234} \BibitemShut {NoStop}%
\bibitem [{\citenamefont {{Heinrich}}\ \emph {et~al.}(2016)\citenamefont
  {{Heinrich}}, \citenamefont {{Burnell}}, \citenamefont {{Fidkowski}},\ and\
  \citenamefont {{Levin}}}]{HL160607816}%
  \BibitemOpen
  \bibfield  {author} {\bibinfo {author} {\bibfnamefont {C.}~\bibnamefont
  {{Heinrich}}}, \bibinfo {author} {\bibfnamefont {F.}~\bibnamefont
  {{Burnell}}}, \bibinfo {author} {\bibfnamefont {L.}~\bibnamefont
  {{Fidkowski}}}, \ and\ \bibinfo {author} {\bibfnamefont {M.}~\bibnamefont
  {{Levin}}},\ }\href@noop {} {\  (\bibinfo {year} {2016})},\ \Eprint
  {http://arxiv.org/abs/arXiv:1606.07816} {arXiv:1606.07816} \BibitemShut
  {NoStop}%
\bibitem [{\citenamefont {Wen}(2002)}]{W0213}%
  \BibitemOpen
  \bibfield  {author} {\bibinfo {author} {\bibfnamefont {X.-G.}\ \bibnamefont
  {Wen}},\ }\href@noop {} {\bibfield  {journal} {\bibinfo  {journal} {Phys.
  Rev. B}\ }\textbf {\bibinfo {volume} {65}},\ \bibinfo {pages} {165113}
  (\bibinfo {year} {2002})},\ \Eprint {http://arxiv.org/abs/cond-mat/0107071}
  {cond-mat/0107071} \BibitemShut {NoStop}%
\bibitem [{\citenamefont {Yao}\ \emph {et~al.}(2010)\citenamefont {Yao},
  \citenamefont {Fu},\ and\ \citenamefont {Qi}}]{YFQ1070}%
  \BibitemOpen
  \bibfield  {author} {\bibinfo {author} {\bibfnamefont {H.}~\bibnamefont
  {Yao}}, \bibinfo {author} {\bibfnamefont {L.}~\bibnamefont {Fu}}, \ and\
  \bibinfo {author} {\bibfnamefont {X.-L.}\ \bibnamefont {Qi}},\ }\href@noop {}
  {\  (\bibinfo {year} {2010})},\ \Eprint
  {http://arxiv.org/abs/arXiv:1012.4470} {arXiv:1012.4470} \BibitemShut
  {NoStop}%
\bibitem [{\citenamefont {Vishwanath}\ and\ \citenamefont
  {Senthil}(2013)}]{VS1306}%
  \BibitemOpen
  \bibfield  {author} {\bibinfo {author} {\bibfnamefont {A.}~\bibnamefont
  {Vishwanath}}\ and\ \bibinfo {author} {\bibfnamefont {T.}~\bibnamefont
  {Senthil}},\ }\href@noop {} {\bibfield  {journal} {\bibinfo  {journal} {Phys.
  Rev. X}\ }\textbf {\bibinfo {volume} {3}},\ \bibinfo {pages} {011016}
  (\bibinfo {year} {2013})},\ \Eprint {http://arxiv.org/abs/arXiv:1209.3058}
  {arXiv:1209.3058} \BibitemShut {NoStop}%
\bibitem [{\citenamefont {{Xu}}(2013)}]{X13078131}%
  \BibitemOpen
  \bibfield  {author} {\bibinfo {author} {\bibfnamefont {C.}~\bibnamefont
  {{Xu}}},\ }\href {\doibase 10.1103/PhysRevB.88.205137} {\bibfield  {journal}
  {\bibinfo  {journal} {\prb}\ }\textbf {\bibinfo {volume} {88}},\ \bibinfo
  {pages} {205137} (\bibinfo {year} {2013})},\ \Eprint
  {http://arxiv.org/abs/arXiv:1307.8131} {arXiv:1307.8131} \BibitemShut
  {NoStop}%
\bibitem [{\citenamefont {Kitaev}(2009)}]{K0986}%
  \BibitemOpen
  \bibfield  {author} {\bibinfo {author} {\bibfnamefont {A.}~\bibnamefont
  {Kitaev}},\ }in\ \href@noop {} {\emph {\bibinfo {booktitle} {Advances in
  Theoretical Physics: Landau Memorial Conference, Chernogolovka, Russia,
  2008}}},\ Vol.\ \bibinfo {volume} {AIP Conf. Proc. No. 1134},\ \bibinfo
  {editor} {edited by\ \bibinfo {editor} {\bibfnamefont {V.}~\bibnamefont
  {Lebedev}}\ and\ \bibinfo {editor} {\bibfnamefont {M.}~\bibnamefont
  {Feigel’man}}}\ (\bibinfo  {publisher} {AIP},\ \bibinfo {address}
  {Melville, NY},\ \bibinfo {year} {2009})\ p.~\bibinfo {pages} {22},\ \Eprint
  {http://arxiv.org/abs/arXiv:0901.2686} {arXiv:0901.2686} \BibitemShut
  {NoStop}%
\bibitem [{\citenamefont {Ryu}\ \emph {et~al.}(2009)\citenamefont {Ryu},
  \citenamefont {Schnyder}, \citenamefont {Furusaki},\ and\ \citenamefont
  {Ludwig}}]{RSF0957}%
  \BibitemOpen
  \bibfield  {author} {\bibinfo {author} {\bibfnamefont {S.}~\bibnamefont
  {Ryu}}, \bibinfo {author} {\bibfnamefont {A.}~\bibnamefont {Schnyder}},
  \bibinfo {author} {\bibfnamefont {A.}~\bibnamefont {Furusaki}}, \ and\
  \bibinfo {author} {\bibfnamefont {A.}~\bibnamefont {Ludwig}},\ }\href
  {\doibase 10.1088/1367-2630/12/6/065010} {\bibfield  {journal} {\bibinfo
  {journal} {New J. Phys.}\ }\textbf {\bibinfo {volume} {12}},\ \bibinfo
  {pages} {065010} (\bibinfo {year} {2009})},\ \Eprint
  {http://arxiv.org/abs/arXiv:0912.2157} {arXiv:0912.2157} \BibitemShut
  {NoStop}%
\bibitem [{\citenamefont {{Wang}}\ and\ \citenamefont
  {{Senthil}}(2014)}]{WS14011142}%
  \BibitemOpen
  \bibfield  {author} {\bibinfo {author} {\bibfnamefont {C.}~\bibnamefont
  {{Wang}}}\ and\ \bibinfo {author} {\bibfnamefont {T.}~\bibnamefont
  {{Senthil}}},\ }\href {\doibase 10.1103/PhysRevB.89.195124} {\bibfield
  {journal} {\bibinfo  {journal} {\prb}\ }\textbf {\bibinfo {volume} {89}},\
  \bibinfo {pages} {195124} (\bibinfo {year} {2014})},\ \Eprint
  {http://arxiv.org/abs/arXiv:1401.1142} {arXiv:1401.1142} \BibitemShut
  {NoStop}%
\bibitem [{\citenamefont {{Metlitski}}\ \emph {et~al.}(2014)\citenamefont
  {{Metlitski}}, \citenamefont {{Fidkowski}}, \citenamefont {{Chen}},\ and\
  \citenamefont {{Vishwanath}}}]{MV14063032}%
  \BibitemOpen
  \bibfield  {author} {\bibinfo {author} {\bibfnamefont {M.~A.}\ \bibnamefont
  {{Metlitski}}}, \bibinfo {author} {\bibfnamefont {L.}~\bibnamefont
  {{Fidkowski}}}, \bibinfo {author} {\bibfnamefont {X.}~\bibnamefont {{Chen}}},
  \ and\ \bibinfo {author} {\bibfnamefont {A.}~\bibnamefont {{Vishwanath}}},\
  }\href@noop {} {\  (\bibinfo {year} {2014})},\ \Eprint
  {http://arxiv.org/abs/arXiv:1406.3032} {arXiv:1406.3032} \BibitemShut
  {NoStop}%
\bibitem [{\citenamefont {{You}}\ and\ \citenamefont
  {{Xu}}(2014)}]{YC14090168}%
  \BibitemOpen
  \bibfield  {author} {\bibinfo {author} {\bibfnamefont {Y.-Z.}\ \bibnamefont
  {{You}}}\ and\ \bibinfo {author} {\bibfnamefont {C.}~\bibnamefont {{Xu}}},\
  }\href {\doibase 10.1103/PhysRevB.90.245120} {\bibfield  {journal} {\bibinfo
  {journal} {\prb}\ }\textbf {\bibinfo {volume} {90}},\ \bibinfo {pages}
  {245120} (\bibinfo {year} {2014})},\ \Eprint
  {http://arxiv.org/abs/arXiv:1409.0168} {arXiv:1409.0168} \BibitemShut
  {NoStop}%
\bibitem [{\citenamefont {{You}}\ \emph {et~al.}(2015)\citenamefont {{You}},
  \citenamefont {{Bi}}, \citenamefont {{Rasmussen}}, \citenamefont {{Cheng}},\
  and\ \citenamefont {{Xu}}}]{YC14046256}%
  \BibitemOpen
  \bibfield  {author} {\bibinfo {author} {\bibfnamefont {Y.-Z.}\ \bibnamefont
  {{You}}}, \bibinfo {author} {\bibfnamefont {Z.}~\bibnamefont {{Bi}}},
  \bibinfo {author} {\bibfnamefont {A.}~\bibnamefont {{Rasmussen}}}, \bibinfo
  {author} {\bibfnamefont {M.}~\bibnamefont {{Cheng}}}, \ and\ \bibinfo
  {author} {\bibfnamefont {C.}~\bibnamefont {{Xu}}},\ }\href {\doibase
  10.1088/1367-2630/17/7/075010} {\bibfield  {journal} {\bibinfo  {journal}
  {New Journal of Physics}\ }\textbf {\bibinfo {volume} {17}},\ \bibinfo
  {pages} {075010} (\bibinfo {year} {2015})},\ \Eprint
  {http://arxiv.org/abs/arXiv:1404.6256} {arXiv:1404.6256} \BibitemShut
  {NoStop}%
\bibitem [{\citenamefont {{Fidkowski}}\ \emph {et~al.}(2013)\citenamefont
  {{Fidkowski}}, \citenamefont {{Chen}},\ and\ \citenamefont
  {{Vishwanath}}}]{FV13055851}%
  \BibitemOpen
  \bibfield  {author} {\bibinfo {author} {\bibfnamefont {L.}~\bibnamefont
  {{Fidkowski}}}, \bibinfo {author} {\bibfnamefont {X.}~\bibnamefont {{Chen}}},
  \ and\ \bibinfo {author} {\bibfnamefont {A.}~\bibnamefont {{Vishwanath}}},\
  }\href {\doibase 10.1103/PhysRevX.3.041016} {\bibfield  {journal} {\bibinfo
  {journal} {Physical Review X}\ }\textbf {\bibinfo {volume} {3}},\ \bibinfo
  {pages} {041016} (\bibinfo {year} {2013})},\ \Eprint
  {http://arxiv.org/abs/arXiv:1305.5851} {arXiv:1305.5851} \BibitemShut
  {NoStop}%
\bibitem [{\citenamefont {{Kapustin}}\ \emph {et~al.}(2015)\citenamefont
  {{Kapustin}}, \citenamefont {{Thorngren}}, \citenamefont {{Turzillo}},\ and\
  \citenamefont {{Wang}}}]{KTT1429}%
  \BibitemOpen
  \bibfield  {author} {\bibinfo {author} {\bibfnamefont {A.}~\bibnamefont
  {{Kapustin}}}, \bibinfo {author} {\bibfnamefont {R.}~\bibnamefont
  {{Thorngren}}}, \bibinfo {author} {\bibfnamefont {A.}~\bibnamefont
  {{Turzillo}}}, \ and\ \bibinfo {author} {\bibfnamefont {Z.}~\bibnamefont
  {{Wang}}},\ }\href@noop {} {\bibfield  {journal} {\bibinfo  {journal}
  {Journal of High Energy Physics}\ }\textbf {\bibinfo {volume} {2015}},\
  \bibinfo {pages} {52} (\bibinfo {year} {2015})},\ \Eprint
  {http://arxiv.org/abs/arXiv:1406.7329} {arXiv:1406.7329} \BibitemShut
  {NoStop}%
\bibitem [{\citenamefont {Read}\ and\ \citenamefont {Sachdev}(1991)}]{RS9173}%
  \BibitemOpen
  \bibfield  {author} {\bibinfo {author} {\bibfnamefont {N.}~\bibnamefont
  {Read}}\ and\ \bibinfo {author} {\bibfnamefont {S.}~\bibnamefont {Sachdev}},\
  }\href@noop {} {\bibfield  {journal} {\bibinfo  {journal} {Phys. Rev. Lett.}\
  }\textbf {\bibinfo {volume} {66}},\ \bibinfo {pages} {1773} (\bibinfo {year}
  {1991})}\BibitemShut {NoStop}%
\bibitem [{\citenamefont {Wen}(1991)}]{W9164}%
  \BibitemOpen
  \bibfield  {author} {\bibinfo {author} {\bibfnamefont {X.-G.}\ \bibnamefont
  {Wen}},\ }\href@noop {} {\bibfield  {journal} {\bibinfo  {journal} {Phys.
  Rev. B}\ }\textbf {\bibinfo {volume} {44}},\ \bibinfo {pages} {2664}
  (\bibinfo {year} {1991})}\BibitemShut {NoStop}%
\bibitem [{\citenamefont {Wang}\ and\ \citenamefont {Wen}(2015)}]{WW1454}%
  \BibitemOpen
  \bibfield  {author} {\bibinfo {author} {\bibfnamefont {J.}~\bibnamefont
  {Wang}}\ and\ \bibinfo {author} {\bibfnamefont {X.-G.}\ \bibnamefont {Wen}},\
  }\href@noop {} {\bibfield  {journal} {\bibinfo  {journal} {Phys. Rev. B}\
  }\textbf {\bibinfo {volume} {91}},\ \bibinfo {pages} {035134} (\bibinfo
  {year} {2015})},\ \Eprint {http://arxiv.org/abs/arXiv:1404.7854}
  {arXiv:1404.7854} \BibitemShut {NoStop}%
\bibitem [{\citenamefont {Kitaev}\ and\ \citenamefont {Kong}(2012)}]{KK1251}%
  \BibitemOpen
  \bibfield  {author} {\bibinfo {author} {\bibfnamefont {A.}~\bibnamefont
  {Kitaev}}\ and\ \bibinfo {author} {\bibfnamefont {L.}~\bibnamefont {Kong}},\
  }\href {\doibase 10.1007/s00220-012-1500-5} {\bibfield  {journal} {\bibinfo
  {journal} {Commun. Math. Phys.}\ }\textbf {\bibinfo {volume} {313}},\
  \bibinfo {pages} {351 } (\bibinfo {year} {2012})},\ \Eprint
  {http://arxiv.org/abs/arXiv:1104.5047} {arXiv:1104.5047} \BibitemShut
  {NoStop}%
\bibitem [{\citenamefont {Hung}\ and\ \citenamefont {Wen}(2012)}]{HW1267}%
  \BibitemOpen
  \bibfield  {author} {\bibinfo {author} {\bibfnamefont {L.-Y.}\ \bibnamefont
  {Hung}}\ and\ \bibinfo {author} {\bibfnamefont {X.-G.}\ \bibnamefont {Wen}},\
  }\href@noop {} {\  (\bibinfo {year} {2012})},\ \Eprint
  {http://arxiv.org/abs/arXiv:1211.2767} {arXiv:1211.2767} \BibitemShut
  {NoStop}%
\bibitem [{\citenamefont {Ye}\ and\ \citenamefont {Wen}(2014)}]{YW1427}%
  \BibitemOpen
  \bibfield  {author} {\bibinfo {author} {\bibfnamefont {P.}~\bibnamefont
  {Ye}}\ and\ \bibinfo {author} {\bibfnamefont {X.-G.}\ \bibnamefont {Wen}},\
  }\href@noop {} {\bibfield  {journal} {\bibinfo  {journal} {Phys. Rev. B}\
  }\textbf {\bibinfo {volume} {89}},\ \bibinfo {pages} {045127} (\bibinfo
  {year} {2014})},\ \Eprint {http://arxiv.org/abs/arXiv:1303.3572}
  {arXiv:1303.3572} \BibitemShut {NoStop}%
\bibitem [{\citenamefont {{Wang}}\ and\ \citenamefont
  {{Levin}}(2015)}]{WL14121781}%
  \BibitemOpen
  \bibfield  {author} {\bibinfo {author} {\bibfnamefont {C.}~\bibnamefont
  {{Wang}}}\ and\ \bibinfo {author} {\bibfnamefont {M.}~\bibnamefont
  {{Levin}}},\ }\href {\doibase 10.1103/PhysRevB.91.165119} {\bibfield
  {journal} {\bibinfo  {journal} {\prb}\ }\textbf {\bibinfo {volume} {91}},\
  \bibinfo {pages} {165119} (\bibinfo {year} {2015})},\ \Eprint
  {http://arxiv.org/abs/arXiv:1412.1781} {arXiv:1412.1781} \BibitemShut
  {NoStop}%
\bibitem [{\citenamefont {Costantino}(2005)}]{C0527}%
  \BibitemOpen
  \bibfield  {author} {\bibinfo {author} {\bibfnamefont {F.}~\bibnamefont
  {Costantino}},\ }\href@noop {} {\bibfield  {journal} {\bibinfo  {journal}
  {Math. Z.}\ }\textbf {\bibinfo {volume} {251}},\ \bibinfo {pages} {427}
  (\bibinfo {year} {2005})},\ \Eprint {http://arxiv.org/abs/math/0403014}
  {math/0403014} \BibitemShut {NoStop}%
\bibitem [{\citenamefont {{Wan}}\ \emph {et~al.}(2015)\citenamefont {{Wan}},
  \citenamefont {{Wang}},\ and\ \citenamefont {{He}}}]{WH14093216}%
  \BibitemOpen
  \bibfield  {author} {\bibinfo {author} {\bibfnamefont {Y.}~\bibnamefont
  {{Wan}}}, \bibinfo {author} {\bibfnamefont {J.~C.}\ \bibnamefont {{Wang}}}, \
  and\ \bibinfo {author} {\bibfnamefont {H.}~\bibnamefont {{He}}},\ }\href
  {\doibase 10.1103/PhysRevB.92.045101} {\bibfield  {journal} {\bibinfo
  {journal} {\prb}\ }\textbf {\bibinfo {volume} {92}},\ \bibinfo {pages}
  {045101} (\bibinfo {year} {2015})},\ \Eprint
  {http://arxiv.org/abs/arXiv:1409.3216} {arXiv:1409.3216} \BibitemShut
  {NoStop}%
\bibitem [{\citenamefont {Moradi}\ and\ \citenamefont {Wen}(2015)}]{MW1514}%
  \BibitemOpen
  \bibfield  {author} {\bibinfo {author} {\bibfnamefont {H.}~\bibnamefont
  {Moradi}}\ and\ \bibinfo {author} {\bibfnamefont {X.-G.}\ \bibnamefont
  {Wen}},\ }\href@noop {} {\bibfield  {journal} {\bibinfo  {journal} {Phys.
  Rev. B}\ }\textbf {\bibinfo {volume} {91}},\ \bibinfo {pages} {075114}
  (\bibinfo {year} {2015})},\ \Eprint {http://arxiv.org/abs/arXiv:1404.4618}
  {arXiv:1404.4618} \BibitemShut {NoStop}%
\bibitem [{\citenamefont {Spanier}(1966)}]{Spa66}%
  \BibitemOpen
  \bibfield  {author} {\bibinfo {author} {\bibfnamefont {E.~H.}\ \bibnamefont
  {Spanier}},\ }\href@noop {} {\emph {\bibinfo {title} {Algebraic Topology}}}\
  (\bibinfo  {publisher} {McGraw-Hill},\ \bibinfo {address} {New York},\
  \bibinfo {year} {1966})\BibitemShut {NoStop}%
\bibitem [{\citenamefont {Wu}(1950)}]{W5008}%
  \BibitemOpen
  \bibfield  {author} {\bibinfo {author} {\bibfnamefont {W.}~\bibnamefont
  {Wu}},\ }\href@noop {} {\bibfield  {journal} {\bibinfo  {journal} {C.R. Acad.
  Sci. Paris}\ }\textbf {\bibinfo {volume} {230}},\ \bibinfo {pages} {508}
  (\bibinfo {year} {1950})}\BibitemShut {NoStop}%
\bibitem [{\citenamefont {Hatcher}(2002)}]{Hat02}%
  \BibitemOpen
  \bibfield  {author} {\bibinfo {author} {\bibfnamefont {A.}~\bibnamefont
  {Hatcher}},\ }\href@noop {} {\emph {\bibinfo {title} {Algebraic Topology}}}\
  (\bibinfo  {publisher} {Cambridge University Press},\ \bibinfo {year}
  {2002})\BibitemShut {NoStop}%
\end{thebibliography}%

\end{document}